\documentclass[aip, amsmath,amssymb, reprint]{revtex4-1}

\usepackage{graphicx}
\usepackage{dcolumn}
\usepackage{bm}
\usepackage{tabularx} 
\usepackage[utf8]{inputenc}
\usepackage[T1]{fontenc}
\usepackage{mathptmx}
\usepackage{verbatim}
\usepackage{xcolor}
\usepackage{upgreek}
\usepackage{amsmath}
\usepackage[subnum]{cases}
\usepackage{soul}
\usepackage{dcolumn}
\usepackage{bm}

\usepackage[subnum]{cases}
\usepackage{colortbl}
\usepackage{subfiles}
\usepackage{braket}
\usepackage{Commands}

\begin{document}

\preprint{AIP/123-QED}


\title{Probing quantum devices with radio-frequency reflectometry}
\author{Florian Vigneau}
\affiliation{University of Oxford, Department of Materials, 16 Parks Road, Oxford OX1 3PH, UK}

\author{Federico Fedele}
\affiliation{University of Oxford, Department of Materials, 16 Parks Road, Oxford OX1 3PH, UK}

\author{Anasua Chatterjee}
\affiliation{Center for Quantum Devices,  Niels Bohr Institute, University of Copenhagen,  2100 Copenhagen,Denmark}

\author{David Reilly}
\affiliation{ARC Centre of Excellence for Engineered Quantum Systems, School of Physics, The University of Sydney, NSW 2006, Australia}
\affiliation{Microsoft Quantum Sydney, The University of Sydney, Sydney, NSW 2006, Australia}

\author{Ferdinand Kuemmeth}
\affiliation{Center for Quantum Devices,  Niels Bohr Institute, University of Copenhagen,  2100 Copenhagen,Denmark}

\author{M.~Fernando~Gonzalez-Zalba}
\email[]{fernando@quantummotion.tech}
\affiliation{Quantum  Motion  Technologies,  Windsor  House,  Cornwall  Road,  Harrogate, HG1  2PW, United Kingdom}

\author{Edward Laird}
\email[]{e.a.laird@lancaster.ac.uk}
\affiliation{Department of Physics, University of Lancaster, Lancaster, LA1 4YB, United Kingdom}

\author{Natalia Ares}
\email[]{natalia.ares@eng.ox.ac.uk}
\affiliation{University of Oxford, Department of Materials, 16 Parks Road, Oxford OX1 3PH, UK}

\date{\today}

\begin{abstract}

Many important phenomena in quantum devices are dynamic, meaning that they cannot be studied using time-averaged measurements alone.
Experiments that measure such transient effects are collectively known as fast readout.
One of the most useful techniques in fast electrical readout is radio-frequency reflectometry, which can measure changes in impedance (both resistive and reactive) even when their duration is extremely short, down to a microsecond or less.
Examples of reflectometry experiments, some of which have been realised and others so far only proposed, include projective measurements of qubits and Majorana devices for quantum computing, real-time measurements of mechanical motion and detection of non-equilibrium temperature fluctuations.
However, all of these experiments must overcome the central challenge of fast readout: the large mismatch between the typical impedance of quantum devices (set by the resistance quantum) and of transmission lines (set by the impedance of free space).
Here, we review the physical principles of radio-frequency reflectometry and its close cousins, measurements of radio-frequency transmission and emission.
We explain how to optimise the speed and sensitivity of a radio-frequency measurement, and  how to incorporate new tools such as superconducting circuit elements and quantum-limited amplifiers into advanced radio-frequency experiments.
Our aim is three-fold: to introduce the readers to the technique, to review the advances to date and to motivate new experiments in fast quantum device dynamics.
Our intended audience includes experimentalists in the field of quantum electronics who want to implement radio-frequency experiments or improve them, together with physicists in related fields who want to understand how the most important radio-frequency measurements work.

\end{abstract}

\maketitle
\tableofcontents

\section{Introduction and motivation}

\subsection{Why use rf measurements?}

The bandwidth of standard electrical measurement setups is typically limited by the \textit{RC} low-pass filter formed by the resistance of the sample, the input impedance of the amplifier and the capacitance of the electrical cables that connect cryogenic devices to the measurement instruments at room temperature. Quantum devices usually have resistances of the order of the quantum of resistance $h/e^2\approx25.8~\mathrm{k\Ohm}$
and the capacitance of the cables is in the range $C_\text{line} = 0.1-1$~nF which bring the cut-off frequency to no more than few kilohertz.

\begin{figure}
    \includegraphics[width=\columnwidth]{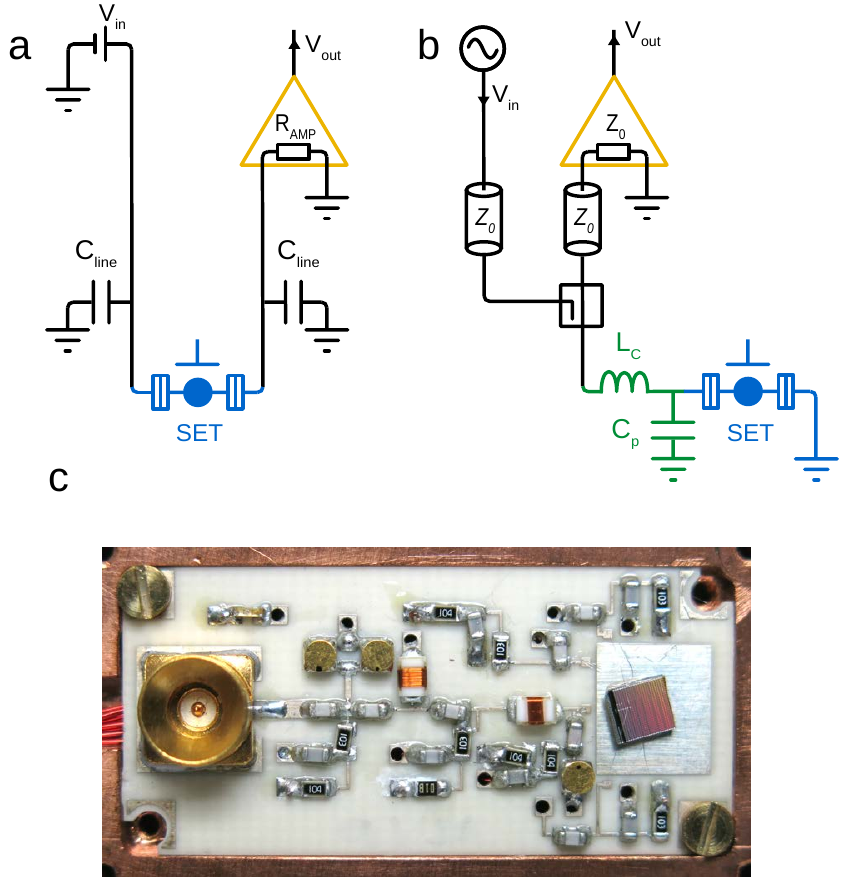}
    \caption{(a) Conventional dc measurement of a SET. The SET is biased with a dc voltage $V_\text{in}$ applied to the source electrode. A second line carries the current to the amplifier, which outputs a voltage $\Vout$. 
    (b) Radio-frequency measurement of the SET embedded in a combination of impedances $\Lc$ and $\Cp$. The SET is illuminated by an rf ac voltage $\Vin$ injected via a transmission line of characteristic impedance $\Zo$. The reflected signal is routed by a directional coupler and a second transmission line to the amplifiers, which outputs $\Vout$. 
    (c) Picture of a sample board for radiofrequency measurements mounted on a copper enclosure~\cite{Gonzalez-Zalba2019tunable}. The PCB hosts a high frequency SMP connector on the left hand side, a horizontally-positioned wirewound ceramic inductor (part of the matching network), three varicap diodes (gold-coloured circular components) for frequency and matching tuning, RC filters on every bias line and a vertically-positioned shunt inductor to provide attenuation of modulation frequencies on the matching varicaps. Finally on the right hand side, a sample is bonded to the matching network.}
    
    \label{fig:Intro}
\end{figure}

An important example of this problem is the single-electron transistor (SET). SET charge sensors \cite{Devoret1991,Dolan1987} are the most sensitive electrometers used to measure the charge occupation of quantum dots (QDs) by monitoring the change of resistance of a closely positioned SET. In principle~\cite{Vleuten1996}, their bandwidth could exceed $10$~GHz, intrinsically limited by the RC filtering due to the resistance of the two tunnel junctions in series ($> 2h/e^2$) and the typical capacitance of few femtofarads between the SET's tunnel junctions. However, in practice, the bandwidth is limited to few kilohertz because of the high capacitance of the cabling that connects the output of the device to room temperature electronics (Fig. \ref{fig:Intro}(a)).

There have been some attempts \cite{pettersson1996,Vleuten1996} to overcome this obstacle by introducing cryogenic amplifiers close to the device.  This has the effect of reducing the capacitance of the cable but also produces a substantial amount of heat near the device. It is difficult to reduce the amplifier resistance, $R_\text{AMP}$, much below the SET resistance since it would create a voltage divider for the signal. Another approach is to replace the voltage amplifier by a current-to-voltage converter, with a lower input impedance, that measures the current at the output of the SET\cite{Delsing1999}, but the improvement is modest. 

In 1998 Schoelkopf $\etal$ introduced the radio-frequency SET (rf-SET) \cite{Schoelkopf1998} which can measure the charge occupation of quantum dots with a bandwidth exceeding 100 MHz. The solution was to place the SET at the end of a transmission line (Fig.~\ref{fig:Intro}(b)) while illuminating the device with an rf signal whose reflected phase and amplitude depend on the impedance  of the SET. The high resistance of the SET is converted to the $\Zo=50~\Ohm$ characteristic impedance of the line by a matching network combining the impedances of an inductor $\Lc$ and a capacitor $\Cp$, in its simplest implementation. Since all the components in the amplification chain, including the amplifier input impedance, are now matched to $\Zo$ the bandwidth of the measurement is greatly enhanced.

Since then, rf techniques for QDs have flourished, motivated in particular by the emergence of quantum computation using the spin of charged particles  confined to real or artificial atoms (QDs) to encode qubits~\cite{DiVincenzo1998}.
Practical quantum computation requires error correction schemes that involve fast high-fidelity single-shot readout of qubits~\cite{Andrew2012}, much faster than their coherence time. Such sensitive and fast readout could be provided by the rf-SET (or related readout devices such as the rf quantum point contact (rf-QPC)), or dispersive readout. 

Radio-frequency techniques are becoming increasingly popular to study other kinds of quantum devices and phenomena. In particular, they have been employed for measuring low-dimensional systems, nanomechanical resonators, superconducting quantum interference devices (SQUIDs), and Majorana devices, and even to perform fast thermometry.
Owing to its high bandwidth, 
rf readout enables measurements of the time evolution of rapid physical effects. In some cases, as we shall see, the input signal induces novel non-equilibrium phenomena, allowing their study.

\subsection{Organisation of the review}

In Section~\ref{Sec:Basics}, we introduce the fundamental concepts essential to understand rf measurements and the subsequent sections of this review. This section is particularly important for readers who are new to high-frequency electronics. We present the basic constituents of rf setups and the principles of propagation of high frequency electronic signals along transmission lines, of impedance matching, of signal composition and demodulation.

In Sections~\ref{sec:resistiveReadout} and~\ref{sec:reactiveReadout}, we present dissipative and reactive readout of quantum devices by focusing on the representative example of charge sensing in QDs. In the case of dissipative readout, the change of the sample resistance due to changes in the electrostatic environment modifies the amplitude of the rf signal. In the case of reactive readout, the phase of the rf signal changes due to variations of the device capacitance or inductance. We describe the examples of charge sensing in QDs as illustration. We explain the working principle of the rf-SET, the rf-QPC and dispersive readout. For each case, we give an overview of the state-of-the-art. 

Technological developments have greatly improved the sensitivity and bandwidth of high-frequency measurements. The engineering of rf cavities is the subject of Section~\ref{section:Optimisation}. Variable capacitors, for example, allow \textit{in-situ} tuning of the cavity resonant frequency in order to optimize impedance matching to the device to be probed. Superconducting circuit elements and optimized circuit topologies can improve the cavity quality factor of the resonator. 
Identifying and reducing sources of noise is key to the readout of weak signals. Low-noise amplifiers, including superconducting amplifiers that reach or exceed the standard quantum limit, will be described in Section~\ref{Sec:amplifiers}. 
Different approaches to scale up measurement setups to read multiple quantum devices are presented in Section \ref{sec:Multiplexing}.

Finally, we focus on the many different quantum phenomena that can be studied using fast readout, and on how they are exploited in quantum technologies and other condensed matter physics experiments. In Section~\ref{sec:Singlet-triplet qubits}, we explain in particular the stimulating application of the fast readout of spin qubits, imperative for fault tolerant quantum computing. Other applications are shown in Section~\ref{sec:Sample}. For example, special symmetries in the effective circuit could allow noise-protected superconducting qubits (Section~\ref{Noise-protected superconducting qubits}). An interesting adaptation allows probing Majorana modes in nanowire devices and could be the basis for topological qubit readout (Section~\ref{subsec:Majorana devices}). The measurement of noise (Section~\ref{sec:noiseMeasrement}) can reveal fundamental properties of a device such as the charge of the carriers and its temperature. High-frequency measurements of nanomechanical resonators (Section~\ref{sec:NanomechanicalResonator}) have proved key for studying fast dynamics and are promising for the generation of quantum states. Rf thermometry (Section~\ref{fast thermometry}) brings a solution for measuring subkelvin temperature and with a speed that could enable to detect out-of-equilibrium phenomena. Rf measurements can also reveal information about the environment of a quantum device (Section~\ref{sensing Environement}). Finally, superconducting quantum interference devices (SQUIDs) allow sensitive magnetic field sensing (Section~\ref{Section9SQUID}). We conclude the review with perspectives on future developments.

We also note the boundaries of this review. Here we focus the regime where the photon energy of the rf signal is much smaller than the quantum level separation, such that resonant excitations do not occur and the system can be described using a semiclassical approach: a classical electric field coupled to a quantum system. The situation when these two energies are comparable and quantum mechanical interactions may occur between the two systems is described in the theory of quantum electrodynamics~\cite{Wallraff2020}.

\section{Basics of high-frequency measurements}
\label{Sec:Basics}

This section is a high-level overview of high-frequency electronic measurements, covering the main elements of the circuit to the final demodulated signal. 
To understand how these measurements work, we need to know what the main constituents of an rf setup are (Section \ref{Sec:measurement setup});  how a signal voltage propagates along a transmission line (Section \ref{sec:transmissionlines}); how it is changed when it scatters off a load impedance (Section \ref{sec:LCRLoad}); and  how information about the load can be extracted from the signal (Section \ref{sec:demodulation}).
From these, the reader should gain a self-contained understanding of how a high-frequency measurement works.
The later sections of this review provide more comprehensive explanations of how these principles are implemented in experiments.

\subsection{High-frequency measurement setups: an overview}
\label{Sec:measurement setup}

\begin{figure}
    \includegraphics[width=\columnwidth]{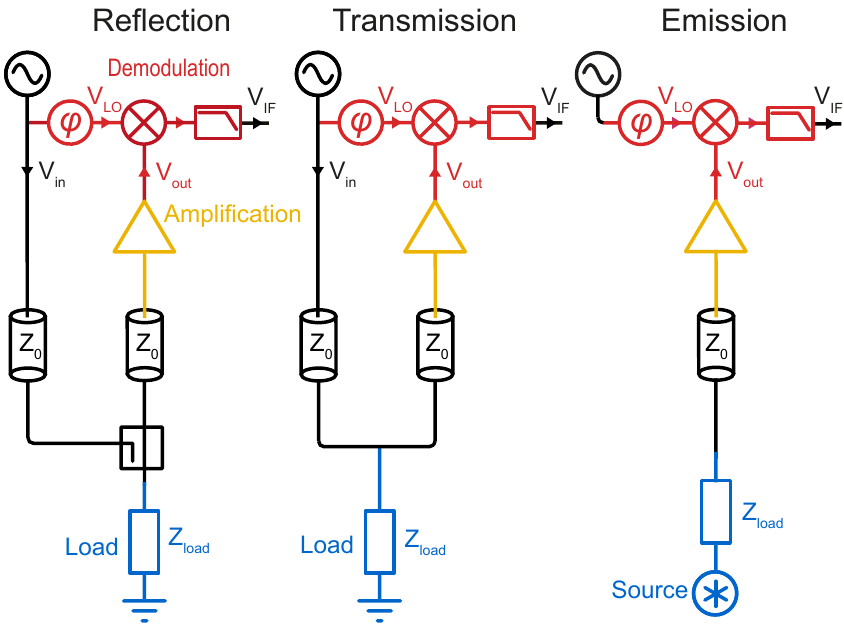}
    \caption{Reflection, transmission and emission measurement setups. In the first two panels, the input signal $\Vin$ is generated by an rf source and travels along transmission lines of characteristic impedance $\Zo$. After illuminating the load the returned signal is amplified into $\Vout$ and homodyne demodulated into $\VIF$. In the third panel, the source emits a signal by itself that is demodulated using an external signal.}
    \label{fig:Reflectometry}
\end{figure}

Circuit diagrams for the three main types of high-frequency measurement are shown in Fig.~\ref{fig:Reflectometry}. In a reflection or transmission measurement (as shown in the first two panels), the aim is to detect changes in the impedance of the device under test by converting them to a voltage.
The device, together with the tank circuit in which it is usually embedded, presents a total impedance $\Zload$, so called because it acts as a load on the transmission line.
When the device impedance changes, $\Zload$ changes.
To measure $\Zload$, it is illuminated by injecting a carrier tone $\Vin(t)$.
The carrier propagates along a transmission line towards the load and is reflected off it (in reflection configuration) or transmitted through or past it (in transmission configuration).
The signal propagating away from the load differs from $\Vin(t)$ in a way that depends on $\Zload$ and therefore on the device impedance.

The outgoing signal is then amplified to boost it well above the noise of subsequent electronics used for analysis.
Finally, it is usually demodulated to shift it away from the carrier frequency and towards a lower frequency, which is usually more convenient to work with.
This is done by multiplying it with a demodulation tone $\VLO(t)$ and low-pass filtering the product, as explained in Section~\ref{sec:demodulation}.
Often $\VLO(t)$ is derived from the original carrier, as in Fig.~\ref{fig:Reflectometry}.
The output $\VIF(t)$ of the demodulation circuit carries the required information about the device impedance.
This is the signal that is recorded.

The third type of high-frequency measurement does not inject a carrier tone at all.
Instead, it treats the device as a voltage source whose emission must be measured.
The circuit used in this kind of measurement (third panel of Fig.~\ref{fig:Reflectometry}) therefore omits the injection path.

\subsection{Wave propagation along transmission lines, the characteristic impedance $\Zo$, and why it is important}
\label{sec:transmissionlines}

To couple a voltage source to the load being measured, as in Fig.~\ref{fig:HFLine}(a), we must provide paths for both signal and ground.
This is done using a coaxial cable, or sometimes using coplanar waveguides.
The general term for any such connection is a transmission line.
The signal propagates as a voltage difference between the inner (or signal) conductor and the outer (or ground).

\subsubsection{Wave propagation along a transmission line}

In dc electronics, two points in a circuit connected by a zero-resistance path will be at the same voltage.
However, this is not generally true at high frequency.
The reason is that the connecting cable has an inductance which presents a high-frequency impedance between its two ends.
Likewise the capacitance between the inner conductor and ground means that the current need not be the same everywhere along the cable.

\begin{figure}
	\includegraphics[width=\columnwidth]{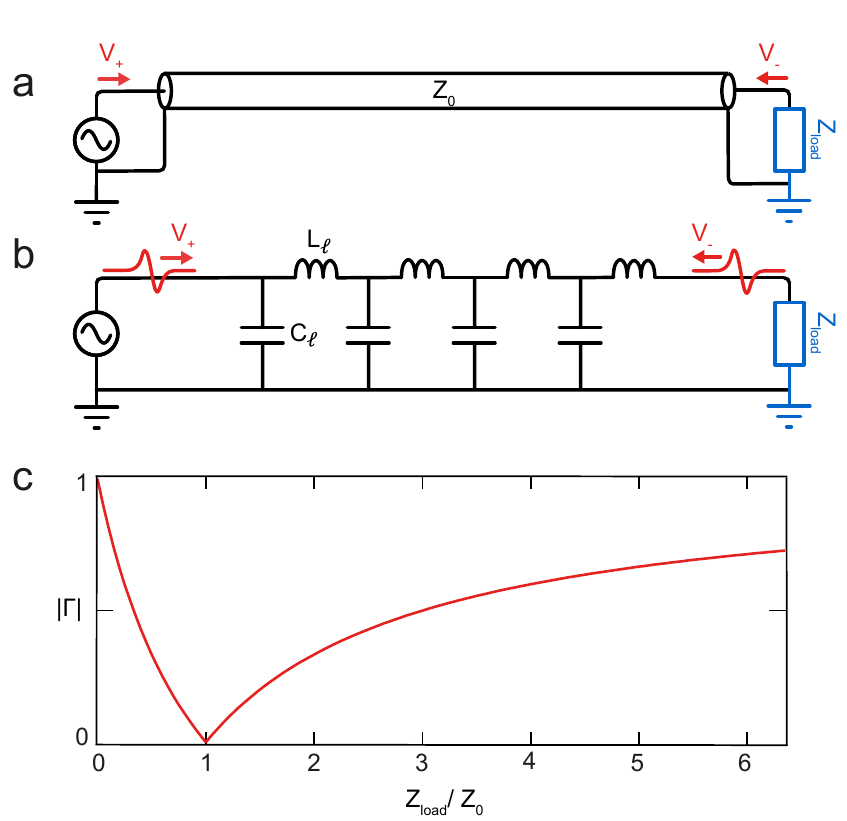}
	\caption{(a) Transmission line with characteristic impedance $\Zo$ linking an rf voltage generator to a load impedance. The signal $V_+$ is output by the generator and transmitted through the line. When it reaches the load, a portion of this signal $V_-$ is reflected back. 
	(b) Lumped-element equivalent of the same circuit.
	The line is represented by short segments of length $\Delta \ell$, each with inductance $\Ll \Delta \ell$ and capacitance $\Cl \Delta \ell$. 
	(c) Magnitude of the reflection coefficient $|\Gamma|$ (Eq. \ref{equ:Reflection2}) as function of the ratio $\Zload/\Zo$. }
	\label{fig:HFLine}
\end{figure}

To describe signal propagation, we must therefore allow the signal voltage $V(x,t)$ and current $I(x,t)$ to depend on location $x$ along the cable as well as on time $t$.
To see how these are related, suppose we have a cable of inductance $\Ll$ and capacitance $\Cl$ per unit length.
We imagine slicing the cable into short segments, each approximated by a single inductor and capacitor as in Fig.~\ref{fig:HFLine}(b).
This is a lumped-element transmission-line model of the cable.
By analysing the voltage and current at each node\cite{Pozar2012}, a pair of coupled equations (the telegraph equations) can be derived:
\begin{align}
\frac{\partial V}{\partial x}	&= -\Ll \frac{\partial I}{\partial t} \\
\frac{\partial I}{\partial x}		&= -\Cl \frac{\partial V}{\partial t}.
\label{equ:telegraphic}
\end{align}
Their solution is:
\begin{align}
 		\label{equ:telegraphicSolnA}
 	V(x,t) &= \phantom{\frac{1}{Z_0}\bigg[} V_+\left(t-\frac{x}{c'}\right) + V_-\left(t+\frac{x}{c'}\right) \\[2ex]
		\label{equ:telegraphicSolnB}
	I(x,t) &= \frac{1}{Z_0} \left[V_+\left(t-\frac{x}{c'}\right) - V_-\left(t+\frac{x}{c'}\right)\right]
\end{align}
where $c'=1/\sqrt{\Ll\Cl}$ is the phase speed of transmission line.
These solutions correspond to waves propagating in the positive direction (described by~$V_+\left(t-\frac{x}{c'}\right)$) and the negative direction (described by $V_-\left(t+\frac{x}{c'}\right)$).

For a wave propagating in a single direction, i.e. either the $V_+$ or the $V_-$ component, there is a fixed ratio between the signal voltage current and the signal current.
This ratio
\begin{equation}
\Zo \equiv \sqrt{\frac{\Ll}{\Cl}}
\end{equation}
is the characteristic impedance of the line.
It is the impedance that a semi-infinite length of line would present at its end, if its internal resistance (which was ignored in the approximation of Fig.~\ref{fig:HFLine}) could be neglected.

\subsubsection{Scattering at an impedance mismatch}

A transmission line's characteristic impedance becomes important when it is connected to a load with a different impedance.
The simplest example is a two-terminal device, such as a resistor, with impedance $\Zload$ (Fig.~\ref{fig:HFLine}(a-b)).
This imposes the boundary condition at the end of the line:
\begin{equation}
\frac{V(x,\omega)}{I(x,\omega)}=\Zload(\omega),
\label{eq:boundarycondition}
\end{equation}
where $V(x,\omega)$ and $I(x,\omega)$ are respectively the time Fourier transforms of $V(x,t)$ and $I(x,t)$ and $\Zload(\omega)$ is the load impedance, which in general depends on the angular frequency $\omega$.

Unless $\Zload=\Zo$, the $V_+$ component of Eqs.~(\ref{equ:telegraphicSolnA}-\ref{equ:telegraphicSolnB}) cannot satisfy Eq.~\eqref{eq:boundarycondition} by itself.
This means that if there is a mismatch between the impedances of the line and the load, part of the signal must be reflected back.
The amount of reflection can be calculated by defining $x=0$ to be the end of the line, and then taking the time Fourier transforms of Eqs.~(\ref{equ:telegraphicSolnA}-\ref{equ:telegraphicSolnB}) to give\cite{Pozar2012}:
\begin{align}
		\label{equ:telegraphicSolnC}
	V(0,\omega) &= V_+(\omega) + V_-(\omega) \\[1ex]
		\label{equ:telegraphicSolnD}
	I(0,\omega) &= \frac{V_+(\omega) - V_-(\omega)}{\Zo}.
\end{align}
Substituting into Eq.~\eqref{eq:boundarycondition} then gives the reflection coefficient $\Gamma(\omega)$ for the component of the incident signal at angular frequency $\omega$:
\begin{align}
	\Gamma(\omega)	&\equiv \frac{V_-(0,\omega)}{V_+(0,\omega)} \\
			&=\frac{\Zload(\omega)-\Zo}{\Zload(\omega)+\Zo}.
			\label{equ:Reflection2}
\end{align}
Figure~\ref{fig:HFLine}(c) plots $\Gamma$ for a purely resistive $\Zload$.
Similar equations hold for the scattered amplitude in a transmission circuit (see supplementary Section~S1).

In one respect, Eq.~\eqref{equ:Reflection2} is good news for measuring an unknown impedance; all we need to do is connect it to a transmission line and see how much power it reflects.
However, Eq.~\eqref{equ:Reflection2} and Fig.~\ref{fig:HFLine}(c) also tell us that if $|\Zload| \gg \Zo$, the reflection barely depends on $\Zload$.
Unfortunately, this is almost always the situation when measuring a quantum device.
This is because the typical resistance of a quantum device, such as an SET, is set by the resistance quantum, i.e.\
\begin{equation}
	\Zload \sim \frac{h}{e^2}\approx25.8~\mathrm{k\Ohm}.
	\label{equ:RQ}
\end{equation}
However, the typical impedance of a transmission line is of the order of magnitude of the impedance of free space $\eta_0$, i.e.\
\begin{equation}
	Z_0 \sim \eta_0 \equiv \sqrt\frac{\mu_0}{\epsilon_0} \approx 377~\Ohm.
	\label{equ:etaZero}	
\end{equation}
For example, a cylindrical coaxial cable has
\begin{equation}
	\Zo=\frac{\eta_0}{2\pi} \sqrt{\frac{\mu_\mathrm{r}}{\epsilon_\mathrm{r}}}~\ln \frac{b}{a},
	\label{equ:ZZero}
\end{equation}
where $\mu_\mathrm{r}$ and $\epsilon_\mathrm{r}$ are the relativity permeability and permittivity of the coaxial insulation and $a$ and $b$ are the diameters of the inner and outer conductor respectively.
For other geometries, similar equations apply\cite{Pozar2012}.
In fact most commercial coaxial cables, and therefore electronics designed to interface with them, use the standard value
\begin{equation}
	\Zo = 50~\Ohm.
\end{equation}

The mismatch between Eqs.~\eqref{equ:RQ} and~\eqref{equ:etaZero} is the fundamental reason why high-speed measurements of quantum devices are so difficult.
One tempting circumvention is to design a transmission line with $\Zo \approx h/e^2$.
Unfortunately, this approach seems doomed to failure\footnote{Its doom may not be final. By inserting Josephson junctions into a coplanar stripline to increase its inductance\cite{Altimiras2013, Stockklauser2017}, the characteristic impedance can be increased to $\Zo \approx 1~\mathrm{k\Ohm}$. Transmission lines exploiting the quantum Hall effect have been proposed as a way to make $\Zo$ even larger\cite{Bosco2019}}.
Equation~\eqref{equ:ZZero} shows that we would need coaxial cable with a diameter ratio of  $b/a\approx 10^{187}$, even using vacuum dielectric!
The key advance that created the field of radio-refrequency reflectometry for quantum devices was to interpose an impedance transformer between the load and transmission line~\cite{Schoelkopf1998}.
This is the topic of Section~\ref{sec:resistiveReadout}.

\subsection{Using an electrical resonator as the load impedance}
\label{sec:LCRLoad}

\begin{figure}
    \includegraphics[width=1\columnwidth]{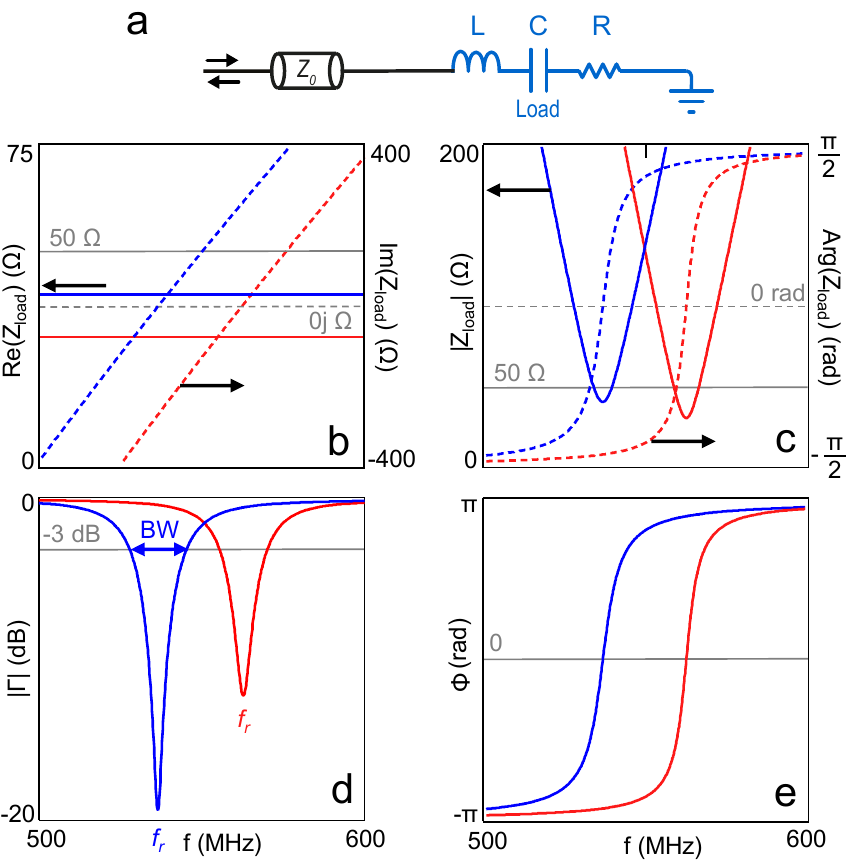}
    \caption{
    (a) Schematic of a transmission line connected to a LCR load. 
    (b) Simulation of the real (solid lines) and imaginary part (dashed lines) of $\Zload$ as function of the frequency $f$ for two slightly different LCR loads.  Blue: $R = 40~\Ohm$, $L = 800~\text{nH}$, $C = 0.11~$pF; red: $R = 30~\Ohm$, $L = 800~$nH, $C = 0.1~$pF.
    (c) Corresponding modulus (solid lines) and phase (dashed lines) of $\Zload$.
    (d) Corresponding reflection amplitude $|\Gamma|$, with $\Zo = 50~\Ohm$. The minimum of each dip marks the resonance frequency $\fr$. The bandwidth $\Bf$ lies approximately -3~dB from the top.
    (e) Corresponding reflection phase.
   }
    \label{fig:RLCspectrum}
\end{figure}

For reasons that will be explained in Section~\ref{sec:resistiveReadout}, the most useful load is usually an electrical resonator.
Near its resonance frequency, such a resonator is well approximated by an equivalent LCR circuit with an inductance $L$, a capacitance $C$ and a resistance $R$ in series (Fig.~\ref{fig:RLCspectrum}(a)). The equivalent complex impedance is:
\begin{equation}
    \Zload(\omega)=j \omega L + \frac{1}{j\omega C} + R
    \label{equ:ZLCR}
\end{equation}
where $j=\sqrt{-1}$. In this subsection, we describe the important properties of such a circuit and their effect on the reflection coefficent $\Gamma(f)$, where as usual the frequency is $f=\omega/2\pi$.

\subsubsection{Resonance frequency}

Figure~\ref{fig:RLCspectrum} shows how the complex impedance $\Zload$ (Eq.~\eqref{equ:ZLCR}) and the reflection spectrum depend on frequency.
Figure~\ref{fig:RLCspectrum}(b) plots the real and imaginary parts of $\Zload(f)$, calculated for two different combinations of $L$, $C$, and $R$.
Figure~\ref{fig:RLCspectrum}(c) shows the same quantities plotted as amplitude and phase.
The resonance frequency is where $\operatorname{Im}(\Zload)$ passes through zero, or equivalently where $\arg(\Zload)=0$.
From Eq.~\eqref{equ:ZLCR}, this frequency is
\begin{equation}
	\fr=\frac{1}{2 \pi \sqrt{L C}}.
	\label{equ:resonancef}
\end{equation}
The resonance also appears clearly in the reflection coefficient $\Gamma(f)$.
It leads to a dip in the amplitude $|\Gamma|$, here expressed in decibels ($|\Gamma|_\mathrm{dB} = 20\log_{10}(|\Gamma|_\mathrm{lin})$) (Fig.~\ref{fig:RLCspectrum}(d)) and a steep change in the reflection phase spectrum $\phi=\mathrm{arg}(\Gamma)$ (Fig.~\ref{fig:RLCspectrum}(e)).
Clearly this is a favourable frequency at which to illuminate the resonator, since a small change in circuit parameters leads to a large change in the amplitude or phase of the reflected signal. A change of $R$ changes the depth of the dip of the $|\Gamma(f)|$ while a change of $C$ or $L$ changes $\fr$ and moves $|\Gamma(f)|$ and $\phi(f)$ horizontally. These two cases are explained in detail in Section \ref{sec:resistiveReadout} and \ref{sec:reactiveReadout}.

\subsubsection{Resonator quality factor and bandwidth}

How fast does the reflected signal respond to a change in circuit parameters?
This is an important question, because it determines whether a transient change can be followed using reflectometry.
The answer is that the reflection will track the circuit parameters provided the rate at which they change is slower than the resonator's bandwidth $\Bf$~\footnote{
	One way to see this is to imagine that while the resonator is illuminated at frequency $\fin$, one of its parameters is modulated at frequency $\fm$.
	The current through the resonator, which without modulation would vary at the illumination frequency, is modulated by the changing impedance and therefore acquires sidebands at frequency $\fin \pm \fm$ (and possibly also at higher multiples of $\fm$).
	The reflected signal $V_-$ arises from the radiation of these sidebands into the transmission line.
	However, if the sidebands fall outside the resonance frequency window, i.e. $\fm \gtrsim \Bf/2$, they couple only weakly to the transmission line and the reflected signal is not modulated.
	
	A way to circumvent this problem, if you want to measure a modulation whose frequency is known but greater than the resonator bandwith, is to illuminate at $\fin+\fm$ so that the sideband appears inside the window~\cite{Regal2008}.
}. 
The resonance, having an inverse Lorentzian shape, has bandwidth corresponding to the full width at half maximum (FWHM) of the reflected \textit{power}
\footnote{since the power is proportional to $|\Gamma|^2$, the half power bandwidth can be found at $1/\sqrt{2}$ of the maximum $\Gamma$ or magnitude when plotted in linear units \cite{Pozar2012}}. 
This approximately corresponds to  $-3$~dB from the top if it is plotted in logarithmic units and the dip is deep (Fig.~\ref{fig:RLCspectrum}(d)). 
This bandwidth is determined by the rate at which energy is lost from the resonator, and includes both internal losses (i.e. dissipation) and external losses (i.e. radiation to the transmission line).

Both channels are conveniently described by an associated quality factor, defined in the conventional way as the inverse of the fraction of energy lost per radian of oscillation.
For the circuit of Fig.~\ref{fig:RLCspectrum}(a), the internal and external quality factors are respectively:
\begin{alignat}{3}
    \label{equ:Qint}
	\Qint&=\frac{1}{R}\sqrt{\frac{L}{C}} &= \frac{2\pi f_\text{r} L}{R}    \\[1ex]
		\Qext&=\frac{1}{\Zo}\sqrt{\frac{L}{C}} &= \frac{2\pi f_\text{r} L}{Z_0}.
	\label{equ:Qext}
\end{alignat}
The loaded (or total) quality factor describes the combination of both mechanisms and is
\begin{align}
	\Qr	&= \left (\Qint^{-1}+\Qext^{-1} \right )^{-1}	\label{equ:Qloaded1}\\
				&= \frac{1}{R+\Zo}\sqrt{\frac{L}{C}}.
\end{align}
In terms of $\Qr$, the bandwidth is
\begin{align}
	\Bf 	&=\frac{\fr}{\Qr} \label{equ:\Bf1}\\
		&=\frac{R+\Zo}{2\pi L}.
		\label{equ:BW1}
\end{align}
The quality factor, and hence the bandwidth, is dominated by whichever loss channel is stronger.

As Fig.~\ref{fig:RLCspectrum} suggests, designing resonators for fast readout involves a trade-off.
A large $\Qr$ is desirable to have a sharp resonance and therefore maximise the sensitivity changing circuit parameters.
However, Eq.~\eqref{equ:BW1} shows that this limits the measurement bandwidth.
This tension between sensitivity and speed is quantified by the Bode-Fano criterion, which states the optimum combination that can be achieved with resonators incorporating particular device impedances~\cite{Pozar2012}.

\subsubsection{Matching and coupling}

\begin{figure}
    \includegraphics[width=1\columnwidth]{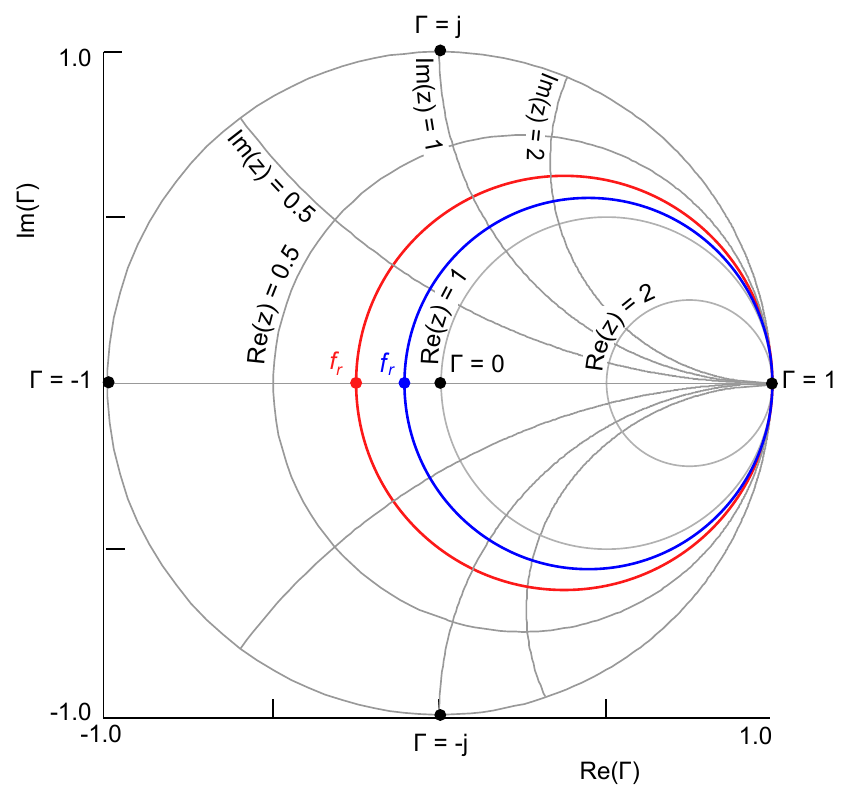}
    \caption{
    Smith chart representation of the same curves as in Fig.~\ref{fig:RLCspectrum}.
    The grey lines represent particular value of the impedance ratio $z=\Zload/\Zo$. Both curves are in the overcoupled regime since they intercept the horizontal line to the left of the point $\Gamma=0$.
   }
    \label{fig:Smith}
\end{figure}

The coupling constant 
\begin{equation}
    \beta = \frac{\Qint}{\Qext} = \frac{\Zo}{R}
    \label{eq:coupling}
\end{equation}

\noindent quantifies the coupling of the load to the line and classifies which part of the circuit dominates the losses. Critical coupling occurs when $\beta=1$, meaning that equal power is dissipated in the load and towards the line.
Usually circuit parameters are chosen to operate near this point because it maximises power transfer between the load and the measurement circuit.
The regime where $\beta<1$, so that internal losses dominate (i.e. $\Qint<\Qext$ and $R>\Zo$ for an LCR circuit), is called \emph{undercoupled}. The opposite regime $\beta>1$, illustrated in Fig. \ref{fig:Smith}, is called \emph{overcoupled}.

A useful way to show the reflection from a resonator and analyse the matching is by using a Smith chart\cite{Pozar2012}.
This is a plot of the reflection coefficient $\Gamma$ in the complex plane.
A graph of $\Gamma$ as a function of frequency, as shown over two panels in Fig.~\ref{fig:RLCspectrum}(d-e), appears as a single curve on the Smith chart (Fig.~\ref{fig:Smith}).
The closer this curve passes to the centre of the chart, i.e. the point $\Gamma=0$, the better the impedance matching.
It can easily be measured using a vector network analyser.

Because Eq.~\eqref{equ:Reflection2} imposes a one-to-one mapping between $\Gamma$ and $\Zload$, each point on the Smith chart also represents a specific value of $\Zload$.
The gridlines of $\Zload$, i.e. contours of constant $\operatorname{Re}(\Zload)$ and $\operatorname{Im}(\Zload)$, appear as circles on the Smith chart.
In Fig.~\ref{fig:Smith}, these gridlines are plotted in terms of the ratio $z=\Zload/\Zo$.
The Smith chart allows the effect of a change in $\Zload$ to be seen graphically.
For example, increasing $R$ moves the horizontal intercept of the $\Gamma(\omega)$ curve to the right. When the intercept lies to the right of the point $\Gamma=0$, the circuit is undercoupled while when it lies to the left, the circuit is overcoupled. 

\subsection{Introduction to demodulation}
\label{sec:demodulation}

We explain here how the information carried by a signal $V(t)$ about the variation of $\Zload$ is contained in two quadrature components and introduce the technique of demodulation to extract them.

\subsubsection{Representing a signal in terms of its quadratures}

A periodic signal $V(t)$ of frequency $f=\omega/2\pi$ can be mathematically expressed using two quadrature components. One common quadrature representation is composed of the amplitude $R$ and phase $\varphi$:
\begin{equation}
    V(t) = V_\text{R}\cos(\omega t + \varphi)
\end{equation}
We can rewrite $V(t)$ as:
\begin{align}
    V(t) 	&= V_\text{R}\cos(\varphi)\cos(\omega t) - V_\text{R}\sin(\varphi)\sin(\omega t)	\\
    V(t)	&= \VI\cos(\omega t) - \VQ\sin(\omega t)
\end{align}
where we define the quadratures $\VI$ and $\VQ$ as:
\begin{align}
    &\VI = V_\text{R}\cos(\varphi)\\
    &\VQ = V_\text{R}\sin(\varphi)
\end{align}

$\VI$ and $\VQ$ are also sometimes labelled "$X$" and "$Y$" or the "in-phase" and "out-of-phase" components. They correspond to two signals shifted in phase by $\pi/2$. The "IQ" quadrature representation is very useful to generate or analyse signals. The relation between ($\VI$,$\VQ$) and ($V_\text{R}$,$\varphi$) is 
\begin{align}
    &V_\text{R} = \sqrt{\VI^2+\VQ^2} \\
    &\varphi = \arctan \left ( {\frac{\VQ}{\VI}} \right )
    \label{eq:defineR}
\end{align}

\begin{figure}[h!]
    \includegraphics[width=\columnwidth]{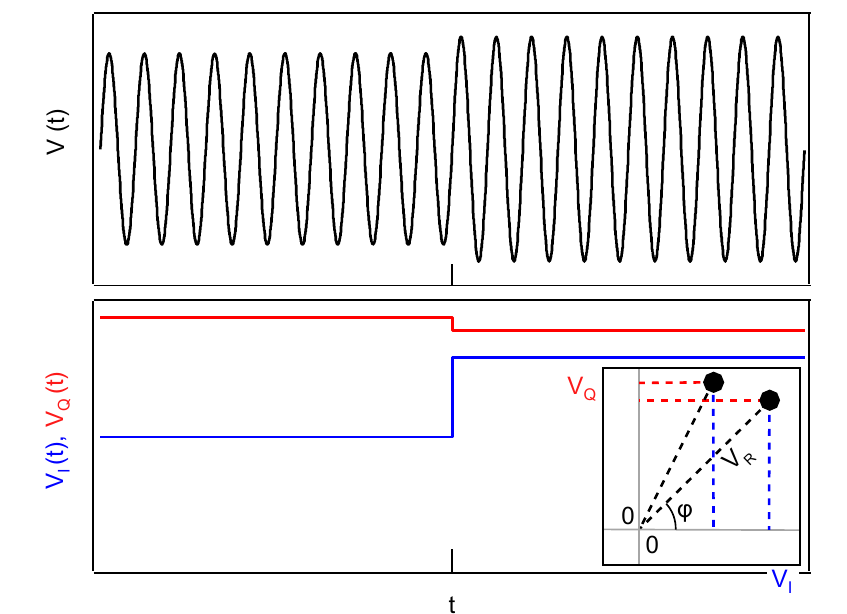}
    \caption{Top: signal $V(t)$ for a device switching between two states. Bottom: $V(t)$  decomposed in $\VI$ and $\VQ$ quadratures. 
   Inset:  representation of the two states in the IQ plane.
    }
    \label{fig:Demodulation&IQ}
\end{figure}

In Fig.~\ref{fig:Demodulation&IQ}(a) the simulated signal $V(t)$ changes phase and amplitude in the middle of the horizontal axis which give two points in the IQ plane (Fig. \ref{fig:Demodulation&IQ}(b)). This change could be due to a switch between two states of the load impedance $\Zload$ with different reflection coefficients $\Gamma$.

\subsubsection{Demodulation}
\label{subsec:demodulation}

We have just seen that the useful information carried by $V(t)$ is embedded in the quadrature components of the signal. A direct measurement of $V(t)$ is complex and inefficient since it would require high-rate acquisition of a huge number of points. A more efficient approach is to demodulate $V(t)$ to an intermediate frequency signal $\VIF(t)$ that has a lower frequency but contains the information in the quadrature components $\VI$ and $\VQ$. 

To demodulate $V(t)=\VI\cos(\omega t)-\VQ\sin(\omega t)$ we need a mixer and a low pass filter. By mixing $V(t)$ with a demodulating signal $\VLO(t) = \cos(\omega t)$ of the same frequency and phase we obtain the product
\begin{equation}
    \VIF(t)=V_\text{out}(t) \cdot \VLO(t),
\end{equation}
which decomposes as
\begin{align}
    &\VI\cos(\omega t) \cdot \cos(\omega t) = \frac{\VI}{2} + \frac{\VI}{2}\cos(2\omega t) \\
    -&\VQ\sin(\omega t) \cdot \cos(\omega t) = 0  - \frac{\VQ}{2}\sin(2\omega t).
\end{align}

This product gives a low-frequency signal proportional to the $\VI$ quadrature ($\VI/2$). The $2\omega$ component of the signal is removed by the low-pass filter. 
We obtain the $\VQ$ quadrature by following the same process using a phase-shifted local oscillator $\VLO(t) = -\sin(\omega t)$:
\begin{align}
    &\VI\cos{(\omega t)} \cdot (-\sin{(\omega t)}) = 0 - \frac{\VI}{2}\sin{(2\omega t)}\\
    -&\VQ\sin{(\omega t)} \cdot (-\sin{(\omega t)}) = \frac{\VQ}{2} - \frac{\VQ}{2}\cos{(2\omega t)}  
\end{align}

If $V$ and $\VLO$ are separated by a phase difference $\varphi$ the demodulated signal is composed of both quadrature components:
\begin{align}
    &\VI\cos{(\omega t)} \cdot \cos{(\omega t + \varphi)} \rightarrow \frac{\VI}{2}\cos{(\varphi)} \\
    -&\VQ\sin{(\omega t)} \cdot\cos{(\omega t + \varphi)} \rightarrow  -\frac{\VQ}{2}\sin{(\varphi)}.
\end{align}
We have removed the $2\omega$ contribution from these two expressions. 

If $V(t)$ and $\VLO(t)$ are not at the same frequency but $\VLO(t) = \cos(\omega_\text{LO} t)$,  the demodulated signal is composed of two angular frequencies $\omega_+=\omega + \omega_\text{LO}$ and $\omega_-=\omega-\omega_\text{LO}$:
\begin{align}
   &\VI\cos {(\omega t)} \cos {(\omega_\text{LO} t)} = \frac{\VI}{2} \left[ \cos{(\omega_-t)} + \cos{(\omega_+t)} \right] \\
    -&\VQ\sin {(\omega t)} \cos {(\omega_\text{LO} t)} = -\frac{\VQ}{2} \left[ \sin{(\omega_-t)} + \sin{(\omega_+t)} \right]  
  \label{equ:demodulation1&2}
\end{align} 

The low-pass time constant $\taum$ has to be carefully chosen. 
A filter with a long time constant passes less noise, but also filters out rapid fluctuations of the signal.
The filter is therefore generally chosen to pass all frequency components of interest in the demodulated signal.

\subsubsection{Homodyne and heterodyne detection}

In homodyne detection, the signal $\Vout(t)$ at frequency $\fout$ is demodulated using the frequency of the input signal so $\fLO=\omega_\text{LO}/2\pi=\fin$ (Fig. \ref{fig:Demodulation&IQ}(c)). This results in two signals at frequencies $(\fout-\fin)=\fm$ and $\fout+\fin$. The second term can be filtered out so only the signal at the frequency $\fm$ remains, which represents the modulations of the sample impedance.
In heterodyne detection, $\Vout(t)$ is demodulated using $\fLO \neq \fin$. The result is two signals at frequencies  $\fout- \fLO$ and $\fout + \fLO$, the second term being usually filtered out.

\section{Measuring a resistive device}
\label{sec:resistiveReadout}

In this section, we detail how radio-frequency measurements are used to probe the resistance of a quantum device. We start by discussing the matching condition between a transmission line (Section \ref{subsec:MatchingR}) and a quantum device and we then focus on two examples which are mostly used in quantum electronic experiments: the quantum point contact (QPC) charge sensor and the single-electron transistor (SET) charge sensor together with its lookalike, the quantum dot (QD) charge sensor (Section \ref{subsec:ChargeSensor}). Later, we describe how these devices can be used as charge sensors (Section \ref{subsec:rfmeasSET}) and which applications arise from the combination of radio-frequency measurements with charge sensing techniques (Section \ref{subsec:ReadPerf}). Next, we present an exemplary phenomenon of dissipation induced by the rf drive, the Sisyphus resistance, which can be used to study dynamic dissipation in  two-level systems (Section \ref{Sec:Sisyphus}). We conclude by discussing the difficulty to scale up for measuring numerous quantum devices such as qubits (Section \ref{subsec:ScalingUp}) 

\subsection{Matching resistive devices with a $LC$ resonator}
\label{subsec:MatchingR}
As shown in Section~\ref{Sec:Basics}, monitoring the reflection coefficient $\Gamma$ can, in particular, reveal changes in the resistance (real part) of a load impedance. Near the resonant frequency, $\Gamma$ becomes highly sensitive to the variations of $\Zload$. However, according to Fig.~\ref{fig:HFLine}, $\Zload$ needs to be close to the characteristic impedance of the transmission line, $\Zo =50~\Ohm$, to ensure good sensitivity to sample resistance changes. This poses a problem since the impedance of quantum devices is typically much larger. To match the transmission line impedance, the quantum device must be embedded in a matching network.

\begin{figure}[h]
  \includegraphics[width=\columnwidth]{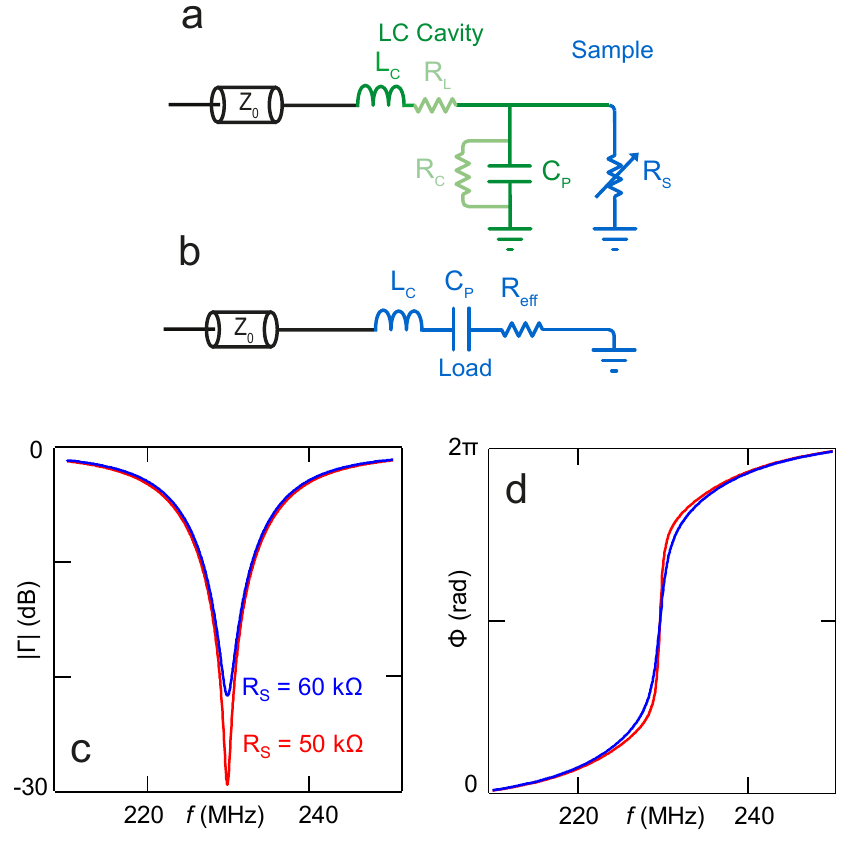}
  \caption{
  (a) Schematic of a reflectometry circuit used to measure a variable resistor $\Rs$. The sample is embedded in an L-matching network made of an inductance $\Lc$ and a capacitance $\Cp$, to match the characteristic impedance $\Zo$ of the line. The resistance $\RL$ represents dissipation in the inductor and the resistance $\Rc$, dielectric losses. 
  (b)~Equivalent RLC series circuit at the resonant frequency.
  Magnitude~(c) and phase spectrum (d) of the reflection coefficient $|\Gamma|$ for two values of $\Rs$. The other circuit parameters are $\Lc = 800~\text{nH}$, $\RL = 20~\Ohm$, $\Rc =100~\text{M}\Ohm$ and $\Cp = 0.6~\text{pF}$.
  }
  \label{fig:LC cavity R}
\end{figure}

L-matching networks, particularly low-pass $LC$ circuits, are widely used because they consist of only two elements. However, more complex matching networks can be employed, especially if independent control of the matching condition and network quality factor is needed. Their main components are an inductor $\Lc$, placed between the line and the sample, and a capacitance $\Cp$ located in parallel to the sample (Fig.~\ref{fig:LC cavity R}(a)). The capacitance can be chosen to be a real or parasitic element. 
We also introduce parasitic resistors to model dissipation in the circuit; here $\RL$ models ohmic losses in the inductor, while $\Rc$ models dielectric losses in the capacitor.
As we shall see in Section~\ref{section:Optimisation}, to optimize the sensitivity to resistance changes, a good matching network should usually minimize $\RL$ and maximize $\Rc$.

The impedance of this circuit, which presents itself as a load on the transmission line, is
\begin{equation}
  \Zload = j\omega \Lc+\RL+\frac{\Req}{1+j\omega \Req\Cp},
  \label{equ:ZloadR}
\end{equation}
where $\Req= \Rs||\Rc$ is the parallel combination if $\Rs$ and $\Rc$.
On resonance, $\mathrm{Im}(\Zload)=0$, which leads to an analytical expression for the resonant frequency.
For typical circuit parameters such that $\Lc/\Req^2\Cp\ll 1$, it reads
 \begin{equation}
  \omegar = \frac{1}{\sqrt{\Lc\Cp} } \label{eq:resonancef}
 \end{equation}
where $\omegar=2\pi\fr$. Substituting Eq.~\eqref{eq:resonancef} into Eq.~\eqref{equ:ZloadR}, the impedance of the circuit at resonance simplifies to that of a series $LCR$ circuit (Fig.~\ref{fig:LC cavity R}(b)), similar to the example in Section~\ref{Sec:Basics}, with an overall impedance given by~\cite{Schoelkopf2003}
\begin{align}
  \Zload    &\approx \Reff+j \omega \Lc +\frac{1}{j \omega \Cp} \\
            &= \Reff + j2\sqrt{\frac{\Lc}{\Cp}}\frac{\Delta\omega}{\omegar},
  \label{eq:LCsimple}
\end{align}
where $\Delta\omega$ is the difference between the probing angular frequency $\omega$ and $\omegar$, and the effective resistance reads
\begin{align}
  \Reff &= \frac{\Lc}{\Cp \Req} + \RL \\
        &\stackrel{\text{ideal}}{\rightarrow} \frac{\Lc}{\Cp \Rs}.
  \label{equ:Reff}
\end{align}
Equation~\eqref{equ:Reff} is a key result of this section, showing that an L-matching network transforms the impedance of the device to a new value that can be more easily matched to the impedance of the line. The matching resistance, which is the device resistance for which the tank circuit matches the line, i.e. $\Reff=\Zo$, is thus ideally

\begin{equation}
  \Rmatch = \frac{\Lc}{\Cp Z_0}.
  \label{equ:Rmatch}
\end{equation}
Hence an rf designer should carefully choose the values of $\Lc$ and $\Cp$ that will make $\Rmatch$ equal to the on-state resistance of the device to be measured. 

The measurement principle relies on the change of the reflection coefficient $\Gamma$ induced by a change in $\Rs$. This is illustrated in Figure~\ref{fig:LC cavity R}(c) in which a change of sample resistance embedded in a matching network manifests as a change in the reflection power near the resonance frequency. Likewise, the reflected phase changes due to the change in loaded quality factor (Fig.~\ref{fig:LC cavity R}(d)). The carrier frequency must be chosen at $\omega \approx \omegar$ to maximise the sensitivity to resistance changes. 
Guidelines on how to optimise the design of the matching network and improve the sensitivity to resistance changes are developed in Section~\ref{section:Optimisation}.

\subsection{Measuring the charge occupation of quantum dots with charge sensors}
\label{subsec:ChargeSensor}

The direct measurement of quantum devices, and specifically QDs, is a challenging task due to large time constants associated with their typical high impedance. This has motivated the use charge detectors coupled to the quantum system as local and sensitive electrometers to investigate a variety of phenomena including detection of single charge occupation in QDs~\cite{Kouwenhoven2003,Gossard2004}, time domain measurements of tunneling events~\cite{Kouwenhoven2004,Gossard2006,Wagner2019}, charge and spin single-shot readout and coherent manipulation~\cite{Elzerman2004,Gossard2005,Petta2005,Amasha2008,Gossard2010_Coherent}. In this section, we analyse the working principle of the most common type of charge detectors: the QPC, the SET and the QD charge sensor.

\subsubsection{Quantum point contact charge sensors}

A QPC is a constriction in which transport occurs through 1-dimensional subbands~\cite{datta1997Book}. For structures with high mobility and at low temperature, the conductance $G_\mathrm{QPC}$, tends to be quantized in plateaus at multiples of the conductance quantum $G_0=2e^2/h=77.5$~$\mu$S (or at $e^2/h$ under high magnetic field)~\cite{Kouwenhoven1988}. In the ballistic limit,
\begin{equation}
  G_\mathrm{QPC}=\frac{e^2}{h}\sum_n{f_e(E_n)g_n}
\end{equation}
with $f_e$ the Fermi probability distribution and $E_n$ and $g_n$ the energy and degeneracy of the $n^{th}$ subband. 

\begin{figure}[h]
  \includegraphics[width=1\columnwidth]{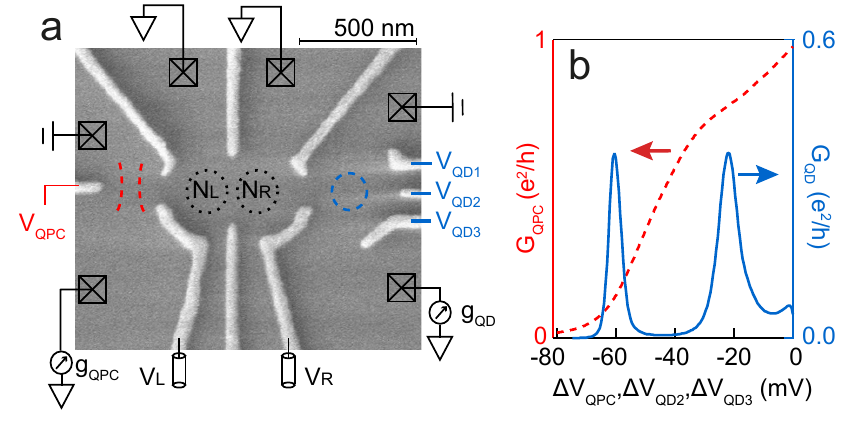}
  \caption{A GaAs/AlGaAs quantum device incorporating a double quantum dot and two charge sensors. 
  (a) SEM micrograph of the device. Metallic gates define a double quantum dot (black dashed lines) which is capacitively coupled to a QPC charge sensor (red dashed lines) and a QD charge sensor (blue dashed lines).
  Leads through which these two charge sensors are measured, are marked by crossed squares.
  (b) dc conductance of the QPC ($G_\text{QPC}$) and the QD charge sensor ($G_\text{QD}$) as a function of their respective control gate voltages. The most sensitive operation points are typically those with highest transconductance, i.e. those with the steepest dependence of conductance on gate voltage. Figure adapted from Ref.~\onlinecite{Gossard2010_fast}.}
  \label{fig:rf-SETandQPC}
\end{figure}

QPCs are easily realised in nanowires whose linear geometry provides a natural one-dimensional confinement, or in two-dimensional electron (hole) gases (2DEGs or 2DHGs) in which the current path is restricted to a narrow channel using depletion gates as in Fig.~\ref{fig:rf-SETandQPC}(a).
The QPC (in red) is created by just a single additional gate $V_\text{QPC}$ creating a constriction against the DQD barriers. The two sides of the constriction are connected to contact leads that allow the measurement of the QPC conductance $G_\mathrm{QPC}$.
As the channel is narrowed using a gate voltage, the conductance decreases (Figs.~\ref{fig:rf-SETandQPC}(b) and \ref{fig:rf-QPC}(b)).
At gate voltage settings for which the conductance changes steeply, the QPC is highly sensitive to its electrostatic environment.

This property makes the QPC an efficient charge sensor for nearby QDs~\cite{Hasko1993,Kouwenhoven2003}. Each additional charge present on the QDs contributes an effective voltage which shifts the conductance-versus-gate voltage curve of the QPC.
Therefore variations in the charge configuration of the QD result in discrete changes in $G_\mathrm{QPC}$. The gate voltage is tuned to the point of maximum derivative of the conductance curve for the best sensitivity, which is often midway between the first conductance plateau and pinch-off~\cite{Smith2007}, setting the QPC resistance to around $25.8$~k$\Ohm$. 

\subsubsection{Single-electron transistor and quantum dot charge sensors}
\label{subsec:SET}

The SET and QD charge sensors are three-terminal devices in which a small region of conducting material (the `island') is connected via tunnel barriers to two charge reservoirs, source and drain. Furthermore, the island is capacitively coupled to a gate electrode that enables changing the charge occupation in the island by means of gate voltage changes. SETs can be realised in metals~\cite{Schoelkopf1998,Delsing2001} or semiconductors, whereas QD charge sensors require quantum confinement, which can usually only be achieved in semiconducting nanostructures~\cite{Dzurak2015,Simmons2015}.
In Fig.~\ref{fig:rf-SETandQPC}(a) a QD charge sensor is realised by the confinement potential of three gates $V_\text{QD1-3}$ (1 and 3 control primarily the tunnel barrier resistance and 2 the QD charge occupation). The sensor is capacitively coupled to a DQD and tunnel-coupled to two (source and drain) reservoirs. 

Electronic transport through SETs and QDs is governed by charge quantization in the island, i.e. Coulomb blockade~\cite{Kastner1992}. For Coulomb blockade to be manifest, the charging energy of the island $E_\mathrm{C}$ needs to be larger than $k_\text{B}T$. Besides, the resistance of each of the tunnel barriers, $R_\text{T}$, needs to be larger than the von Klitzing resistance $R_\text{K} \approx 25.8$ k$\Ohm$ to ensure that the energy uncertainty of each charge state is smaller than $E_\text{C}$. 
At finite bias, Coulomb blockade gives rise to regular sharp conductance peaks as a function of the gate voltage~\cite{Hanson2007} (Fig.~\ref{fig:rf-SETandQPC}(b)). 

In an SET, transport is considered through a quasi-continuum of states~\cite{Beenakker1991}. The SET conductance as function of the gate voltage $\Vg$, close to a charge degeneracy point $\Vg^0$, can be described as
\begin{equation}
    G_\text{SET} = G_\mathrm{max}\cosh^{-2}\left [\frac{\alpha(\Vg-\Vg^0)}{2.5k_\mathrm{B}T}\right ].
\end{equation}
Here $G_\mathrm{max}$ represents the conductance at the charge degeneracy point ($\Vg=\Vg^0$), and $\alpha$ is the ratio of the gate and total capacitance (lever arm). 

If the island is made sufficiently small, quantum confinement can lead to electronic transport through discrete energy levels once the energy level spacing, $\Delta E$ is larger than $k_\text{B}T$~\cite{Marcus2007,Marcus2009,churchill2009, Marcus2012,Marcus2014,kurzmann2019}. In this case, charge tunnelling occurs through a single level. We refer to these devices as QDs. The QD conductance can be expressed as~\cite{Beenakker1991}: 
\begin{equation}
    G_\text{QD} = G_\mathrm{max}\frac{\Delta E}{4k_\mathrm{B}T}\cosh^{-2}\left [\frac{\alpha(\Vg-\Vg^0)}{2k_\mathrm{B}T}\right ].
\end{equation}
Because QD charge sensors present sharper conductance peaks, they can reach higher sensitivity than SETs. Charge sensing with SETs or QD charge sensors works on the same principle as with QPCs~\cite{Dolan1987,Devoret1991,devoret2000Review}.
Charge sensing is realised by monitoring the conductance of the island at a constant gate voltage, chosen on the flank of a Coulomb peak so that the conductance depends steeply on the electrical potential. When a charge is added to or removed from a nearby device, the small variation of electric field shifts the position of the Coulomb peak on the gate voltage axis resulting in a different current.

Both SETs and QD charge sensors are now commonly used for charge sensing, with the choice of one or the other being mostly dependent on the geometry and the material of the experiment. 
QDs tend to be used in systems where quantum confinement can be routinely be achieved. Materials with low effective mass like AlGaAs/GaAs heterostructures~\cite{Gossard2010_fast} or nanowires~\cite{churchill2009}, that provide natural confinement, are typical examples. On the other hand, SETs are more common in materials with higher effective mass like silicon~\cite{Angus2007a, GonzalezZalba2012a, Simmons2015}.  
SETs and QDs are technologically more complex to fabricate than QPCs because of the additional number of gates needed, but provide in general a better sensitivity because of their steeper slope. However, QPCs work over a wider range of gate voltages and often have a greater dynamic range to charge sensing signals than SETs and QD charge sensors whose sensitivities drop away from a Coulomb oscillation. Voltage cross-talk from neighbouring gates also affects the bias point and hence the conductance of SETs and QD charge sensors, thus requiring re-adjustment of the gate voltages to maintain the sensors at the bias point for maximum sensitivity. This has recently motivated the use of advanced compensation strategies based on fast feedback~\cite{Tarucha2021}.

\subsubsection{Sensitivity and limits of low-frequency charge sensors}

We have just seen that devices such as the QPC and the SET can act as fine electrometers due to their sharp transconductance at low temperatures. However, when measured in the setup Fig.~\ref{fig:Intro}(a), their measurement bandwidth is limited to a few tens of kilohertz, due to the $RC$ constant formed between the impedance of the current amplifier and the capacitance of the cabling leading to the first amplifier stage ($C_\text{line}=0.1-1$~nF), usually sitting at room temperature, $\sim 1.5$~m away from the device. This measurement bandwidth is far below the intrinsic maximum bandwidth of the charge sensors, which in the case of the SET is set by the intrinsic $RC$ constant of the tunnel barriers ($R\approx R_\text{K}$ and $C\approx 1$~fF) and can exceed 10~GHz. The limited bandwidth of conventional low-frequency measurements has a knock-on detrimental effect on the sensor's charge sensitivity. The charge sensitivity of a sensor is not exclusively determined by the sharpness of its transconductance but also by the noise level at the measurement frequency which, as we shall see later, can be substantial at low frequencies. 

To quantify the sensitivity of a sensor and discuss its ultimate limits, we resort to the definition of charge sensitivity. The charge sensitivity $\Sqq$ of a charge detector tells us the amount of charge $Q$ that can be discerned in a measurement lasting a second. It is defined as 
\begin{equation}\label{eq:deltaq}
   \Sqq(f)=\frac{\sqrt{\SII(f)}}{|\partial I/ \partial Q|}
\end{equation}

\noindent where $\SII(f)$ is the current noise spectral density
\footnote{All spectral densities in this Review are one-sided. Most experimentalists follow this convention, but a two-sided convention is widespread in theory-land~\cite{Clerk2010}. To make a double-sided spectral density single-sided, multiply it by two; more details are given in the Supplementary Information.}
and $\partial I/\partial Q$ is the change in device current $I$ induced due to a change in the charge $Q$ on the device, a magnitude proportional to the transconductance. This figure is the noise-limited charge sensitivity. Charge sensitivities as good as 20 $\mu$e$/\sqrt{\text{Hz}}$ have been measured for SETs in the normal state at 4.4~kHz~\cite{starmark1999}, outperforming state-of-the-art conventional transistors by three orders of magnitude~\cite{Mar1994}. However, this number is still far from the theoretical limit of the SET, which is dominated by shot noise (at dilution refrigerator temperatures Johnson-Nyquist noise is typically much smaller). Shot noise has a current noise spectral density given by~\cite{Buttikker2000}
\begin{equation}
   \SII=F2eI
   \label{equ:SII}
\end{equation}

\noindent where $F$ is the Fano factor, which varies between 0.5 and 1 in the Coulomb blockade regime 	\footnote{Expressions for the Fano factor in different kinds of device are given in Ref.~\onlinecite{Buttikker2000}.
The most important is
	\begin{equation}
	    F=\frac{\sum_n T_n (1-T_n)}{\sum_n T_n},
	\end{equation}
which applies when non-interacting electrons tunnel through a barrier, each of whose conductance channels has an energy-independent transmission $T_n$.}
and accounts for the correlation of between charge tunneling events. $I$ is the average current flowing through the device. In this limit, the ultimate charge sensitivity reads~\cite{korotkov1999}
\begin{equation}\label{Eq:limit}
   \Sqq= 1.9e\left(R_\text{T}C_\Sigma\right)^{1/2}\left(k_\text{B}TC_\Sigma/e^2\right)^{1/2}.
\end{equation}

For common experimental values $C_\Sigma=0.45$~fF, $R_\text{T}=100$~k$\Omega$ and $T=100$~mK, Eq.~\eqref{Eq:limit} gives 1 $\mu$e$/\sqrt{\text{Hz}}$. As we can see, experimental values are far from this ultimate limit. The reason for this loss of sensitivity is additional sources of noise that appear at the low frequencies of the measurements. For example 1/$f$ noise, which originates from time-dependent occupation of charge trap centres in the neighbourhood of the charge sensor, can be substantial below 10~kHz. Several solutions were proposed to increase the bandwidth such as the use of superconducting SETs that have a lower tunnel barrier resistance~\cite{Zorin1996} and bringing the amplifier closer to the SET. However, these approaches only improved the bandwidth moderately up to 700~kHz~\cite{pettersson1996,Curry2019}. 

The measurement bandwidth limitation can be overcome by the use of rf reflectometry, which removes the effect of $C_\text{line}$ by impedance matching the device to a high-frequency line. Typical operation frequencies are in the few megahertz to 2~GHz regime (set by the first amplifier stage, see Section~\ref{Sec:amplifiers}) and the measurement bandwidth can reach values as high as 100~MHz~\cite{Schoelkopf1998}. This phenomenal increase in operational frequency, closer to the intrinsic limit of the devices, allows operation well above the $1/f$ noise and, as we shall see later, leads to a subsequent improvement of the charge sensitivity and measurement speed.

\subsection{Reflectometry of charge sensors}
\label{subsec:rfmeasSET}

In this subsection, we present examples from the literature where rf-QPCs and rf-SETs have been used to measure QDs in their periphery. 

\subsubsection{Radio-frequency measurement of a QPC}

Radio frequency measurements of a QPC~\cite{Williams2006,Gossard2007,Smith2007,Muller2007,Kycia2010,Muller2010,Muller2012,Muller2013} and of an SET/QD~\cite{Hirayama2000,Delsing2001,Delsing2006,Clark2007,Tarucha2015,Simmons2016} are similar in principle. The charge sensor is embedded as a resistive element in a matching network at the end of a transmission line in a setup similar to Fig.~\ref{fig:LC cavity R}.

\begin{figure}
  \includegraphics[width=1\columnwidth]{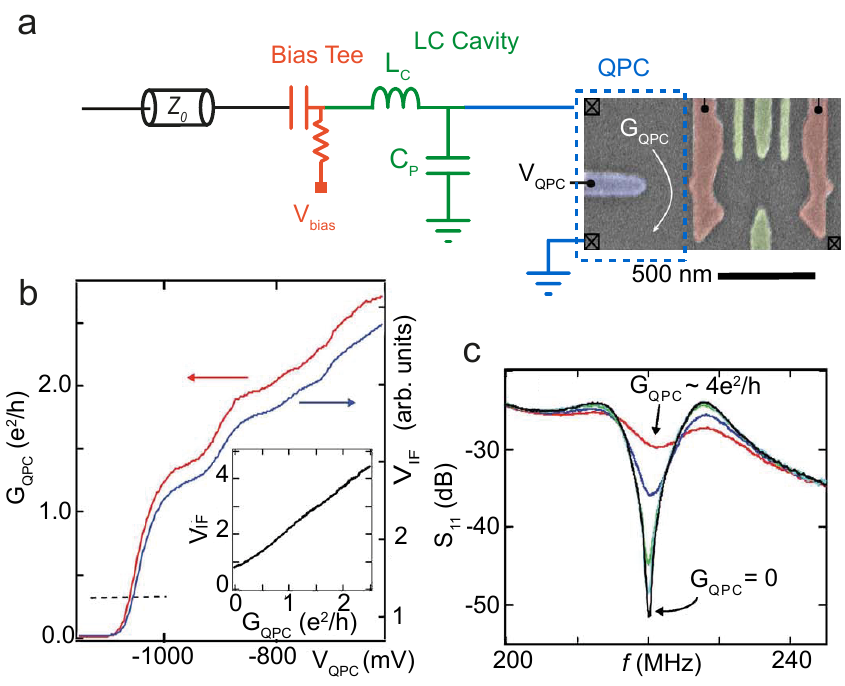}
  \caption{
  (a)  An $LC$ matching network is attached to one of the contact leads of a QPC. In this example~\cite{Gossard2007}, $\Lc=820$~nH and $\Cp=0.63$~pF. 
  The bias tee, made of a 100~pF capacitor in the ac path and a 5~k$\Ohm$ resistor in series with a 100~nH inductor (not shown) in the dc path, allows a dc bias to be applied across the QPC.
  (b) Demodulated response $\VLF$ (right) and dc conductance of the QPC $G_\mathrm{QPC}$ (left) versus gate voltage ($V_\text{QPC}$). The dashed line indicates the operation point for charge sensing. The inset shows the transfer function: $\VLF$ vs conductance. 
  (c) Reflection ratio $S_{11}$ versus frequency for different values of $G_\mathrm{QPC}$. Panel (a) adapted from Ref.~\onlinecite{Gossard2007}.}
  \label{fig:rf-QPC}
\end{figure}

The equivalent circuit for an rf-QPC shown in Fig.~\ref{fig:rf-QPC}(a) is similar to the one presented in Fig.~\ref{fig:LC cavity R}, with $\Rs=1/G_\mathrm{QPC}$ being the QPC resistance (Fig.~\ref{fig:rf-QPC}). In a reflectometry setup, the matching network is connected to one of the contacts of the charge sensor while the other is grounded. 
Further, a bias tee allows a dc voltage to be added to the source-drain bias, so that the dc conductance can be measured simultaneously with the rf response.
The load impedance can be expressed using Eq.~\eqref{equ:ZloadR} allowing the reflection coefficient to be calculated.

The measurement detects the change of rf reflection induced by variations in $\Rs$. In the example of Fig.~\ref{fig:rf-QPC}(c), the reflectance $S_{11}$ shows a trough at the resonance frequency, which is deepest when the QPC is pinched off ($G_\text{QPC}=0$). 
At higher gate voltages the resistance $\Rs$ of the QPC decreases and the circuit becomes undercoupled, resulting in a higher reflection. As shown in Fig.~\ref{fig:rf-QPC}(b), this allows the QPC conductance to be probed over a wide gate voltage range. When $G_\text{QPC}\approx 4e^2/h$, the trough at the resonant frequency is barely evident. 
\footnote{In this example, the matching condition occurs when the QPC is pinched off ($G_\text{QPC}=0$) due to the relatively high dielectric losses (low $\Rc$). As we shall see in Section~\ref{section:Optimisation}, when trying to maximise sensitivity to changes in device resistance, low $\Rc$ is detrimental and matching should preferentially be achieved in the bias condition under which the device is mostly used.}

\subsubsection{Radio-frequency measurement of the SET (and QD)}

\begin{figure}[h]
  \includegraphics[width=1\columnwidth]{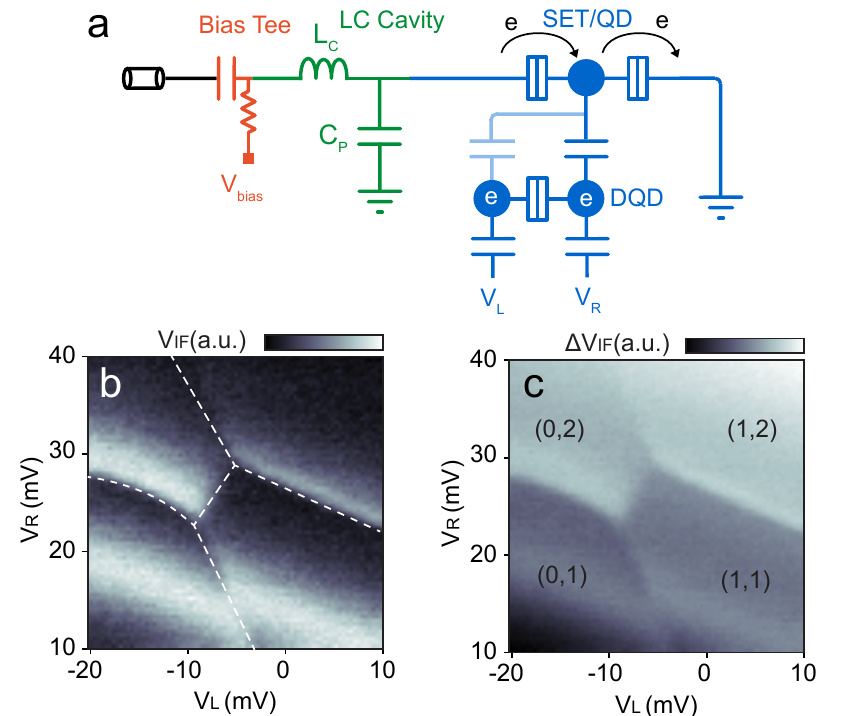}
  \caption{Radio-frequency charge sensing a DQD with a QD charge sensor. 
  (a) Schematic of an rf-QD/SET charge sensor probing the charge occupation of a DQD. The charge sensor has a strong capacitive coupling with the right QD and a smaller coupling with the left one.
    (b) Demodulated voltage $\VIF$ as a function of QD plunger gates $V_\text{L}$ and $V_\text{R}$.
    Dark regions correspond to the QD Coulomb valley where the sensitivity is low, while the bright regions corresponds to the QD Coulomb peak which is highly sensitive to the DQD charge configuration. Abrupt shifts of the Coulomb peak occur each time the occupation of the DQD changes, giving rise to the honeycomb pattern typical of a DQD charge stability diagram.
    (c)  The same data, plotted with a plane background subtracted to make the honeycomb clearer. The four charge configurations are labelled according to the occupation of the DQD.
   Panels (b) and (c) adapted from Ref.~\onlinecite{Fedele2021}.}
  \label{fig:ChargeSensingMeasure}
\end{figure}

Similar to the rf-QPC, the rf-SET and rf-QD charge sensors are connected to an $LC$ matching network via one of the contact leads. Readout of the SET (QD) resistance is then accomplished by monitoring the reflected amplitude of a high-frequency signal, see Fig.~\ref{fig:ChargeSensingMeasure}(a). Compared to the rf-QPC, rf-SET and rf-QD are more sensitive~\cite{Gossard2010_fast}, but their higher resistance (see Fig.~\ref{fig:rf-SETandQPC}(b)) makes them more challenging to match to a 50 $\Ohm$ line~\cite{Williams2006}. 

In Fig.~\ref{fig:ChargeSensingMeasure}(b) we show the demodulated voltage $\VIF$, which is proportional to the sensor conductance, measured as a function of gates voltages $V_\text{R}$ and $V_\text{L}$, which control the number of electrons in a DQD device similar to the one shown in Fig.~\ref{fig:rf-SETandQPC}.
In this plot, the color scale can be directly associated with variations in the absorption of the rf signal by the resonant circuit, where the rf-QD charge sensor is embedded. Dark regions correspond to Coulomb valleys of the charge sensor. Here the charge sensitivity is low, since changes in the charge neighbourhood produce minor changes in the QD resistance. Further, the high QD resistance places the system off the matching condition and most of the rf carrier power gets reflected. Conversely, bright regions correspond to the charge sensor being biased near a Coulomb peak setting the system closer to the matching condition where the sensitivity is best. Further, in this bias condition, the charge sensor's conductance is strongly dependent on the surrounding charge. Sudden jumps in the position of the Coulomb peak are caused by charging and discharging of the neighbouring QDs, which suddenly detune the QD charge sensor and allow the DQD's charge stability diagram (highlighted with white dashed lines) to be mapped down to the very last electrons. 

Figure~\ref{fig:ChargeSensingMeasure}(c) shows a common way to present these data, obtained from Fig.~\ref{fig:ChargeSensingMeasure}(b) after the subtraction of a background plane $\Delta \VLF$. This measurement reveals four distinct regions associated with four different charge configurations where the number of charges within the dots is stable and described by the numbers in parentheses, corresponding to the left and right dot. Another typical way to present these data in the literature (not shown), is to plot the derivative of the raw data to better highlight the charge transitions.
Note how charge sensing allows a clear distinction between the two-electron configurations (0,2) and (1,1). The corresponding interdot transition is primarily of importance for measuring spin states via spin-to-charge conversion~\cite{Gossard2005PSB} and spin qubits in general~\cite{Petta2005}.

\subsection{Readout performance}
\label{subsec:ReadPerf}

The first radiofrequency measurement of a SET in 1998~\cite{Schoelkopf1998} obtained a charge sensitivity of $12~\mu e/\sqrt{\text{Hz}}$ with 1.1 MHz bandwidth. In this case, the charge sensitivity $\Sqq$ refers to probing the charge occupation in the SET island itself (see Section \ref{sec:sensitivity} for more explanation of charge sensitivity and how to measure it.). Since then, devices based on Al/AlO$_x$-based tunnel junctions have demonstrated sensitivities as good as $1~
\mu e/\sqrt{\text{Hz}}$ in the normal state and 0.9~$\mu e/\sqrt{\text{Hz}}$ in the superconducting state~\cite{Delsing2006}. These numbers are more than an order of magnitude better than their low-frequency counterparts but are still not at the theoretical limit, which, for rf-SETs, is only 1.4 times worse than that of Eq.~\eqref{Eq:limit}~\cite{korotkov1999}. The reason is that the noise in these experimental demonstrations contains contributions from the first amplifying stage as well as shot noise, as we shall see in Section~\ref{Sec:amplifiers}. These charge sensors have been used or proposed for readout of charge qubits~\cite{Delsing2001}, lifetime measurements of Cooper-pair box states~\cite{Schoelkopf2003}, and real-time measurements of tunneling events in quantum dots~\cite{Rimberg2003}. 

Later, rf-QPC and rf-SET charge sensors made of semiconductor materials were used to measure the charge occupation of DQDs.
To date, in semiconductors, the most sensitive dissipative electrometers are the rf-QPC and the rf-SET providing charge sensitivities in the $\mu e/\sqrt{\text{Hz}}$ range. The best reported sensitivity for the rf-QPC is $146~\mu e/\sqrt{\text{Hz}}$~\cite{Mason2010} with a bandwidth of at least 1~MHz, demonstrated on a GaAs/AlGaAs heterostructure. 
In silicon, the best reported sensitivity for an rf-SET is 10~$\micro\eHz$~\cite{Clark2007}. As we shall discuss in Section~\ref{sec:sensitivity}, the figures for rf-QPCs and rf-SETs are typically reported under different benchmarking conditions: QPC sensitivities are usually specified in terms of charge on the object being sensed, but SET sensitivities are in terms of the charge on their own island, the two being linked by the capacitive coupling ratio to the system to be sensed (see Eq.~\eqref{Eq:tminSET}). Hence a direct comparison can only be made if the ratio is known.

At present, charge sensing via rf-QPCs and rf-SETs is routinely used in, but not limited to, spin-qubit and quantum information processing experiments to achieve rapid spin-to-charge conversion measurements. These are key to spin-qubit readout, ultimately leading to  single-shot-readout~\cite{Gossard2009,Tarucha2016,Nichol2020}, quantum non-demolition measurements~\cite{Tarucha2019}, and fast qubit-gate operations with demonstrated fidelities above the fault-tolerant threshold~\cite{Nakajima2017,yoneda2018}.
Recently, rf-SETs have also been implemented as a fast characterization tool for nanowires~\cite{Petta2012}, nanowires coupled to superconducting resonators~\cite{Petta2015} and hybrid semiconductor-superconductor nanowire systems (InAs/Al)~\cite{Marcus2019}. The latter example is relevant for the search of Majorana zero modes, (see Section~\ref{subsec:Majorana devices}). In Supplementary Table~SI, we have summarised the charge sensitivity obtained in various experiments in the literature.

\subsection{The Sisyphus resistance}
\label{Sec:Sisyphus}

In the previous subsections, we have learned how rf reflectometry can be used to probe the resistance of nanoelectronic devices on short timescales, for example for fast charge sensing. Beyond this possibility, rf reflectometry offers the opportunity to induce dynamical effects on the sensed nonelectric devices themselves. In this subsection, we deal with a prototypical example of an induced dissipative phenomenon, that of excess dissipation induced by an rf drive: the Sisyphus resistance. 

\begin{figure}
	\includegraphics{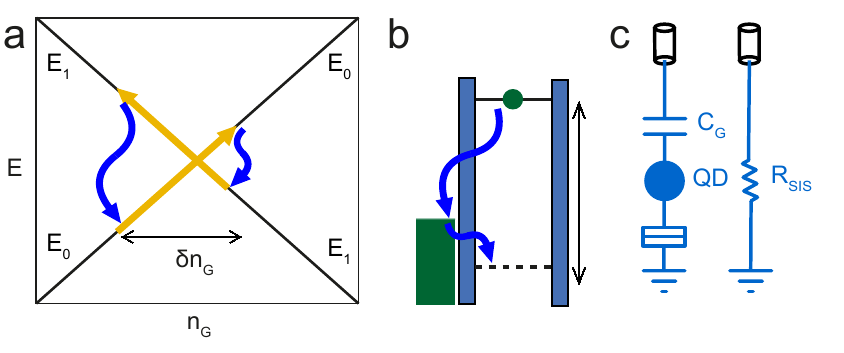}
	\caption{Process associated with Sisyphus resistance.
	(a)  Energy diagram of an uncoupled two-level system representing an (un)occupied QD $E_1$($E_0$) as a function of the reduced gate voltage. The yellow arrows indicate the work done by the rf voltage source (non-adiabatic transitions) and the blue wiggling arrows indicate phonon emission by relaxation.
	(b) Schematic representation of a QD (green circle) tunnel-coupled to a charge reservoir (in green) in the situation described in panel (a). The rf voltage source varies the relative position of the QD electrochemical level with respect to the Fermi energy of the reservoir (amplitude indicated by the double arrow). The blue arrows indicate the inelastic tunneling events. 
	(c) dc (left) and ac (right) small-signal equivalent circuits of the QD as seen from the gate electrode. 
}
	\label{fig:Sisyphus}
\end{figure}

We focus on devices that can be modeled by a two-level system and are driven at a rate comparable to the the their relaxation rate. Under those conditions dynamic power dissipation occurs. Understanding dissipation in these systems is important since two-level systems are the basis of quantum bits and one of the detrimental elements in achieving high quality factors in electrical resonators. We choose the example of a single-charge QD capacitively coupled to a gate electrode and tunnel-coupled to a single charge reservoir to allow particle exchange (Fig. \ref{fig:Sisyphus}). The two energy levels involved correspond to the dot having none ($E_0$) or one ($E_1$) excess electron whose energy separation can be controlled by a parameter $n_\mathrm{G}$, the reduced gate voltage, in the following way, $\Delta E= E_1-E_0=E_\mathrm{C}(1-2n_\mathrm{G})$~\cite{persson2010Thesis}. Here $E_\mathrm{C}$ is the charging energy of the device. 

To explain the physics, let us assume the system is biased so that it has an equilibrium reduced gate voltage, $n_\mathrm{G}^0$, away from the degeneracy point, and is then driven by an rf gate voltage so that the voltage varies as  $n_\mathrm{G}(t)=n_\mathrm{G}^0+\delta n_\mathrm{G}\sin(\omega t)$,  where $\delta n_\mathrm{G}$ is sufficiently large to bring the system past the degeneracy point. In the first half of the cycle, the system is driven past the degeneracy point non-adiabatically. At some point, due to the finite relaxation rates, it relaxes to the ground state, dissipating energy that had been provided by the rf generator. This excitation followed by relaxation occurs indefinitely generating an excess power dissipation in the system that can be modeled by a single resistor, i.e. the Sisyphus resistance $\Rsis$. The Sisyphus resistance can be calculated by solving the dynamics of the system using a master-equation formalism.
The probabilities $P_{0(1)}$ to be in state $E_{0(1)}$ obey

\begin{align}\label{ME}
  \dot{P_0} & = -\Gp P_0 + \Gm P_1 \\
	 \dot{P_1} & = \phantom{-} \Gp P_0 - \Gm P_1 \nonumber
\end{align}

\noindent where $\gamma_{\pm}$ are the relaxation rates. For transitions between a charge reservoir with a 3D density of states and a QD with a discrete density of energy states, the tunnel rates take the following form~\cite{Gonzalez-Zalba2015_limits}:

\begin{equation}
	\gamma_{\pm}=\frac{\Go}{1+\exp\left(\pm\Delta E/k_\mathrm{B}T\right)}
\end{equation}

\noindent where $\Go$ is the maximum relaxation rate that applies away from the degeneracy point. We solve Eq.~\eqref{ME} to first order in $\delta n_\mathrm{G}$ to obtain the quasi-static probabilities

\begin{equation}\label{prob}
	P_{0(1)}=P_{0(1)}^0-(+)\frac{\eta\delta n_\mathrm{G}}{\omega^2+\Gtot^2}[\Gtot\sin(\omega t)-\omega\cos(\omega t)]
\end{equation}

\noindent where $P_{0(1)}^0$ are the thermal probabilities at $n_\mathrm{G}^0$,  
\begin{equation}
  \Gtot=\Gp+\Gm
  \label{equ:gammatot}
\end{equation}
is the effective tunnel rate of the system and $\eta$ is the induced probability oscillation~\cite{persson2010Thesis}
\begin{equation}
  \eta=
  P_1^0\left.\frac{\partial\Gm}{\partial n_\mathrm{G}}\right|_{n_\mathrm{G}^0}-
  P_0^0\left.\frac{\partial\Gp}{\partial n_\mathrm{G}}\right|_{n_\mathrm{G}^0}. 
  \label{equ:eta}
\end{equation} 
The average power dissipation over one period $T$ can then be calculated as~\cite{Delsing2010,Ferguson2011,Gonzalez-Zalba2015_limits}
\begin{equation}
	P_\mathrm{Sis}=\frac{1}{T}\int_0^T\left[P_1\Gm\Delta E-P_0\Gp\Delta E\right]dt.
\end{equation}

By comparing the averaged power dissipation with that of a resistor driven by an oscillatory voltage, we obtain the Sisyphus resistance:

\begin{equation}
	\Rsis^{0D}=\frac{2k_\mathrm{B}T}{(e\alpha)^2}\left(\frac{1+\Go^2/\omega^2}{\Go}\right)\cosh^2\left(\frac{\Delta E}{2k_\mathrm{B}T}\right).
\end{equation}
Here $\alpha$ is the ratio between the gate capacitance and total capacitance of the device. For the case of a 3D density of states in the island, the relaxation rates present a different expression,

\begin{equation}
	\gamma_{\pm}=\frac{R_\mathrm{K}}{R_\mathrm{T}}\frac{\mp \Delta E/h}{1-\exp\left(\pm\Delta E/k_\mathrm{B}T\right)}
\end{equation}

\noindent where $R_\mathrm{T}$ is the resistance of the tunnel barrier. Hence the Sisyphus resistance reads
\begin{equation}
	\Rsis^{3D}=\frac{R_\mathrm{T}}{\alpha^2}\frac{2k_\mathrm{B}T}{\Delta E}\left(1+\frac{\Gtot^2}{\omega^2}\right)\sinh\left(\frac{\Delta E}{k_\mathrm{B}T}\right).
\end{equation}

Radio-frequency reflectometry techniques have been instrumental in detecting excess dissipation in single-electron devices such as superconducting single-electron boxes~\cite{Petersson2010}, and silicon QDs~\cite{Gonzalez-Zalba2015_limits}. 

\subsection{Scaling up}
\label{subsec:ScalingUp}

Charge sensing has been instrumental in some of the most comprehensive studies of QD static and dynamical properties, however it suffers form a potential downside: It is an indirect measurement, i.e. it requires reading the state of a charge sensor that is placed in close proximity to the quantum device of interest. This becomes especially demanding in spin-qubits devices where the growing complexity of the modern geometries~\cite{Volk2019,Petta2019,Meunier2021,Fedele2021} poses substantial spatial constraints.
A possible solution, that leverages the benefits of rf-reflectometry, is to measure the dispersive signal generated by the shifts of the QD's quantum capacitance~\cite{Petersson2010,Reilly2013}. This allows electron tunneling to be measured with rf reflectometry by embedding the QD in the matching network either via the dot's leads or via one of the existing gate electrodes. The latter is especially convenient to realize a compact sensing technique since it alleviates the need for external electrometers~\cite{Reilly2013}. Due to its relevance and substantial technological development, in Section~\ref{sec:reactiveReadout} we present the theory and analyse the details about using rf measurements to probe directly the quantum capacitance of quantum devices.

\section{Measuring a reactive device}
\label{sec:reactiveReadout}

Radio-frequency measurements can be used to detect changes of capacitance or inductance in quantum devices. In this section, we detail the case of variable capacitors (Section \ref{subsec:measuringcap}), which covers in particular the effective capacitance of gated semiconducting devices. However, this discussion is also applicable to variable effective inductance devices. We explain also the concept of \textit{quantum capacitance} (Section \ref{subsec:QuantumCapacitance}), which offers a means to measure various physical phenomena in quantum devices. Finally we explain the techniques of dispersive readout (Section \ref{sec:dispersiveQD}) to measure the charge occupation of quantum dots.

\subsection{Measuring a capacitance}
\label{subsec:measuringcap}

Consider a sample represented by a capacitance $\Cs$. As with a resistive sample (Section \ref{sec:resistiveReadout}), $LC$ resonators are used to match the characteristic impedance of a transmission line and translate the change of sample capacitance to a change of the reflection coefficient $\Gamma$.

\begin{figure}[h!]
  \includegraphics[width=\columnwidth]{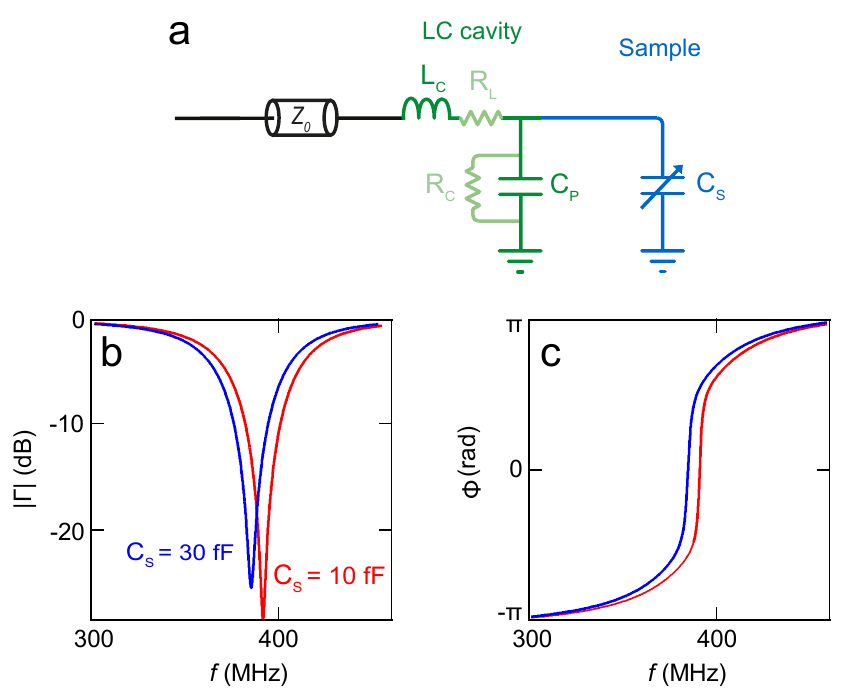}
  \caption{
  (a) Schematic of a reflectometry circuit to measure a capacitive device $\Cs$ using an $LC$ cavity. 
  (b) Reflection as function of the frequency $f$ for the circuit in (a) with two different loads: $\Cs = 10~\mathrm{fF}$ (red) and $\Cs = 30~\mathrm{fF}$ (blue). 
  (c) Corresponding phase spectrum.
  Simulation parameters: $\Lc = 270~\mathrm{nH}$, $\RL = 2~\Ohm$, $\Cp = 0.6~\mathrm{pF}$, $\Rc =10~\mathrm{k}\Ohm$. The circuit is in the overcoupled regime $\Qint > \Qext$.}
  \label{fig:LCcavityC}
\end{figure}

The $LC$ circuit used to measure a device with capacitance $\Cs$ is shown in Fig.~\ref{fig:LCcavityC}(a). We represent the device as dissipationless (which as we shall see may be the case in some limits) and include in the circuit model sources of external dissipation. 
Here we assume a resistor $\RL$ in series with the matching inductor $\Lc$ and a resistor $\Rc$ in parallel with the capacitor $\Cp$.
In this approximation, the impedance presented by the resonator to the transmission line is 

\begin{equation}
\Zload = j\omega\Lc+\RL+\frac{\Rc}{1+j\omega\Rc(\Cp+\Cs)},
\end{equation} 
giving the resonance frequency
\begin{equation}
  \fr=\frac{1}{2\pi\sqrt{\Lc(\Cp+\Cs)}}.
\end{equation}

The main effect of a change of $\Cs$ is to shift $\fr$, which changes 
the reflected signal as seen in Fig. \ref{fig:LCcavityC}(b-c).
The phase change is particularly large near $\fr$ and so it is sometimes used as output of the measurement. In the overcoupled regime $\Qint > \Qext$ and when $\Qext \Delta\CS/(\Cp+\Cs)\ll 1$ , the phase change is
\cite{Buitelaar2012,Gonzalez-Zalba2015_limits}:
\begin{equation}
  \Delta \Phi \equiv \Delta \, \mathrm{arg}(\Gamma(f)) \approx -2 \Qext \frac{\Delta \CS}{(\Cp+\Cs)}.
\end{equation}

Although the circuit equivalent appears similar to that of the resistive case, the sensitivity optimisation strategies are different and will be explained in Section~\ref{section:Optimisation}.

\subsection{Quantum capacitance}
\label{subsec:QuantumCapacitance}

Quantum capacitance is a quantum correction to the capacitance of a system, $\Cs$, that arises due to the additional kinetic energy, in excess of the electrostatic energy, required to add charges to a material containing $N$ charged fermions. This additional energy reflects the fact that since the particles are fermions they must enter unique quantum states with corresponding eigenergies as they fill the system.

\subsubsection{Quantum  capacitance  in  low-dimensional systems}

 The concept of quantum capacitance can be easily understood in the context of a capacitor with a geometrical capacitance $\Cgeo$ formed by a metal gate electrode and a mesoscopic conductor separated by a thin dielectric layer, for example a metal-oxide-semiconductor capacitor~\cite{Serge1988}. Due to its metallic nature, the density of states in the gate is comparatively large compared to the relatively small capacitance in the mesoscopic conductor.

In such systems, a voltage $\Delta \Vg$ applied to the metallic electrode produces an electrostatic ($\Delta V_\mathrm{ELECT}$) as well a chemical ($\Delta V_\mathrm{CHEM}$) potential change~\cite{buttiker1993}

\begin{equation}
\Delta \Vg= \Delta V_\mathrm{ELECT} +\Delta V_\mathrm{CHEM} \text{   and  } \Delta V_\mathrm{ELECT}=\frac{e\Delta N}{\Cgeo}. 
\end{equation}

The contribution of the chemical potential $\mu$ can be expressed in terms of the induced change in charged particles in the mesoscopic conductor:
\begin{equation}
	\Delta V_\mathrm{CHEM}=\frac{\Delta\mu}{e}=\frac{1}{e}\frac{d\mu}{dN}\Delta N.
\end{equation}
Combining the equations above we arrive at the following expression:
\begin{equation}
	\Delta \Vg=e\Delta N\left(\frac{1}{\Cgeo}+\frac{1}{e^2dN/d\mu}\right).
\end{equation}
We see that the total capacitance of the system is composed by the geometrical capacitance in series with a correction that is exclusively dependent on the band structure, the quantum capacitance $\CQ$:
\begin{equation}
  \Cs^{-1}=\Cgeo^{-1}+\CQ^{-1},
  \label{eq:mesocap}
\end{equation}
with 
\begin{equation}
    \CQ=e^2\frac{dN}{d\mu}.
    \label{EqCs}
\end{equation}

To gain more insight into the origin of the quantum capacitance, it is useful to express the definition of the total number of particles in terms of the density of states $\rho(E)$ and the Fermi function $f_e(E)$ at the energy $E$
\begin{equation}
		N=\int \rho(E)f_e(E)dE
\end{equation}
where
\begin{equation}
f_e(E)=\frac{1}{e^{(E-\mu)/k_\mathrm{B}T}+1}.
\end{equation}

Because $\mu$ appears only in $f_e$, the quantum capacitance is proportional to the thermal average of the density of states around the chemical potential. In the limit of zero temperature,
the quantum capacitance can be expressed as being proportional to the density of states at the Fermi energy, $E_\text{F}$:
\begin{equation}
	C_\mathrm{Q}=e^2\frac{dN}{d\mu}\overset{T\rightarrow 0}{=}e^2\rho(E_\text{F}).
	\label{QuantumCap}
\end{equation}

For metallic devices and structures with negligible level spacing, the large density of states means that the quantum capacitance in Eq.~\eqref{EqCs} can be considered infinite, i.e. the total capacitance is simply equal to the geometric capacitance.

Since quantum capacitance is related to the thermodynamic electron compressibility $K=\frac{1}{N^2}\frac{dN}{d\mu}$, it is sometimes called \textit{electron compressibility} when measured at finite temperatures and \textit{quantum capacitance} strictly at $T=0$~K~\cite{sulpizio2011book}. Radio-frequency reflectometry can be an efficient way to measure the quantum capacitance of mesoscopic devices, especially low-dimensional systems where the density of states is low. The quantum capacitance can be computed explicitly according to the dimensionality of the system (2D, 1D or 0D) taking into account the specific density of states $\rho(E)$:
\begin{align}
  & \rho_{2D}(E) = \sum_n \frac{g m^*}{ 2 \pi \hbar^2} H(E-E_\mathrm{n}) \\
  & \rho_{1D}(E) = \sum_n \frac{g}{2\pi\hbar}\sqrt{\frac{m^*}{2(E-E_\mathrm{n})}} \\
  & \rho_{0D}(E) = \sum_n g \delta (E-E_\mathrm{n})
\end{align}
where $g$ is the degeneracy (valley, spin, orbital degree of freedom...), $m^*$ is the effective mass, $H$ is the Heaviside step function and $E_n$ are the subband energy offsets. Examples of quantum capacitance measurements performed in 2D systems involve measurements of 2DEGs~\cite{Eisenstein1994} as well as graphene~\cite{Nongjian2009,Henriksen2010,Ihn2010} including magic-angle twisted bilayer graphene~\cite{Ashoori2019}. In 1D system there are carbon nanotubes~\cite{ilani2006} and quantum point contacts~\cite{Reilly2020QPC}. Some examples of 0D systems include GaAs QDs~\cite{Petersson2010}, InAs QDs ~\cite{Petta2012} and Si QDs~\cite{Gonzalez2020Quintet, Duanambipolar2021}.

\subsubsection{Quantum and tunneling capacitance in quantum dots}
\label{sec:CQandCtuinQD}

We have seen in Section~\ref{Sec:Sisyphus} that the ac response of low-dimensional systems may differ from the classical expectation. In this section, we show how the aforementioned ac component, the Sisyphus resistance (see Section \ref{Sec:Sisyphus}) and now the quantum capacitance can manifest simultaneously in coupled two-level systems. More particularly, we will be able to make a further distinction about the origin of the capacitance term, and separate it into two components associated with reversible and irreversible charge tunneling~\cite{Gonzalez-Zalba2019}, i.e. the pure quantum capacitance $\CQ$~\cite{Gonzalez-Zalba2017, Ahmed2020} as strictly defined in Eq.~\eqref{QuantumCap} and the tunneling capacitance $\Ctu$~\cite{West1992,ashoori1993}, respectively.

\begin{figure}
	\includegraphics{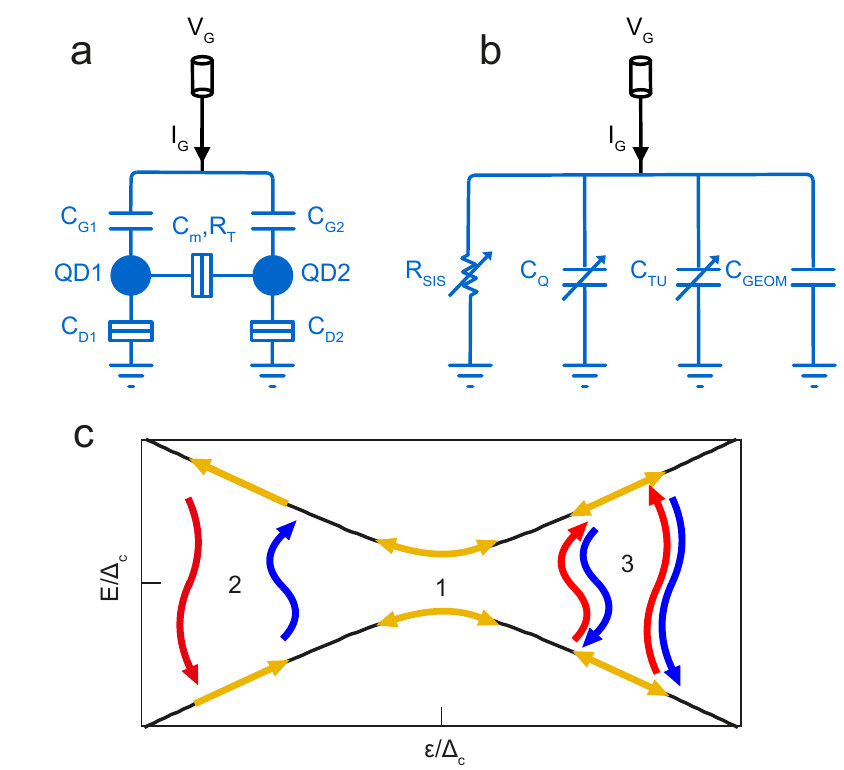}
	\caption{Double quantum dot equivalent circuit and physical processes. 
	(a) dc equivalent circuit of a DQD. The tunnel barriers, indicated by rectangles, consist of a capacitor in parallel with a resistor.
	(b) ac small-signal equivalent circuit of the DQD as seen from the gate electrode (G). The arrows indicate variable impedances. 
	(c) Ground state and excited state energy of the DQD as a function of reduced detuning (black lines). The yellow arrows indicate the work done by the ac voltage source and the red and blue wiggling lines indicate phonon emission and absorption. The process associated with quantum capacitance is marked (1); that with Sisyphus resistance is marked (2) and that with tunneling capacitance  is marked (3).
}
	\label{fig:QDCapacitance}
\end{figure}

In particular, we consider a tunnel-coupled DQD where the two dots QD$i$ $i=1,2$ are connected to an rf gate electrode via gate capacitances $C_\mathrm{Gi}$ and to grounded charge reservoirs at temperature $T$ via $C_{\mathrm{D}i}$ (Fig.~\ref{fig:QDCapacitance}(a)). 
The interdot tunnel barrier has a mutual capacitance $C_\mathrm{m}$ and tunnel resistance $R_\mathrm{T}$. 
The system can be described by an equivalent impedance $Z_\mathrm{eq}$ such that $\Vg=Z_\mathrm{eq}I_\mathrm{G}$, where $\Vg$ and $I_\mathrm{G}$ are the gate voltage and the gate current, respectively. 
Here, we consider the system driven by a small-amplitude excitation, $\Vg= \delta \Vg\sin(\omega t)$, where the excitation frequency is much smaller than the DQD frequency (i.e. $\omega\ll\Delta_\mathrm{C}/\hbar$), the rate of transit through the anticrossing is small and the DQD is weakly coupled to the reservoirs. 
In this limit, as we shall see later, the DQD impedance is $Z_\mathrm{eq}=\left(j\omega \Cs+1/\Rsis\right)^{-1}$ where $\Cs$ is the total equivalent capacitance of the system and $\Rsis$ is the Sisyphus resistance of the DQD. 

To obtain an analytical expression for $Z_\mathrm{eq}$, we take the definition of the gate current 
\begin{equation}\label{current}
	I_\mathrm{G}=\frac{d(Q_1+Q_2)}{dt},
\end{equation}

\noindent where $Q_i$ is the gate charge on the respective QD$i$. We expand the total gate charge in the DQD as a function of the gate coupling factors, $\alpha_i=C_{\text{G}i}/(C_{\mathrm{D}i}+C_{\text{G}i}+C_\text{m})$ and the average electron occupation probability in QD$i$, $P_i$. We further assume the weak DQD coupling limit $C_\mathrm{m}\ll C_{\mathrm{D}i}+C_{\text{G}i}$ and obtain
\begin{equation}\label{charge}
	Q_1+Q_2=\sum_i \alpha_i(C_{\mathrm{D}i}\Vg+eP_i).
\end{equation}

Using Eqs.~(\ref{current} - \ref{charge}) for inter-dot charge transitions, and noting that in that case $dP_2/dt=-dP_1/dt$, we obtain
\begin{equation}\label{dqdt}
	I_\mathrm{G}=\Cgeo\frac{d\Vg}{dt}+e\alpha'\frac{dP_2}{dt}.
\end{equation}
Here, $\Cgeo=\sum_i\alpha_i C_{\mathrm{D}i}$ and $\alpha'=\alpha_2-\alpha_1$. We further note that the gate voltage induces an electrochemical potential energy difference between the QDs, i.e. the detuning $\varepsilon=\mu_2-\mu_1=e\alpha'(\Vg-\Vg^0)$, where $\Vg^0$ is the gate voltage offset at which the difference is zero. Hence, Eq.~\eqref{dqdt} can be further expanded into:
\begin{equation}\label{dqdt2}
	I_\mathrm{G}=\Cgeo\frac{d\Vg}{dt}+e\alpha'\frac{dP_2}{dt}=\left[\Cgeo+(e\alpha')^2\frac{dP_2}{d\varepsilon}\right]\frac{d\Vg}{dt}.
\end{equation}

In Eq.~\eqref{dqdt2}, the semi-classical nature of our system becomes apparent.
The first term is the geometrical capacitance of the DQD, whereas the second term, which appears as if it were a second capacitance in parallel, is related to the electron compressibility as defined below Eq.~\eqref{QuantumCap}. It is linked to changes in charge occupation caused by time-dependent changes in detuning. However, as we shall see, in-depth investigation of this second term reveals two distinct physical mechanisms leading to charge redistribution. For now, the problem boils down to calculating the time-dependent occupation of QD2.

In order to understand the nature of this second term, we revert to the quantum description of the DQD. In the single-charge regime the DQD is described by the Hamiltonian 
\begin{equation}\label{eq:DQDHamilt}
	H=-\frac{\Delta_\mathrm{C}}{2}\sigma_x-\frac{\varepsilon}{2}\sigma_z,
\end{equation}

\noindent where $\Delta_\mathrm{C}$ is the tunnel coupling energy and $\sigma_{x(z)}$ are the Pauli matrices. The eigenenergies are
\begin{equation}\label{eigenenergies}
	E_\pm=\pm\frac{1}{2}\sqrt{\varepsilon^2+\Delta_\mathrm{C}^2},
\end{equation}

\noindent and the energy difference between the excited and the ground state is $\Delta E=E_+-E_-$ (Fig.~\ref{fig:QDCapacitance}(c)). At large detunings, the eigenstates coincide with the charge states of the DQD. In general, the probability in the charge basis ($P_2$) can be expressed in terms of the probabilities in the ground (GS) and excited state (ES) energy basis, $P_\pm$,
\begin{equation}\label{p2}
	P_2=P_2^-P_-+P_2^+P_+=\frac{1}{2}\left[1+\frac{\varepsilon}{\Delta E}\chi\right]
\end{equation}

\noindent where $P_2^\pm=(1\mp\varepsilon/\Delta E)/2$~\cite{Gonzalez-Zalba2016} and $\chi=P_--P_+$ is the polarization of the system in the energy basis. If the system is driven at a finite rate $\varepsilon(t)=\varepsilon_0+\delta\varepsilon\sin(\omega t)$ (where $\varepsilon_0$ is the bias point) and the excitation rate is low $\omega\ll\Delta_\mathrm{C}^2/(\hbar\delta\varepsilon)$ to avoid Landau-Zener transitions~\cite{Shevchenko2010}, an electron can change its probability distribution in the DQD in two different ways~\cite{Otxoa2019}: (i) via adiabatic charge tunneling (process 1 in Fig.~\ref{fig:QDCapacitance}(c) associated with the derivative of $\varepsilon/\Delta E$), or (ii) irreversibly via phonon absorption and emission (processes 2 and 3, associated with the derivative of $\chi$). By expanding the second term in Eq.~\eqref{dqdt2}, we can extract our first conclusion:
\begin{equation}\label{dP}
	(e\alpha')^2\frac{dP_2}{d\varepsilon}=\frac{(e\alpha')^2}{2}\left[\frac{\partial^2E_+}{\partial\varepsilon^2}\chi+\frac{\varepsilon}{\Delta E}\frac{\partial\chi}{\partial\varepsilon}\right].
\end{equation}

The first term on the right can be associated with the description of quantum capacitance in QDs in the literature as originating from the second derivative of the eigenenergies with respect to detuning~\cite{Petersson2010,urdampilleta2015,Simmons2015,West2019}. It coincides with the strict definition of quantum capacitance being the electron compressibility at $T=0$~K. The second term is linked to irreversible redistribution processes that, as we shall see, lead to Sisyphus dissipation and also to an additional source of capacitance, the tunneling capacitance. To gain further insight into the second term, we calculate changes in $\chi$ using the master equation formalism introduced in Section~\ref{Sec:Sisyphus} to first order approximation in $\delta\varepsilon/\Delta_\mathrm{C}$ and find
\begin{equation}
	\delta\chi=\frac{-2\eta\delta\varepsilon}{\omega^2+\Gtot^2}[\Gtot\sin(\omega t)-\omega\cos(\omega t)].
\end{equation}

Here, $\Gtot$ is the characteristic relaxation rate of the system and $\eta$ relates to the amplitude of the induced probability oscillations. More concretely, $\Gtot=\Gp+\Gm$ where $\Gp=\Gc n_\mathrm{p}$ is the phonon absorption rate, $\Gm=\Gc(1+n_\mathrm{p})$ is the phonon emission rate, $\Gc$ is a material-dependent charge relaxation rate and $n_\mathrm{p}=(\exp(\Delta E/k_\mathrm{B}T)-1)^{-1}$ is the phonon occupation number. Hence, $\Gtot$ can be expressed as  
\begin{equation}
  \Gtot=\Gc\coth(\Delta E_0/2k_\mathrm{B}T).
\end{equation}

Further, $\eta$,  according to Eq.~\eqref{equ:eta} can be written as
\begin{equation}
\eta=\frac{\Gtot}{4k_\mathrm{B}T}
\frac{\varepsilon_0}{\Delta E_0}
\,\cosh^{-2}\left(
\frac{\Delta E_0}{2k_\mathrm{B}T}\right).
\end{equation}

By inserting the detuning derivative of the change in energy polarization into Eq.~\eqref{dP} and averaging over a cycle of the rf signal we get
\begin{equation}\label{all}
	\begin{aligned}
	I_\mathrm{G}&=\Cgeo\frac{d\Vg}{dt}+\frac{(e\alpha')^2}{2}\frac{\Delta_\mathrm{C}^2}{(\Delta E_0)^3}\chi^0\frac{d\Vg}{dt} \\
	&+\frac{(e\alpha')^2}{2}\frac{\varepsilon_0}{\Delta E_0}\frac{2\eta\Gtot}{\omega^2+\Gtot^2}\frac{d\Vg}{dt}+\frac{(e\alpha')^2}{2}\frac{\varepsilon_0}{\Delta E_0}\frac{2\eta\omega^2}{\omega^2+\Gtot^2}\Vg,
	\end{aligned}
\end{equation}
where $\chi^0=\tanh(\Delta E_0/2k_\mathrm{B}T)$ is the equilibrium polarization and $\Delta E_0=\Delta E(\varepsilon=\varepsilon_0)$. From Eq.~\eqref{all} we find the form of the equivalent impedance of the system, $Z_\mathrm{eq}$. The prefactors in the terms linear in $d\Vg/dt$ correspond to capacitances, whereas the prefactor in the linear term in $\Vg$ is a conductance. The reactive terms add up to a total sample capacitance $\Cs$, see Fig. \ref{fig:QDCapacitance}(b), 
\begin{equation}
  \Cs = \Cgeo + C_\mathrm{Q} + \Ctu.
  \label{eq:CgeoCQCtu}
\end{equation}
\noindent which corresponds to the sum
\footnote{We note that in Eq.~\eqref{eq:CgeoCQCtu}, the capacitances appear in parallel as opposed to the example of the mesoscopic capacitor where the capacitances appear in series (see Eq.~\eqref{eq:mesocap}). This is a result of the different geometry of the two examples, where the DQD system contains additional capacitances to the charge reservoirs.}
of the geometrical capacitance
\begin{equation}
	\Cgeo= \sum_i \alpha_i C_{\text{S}i},
\end{equation}

\noindent the quantum capacitance 
\begin{equation}
	C_\mathrm{Q}=\frac{(e\alpha')^2}{2}\frac{\Delta_\mathrm{C}^2}{(\Delta E_0)^3}\chi^0,
	\label{QuantumCapDQD}
\end{equation}
and the tunneling capacitance
\begin{equation}
    \label{Eq:tunnel}
	\Ctu=\frac{(e\alpha')^2}{2}\frac{1}{2k_\mathrm{B}T}\left(\frac{\varepsilon_0}{\Delta E_0}\right)^2\frac{\Gtot^2}{\omega^2+\Gtot^2}
	\,\cosh^{-2}\left(
	\frac{\Delta E_0}{2k_\mathrm{B}T}\right).
\end{equation}

The dissipative term, which appears in parallel with $\Cs$ , is the Sisyphus resistance
\begin{equation}
	\Rsis=\frac{4R_\mathrm{K}}{\alpha'^{2}}\frac{k_\mathrm{B}T}{h\Gtot}\left(\frac{\Delta E_0}{\varepsilon_0}\right)^2\frac{\omega^2+\Gtot^2}{\omega^2}
		\,\cosh^{2}\left(
	\frac{\Delta E_0}{2k_\mathrm{B}T}\right).
\end{equation}

\noindent For comparison, the resistance of the inter-dot tunnel barrier is $R_\mathrm{T}=2R_\mathrm{K}k_\mathrm{B}T/(h\Gc)$~\cite{Gonzalez-Zalba2015_limits}.

\begin{figure}
	\includegraphics{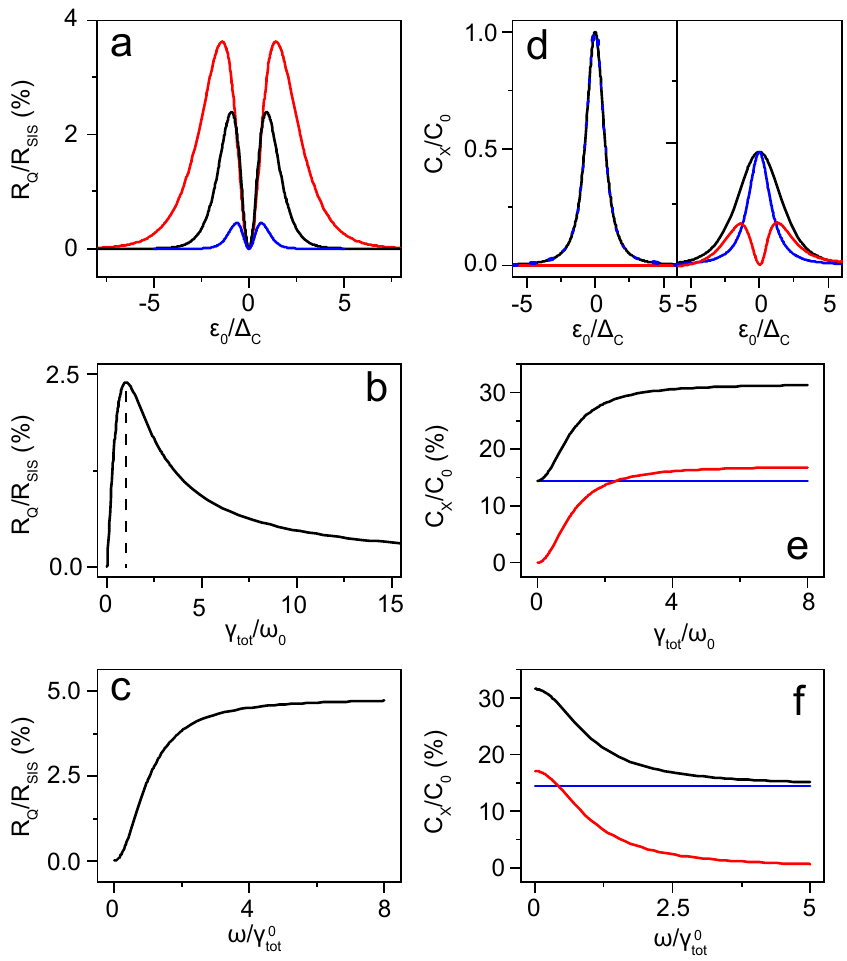}
	\caption{Parametric impedances. 
	(a) Relative change of the normalized inverse of the Sisyphus resistance versus reduced detuning for $k_\text{B}T/\Delta_\mathrm{C} = 0.25,0.5$ and 1 (blue, black and red traces respectively) and $\Gc = \omega$. 
	(b) $R_\text{Q}/\Rsis$ as a function of reduced relaxation rate for a given operation angular frequency $\omega_0$ and 
	(c) as a function of reduced operation angular frequency for $k_\text{B}T/\Delta_\mathrm{C} = 0.5$ for a given relaxation rate $\Gtot^0$ and $\varepsilon_0/\Delta_\mathrm{C} = 1$. 
	(d) Normalized parametric (black), quantum (blue) and tunneling capacitance (red) as a function of reduced detuning for $k_\text{B}T/\Delta_\mathrm{C} = 0.01, 1$ (left and right panels, respectively) and $\Gc/\omega = 10$. Here, $C_0 = (e\alpha')^2/2\Delta_\mathrm{C}$ and we set $\alpha' = 1$. $C_\text{x}/C_0$ as a function of reduced relaxation rate 
	(e) and reduced operation angular frequency 
	(f) for $k_\text{B}T/\Delta_\mathrm{C} = 1$ and $\varepsilon_0/\Delta_\mathrm{C} = 1$. 
	}
	\label{fig:parametricimpedances}
\end{figure}

In Fig.~\ref{fig:parametricimpedances}, we show the functional dependence of these different components. We start with the Sisyphus dissipation which is proportional to $\Rsis^{-1}$ and see that it presents two symmetric maxima at finite detuning that increase with temperature (panel (a)). Furthermore, when the system is driven at constant frequency, $\omega=\omega_0$, the dissipation presents a maximum when the effective relaxation rate coincides with the driving frequency (panel (b)). Finally, at a fixed relaxation rate, $\Gtot=\Gtot^0$, the dissipation increases asymptotically as the driving frequency increases. The asymmetry between $\omega$ and $\Gtot$ can be understood by noting that although dissipation in each cycle decreases as $\omega$ is increased, the overall number of cycles increases, exactly matching the reduction of energy dissipation per cycle. The Sisyphus cycle, as explained in Section~\ref{Sec:Sisyphus}, is driven by phonon pumping (Fig.~\ref{fig:QDCapacitance}(c), process 2).

Now, we focus on the quantum and tunneling capacitance and their sum, the parametric capacitance $C_\text{par}$. In panel (d), we show how they depend on detuning in the low and high temperature limits (left and right panels, respectively). In the low-$T$ limit, the parametric capacitance (black) has a single peak centered at $\varepsilon_0 = 0$ and contains exclusively contributions from the quantum capacitance (dashed blue). In the high-$T$ regime, the parametric capacitance (black) still has a single peak, although of reduced height due to the reduced equilibrium polarization in the energy basis. However, the peak now consist of contributions from both $\CQ$ and $\Ctu$ in blue and red, respectively, the latter is responsible for the increased linewidth. The lineshape of $\Ctu$ coincides with that of the Sisyphus dissipation indicating that the same mechanism, phonon pumping, drives the process. However, when we explore the dependence of the capacitance on $\Gtot/\omega_0$ and $\omega/\Gtot^0$, we observe subtle differences. In panels e and f, we see that $\CQ$ (blue) does not depend on the drive frequency. On the
other hand, $\Ctu$ (red), and hence $C_\text{par}$, increases with increasing $\Gtot/\omega$ in a symmetric way, see panels (e-f). With these three plots, we can get a comprehensive picture of the dispersive response. The quantum capacitance is linked to isentropic charge polarization due to the nonlinearity of the discrete energy levels of the DQD, whereas the tunneling capacitance is linked to thermal probability redistribution (maximal entropy production). The latter depends strongly on the system dynamics, i.e., it only manifests when $\Gtot$ is comparable to or larger than $\omega$; this is when tunneling occurs either nonadiabatically (as in the case of the Sisyphus heating) or adiabatically. In the specific case that $\Gtot$ and $\omega$ are comparable, the Sisyphus and tunneling capacitance processes are linked to phonon pumping~\cite{Pritiraj2003,Schon2008} and lead to net power dissipation. 

In short, radio-frequency reflectometry can be used to probe additional components in the high-frequency response of low-dimensional systems. More particularly, we have learned that near the charge degeneracy point, a DQD behaves effectively as a variable capacitor (composed of the parallel sum of the quantum, tunneling capacitance and constant geometrical capacitance) in parallel with a variable resistor (the Sisyphus resistance).

\subsection{Dispersive readout of QDs}
\label{sec:dispersiveQD}

\begin{figure}
    \includegraphics{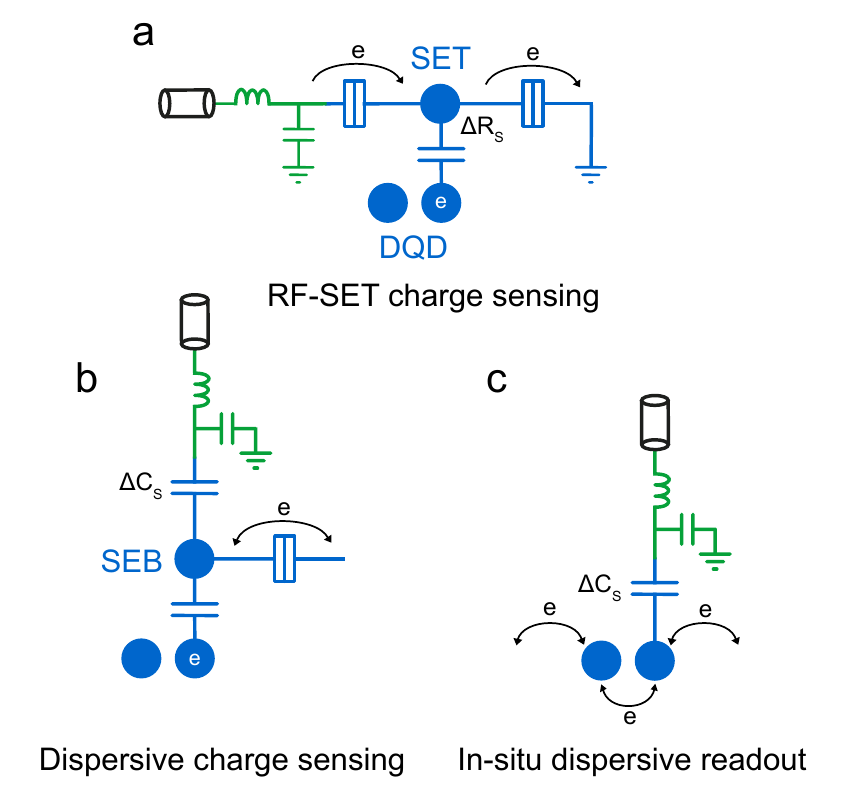}
    \caption{Schematics of the three main reflectometry sensing techniques to probe the charge occupation of DQDs. 
    (a) An rf-SET charge sensor detects the charge occupation of a DQD by measuring changes in the SET channel resistance $\Delta \Rs$.
    (b) Dispersive charge sensing detects the charge occupation of a DQD by measuring changes in the tunneling capacitance $\Ctu$ of a SEB induced by changes in the charge configuration of the DQD.
    (c) \textit{In-situ} dispersive readout measures directly changes in the capacitance of the DQD due to bistable tunneling between QDs or between a QD and charge reservoir.
    }
    \label{fig:SuperSketch}
\end{figure}

Dispersive readout is based on measuring capacitance changes in quantum devices via rf-reflectometry techniques. It is ideal for applications where electrical transport measurements may not be possible. Dispersive techniques can be implemented with fewer electrodes than required for conventional dissipative sensors based on measuring two terminal conductance, such as the rf-QPC or the rf-SET. For that reason, they have gained considerable traction in spin-based quantum computing, where scaling is an important challenge. We note that, although motivated by developments in QD science and technology, dispersive readout techniques can be used to measure varying capacitance in generic quantum devices.   
To put dispersive readout techniques in perspectives, we present Fig.~\ref{fig:SuperSketch}, summarising the three main techniques used to probe the quantum state in QD systems using rf-reflectometry: dissipative and dispersive charge sensing (panel a and b, respectively) and \textit{in-situ} dispersive readout (panel c). Dissipative charge sensing, exemplified by the rf-SET, utilizes the variable resistance of the SET to detect the charge states of a coupled DQD. The sensor is coupled to two charge reservoirs. Dispersive charge sensing uses the variable capacitance, in this case of a single-electron box (SEB), to detect the charge state of a coupled DQD. In this case, the sensor needs to be coupled only to one charge reservoir. The resonator can be connected to gate or reservoir of the SEB. And finally, \textit{in-situ} dispersive readout detects directly the state-dependent capacitance of the DQD and requires no sensor apart from the coupling gate. In Supplementary Table SI, we have benchmarked the three different methods utilizing available data in the literature. Here, we explain dispersive charge sensing and \textit{in-situ} dispersive readout using some examples from the literature. 

\subsubsection{Dispersive charge sensing}

An interesting approach to measure the charge occupation in QD arrays is to combine charge sensing techniques (see Section \ref{subsec:SET}) with dispersive readout. This is the case of the single-electron box (SEB) \cite{Sillanpaeae2006,Persson2010a,Urdampilleta2020,Kuemmeth2020array,Kuemmeth2020quadruple,Morton2020array, CirianoTejel2021}. The SEB (or single-lead quantum dot) is a charged island with only one connection to a lead (rather than two for SETs) and is capacitively coupled to one or more gates~\cite{Tarucha2015, Simmons2016}. Cyclic tunneling between the island and the reservoir results in an effective capacitance that can be calculated analytically for the case in which the island and the reservoir present a zero- and three-dimensional density of states, respectively ~\cite{Gonzalez-Zalba2015_limits,Gonzalez-Zalba2018_thermo}

\begin{equation}
	\Ctu^{0D}=\frac{(e\alpha)^2}{4k_\text{B}T}
	\,\frac{\gamma_0^2}{\gamma_0^2+\omega^2}
	\,\cosh^{-2}
	\left (\frac{\epsilon_0}{2k_\text{B}T} \right ),
\end{equation}

\noindent where $\gamma_0$ is the dot-reservoir tunneling rate at the charge degeneracy point, $\varepsilon_0$ is the bias point and $\omega$ is the probing angular frequency. Recall the relation, $\epsilon=\alpha(\Vg-\Vg^0)$. Equally, the capacitive response of a charged island with a 3D density of states can be calculated analytically,

\begin{equation}
	\Ctu^{3D}=\frac{(e\alpha)^2}{k_\text{B}T}
	\,\frac{R_\text{K}}{R_\text{T}}
	\,\frac{\epsilon_0}{h\gamma}
	\,\frac{\gamma^2}{\gamma^2+\omega^2}
	\, \sinh^{-1}
	\left (\frac{\epsilon_0}{k_\text{B}T} \right ),
\end{equation}

\noindent where

\begin{equation}
    \gamma=\frac{R_\text{K}}{R_\text{T}}\frac{\epsilon_0}{h}\coth\left (\frac{\varepsilon_0}{2k_\text{B}T} \right ).
\end{equation}

\begin{figure}
  \includegraphics[width=\columnwidth]{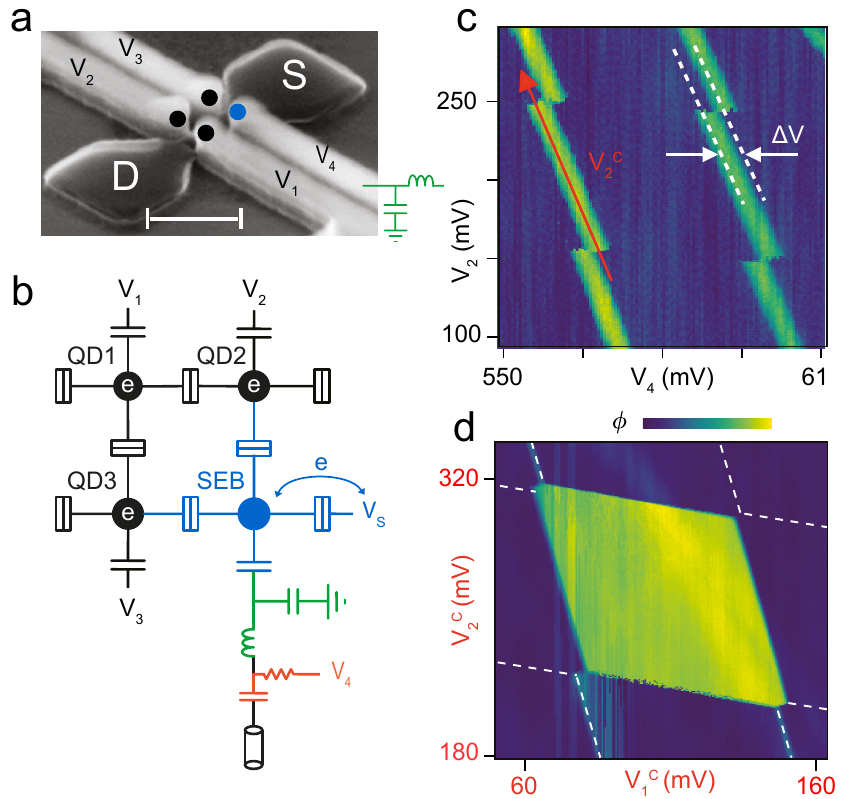}
  \caption{
  (a) SEM of a $2\times 2$ QD array in a double split-gate fully-depleted silicon-on-insulator (FD-SOI) transistor. Gate 4 is wirebonded to an rf resonator. The scale bar is 200 nm. The position of each QD is indicated by circle, with SEB in blue, sensed dots in black.
  (b) Circuit schematic showing the $LC$ resonator attached to the SEB charge sensor and probing the other QDs of the array. The gate voltage $V_\mathrm{4}$ is applied via a bias tee. The arrows represent the tunneling of charges.
  (c) Phase response $\phi$ of the gate reflectometry measurement as gates $V_\mathrm{2}$ and $V_\mathrm{4}$ are swept, revealing peaks at the dot-to-reservoir transitions of the SEB. $\Delta V$ indicates the shift of the SEB peaks due the change of occupation of QD2. We also indicate the definition of a compensated control voltage, $V_2^C$, that changes the potential of QD2 without changing the potential of the SEB.
  (d) Phase response of the SEB as a function of two compensated control voltages $V_2^C$ and $V_1^C$ defined similar to that in panel (c). In this case, a hexagonal charge-stable region of the QD1-QD2 DQD is clearly visible.
  Panels (a),(c) and (d) adapted from Ref.~\onlinecite{Kuemmeth2020array}.
  }
  \label{fig:Array}
\end{figure}

SEBs find applications in QD arrays where for example, the dots at the edge of the array in proximity to contact leads can be used as SEB charge sensors~\cite{Hutin2019,Urdampilleta2020,Kuemmeth2020array,Kuemmeth2020quadruple,Morton2020array} (Fig.~\ref{fig:Array}(a)). SEB charge sensors are sensitive detectors of the electrostatic environment. When the occupation of a nearby QD changes, the SEB island potential shifts, which in turn produces a change in capacitance that can be detected by gate reflectometry measurements. Whereas QPCs or SETs require two leads, SEB charge sensors require only one. This  means they take up less space, which is an advantage in quantum circuits~\cite{Tristan2012,Tristan2017,Vandersypen2018Array,Dzurak2020,Tristan2020,Veldhorst2020Array,Veldhorst2020HoleArray} and provide an interesting direction forward for scalable quantum electronics circuits.

Figure~\ref{fig:Array} shows a $2\times2$ array of QDs in which each dot is primarily controlled by one gate~\cite{Kuemmeth2020array}. The readout technique employs the QD tunnel-coupled to the source electrode at the edge of the array, as a SEB charge sensor. The $LC$ resonator is coupled to this QD (Fig.~\ref{fig:Array}(b)). Gate reflectometry readout detects a phase change at the dot-to-reservoir transition due to the increased tunneling capacitance (Fig.~\ref{fig:Array}(c)). When the occupation of one of the neighbour dots changes, the dot-to-reservoir transition line shifts by a small amount $\Delta V$ in the gate voltage space. By tuning the SEB near a dot-to-reservoir transition, the reflectometry phase signal becomes highly sensitive to the change of occupation of the nearby dot in a similar manner to SET charge sensors (Fig.~\ref{fig:Array}(d)).

Going forward, with the number of QDs in arrays increasing, it may become difficult to bring a sizable number of lead electrodes to create SEB charge sensors without disturbing the connectivity of the QD array. An approach that could overcome this challenge in the short term could be the use of floating gates to capacitively couple one of the QDs in an array to a remote sensor~\cite{Morton2020array}. We note that this approach can be applied to any type of charge sensor described in this review. 

\subsubsection{In-situ dispersive readout}
\label{sec:insitudispersive}

\begin{figure}
	\includegraphics{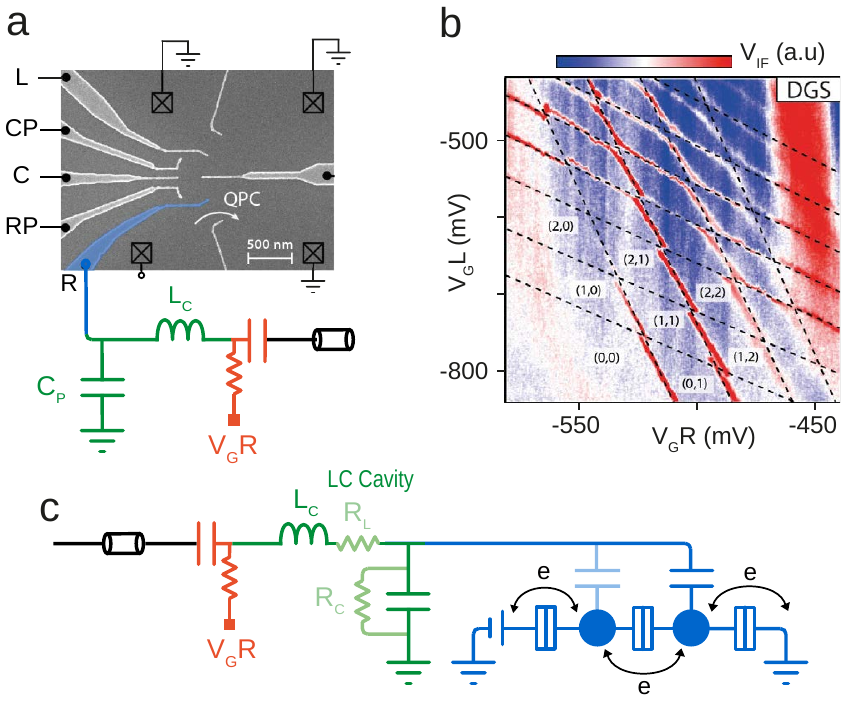}
	\caption{ 
	(a) Micrograph of a GaAs DQD with an $LC$ resonator attached to one of its gates. The $LC$ resonator is composed of a superconducting inductor ($\Lc= 210$~nH) and the parasitic capacitance ($\Cp=0.2$~pF).
	(b) Reflected voltage $V_\text{out}$ as a function of voltages $\Vg\mathrm{L}$ and $\Vg\mathrm{R}$ applied to gates L and R.
	The charge occupation of each dot is indicated in brackets.
	Bright features indicate the regions of charge bistability. .
	(c) Circuit equivalent of $LC$ resonator attached to one gate of a DQD. The arrows represent the types of tunneling to which the measurement is sensitive: tunneling between to and from the leads, and tunneling between dots. 
	Panels (a) and (b) adapted from Ref.~\onlinecite{Reilly2013}.
}
	\label{fig:DispersiveReadout}
\end{figure}

Figure~\ref{fig:DispersiveReadout} shows an example of \textit{in-situ} dispersive readout performed on a GaAs gate-defined DQD~\cite{Reilly2013}. The rf resonator is connected directly to one of the gate electrodes that controls the electrostatic potential of a QD. The measurement circuit (Fig.~\ref{fig:DispersiveReadout}(a)), works on the principle explained in Section~\ref{subsec:measuringcap} and can be sensitive to both the capacitive and resistive contributions from the device.
Measuring the demodulated signal as a function of the gate voltages that control the energy levels in the two dots shows the characteristic honeycomb pattern of double dot Coulomb blockade (Fig.~\ref{fig:DispersiveReadout}(b)).
Each type of charge transition (top dot-to-lead, bottom dot-to-lead, and dot-to-dot; see Fig.~\ref{fig:DispersiveReadout}(c)) appears as a series of high-intensity lines in the plot, with a slope determined by the relative lever arms to the two gates.
These lines mark locations in gate space at which electrons can tunnel cyclically in response to the rf gate voltage, which through the combination of $\CQ$, $\Ctu$, and $\Rsis$ loads the resonator and therefore changes the reflected voltage.
The intensities of different lines arise from the different response of particular transitions to the rf voltage; as expected, the transitions of the bottom dot are stronger because of the larger gate coupling and hence lever arm.

\subsubsection{Sensitivity and state of the art}

The first demonstration of \textit{in-situ} dispersive readout was reported in 2010 with an $LC$ resonator attached to a lead of a GaAs DQD~\cite{Petersson2010}. A minimum integration time of $\taum=4$~ms was needed to discern the charge of a single electron tunneling between QDs with a signal to noise ratio of 1. Since then, several works have demonstrated a similar methodology~\cite{Buitelaar2012,GonzalezZalba2014,Simmons2015} showing detection of regions of charge bistability when a tunnel rate, either of a dot-to-lead or dot-to-dot transition, is larger than the overall tunnel rate through the structure.

In 2013, dispersive readout with a $LC$ resonator attached to a gate was first demonstrated using a GaAs DQD~\cite{Reilly2013}. To discern dot-to-lead charge transitions, the authors required a minimum integration time of $\taum=5$~$\micro$s. From these results, it becomes apparent that the lever arm of the sensing gate to the specific transition to be sensed is a primary factor in $\taum$, as shall be discussed in Section~\ref{section:Optimisation}. The latter work opened the path to more advanced \textit{in-situ} dispersive readout demonstrations based on an enhanced gate lever arm~\cite{Gonzalez-Zalba2015_limits} and optimized resonator topologies~\cite{Gonzalez-Zalba2018_RFgate} meaning that a dot-to-lead transition could be discerned within a time $\taum=12.5$~ns.

After that, single-shot readout of spin qubits in silicon QDs via \textit{in-situ} dispersive readout~\cite{Simmons2018, West2019} of dot-to-dot charge transitions was performed. More recently, a DQD was coupled to a superconducting microwave cavity to obtain $\taum=170$~ns~\cite{Vandersypen2019}, see Section~\ref{sec:Singlet-triplet qubits}. Finally, readout of interdot charge transitions has been further improved to allow $\taum=80$~ns~\cite{Morton2020_JPA} and then $\taum=10$~ns~\cite{Gonzalez-Zalba2021Interaction} by using a Josephson parametric amplifier (see Section~\ref{Sec:amplifiers}) and by adapting the resonator topology to increase the quality factor (see Section~\ref{sec:Optimisation:dispersive}), respectively. 

With regard to dispersive charge sensing, the SEB's capacitance is read using the same methodology as \textit{in-situ} dispersive readout. It is hence expected that the aforementioned approaches could be used to achieve similar $\taum$ when charge sensing events shift the sensor from a capacitance peak to the background. So far, SEBs charge sensors have allowed measurements of charge occupation with SNR = 1 in 550~ns integration time~\cite{Simmons2016}, charge detection in silicon nanowire QDs~\cite{Urdampilleta2019,Kuemmeth2020array} and single-shot spin readout in less than  100~$\micro$s with a SNR > 1~\cite{Urdampilleta2019}. 

The recent progress on dispersive readout shows that, with an adequate resonator design, dispersive signals (either for \textit{in-situ} dispersive readout or dispersive charge sensors) can approach the signal levels of dissipative sensors (see Sec~\ref{section:Optimisation}). Under certain conditions, \textit{in-situ} dispersive sensing and SEBs could achieve comparable or even faster readout. One reason is that SETs are intrinsically shot noise-limited with values for typical bias conditions approaching or even exceeding the noise of cryogenic amplifiers (see Section~\ref{Sec:amplifiers}). Dispersive readout is Sisyphus noise-limited~\cite{Gonzalez-Zalba2015_limits}, noise that can be made comparatively smaller than the noise temperature of cryoamps and hence allows quantum-limited amplifiers to achieve lower noise temperatures~\cite{Morton2020_JPA} (see Section~\ref{Sec:amplifiers}).

\section{Optimization of radio-frequency resonators}
\label{section:Optimisation}

The optimisation of a radio-frequency resonator has the purpose of reducing the time required to perform a measurement or in other words, of maximizing the SNR. In rf reflectometry, the signal corresponds to the difference in reflected voltage between the two states to be measured, so that
\begin{align}
    \textrm{SNR}&=\left|(\Gamma_\mathrm{b}-\Gamma_\mathrm{a})\frac{V_{0}}{\VN}\right|^2 \nonumber
    \\
                &=|\Delta \Gamma|^2 \frac{P_{0}}{P_\mathrm{N}} ,
    \label{eq:SNR}
\end{align}
where $V_{0}$~($P_0$) and $\VN$~($P_\mathrm{N}$) are the input and noise voltage (power) respectively, and $\Gamma_\mathrm{a(b)}$ are the reflection coefficients corresponding to the two states to be measured.
Maximizing the SNR entails two objectives: (i) maximizing the change in reflection coefficient between states for a given input power, (ii) minimising the noise power.

The task of minimising the noise  is discussed in Section~\ref{Sec:amplifiers}. In this section, we shall describe strategies to maximise the signal by optimising the radio-frequency circuit. From Eq.~\eqref{eq:SNR}, it is clear that the SNR can be increased by maximizing $|\Delta\Gamma|$ at a given input power. The strategy that should be followed to optimise $|\Delta\Gamma|$ depends on the type of device to be measured (resistive or reactive) and on the size of the signal change (small or large-signal regime).

\subsection{Optimising for changes in resistance}

\begin{figure}
    \includegraphics[width=\columnwidth]{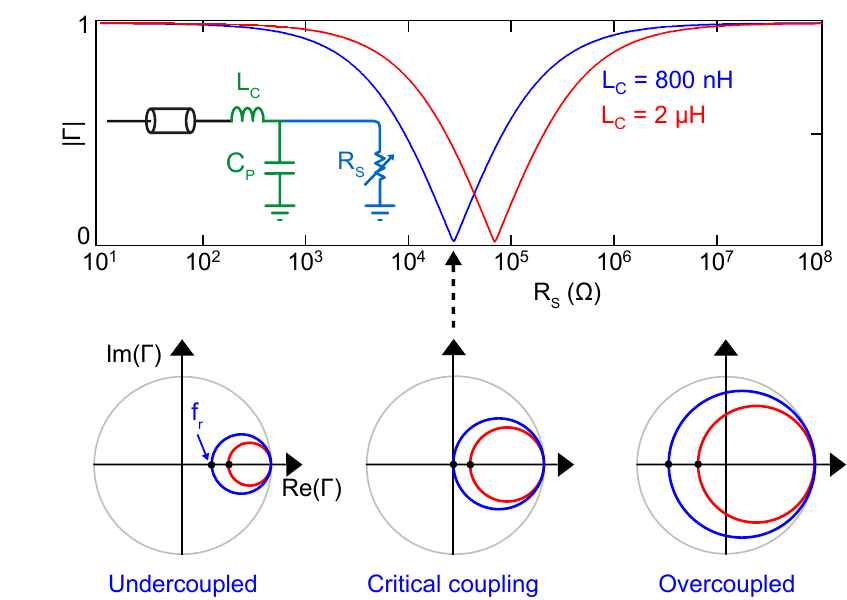}
    \caption{Top: Reflection coefficient amplitude $|\Gamma(\Rs)|$ simulated for the $LC$ circuit in the inset considering $\Lc=800$~nH, and $f=229.4$~MHz (blue trace), $\Lc=2~\micro $H and $f=145.2$~MHz (red trace); $\Cp=0.6$~pF. Critical coupling is reached at a specific value of $\Rs$ that depends on the circuit parameters.
    Bottom: Smith charts of the reflection coefficient spectrum (see Fig. \ref{fig:Smith}). The black points indicate the resonance frequency. For low  values of $\Rs$ (left), the circuit is undercoupled to the line. For high  values of $\Rs$ (right), the circuit is overcoupled. At critical coupling (blue curve middle), the curve meets the origin of the graph.
        \label{fig:LC cavity (R)}
    }
\end{figure}

A simulation of $|\Gamma(\Rs)|$ as a function of a variable resistance $\Rs$ is presented in Fig.~\ref{fig:LC cavity (R)} for an ideal $LC$ resonator ($\Lc$, $\Cp$) attached to the variable resistance (see Section~\ref{sec:resistiveReadout}).
The point $|\Gamma(\Rs)|=0$ marks the critical coupling condition.
The optimisation of the resonator consists of maximising the variation of $\Delta\Gamma$ for a given change of sample resistance $\Delta \Rs$. In the analysis of this problem, we need to distinguish two cases: when $\Delta\Rs/\Rs\ll 1$ (the small-signal regime) or when $\Delta\Rs/Rs\approx 1$ (the large-signal regime).

\subsubsection{Resistive readout - The small-signal regime}

For small changes of device resistance $\Delta \Rs$,  the change in reflection coefficient can be calculated to first order as 
\begin{equation}
    \Delta\Gamma = \left.\frac{\partial\Gamma}{\partial \Rs}\right|_{\omega=\omega_\text{r}}\Delta \Rs
    \label{eq:smallcapa2}
\end{equation}

For the circuit in Fig.~\ref{fig:LC cavity (R)}, the change of reflection coefficient takes the form:
\begin{equation}
    \Delta\Gamma \approx\frac{2\Zload\Zo}{(\Zload+\Zo)^2}\frac{\Delta \Rs}{\Rs}.
    \label{eq:smallcapa3}
\end{equation}

The first ratio in Eq.~\eqref{eq:smallcapa3} relates to the circuit coupling, which is maximal when the equivalent impedance of the circuit at resonance, $\Zload$, is equal to the impedance of the line $\Zo$. The second ratio, $\Delta \Rs/\Rs$ is maximal when the fractional changes in resistance are maximised. Figure \ref{fig:LC cavity (R)} illustrates that $\Delta\Gamma$ is maximal near the critical coupling condition, as expected from Eq.~\eqref{eq:smallcapa3}.

\subsubsection{The matching capacitor and in-situ tuneable resonators to achieve critical coupling}
\label{sec:insitu}

As explained in Section~\ref{sec:resistiveReadout}, critical coupling for an $LC$ resonator attached to a resistive device is achieved when $\Rs=\Rmatch = \Lc/\Cp \Zo$ (Eq. \eqref{equ:Rmatch}).
If $\Rs$ and $\Cp$ are known, a user can choose $\Lc$ to obtain the best sensitivity. 
However, in practice it is not always easy to know these values with sufficient precision due to temperature dependence of the circuit parameters and uncertainty over parasitic capacitance. Moroever, Eq.~\eqref{equ:Rmatch} implies that a large $\Lc$ is required to match samples with large resistances and parasitic capacitances. Increasing the value of $\Lc$ reduces the readout bandwidth; more problematically, large surface mounted inductors introduce self-resonances close to the operating frequency~\cite{Pepper2008}.
This is a difficulty for many quantum devices, which typically have resistance $\Rs \gtrsim $100~k$\Omega$. Even with careful engineering, sample wiring typically contributes a sample capacitance $\gtrsim 0.3$~pF in parallel with the device~\cite{Reilly2014}. 
A matching capacitor $\CM$, in parallel with the circuit, allows us to shift the critical coupling to a higher value of $\Rs$ to enable $\Zload=\Zo$.

\begin{figure}
    \includegraphics[width=\columnwidth]{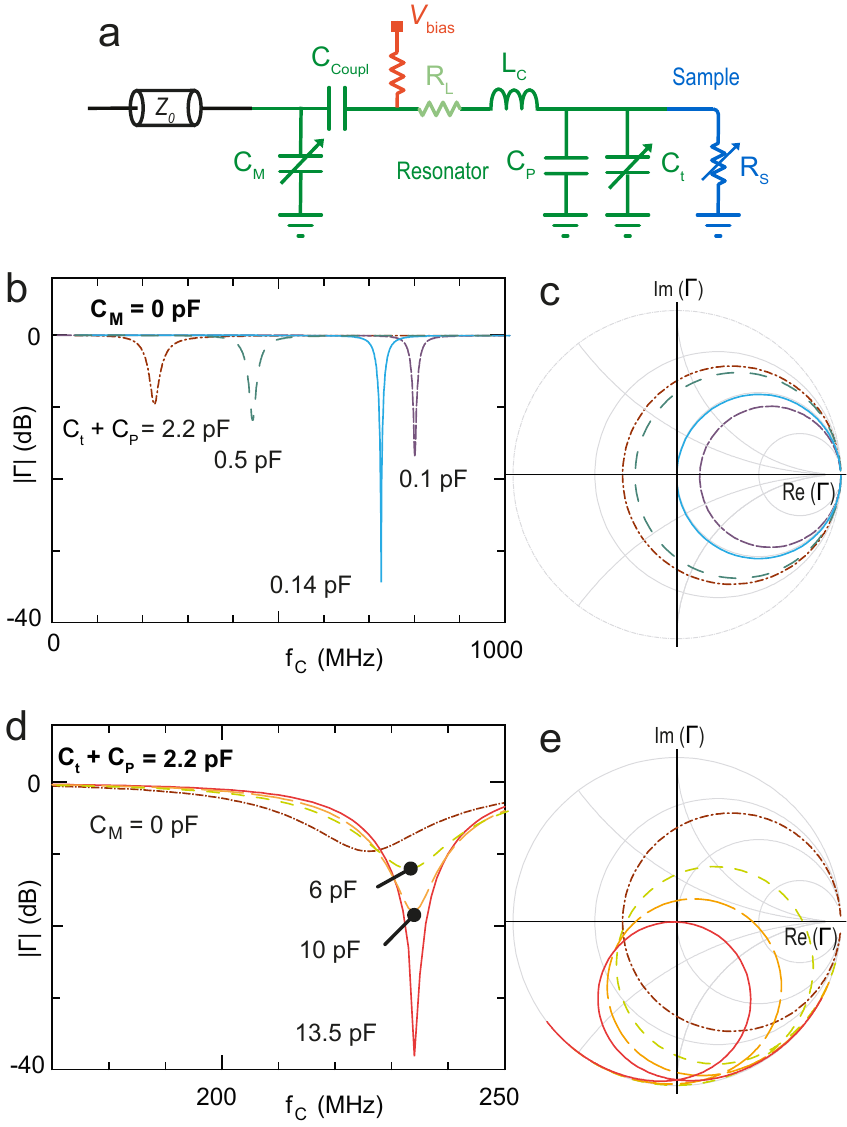}
    \caption{(a) A device is coupled to an impedance matching network formed from an inductor, variable capacitors $\Ct$ and $\CM$, and fixed capacitor. Parasitic losses are parameterized by an effective resistance $\RL$.
    (b) Simulation with no matching capacitor ($\CM=0$)~\cite{Laird2016matching}. Voltage reflection coefficient $\Gamma$ is plotted as a function of frequency for different values of sample capacitance $\Cc$, as magnitude, left , and as a Smith chart,~right.  $\RL=20~\Omega$ takes into account the losses in the resonator. The sample resistance is $\Rs=1~\mathrm{G}\Ohm$, and the capacitance of the device is included in $\Cc$. The inductor value is $\Lc=223$~nH and $C_\mathrm{Coupl}$ is 87~pF.
    (c) Simulated reflection for varying the matching capacitor $\CM$~\cite{Laird2016matching}. Adapted from Ref.~\onlinecite{Laird2016matching}.
        \label{fig:Fig_S6_9}
    }
\end{figure}

Voltage-tunable capacitors (varactors) allow for \emph{in situ} tuning of the matching condition~\cite{Muller2010}. Varactors in parallel with the sample can also be used to tune the resonance frequency.
An example of \emph{in situ} tunable resonator with a matching capacitor~\cite{Laird2016matching} is illustrated in Fig.~\ref{fig:Fig_S6_9}(a). This circuit incorporates two varactors: one primarily for impedance matching ($\CM$) and one primarily for frequency tuning ($\Ct$). The varactors allow the capacitance to be tuned with a dc voltage. Fig.~\ref{fig:Fig_S6_9}(b) shows a simulation of the reflection coefficient $\Gamma$ as a function of frequency for typical device parameters with no matching capacitor ($\CM=0$).  
Tuning $\Ct$ allows changing the resonance frequency
\begin{align}
    \fr=\frac{1}{2\pi\sqrt{\Lc(\Ct+\Cp)}}
\end{align}
as well as modifying the coupling. In this example, critical coupling is achieved at $\Ct+\Cp= 0.14$~pF which is below the typical parasitic capacitance of the measurement set up. Hence $\Ct$ on its own does not allow for critical coupling to be achieved. However, by increasing the capacitance of the shunt capacitor $\CM$, critical coupling can be achieved for a larger range of $\Ct$ and $\RS$ values. The shunt capacitor has little impact on the resonance frequency, but modifies the resonator's impedance and thus the circuit coupling in the circuit of Fig.~\ref{fig:Fig_S6_9} as follows:
\begin{align}
\Zload=\frac{\Ctot^2 \Lc \Rs}{\CM^2L+\Ctot^3 \Rs^2}
\end{align}
where $\Ctot = \CM + \Ct + \Cp$. 
Controllable perfect matching with resistive devices such as rf-QPCs \cite{Muller2010,Muller2012,hellmuller2012,Muller2013} and rf-SETs  \cite{Laird2016matching} is thus possible. 
Typically, GaAs varactor have been used as tunable elements~\cite{Laird2016matching,Gonzalez-Zalba2019tunable}. Although they provide a wide tuning range for temperatures above  1~K, the tunability range of such components is reduced drastically as the temperature is lowered~\cite{Gonzalez-Zalba2019tunable}. A list of components used in various studies cited in this review is available as Table~SI in the supplementary.

\subsubsection{The large-signal regime}

For large resistive changes, for example when a charge sensing event shifts an SET from a Coulomb peak to a valley, the first order approximation of Eq.~\eqref{eq:smallcapa2} breaks down. Instead we must consider $\Delta\Gamma=\Gamma(R_\mathrm{b})-\Gamma(R_\mathrm{a})$, the change in the reflection coefficient given a resistance change from $R_\mathrm{b}$ to $R_\mathrm{a}$. Circuit losses are generally detrimental, but become particularly important in the large-signal regime. 

\begin{figure}
    \includegraphics[width=\columnwidth]{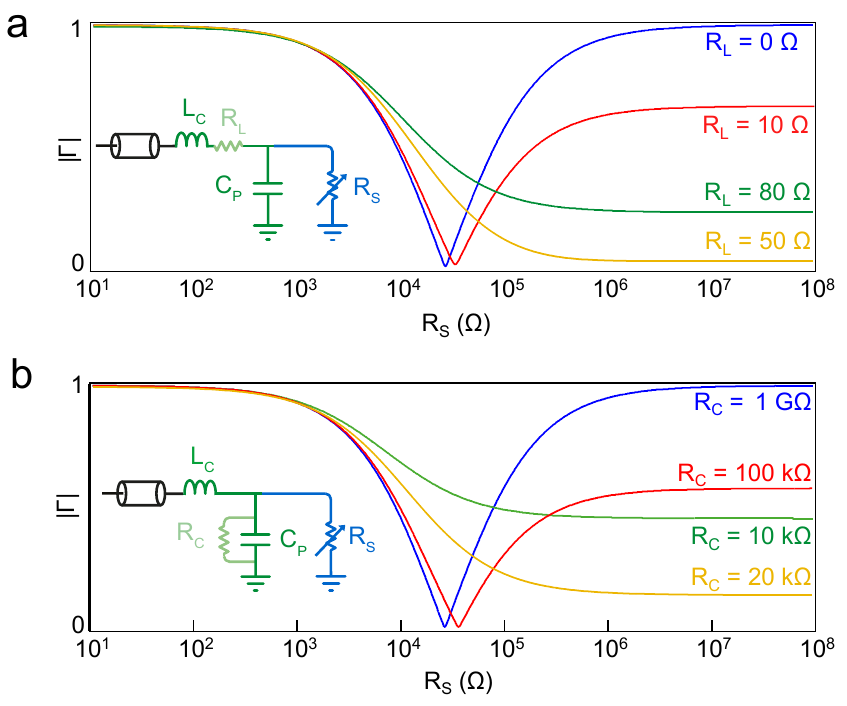}
    \caption{Reflection amplitude $|\Gamma|$ as a function of sample resistance $\Rs$ considering losses in the inductor (a) or to the ground~(b). The simulation parameters are $\Cc=0.6$~pF, $\Lc=800$~nH and $f=229.4$~MHz. The blue curves correspond to an ideal $LC$ resonator $\RL= 0~\Ohm$, $\Rc= +\infty$. The red curves correspond to modest losses that reduce the sensitivity of the circuit while the green and orange curves correspond to losses so high that the circuit is always undercoupled.
    \label{fig:LC cavity (R) unperfect}
    }
\end{figure}

Losses in the inductor (caused by its resistance $\RL$) reduce $|\Gamma|$ when the device is in a highly resistive state (Fig.~\ref{fig:LC cavity (R) unperfect}). As a result, the maximum $|\Delta\Gamma|$ when the device resistance increases above the match value is reduced to 
\begin{align}
|\Delta\Gamma|_\mathrm{max}=\left|\frac{\RL-\Zo}{\RL+\Zo}\right|
\end{align}

\noindent for $\RL<\Zo$. For $\RL>\Zo$ achieving critical coupling is not possible. The goal is thus to minimize $\RL$. Superconducting inductors are a way forward (see Section \ref{sec:Optimisation:superconductingL}).

Capacitive losses ($\Rc$) have a similar effect (Fig.~\ref{fig:LC cavity (R) unperfect}). These losses, as they are in parallel with the device resistance, reduce the maximum achievable $|\Delta\Gamma|$ between the two resistive states of the quantum device to
\begin{align}
|\Delta\Gamma|_\mathrm{max}=\left|\frac{\frac{\Lc}{C_\text{p}\Rc}-\Zo}{\frac{\Lc}{C_\text{p}\Rc}+\Zo}\right|
\end{align}

\noindent for $\Rc>\Rmatch$. For $\Rc<\Rmatch$ achieving critical coupling is not possible. In order to minimize $\Rc$, low-loss dielectric materials and high resistance device gate oxides can be used to avoid unintentional leakage paths to ground~\cite{Gonzalez-Zalba2015_Dispersive}.

\subsubsection{Measurement back-action: Relaxation and dephasing}

Using rf-SETs to measure the state of a qubit causes two types of back-action: relaxation (i.e. random transitions between eigenstates) and dephasing (i.e. randomisation of the phase in superpositions)~\cite{devoret2000Review}.
Various processes contribute to measurement-induced relaxation but the most widely considered in the literature are shot noise in the SET and quantum fluctuations in the qubit’s environment~\cite{Delsing2001}. The measurement-induced relaxation rate $\Gamma_1$ is proportional to the spectral density $S_\text{VV}(\omega)$ of the voltage fluctuations on the SET island at the qubit frequency
\footnote{The expressions for the relaxation and dephasing rates from which Eqs.~\eqref{eq:backactionSET} and \eqref{eq:T2spectraldensity} follow are:
\begin{align}
    \Gamma_1 &= \frac{1}{T_1}       = \frac{1}{2\hbar^2} \left|
    \braket{0_\mathrm{q} | \frac{\partial H}{\partial \lambda} | 1_\mathrm{q}}
    \right|^2 \, 
    S_{\lambda \lambda}\left(\frac{\omega_\mathrm{q}}{2\pi}\right)\\
    \Gamma_2 &= \frac{1}{T_\phi}     = \frac{1}{4} \left( \frac{\partial \omega_\mathrm{q}}{\partial \lambda} \right)^2 \, 
    S_{\lambda \lambda}(0)      \label{eq:footnoteGamma2}
\end{align}
where $\ket{0_\mathrm{q}}$ and $\ket{1_\mathrm{q}}$ are the qubit eigenstates, $\omega_\mathrm{q}$ is its angular frequency, and $\lambda(t)$ is a noisy environmental parameter.
Equation~\eqref{eq:footnoteGamma2} applies to a Ramsey-type dephasing measurement, provided that the spectral density is approximately constant at low frequency, i.e.\ that its cutoff frequency is higher than the inverse measurement duration. Often $T_\phi$ is called $T_2^*$.

These equations can be derived using the method of Ref.~\onlinecite{Clerk2014}.
A nice pedagogical motivation appears in Ref.~\onlinecite{William2019SuperQubits}, although the expression corresponding to Eq.~\eqref{eq:footnoteGamma2} contains a mistake; it appears correctly in Ref.~\onlinecite{Bylander2011}.
}:
\begin{equation}\label{eq:backactionSET}
    \Gamma_1= \frac{1}{8} \left(\frac{e}{\hbar}\right)^2\kappa^2\frac{\Delta_\text{C}^2}{\Delta E^2}~ \SVV\left(\frac{\Delta E}{h}\right),
\end{equation}

\noindent where the qubit Hamiltonian is Eq.~\eqref{eq:DQDHamilt} and the lever arm $\kappa$ is the ratio between the qubit-SET capacitance and the total capacitance of the qubit, which determines how strongly the SET island voltage fluctuations couple to the qubit.

The shot-noise relaxation process dominates for low qubit frequencies, i.e. when $\Delta E/\hbar\ll I/e$ where $I$ is the current through the SET.
Using “orthodox” SET theory~\cite{Korotkov1994}, the corresponding spectral density is
\begin{equation}\label{eq:voltfluct}
    S_\text{VV}^\mathrm{o}(f)=4\frac{E_\text{C}^2}{e^2}\frac{4\omega_\text{I}}{(2\pi f)^2+16\omega_\text{I}^2} 
\end{equation}

\noindent where $\omega_\text{I}=I/e$ is the tunneling rate through the SET and $E_\text{C}$ the SET charging energy. Equation~\eqref{eq:voltfluct} assumes a symmetric SET with no cotunneling.

The quantum-fluctuation process dominates at high frequency where $\Delta E\gg E_\text{C}$.
This process is driven by the fact that every electromagnetic mode containing the SET has associated quantum fluctuations. 
Their total spectral density can be modelled by considering the impedance of the SET island to ground as two parallel tunnel junctions each with resistance $R_\text{T}$, giving
\begin{equation}\label{eq:voltfluct2}
    S_\text{VV}^e(f)= hf \frac{R_\text{T}}{1+\left(\frac{hf}{E_\text{C}}\frac{\pi R_\text{T}}{2R_\text{K}}\right)^2}.
\end{equation}
To evaluate the noise spectral density of the SET island in all frequency regimes, a full quantum mechanical calculation is necessary~\cite{Buttikker2000}.

The other type of back-action is measurement-induced dephasing, caused when voltage fluctuations modify the energy splitting between qubit states without driving transitions between them.
The general expression for the dephasing rate is
\begin{equation}
    \Gamma_\phi=\frac{1}{4} \left(\frac{e}{\hbar}\right)^2\kappa^2\frac{\varepsilon^2}{\Delta E^2}~S_\text{VV}(0). 
    \label{eq:T2spectraldensity}
\end{equation}

The SET approaches the quantum limit for qubit readout (given by the equality in the Heisenberg uncertainty principle, $\Gamma_\phi\tau_\text{min}\geq 1/2$) but so far has not reached it~\cite{Rimberg2009}. Here $\tau_\text{min}$ is the measurement time needed to discern the state of the qubit with a signal-to-noise ratio of 1. 

\subsection{Optimising for changes in capacitance}
\label{sec:Optimisation:dispersive}

In this subsection, we describe analytically the SNR optimization problem to changes in device capacitance ($\Delta \CS$) and provide experimental strategies to achieve higher SNRs. We show that the strategies are different depending on the whether one is in the small-signal or the large-signal limit. The figure of merit that defines in which limit one operates is $F_\text{Q}=\Qr\Delta \Cs/\Ctot$, with $F_\text{Q}\ll 1$ being the small-signal regime and $F_\text{Q}\approx 1$ the large-signal regime.

\subsubsection{Capacitive readout - The small-signal regime}

In the small-signal regime, we consider the effect on $\Gamma$ of changes in device capacitance only to first order,
\begin{equation}
    \Delta\Gamma = \left.\frac{\partial\Gamma}{\partial \CS}\right|_{\omega=\omega_\text{r}}\Delta \CS,
    \label{eq:smallcapa}
\end{equation}
where $\Delta \CS$ is the change in the device capacitance.
Considering the circuit topology of Fig.~\ref{fig:superconductingL}(a), we obtain:
\begin{equation}\label{deltagamma_cap}
    \Delta\Gamma \approx j \frac{2\Zload \Zo}{(\Zload+\Zo)^2} \Qint\frac{\Delta \CS}{\Ctot},
\end{equation}

\noindent where $\Zload=\Lc/(\Ctot\Rc)$ is the equivalent impedance of the circuit at resonance and $\Ctot=\Cp+\CS$ is the total capacitance. The first ratio corresponds to the matching condition and is maximal when $\Zload=\Zo$, as in the resistive case. The second factor is the internal quality factor of the resonator $\Qint=\Rc\sqrt{\Ctot/\Lc}$ and the third to the fractional change in capacitance. Unlike in the case of resistive readout (Eq.~\eqref{eq:smallcapa3}), the internal quality factor plays a significant role~\cite{Gonzalez-Zalba2018_RFgate}. Equation~\eqref{deltagamma_cap} sets the first guidelines for SNR optimization to capacitance changes: (i) good matching to the line, (ii) high internal quality factor and (iii) large fractional changes in capacitance. In other words, both the parasitic capacitance and the internal circuit losses need to be minimised (increase $\Rc$) while achieving good coupling to the line. This can be equivalently seen as designing a high-$Q$, high impedance resonator. In the following, we explain possible strategies to achieve those goals.

\subsubsection{Resonator topology}
\label{sec:parallelconfig}

Considering the critical coupling condition $\Qint =\Qext$ and the requirement to achieve high internal quality factors, improving the SNR to capacitance changes requires increasing $\Qext$. The standard $LC$ resonators used to couple to resistive devices have $\Qext=\sqrt{\Lc/\Ctot}/\Zo$, which is well below 100 for typical circuit parameters~\cite{Sillanpaa2005}. A new circuit topology is needed to allow the necessary degrees of freedom to achieve critical coupling while maintaining high quality factors. In a different design~\cite{Gonzalez-Zalba2018_RFgate,Morton2020_JPA}, the inductor is placed in parallel with the sample and coupled though a coupling capacitor $\Ccoupl$ to a coplanar waveguide (Fig.~\ref{fig:superconductingL}(c)) to increase the quality factor and thus the readout sensitivity. In this configuration, the external and internal quality factors are:
\begin{align}
    \Qext &= \left (\frac{\Ccoupl+\Ctot}{\Ccoupl}\right)\frac{1}{\Zo}\sqrt{\frac{\Lc(\Ccoupl+\Ctot)}{\Ccoupl^2}} \\
    \Qint &= \sqrt{\frac{\Ccoupl+\Ctot}{\Lc}}\Rc. 
\end{align}
By introducing the extra degree of freedom of $\Ccoupl$, the circuit topology enables increasing the external quality factor while maintaining similar $\Qint$. A circuit that introduced this design reached $\Qext=680$ and $\Qint= 943$ ($\Qr \approx 400$) with $\Ccoupl = 90$~fF, $\Cp=0.48$~pF and a superconducting inductor $\Lc=405$~nH~\cite{Gonzalez-Zalba2018_RFgate}. The same paper reports $\Qr=790$ with another resonator. Later, a loaded quality factor of $\Qr=966$ was obtained~\cite{Morton2020_JPA} resulting in a consequent improvement of the sensitivity. Further improvements have been achieved by using inductive coupling rather than capacitive coupling as demonstrated in Ref.~\onlinecite{Gonzalez-Zalba2021Interaction}. Inductive coupling removes the need to add $\Ccoupl$, further increasing the fractional changes in capacitance.

\subsubsection{Reducing the parasitic capacitance}
\label{sec:Optimisation:superconductingL}

The analysis of the SNR to capacitance changes concluded that is necessary to reduce both parasitic losses and capacitance. Superconducting inductors have two main advantages. Firstly, they minimise dissipative losses. Secondly, their planar geometry allows significantly smaller parasitic capacitance than in wire-wound surface-mount inductors. These two advantages both increase the internal quality factor $\Qint$ and increase the fractional changes in capacitance. Conventional wirewound surface-mount inductors~\cite{Simmons2018} do not exceed quality factors of 100 while air-core inductors go just above~\cite{Gonzalez-Zalba2019tunable}.

 \begin{figure}
    \includegraphics[width=\columnwidth]{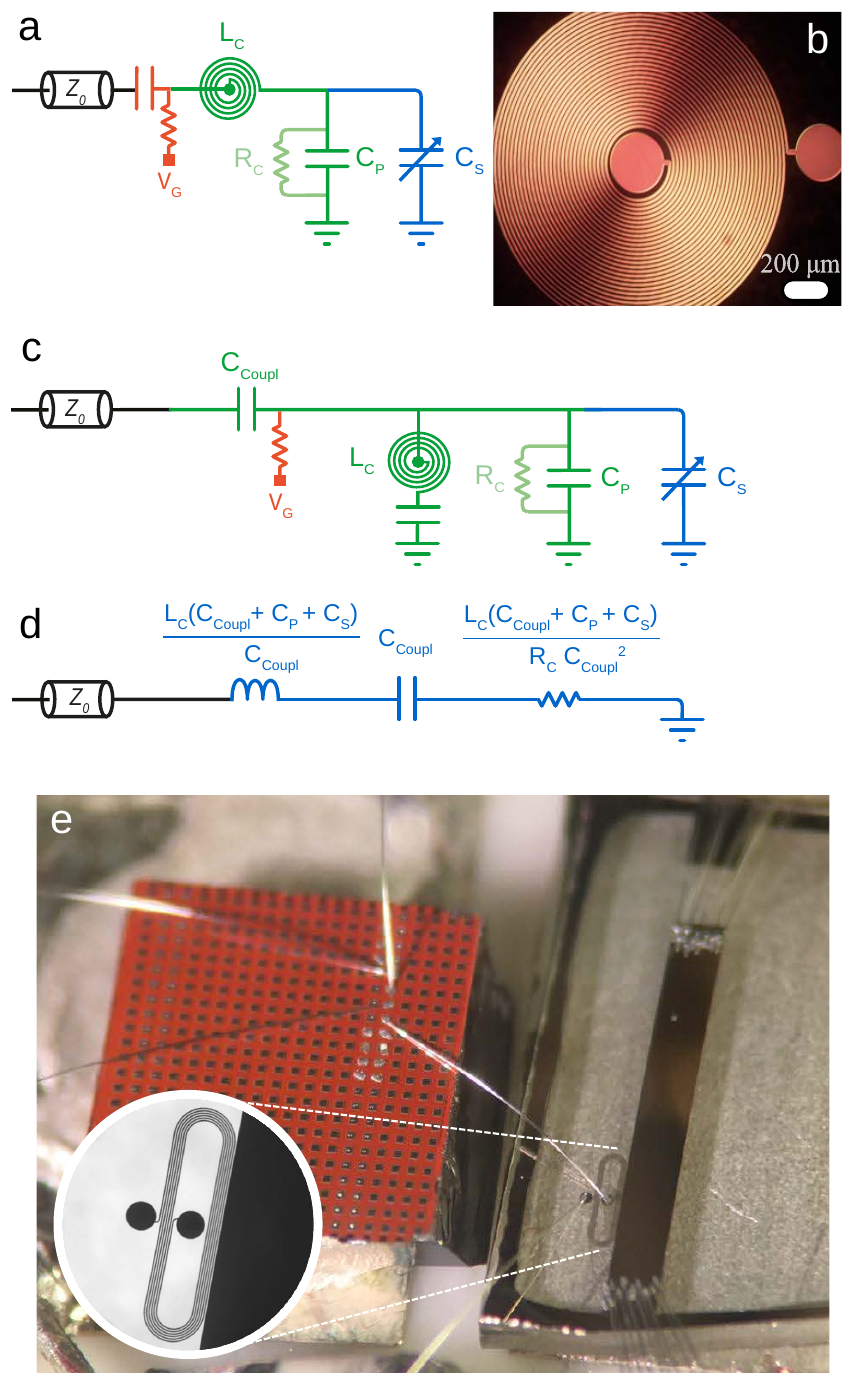}
    \caption{(a) Conventional $LC$ resonator circuit to measure a capacitance, similar to Fig.~\ref{fig:LCcavityC}(a) but using a superconducting inductor $\Lc$. 
    (b) Photograph of a superconducting planar inductor.
    (c) Schematic of a resonator with superconducting inductor in the parallel configuration.
    (d) Equivalent circuit at resonance. The resistance $\Rc$ represents the losses in the resonator.
    (e) Photograph of a multi-module set-up. The silicon chip with an array of square bond-pads is seen to the left, and to the right is the NbN-on-sapphire substrate. The two modules are positioned on a printed circuit board. The inductor, an elongated spiral (shown magnified in inset), is inductively coupled to a 50$~\Omega$ waveguide. This circuit provided an inductance $\Lc = 47$~nH and a parasitic capacitance $\Cp = 0.15 $~pF.
    Panel (b) adapted  from~\onlinecite{Gonzalez-Zalba2018_RFgate}. Panel (e) adapted from Ref.~\onlinecite{Gonzalez-Zalba2021Interaction}.
    \label{fig:superconductingL}}
\end{figure}

Superconducting inductors take, for example, the shape of a planar spiral with wirebonding pads at each end (Fig.~\ref{fig:superconductingL}(a,b)). They can be fabricated on a dedicated chip separate from the sample in order to allow for different fabrication strategies for each chip. They are typically made from thin films of the Type II superconductors Nb \cite{Reilly2013}, NbN~\cite{Simmons2018} or NbTiN\cite{Morton2020_JPA}. Important considerations are the critical temperature and critical magnetic field of the thin film. For experiments requiring high magnetic fields, such as for spin qubits, NbN and NbTiN are suitable provided the field is in the plane of the film~\cite{Gonzalez-Zalba2018_RFgate}.

Multi-module assemblies in which a semiconductor and superconducting chip are connected via wirebonds have been demonstrated (Fig.~\ref{fig:superconductingL}(d)). For example, in Refs.~\onlinecite{Gonzalez-Zalba2021Interaction,lundberg2021nonreciprocal} the superconducting chip contains an elongated spiral inductor that is inductively coupled to a 50~$\Omega$ microstrip waveguide, fabricated using optical lithography from an 80~nm thick sputter-deposited NbN film on a sapphire substrate. 

To reduce parasitic capacitance further, careful rf engineering of the circuit board is essential. High-frequency signals should be delivered by PCB waveguides with $50~\Ohm$ characteristic impedance. Parasitic capacitances can be further reduced by fabricating the PCB board from low-loss dielectrics such as the RT/duroid 4000 and 5000 families~\cite{Gonzalez-Zalba2015_Dispersive}.

\subsubsection{On-chip superconducting microwave resonators}
\label{sec:superconducting cavity}

The quality factors of resonators mounted on a printed circuit board are ultimately limited by dielectric losses and by parasitic capacitance to the ground, which in turn is set by the size of the components and of wirebonds.
To reduce $\Cp$ further, one must mount the resonator on the chip itself.
This adds fabrication and integration complexity.
However, ultimately this approach is limited by the internal quality factor, which for superconducting on-chip microwave resonators can be as large as a million \cite{Cleland2012,DiCarlo2015}. 
When such resonators are incorporated into a spin qubit device, the quality factors are smaller than in fully superconducting systems because of losses in the semiconductor substrate~($\Qr\approx 2000$)~\cite{Petta2012cirQED,Petta2015,Kontos2015,Vandersypen2019}.
However, they still benefit from substantially lower parasitic capacitance than in multi-module assemblies. As well as for spin qubits, they have also been used  to measure nanomechanical resonators~\cite{LaHaye2009, Konrad2013, Lehnert2017, Huttel2020}. The reduction in parasitic capacitance allows operation in the microwave range, i.e. $\fr>1$~GHz, where resonant interactions between the resonator and the device can occur. This regime where the energy of the resonator photons and the system to be probed are similar (also known as the resonant regime of circuit QED) is out of the scope of this review, but interested readers can find information in Ref.~\onlinecite{Wallraff2020}.

\subsubsection{Device capacitance}

In Section~\ref{sec:reactiveReadout}, we discussed the origin of quantum capacitance (Eq.~\eqref{QuantumCap}) and its manifestation in DQDs (Eq.~\eqref{QuantumCapDQD}). The reader can see that the device capacitance depends on the lever arm $\alpha$, which quantifies the efficiency of  a gate in modifying the electrochemical potential of the quantum system to be tested. In semiconductor devices, $\alpha$ can be increased by: (i) using small equivalent gate oxide thicknesses, i.e. by using thin high-k dielectrics, (ii)  using devices with a thin active region, such as on-insulator substrates~\cite{Laucht2021} and (iii) using non-planar gate geometries as in carbon nanotubes~\cite{Buitelaar2012}, InAs nanowires~\cite{Petta2012cirQED}, or silicon nanowire transistors~\cite{Betz2014}. Finally, quantum capacitance changes can be made more pronounced in low-dimensional devices by decreasing the temperature. Lower temperatures result in larger changes in the number of states with respect to changes in the chemical potential (in 1D and 2D systems).

\subsubsection{Capacitive readout - The large-signal regime}

Once the above strategies have been implemented, $F_\text{Q}$ may approach 1. So far, we have studied the small-signal response, i.e. the first-order changes of the reflection coefficient with respect to capacitance. However, for large signals, we must consider $\Delta\Gamma=\Gamma(C_\mathrm{b})-\Gamma(C_\mathrm{a})$, the change in reflection coefficient when the capacitance changes from state $C_\mathrm{b}$ to $C_\mathrm{a}$. To calculate this difference, we consider the reflection coefficient of a parallel resonator geometry (Fig.~\ref{fig:superconductingL}(c)):

\begin{equation}
    \Gamma=\frac{\Gamma_\text{min}+2j\Qr\Delta\omega/\omega_\text{r}}{1+2j\Qr\Delta\omega/\omega_\text{r}}, 
\end{equation}
\noindent where $\omegar$ is the angular resonance frequency, $\Delta\omega$ is the difference from the resonance frequency and $\Gamma_\text{min}$ is the minimum value of the reflection coefficient.   
It can be shown that the difference in $\Gamma$ between two states is maximal when Re$(\Delta\Gamma)=0$ and that this condition occurs when the operation frequency is chosen equal to average resonant frequency of the two capacitive states to be discerned~\cite{Gonzalez-Zalba2021Interaction}. In this case 
\begin{equation}
    \Delta\Gamma=j\frac{2\Qr\frac{\Delta\omega}{\omega_\text{r}}(1-\Gamma_\text{min})}{1+\left(2\Qr\frac{\Delta\omega}{\omega_\text{r}}\right)^2}.
\end{equation}
Taking into account that $\Qr=\Qint/(1+\beta)$, $1-\Gamma_\text{min}=2\beta/(1+\beta)$ and $\Delta\omega/\omega_\text{r}=\Delta \CS/(2\Ctot)$ where $\beta$ is the coupling coefficient and $\Ctot$ now includes $\Ccoupl$, we arrive at the general expression for $\Delta\Gamma$:
%
\begin{equation}
    \Delta\Gamma=j\frac{\frac{2\beta}{(1+\beta)^2}\Qint\frac{\Delta \CS}{\Ctot}}{1+\left(\Qr\frac{\Delta \CS}{\Ctot}\right)^2}.
\end{equation}
In the limit $\Qr\Delta \CS/\Ctot\ll 1$, we recover Eq.~\eqref{deltagamma_cap}, the small-signal limit. Furthermore, we see that $\Delta\Gamma$ becomes maximal when $\Qr\Delta \CS/\Ctot=1$. This condition translates into
\begin{equation}\label{eq:bigcapa4}
    \Delta\omega=\frac{\omegar}{\Qr}=2\pi\Bf,
\end{equation}
\noindent where we have to consider that the system is probed at the average resonant frequency of the two measurement outcomes~\cite{Gonzalez-Zalba2021Interaction}. Given a device-induced frequency shift, the best strategy is to couple the resonator such as its bandwidth will match such frequency shift. This is known as the condition for maximum state visibility. Under these circumstances:
\begin{equation}\label{eq:bigcapa5}
    \Delta\Gamma=j\frac{2\beta}{1+\beta},
\end{equation}
\noindent and hence, if a large enough frequency shift is available, overcoupling the resonator to the line will result in higher $\Delta\Gamma$ compared to the low-signal regime, where the optimal coupling condition is that critical coupling. 

\subsection{Large gated semiconductor devices}

Large gated semiconductor devices are difficult to match because of their large capacitance, but nevertheless, dissipative rf measurement measurements of 2D systems have been achieved~\cite{Pepper2008}. 
However,  accumulation-mode quantum dots remain a challenge because the resistance of the contact leads and capacitance of the large accumulation gate form an RC filter that prevent the signal from reaching the quantum dot~\cite{Yacoby2020}. This problem can be mitigated with adapted device designs that minimise the accumulation region~\cite{Nichol2020}
or by using doping rather than gates to fabricate the lead~\cite{Tarucha2020}.

Another approach, which requires less optimisation, is to connect the resonator to an accumulation gate. Thanks to the high gate capacitance, the reflected signal is sensitive to the resistance of the quantum dot rather than only its capacitance measurement~\cite{Kuemmeth2019,Yacoby2020}. In this configurations the path that would allow the signal to leak directly from the contact lead to ground needs to be blocked by a resistor \cite{Kuemmeth2019} or using gates~\cite{Yacoby2020}.

In dispersive measurements, the accumulation of charges in the surroundings of the quantum dot creates a voltage-dependent change of $\Cp$ that degrades the sensitivity \cite{Gonzalez2017accumulation,Dzurak2019}.
This unwanted accumulation of charge in the areas surrounding the quantum dot can be reduced by using depletion gates~\cite{Dzurak2019}. 

\subsection{Optimal SNR and back action}

From Eq.~\eqref{eq:SNR}, one might conclude that increasing the input power $P_0$ results in an indefinite increase in SNR. However, $P_0$ cannot be arbitrarily large since eventually the large voltage swing across the device will broaden the lineshape of the feature under study, i.e. create back-action by over-driving the system. This voltage scale might correspond, for example, to an energy swing equivalent to an energy-level splitting in a DQD, or to the energy associated with the tunneling rate or the electron temperature for a SEB. To study this problem, we divide the task in two: (i) understanding the effect of the voltage drop at the device $V_\text{dev}$ on the observable capacitance $\Delta \Cs$ and (ii) determining $V_\text{dev}$ given an input power $P_0$.

Point (i) has been considered in literature for charge, spin and Majorana devices using the adiabatic approximation~\cite{Maman2020a}. Here, we use the simplest example: a charge qubit, i.e. a coupled DQD as in Section~\ref{sec:CQandCtuinQD}. In the adiabatic limit, where probe-induced excitations and inelastic relaxation processes can be neglected, the charge on QD2 (the dot which we take as a reference) can be expressed as:
\begin{equation}
\label{eq:chargepolarization}
   n_2=\frac{1}{2}\left(1+\frac{\epsilon}{\Delta E}\right)
\end{equation}
\noindent where $\epsilon$ is the energy detuning between QDs and $\Delta E$ the DQD energy difference. The effective parametric capacitance of the DQD is the ratio between the in-phase Fourier component of the charge response and the Fourier component of the probing voltage during the time $T$, both taken at the probe frequency and weighted by the probing gate lever arm ($\alpha$):
\begin{equation}
\label{eq:FourierCp}
  \CQ=\frac{\frac{1}{T}\int^T_0 \alpha e n_2(t)\sin{(\omega t)}dt}{\frac{1}{T}\int^T_0V_\text{dev}\sin{(\omega t)}\sin{(\omega t)}dt}.
\end{equation}

The denominator can be readily evaluated: $V_\text{dev}/2$. Considering that $n_2(t)$ is periodic in time, we find
\begin{equation}
\label{eq:FourierCp2}
  \CQ=\frac{2\alpha e}{V_\text{dev}}\frac{1}{T}\int^T_0 n_2(t)\sin{(\omega t)}dt.
\end{equation}

Equation~\eqref{eq:FourierCp2} can be solved analytically after inserting Eq.~\eqref{eq:chargepolarization} to yield
\begin{equation}
\label{eq:FourierCp3}
  \CQ=\frac{2\alpha e}{\pi V_\text{dev}}f_C(x)
\end{equation}

\noindent where the dimensionless function characterizing the capacitance is defined as
\begin{equation}
\label{eq:fcx}
    \fC(x)=\frac{\left(1+x^2\right)E\left(\frac{x^2}{1+x^2}\right)-K\left(\frac{x^2}{1+x^2}\right)}{x\sqrt{1+x^2}},
\end{equation}

\noindent and $x=\alpha e V_\text{dev}/\Delta_\text{C}$, $E(x)$ is the complete elliptic integral of the second kind and $K(x)$ is the complete elliptic integral of the first kind. We can now evaluate the effect of increasing $V_\text{dev}$ on the capacitance amplitude. At low voltages, $f_C(x)\rightarrow\frac{\pi}{4}x$ and we recover the expected function for the ground state capacitance in the small excitation regime (see Eq.~\eqref{QuantumCapDQD})
\begin{equation}
\label{eq:CQ}
    \CQ=\frac{\left(\alpha e\right)^2}{2\Delta_\text{C}}
\end{equation}
\noindent whereas for large voltages, $f_C(x)\rightarrow 1$, and the capacitance becomes a decreasing function of $V_\text{dev}$. We can see how overdriving leads to back-action by reducing the measured capacitance. 

Next, we move to point (ii), which requires calculating the relationship between $P_0$ and $V_\text{dev}$. We consider the parallel circuit configuration (see Section~\ref{sec:parallelconfig}). When applying a voltage $V_\text{in}$ to a transmission line with characteristic impedance $\Zo$ and a load $\Zload$, a fraction $\lambda$ of that power will be transferred to the device producing a voltage drop $V_\text{dev}$ across the resistance~$\Rc$,
\begin{equation}
\label{eq:PB1}
   \lambda\frac{V_\mathrm{in}^2}{\Zo}=\frac{V_\text{dev}^2}{\Rc}.
\end{equation}
Here we neglect the dissipation in the inductor resistance $\RL$. The fraction of available power is then given by
\begin{equation}
\label{eq:PB2}
   \lambda=1-|\Gamma|^2=\frac{4\Zload\Zo}{(\Zload+\Zo)^2},
\end{equation}

\noindent which allows us to calculate the relation
\begin{equation}
\label{eq:PB3}
   P_\mathrm{0}=\frac{V_\text{dev}^2}{\Rc}\frac{(\Zload+\Zo)^2}{4\Zload\Zo}.
\end{equation}

We can now go back to our definition of SNR (Eq.~\eqref{eq:SNR}) and insert Eqs.~\eqref{eq:PB3} and \eqref{deltagamma_cap} to find an explicit expression for the SNR.
\begin{equation}
\label{eq:SNRfinal}
    \text{SNR}=\frac{8}{\pi^2}\frac{\beta}{\left(1+\beta\right)^2}\frac{\left(\alpha e\right)^2}{k_\text{B}T_\text{N}}\frac{\Qint\omegar}{\Ctot}
    \,\fC^2\left(\frac{\alpha eV_\text{dev}}{\Delta_\text{C}}\right)\taum,
\end{equation}

\noindent where we have used the fact that $P_\mathrm{N}=k_\text{B}T_\text{N}/2\taum$, with $\taum$ being the integration time. Equation~\eqref{eq:SNRfinal} provides clear guidelines on the optimal steps to maximise the SNR: (i) Achieve critical coupling, (ii) increase the internal quality factor i.e. reduce internal losses, (iii) reduce parasitic capacitance, (iv) operate at high frequency, (v) maximise the lever arm $\alpha$, (vi) reduce the noise temperature and obviously (vii) increase the illumination level $\Vin$ and (viii) increase the integration time. 
One should keep in mind that some of these parameters affect the voltage drop across the device, i.e. $V_\text{dev}=2\Ccoupl \Qr \Vin/(\Ccoupl+\Cp)$, and $\Vin$ may need to be readjusted to avoid overdriving the system. 

For singlet-triplet spin qubits (Section~\ref{sec:Sample}), Ref.~\onlinecite{Maman2020a} shows that the SNR in a dispersive readout experiment also saturates as $V_\text{in}$ is increased. 
For Majorana qubits (see Section~\ref{sec:Sample}), the model predicts an optimal $V_\text{in}$, with a decreasing readout fidelity as $V_\text{in}$ is increased beyond this optimum.

\subsection{Resonator-induced dephasing}

Any readout method is bound by the Heisenberg uncertainty principle that poses constraints on sensitivity and back-action. For qubit readout, the measurement time needed to acquire the state of a qubit with a signal-to-noise ratio of 1, $\tau_\text{min}$, is related to the induced rate of dephasing $\Gamma_\phi$ by the following relation

\begin{equation}
    \Gamma_\phi\tau_\text{min}\geq 1/2,
\end{equation}

\noindent meaning that a measurement completely dephases the qubit and the rate at which it does so is at least $1/(2\tau_\text{min})$. If the readout method follows the equality, it is said to have a quantum efficiency of 1. The problem of induced dephasing using dispersive readout has been analysed in Ref.~\onlinecite{johansson2006} for a slow oscillator ($\omega_\text{r}$ much smaller than the characteristic discrete energy level spacing in the qubit) which is the common case for rf reflectometry. For a wide range of parameters, dispersive readout is found to have unit quantum efficiency. 

Another important consideration is the rate of dephasing induced by the measurement system when it is not measuring $\Gamma_\phi^\text{off}$, i.e. not being driven by an external rf tone. The rate of induced dephasing for a thermally occupied oscillator is
\begin{equation}
\label{eq:dephasing}
   \Gamma_\phi^\text{off}=n(\omegar)\left[1+n(\omegar)\right]\Qr\frac{\Delta \Cs^2}{\Ctot^2}\omegar,
\end{equation}

\noindent where $n(\omegar)=1/\left(e^{\hbar\omegar/k_\text{B}T}-1\right)$ is the thermal occupation number at the resonator frequency. To minimize off-state dephasing, it is advantageous to cool down the resonator -- either by increasing the frequency or lowering its physical temperature -- but it also to reduce the fractional change in capacitance and to use a low-$Q$ resonator. Since some of these conditions compete against the SNR optimization strategies presented above, SNR and $\Gamma_\phi^\text{off}$ need to be evaluated simultaneously to reduce the readout time while maintaining low dephasing rates.  

\subsection{Measures of sensitivity}
\label{sec:sensitivity}
\subsubsection{The charge sensitivity}

Comparing the performance of different sensing devices and methodologies is essential in assessing the quality of a particular readout technology. Different figures of merit have been used in the literature to benchmark readout sensors but all can be related to a single magnitude, the minimum measurement time $\tmin$, defined as the integration time needed to discern two states with a SNR of 1. 

\begin{figure}
    \includegraphics[width=\columnwidth]{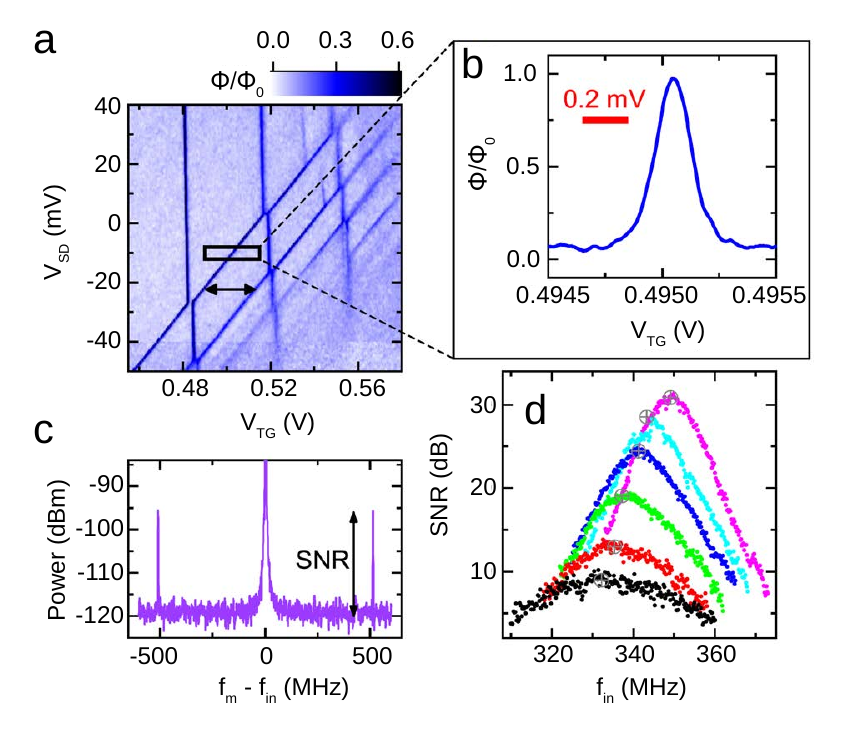}
    \caption{Charge sensitivity in the frequency domain. 
    (a) Reduced phase response of the resonator as a function of source-drain $V_\text{SD}$ and top-gate voltage $V_\text{TG}$ at 50 mK, measured with an rf power of -93 dBm. The black box indicates the dot-to-reservoir transition used to measure sensitivity and the arrow the gate voltage period. 
    (b) The same transition as indicated in (a), measured at $V_\text{SD} = -10$~mV. The input rf power here is $-103$~dBm. The red line indicates the peak-to-peak amplitude of the top-gate modulation signal. 
    (c) Spectrum of the reflected RF, showing sidebands at the gate modulation frequency of 511 Hz. 
    (d) SNR as a function of RF carrier frequency ($\fin$) for a frequency tunable resonator including a variable capacitor as in Fig.~\ref{fig:Fig_S6_9}(a). Black, red, green, blue, cyan and pink correspond to voltages across $C_\text{t}$ equal to 0, 1.5, 3, 4.6, 6.5, 15 V, respectively. Measurements taken with an rf power of $-90$~dBm and a spectrum analyzer resolution bandwidth of 2 Hz. The maximum SNR is marked for each data set with a grey circle. figures adapted from Ref.~\onlinecite{Gonzalez-Zalba2019tunable}.
    \label{fig:SNR1}
    }
\end{figure}

For charge sensors, the most commonly used figure of merit is the charge sensitivity, $\Sqq$ which can be understood as the amount of charge that can be discerned in a measurement lasting a second (see Eq.~\ref{eq:deltaq}). In this case, the minimum measurement time corresponds to
\begin{equation}
    \tmin=\frac{S_{QQ}^\mathrm{N}}{2e^2}.
\end{equation}
In radio-frequency mode, the charge sensitivity of a sensor can be extracted in two ways: 

(i) In the frequency domain (see Fig.~\ref{fig:SNR1}(a,b)), by applying to a device electrode a small sinusoidal signal of frequency $\fm$ and calibrated charge rms amplitude (typically $\Delta q_\text{rms}=0.01e$ or less to guarantee that the sensor operates in the linear regime). This method is particularly useful for SETs where the gate voltage period is a direct measurement of the addition of one unit of charge. If the device is biased at a point of finite transconductance, the ac signal modulates the carrier frequency $\fin$ producing sidebands in the power spectrum of the reflected signal at $\fin \pm \fm$ (Fig.~\ref{fig:SNR1}(c)). The sensitivity is then calculated from the SNR of the side bands (Fig.~\ref{fig:SNR1}(d)) according to~\cite{Aasime2001, Brenning2006} (see Supplementary for a derivation):
\begin{equation}
    \Sqq=\frac{\Delta q_\text{rms}}{\sqrt{2\Rf}\times10^{\text{SNR}_\text{dB}/20}}
    \label{eq:chargesensitivity}
\end{equation}
where $\Rf$ is the resolution bandwidth of the measurement and SNR$_\text{dB}$ is the power signal-to-noise ratio in decibels
\footnote{The resolution bandwidth is the width of each frequency bin in the numerically calculated power spectrum. A resolution bandwith $\Rf$ admits the same amount of white noise as an average over an integration time $\taum=1/2\Rf$, as proved in Supplementary Section S3.B.3}.
The factor of $\sqrt{2}$ takes into account that information can be extracted from both sidebands by homodyne detection.

It is important to distinguish between sensitivity to charge on the charge sensor (for example on the island of an SET) and on the target (i.e. the object being sensed, such as a qubit).
The sensitivity to charge on the target is generally worse, because one electron on the target induces less than one electron on the sensor. 
If $\Delta q_\text{rms}$ is a charge induced on the sensor, then
Eq.~\eqref{eq:chargesensitivity} gives the sensitivity to charge on this sensor, and can be used to compare sensors without any need to measure a target.  
However, the time taken to resolve a charge of one electron on the target is:
\begin{equation}
\label{Eq:tminSET}
    \tmin=\frac{S_{QQ}^\mathrm{N}}{2e^2} \left (\frac{C_\Sigma}{C_\text{m}} \right)^2 
\end{equation}
where $C_\text{m}$ is the mutual capacitance and $C_\Sigma$ is the total capacitance of the system to be sensed~\cite{Aasime2001}.  

\begin{figure}[!ht]
    \includegraphics[width=\columnwidth]{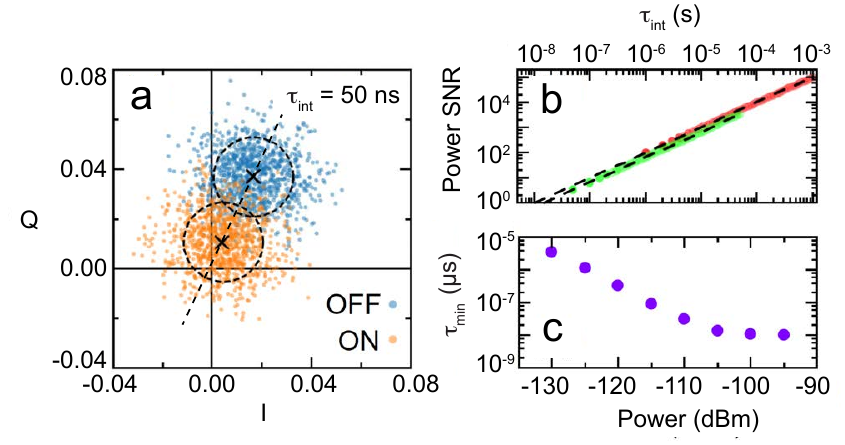}
     \caption{Signal-to-noise ratio in the IQ plane. (a) Distribution of the reflected signal in quadrature space, collected at gate settings on and off an interdot charge transition. Each point is collected with an integration time $\tauM = 50$~ns. 
    For each distribution, the black cross marks the centre (mean), the dashed circle indicates the standard deviation of distance to the centre. The dashed line markes the axis that joints the two centres. 
    (b) SNR dependence on $\tauM$ for input power $= -100$~dBm. Red points are taken with a 1~MHz low-pass filter and green points are taken with a 20~MHz low-pass filter. The dashed lines extrapolate the data to SNR = 1, from which the minimum integration time $\tmin$ can be extracted. 
    (c) Decrease in $\tmin$ with increasing input power, showing saturation due to power broadening at approximately $-100$~dBm. 
    Figure adapted from Ref.~\onlinecite{Gonzalez-Zalba2021Interaction}.
    \label{fig:SNR2}}
\end{figure}

(ii) In the time domain, by monitoring the sensor response with and averaging it over bins of duration $\taum$ while the target system to be sensed changes state (either actively driven by voltage pulses, or passively when it fluctuates between states). Then the data is collected in a 2D histogram in the IQ plane and the SNR$=(S/\sigma)^2$ calculated from the voltage distance between centers of the clusters $(S)$ and their average standard deviation ($\sigma$) along the axis that joints the two centres, see Fig.~\ref{fig:SNR2}(a). The charge sensitivity is
\begin{equation}
    \Sqq=\frac{\sqrt{2 \taum} \, e}{\sqrt{\text{SNR}}}
    \label{eq:chargesensitivity2}
\end{equation}

For dispersive sensing, the readout resonator probes directly the system to be sensed rather that detecting it via an intermediate charge sensor. For that reason, the preferred figure of merit has been the SNR of the two possible outcomes of the measured system with given measurement duration $\tauM$ (Fig.~\ref{fig:SNR2}(b)). The methodology that has been followed to extract the SNR is identical to the time-domain case (ii) above. Given the relation
\begin{equation}
    \text{SNR}=\frac{\tauM}{\tmin},
\end{equation}

\noindent from the SNR at a given $\tauM$, one can directly extract $\tmin$ (Fig.~\ref{fig:SNR2}(b,c)), enabling a direct comparison between the measurement time for charge sensors and dispersive readout. 

\subsection{Opportunities and challenges}

In Section \ref{sec:insitu} we discussed how voltage-controlled capacitors can be used to optimise the impedance matching \textit{in situ}. However, semiconductor-based varactors are lossy, degrading the quality factor of the circuit and thus its sensitivity. Another challenge is the small tuning range of these varactors at cryogenic temperatures~\cite{Laird2016matching,Gonzalez-Zalba2019tunable}. 
To overcome these limitations, we could use varactors based on ferroelectric materials, such as the lead titanate and barium strontium titanate families of solid solutions. The highly non-linear dielectric permittivity enables control of the capacitance via an electric field and, at temperatures at which the material is in its paraelectric state, low dissipation can be achieved. However, at low temperatures, ferroelectricity affects the tunability and loss tangent of these varactors. Opportunities are therefore open for the improvement of varactors.

Coplanar waveguide architectures can also benefit from tunable capacitances. Quantum paraelectric materials, such as SrTiO$_3$, KTaO$_3$, and CaTiO$_3$, allow for such capability. In these materials, quantum fluctuations suppress ferroelectricity at low temperatures. In particular, SrTiO$_3$ has a very high relative permittivity at mK temperatures~\cite{Saifi1970,Sakudo1971,Neville1972,Rowley2014,Davidovikj2017}, which is tunable using electric fields. A SrTiO$_3$ varactor was integrated in an rf circuit, allowing for perfect impedance matching down to 6 mK~\cite{Buitelaar2021}. Other quantum paraelectrics, such as KTaO$_3$, may reduce losses further, at the cost of less tunability~\cite{Geyer2005}. 
Tunable microwave impedance matching can also be achieved using a coplanar resonator whose inner conductor contains a high kinetic inductance metamaterial, in this case a series array of SQUIDs~\cite{Altimiras2013}. The matching frequency of such circuits was demonstrated to be tunable between 4 and 6~GHz. This approach has not been yet applied to the rf readout of quantum devices.

A potential avenue for improving the readout of resistive devices is to design impedance matching networks with a matching resistance larger than the on-state resistance of the device. In that scenario, by moving from the overcoupled (high resistance state) to the undercoupled regime (low resistance state), $|\Delta\Gamma|>1$ could be achieved. Note that $|\Delta\Gamma|=1$ for the case where the inductive and dielectric losses are negligible and critical coupling is achieved for the on-state of the device.   

Going beyond varactors, which are essential elements for optimal readout of resistive devices, dispersive readout of reactive devices will benefit from further improvements. At the device level, structures with high lever arm are desirable since they results in higher quantum capacitance changes (see Eq.~\ref{eq:CQ}). Using thin gate oxides or high-k dielectrics will facilitate that goal. Also thin layers of material, like thin silicon-on-insulator, or wrap-around gates can increase the lever arm further. 

At the resonator level, the directions to go are towards high-impedance, high-$Q$ and high-frequency resonators. High-frequency, high-impedance resonators can be achieved by minimising the effect of parasitic capacitance. Planar circuit elements, either capacitors or inductors, have less parasitic capacitance than surface mount components. On-chip resonators reduce the effect of parasitics further~\cite{Zheng2019}. Besides, inductive coupling results in even lower total capacitance than capacitive coupling~\cite{Gonzalez-Zalba2021Interaction}. High-frequency operation is also favourable for minimising back-action due to the reduced thermal photon shot noise (Eq.~\ref{eq:dephasing}). A potential drawback of operating at higher frequencies is that quantum capacitance effects are governed by charge reconfiguration due to the high-frequency electric field excitation. If the characteristic charge tunneling times are comparable or slower that the probe frequency, the magnitude of the quantum capacitance change is reduced~\cite{Gonzalez-Zalba2017}. To reduce non-radiative losses in the resonator and hence increase the internal quality factor, resonators will need to be manufactured using superconducting materials on low-loss substrates with high quality interfaces such as sapphire or quartz. Even further advances may be possible by changing paradigm to longitudinal coupling, by modulation of the resonator-qubit coupling at the frequency of the resonator, an approach considered to be generally quantum-limited~\cite{Didierlongitudinal2015}.

\section{Amplifiers and noise}
\label{Sec:amplifiers}

In a typical quantum electronic experiment, the signal of interest is tiny, with the useful information often contained within a total signal amplitude of 1~$\mu$V or less~\cite{Laird2016matching}.
Inevitably, this signal is accompanied by noise.
To extract the information, the signal must usually be increased to a level where it can be analysed by digital electronics, which typically operates at logic levels above 1~V.
A central challenge in quantum electronics is to do this with as little noise as possible.
Unfortunately, on top of noise intrinsic to the experimental device, there are noise contributions (which are often much larger) from almost every component in the measurement chain.
The topic of this section is how to quantify and reduce these, in order to minimise the effects of noise in an experiment.

\subsection{Quantifying noise in an electrical measurement}

\begin{figure*}
    \centering
    \includegraphics[width=\textwidth]{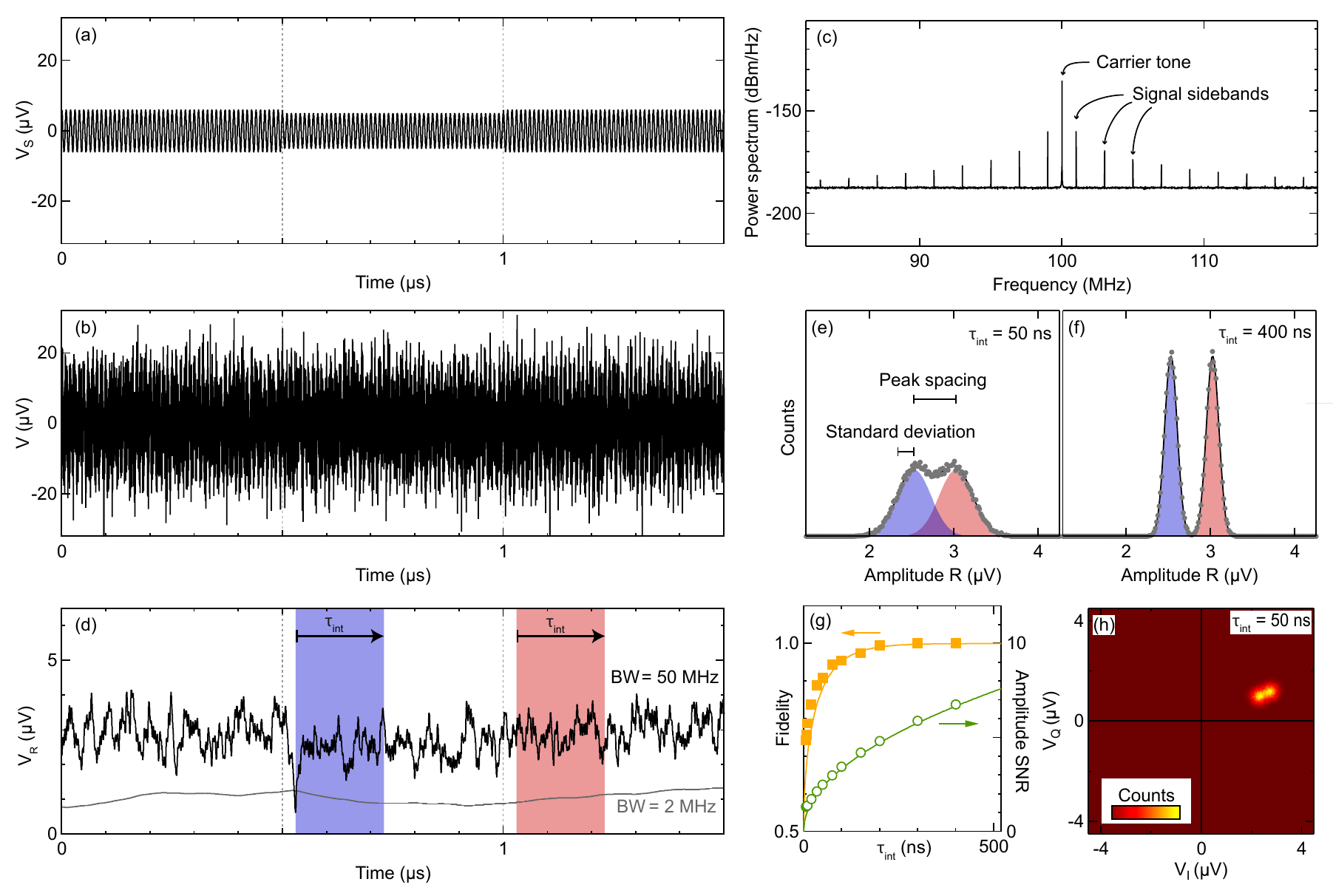}
    \caption{Simulation of the effect of noise on raw data and on the processed signal.
    (a) Reflected signal $\VS(t)$ switching regularly  between two amplitude levels, as caused by a device switching between two states.
    (b) The same signal with added white noise of spectral density $\SVVN=10^{-20}~\mathrm{V^2}/\mathrm{Hz}$, corresponding to a noise temperature of 14.5~K.
    (c) The corresponding power spectral density $\SVV(f)/Z_0$ (from Eq.~(S21)).
    %
    (d) Upper trace: Signal amplitude $V_\text{R}(t)$, defined by Eq.~\eqref{eq:defineR} and obtained by demodulating the trace in (a) and applying a 50~MHz low-pass filter.
    The shaded regions mark two intervals, each of duration $\taum$, during which $V_\text{R}(t)$ is averaged in order to determine whether it is high or low.
    Each interval begins shortly after the transition, with a short delay to cut out the response time of the filter.
    Lower trace: the same data, with the filter bandwidth set to 2 MHz (and vertically shifted for clarity).
    This filter eliminates most of the noise, but means that averaging overlaps with the rise time of the filter.
    It is therefore a bad choice.
    (e,f) Symbols: Histogram of amplitude measurements obtained from many averages as in (d).
    Shaded areas: inferred gaussian contributions from the low and high portions of the signal.
    (g) Right axis: Signal-to-noise ratio (SNR) as a function of averaging time $\taum$, extracted from histograms as above. Signal is defined as the spacing between peaks, noise as the standard deviation.
    Left axis: Fidelity, defined as the probability of deducing the correct amplitude level based on a single averaging interval.
    In both cases, symbols are values extracted from the simulation and curves are analytical predictions, using Eq.~(\ref{eq:sensitivity}(b)) for SNR and Eq.~\eqref{eq:fidelity} for fidelity.
    (h) Similar averaged data as in (e), represented as a two-dimensional histogram over in-phase and quadrature voltages.
    \label{fig:noise}
    }
\end{figure*}

Suppose we want to measure the voltage being reflected from a radio-frequency resonator as in Section~\ref{Sec:Basics}.
The signal that we want to measure is $\VS(t)$.
For example, Fig.~\ref{fig:noise}(a) shows a simulated voltage trace from a device that is switching regularly between two states.
Instead, we measure something like Fig.~\ref{fig:noise}(b).
Our measured signal is
\begin{equation}
    V(t) = \VS(t) + \VN(t).
    \label{eq:VSN}
\end{equation}
The second term, which by assumption carries no information about the signal, is the noise.

To understand the effect of the noise on our experiment, we need to answer two questions:
\begin{enumerate}
\item
How should we describe the noise contained in a voltage trace $V(t)$?
This question is answered in Sections~\ref{sec:spectraldensity} and~\ref{sec:noisepower}.
\item
What uncertainty will this noise introduce in an estimate of $\VS(t)$, or of a quantity derived from it?
This question is answered in Section~\ref{sec:predictinguncertainty}.
\end{enumerate}

\subsubsection{Quantifying noise using the spectral density}
\label{sec:spectraldensity}
How should we characterise a noisy voltage trace $V(t)$?
Since on average the noise is zero, we should quantify the variance.
Suppose that we construct a filter that passes only those components within a frequency bandwidth $\fBW$ centered at frequency $f$.
The magnitude of the filtered signal $\VF(t)$ depends on which components are passed, i.e. on how wide we choose $\fBW$.
We therefore quantify it by means of the one-sided spectral density $S_{VV}(f)$, which in almost all circumstances is given by
\begin{align}
     S_{VV}(f)  &= \lim_{\fBW \rightarrow 0} \frac{\EEE{\VF^2(t)} }{\fBW}
     \label{eq:SVV_defn}
\end{align}
where $\langle\cdot\rangle$ denotes an expectation value and $\lceil \cdot \rceil$ is a time average.
The spectral density is a measure of how strongly $V(t)$ fluctuates near frequency $f$.
For a precise definition of $\SVV(f)$ and instructions how to calculate it, see Supplementary Information~S3.

If the signal and the noise are uncorrelated, which is usually the case, the spectral density can be separated into a signal contribution $\SVVS(f)$ and a noise contribution $\SVVN(f)$:
\begin{equation}
    S_{VV}(f)   = \SVVS(f) + \SVVN(f).
\end{equation}
We therefore describe the noise quantitatively by specifying the noise spectral density $\SVVN(f)$, which is the spectral density in the absence of signal, i.e. when $\VS=0$.
Generally we want $\SVVS(f)$ to be large and $\SVVN(f)$ to be small.

To allow comparison between measurements, $\SVVN$ is usually quoted as an input-referred noise, which means that it is based on the inferred signal $V(t)$ at the input to the first amplifier encountered by the signal.
To calculate the input-referred voltage $V(t)$ from a voltage record such as an oscilloscope trace, you should divide the recorded voltage by the total gain of the amplifier chain before the recording device.

\subsubsection{Other ways to specify noise: Noise power, noise temperature, and noise quanta}
\label{sec:noisepower}
The noise spectral density can be expressed in three equivalent ways.
Firstly, it can be written as a noise power density $\PN(f)$, which is the power per unit bandwidth that the noise delivers to a matched load:
\begin{equation}
	\PN(f)  \equiv  \frac{\SVVN(f)}{Z_0}
	\label{eq:powerspectrumdefinition}
\end{equation}
where $Z_0$ is the input impedance of the measurement circuit.
The power density has units of $\mathrm{W}/\mathrm{Hz}$, or equivalently $\mathrm{dBm}/\mathrm{Hz}$.

Secondly, it can be written as a noise temperature
\begin{equation}
	\TN(f) \equiv \frac{\PN(f)}{\kB}=\frac{\SVVN(f)}{\kB Z_0}.
	\label{eq:noisetemperature}
\end{equation}
This is the temperature of a fictitious classical resistor~\cite{foot:noisetemperature} with resistance equal to the amplifier's input impedance, that when connected to the amplifier would generate a thermal noise spectrum equal to $\SVVN(f)$.

Thirdly, a noise spectrum is occasionally~\cite{Caves1982,Castellanos-Beltran2008} expressed as a number of noise quanta
\begin{equation}
	N_\mathrm{N}(f) \equiv \frac{\PN(f)}{hf} = \frac{\kB \TN}{hf}.
\end{equation}
The physical interpretation~\cite{Clerk2010} is that a measurement with bandwidth $B_f$ detects a noise power equivalent to quanta incident at a rate $N_\mathrm{N}B_f$.

\subsubsection{Predicting measurement uncertainty; sensitivity}
\label{sec:predictinguncertainty}

As seen from Fig.~\ref{fig:noise}, the noise voltage $\VN(t)$ obscures the signal $\VS(t)$.
In an experiment, we must try to estimate what $V(t)$ would have been had the noise not been present.
To be concrete, suppose the signal is
\begin{equation}
\VS(t)=V_\text{m} \cos(2\pi \fm t).
\label{eq:CosineVoltage}
\end{equation}
For example, $V_\text{m}$ might take one value if a qubit has state 0 and a different value if the qubit has state 1.
After acquiring a voltage record of duration $\taum$, which necessarily includes the noise, we want to estimate $V_\text{m}$ by taking the average (if $\fm=0$) or a Fourier integral (if $\fm \neq 0$). What error do we expect in this estimate?

To answer this, we must calculate the variance in our estimate over different random values of the noise.
This calculation (see Supplementary Section S3 B 1) gives for the expected error, i.e. the standard deviation in the estimate of $V_\text{m}$:
    \begin{numcases}{\sigma(V_\text{m})=\label{eq:sensitivity}}
    \sqrt{\frac{\SVVN(0)}{2\taum}} & if $\fm=0$
    \label{eq:sensitivity1}
    \\
    \sqrt{\frac{\SVVN(\fm)}{\taum}} & if $\fm \taum \gg 1$
    \label{eq:sensitivity2}
    \end{numcases}
This is the minimum uncertainty in our estimate of $V_\text{m}$. It is the reason why it is important to suppress the noise spectral density at the frequency of the signal. 

Because Eqs.~(\ref{eq:sensitivity1}-\ref{eq:sensitivity2}) determine the smallest signal that can be resolved in a measurement of duration $\taum$, $\sqrt{\SVVN(f)}$ is called the sensitivity of the voltage measurement.
In an experiment in which another quantity $X$ is transduced to a voltage, the sensitivity of the measurement of $X$ is
\begin{equation}
\sqrt{S_{XX}^\mathrm{N}(f)} = \left|\frac{\partial X}{\partial V}\right| \sqrt{\SVVN(f)}.
\label{eq:sensitivityX}
\end{equation}
provided that $\partial X/\partial V$ is constant over the range of the noise.

\subsection{The effects of noise}

\subsubsection{How noise appears in different types of measurement}

Let us now see how noise affects the data recorded in a reflectometry experiment, and how this changes when the data are represented in different ways.
Suppose we have a device, for example a qubit, which changes regularly between two states in such a way that the reflected signal switches between two amplitudes, ideally as in Fig.~\ref{fig:noise}(a).
A more realistic simulation must include noise (Fig.~\ref{fig:noise}(b)).
Here this is taken as white noise, meaning that $\SVVN (f)$ is independent of $f$ within the frequency range to which the experiment is sensitive.
In the time domain, the effect of noise is to increase the scatter of the data points.
In the frequency domain (Fig~\ref{fig:noise}(c)), the noise appears as a nearly uniform background in the power density, between the sharp signal sidebands which contain the useful information.

Our typical task is to deduce the device state based on a segment of the time trace.
As explained in Section~\ref{sec:demodulation}, we begin by demodulating the signal and low-pass filtering it to keep only the spectral range of interest.
The top trace in Fig.~\ref{fig:noise}(d) shows the amplitude of such a demodulated filtered signal.
The two levels are barely evident, and obscured by noise near the carrier frequency that has been shifted downwards by demodulation and survives the filter.
To identify the device state, the trace is therefore averaged over an interval $\taum$, beginning just after the switching event.

When the averaged data are plotted as a histogram (Fig.~\ref{fig:noise}(e-f)), the two levels become evident.
With sufficiently long $\taum$, the distribution separates clearly into two peaks, whose width is set by Eq.~\eqref{eq:sensitivity2}.
To assign the device state based on the record from a single measurement interval, the criterion is obvious: if the average signal is above the midpoint threshold, the device is in the high-reflection state, otherwise it is in the low-reflection state.
The probability to assign the state correctly is called the fidelity
\footnote{In other words, the fidelity is 
\begin{equation}
	\mathcal{F} \equiv 1-\frac{P(`0\text{'}|1)+P(`1\text{'}|0)}{2},
	\label{eq:footFidelity}
\end{equation}
where $P($`$0$'$|1)$ is the probability to label the state as `$0$' when the true state is 1, and vice versa.
A related quantity is the visibility~\cite{Elzerman2004,Gossard2009}, defined as 
\begin{align}
	\mathcal{V}	&\equiv 	1-(P(`0\text{'}|1)+P(`1\text{'}|0))	\\
				&=		2\mathcal{F}-1.
\end{align}
Occasionally one sees $\mathcal{V}$ called the fidelity~\cite{Gambetta2007,Touzard2019}.
\newline
In terms of the signal-to-noise ratio $\mathrm{SNR} \equiv \frac{\mathrm{Peak~spacing}}{\mathrm{Standard~deviation}}$, the fidelity is
\begin{equation}
	\mathcal{F} = \frac{1}{2}\left(1+\mathrm{erf} \left(\frac{\mathrm{SNR}}{2\sqrt{2}}\right)\right),
	\label{eq:fidelity}
\end{equation}
provided that the peaks in a histogram such as Fig.~\ref{fig:noise}(e-f) are Gaussian and that the qubit does not decay during measurement~\cite{Gambetta2007}.
}
and is plotted in Fig.~\ref{fig:noise}(g).
Increasing the integration time or decreasing the noise allows higher-fidelity readout.
The ability to distinguish the two states can also be expressed as the amplitude signal-to-noise ratio (SNR), which here is defined as the ratio between the peak spacing and the standard deviation
\footnote{Sometimes SNR is defined as the ratio of signal and noise power, instead of amplitude. With this definition, the SNR plotted in ~\ref{fig:noise}(g) is squared.} 
.
Once the two histograms become distinct, i.e.\ $\mathrm{SNR} \gtrsim 2$, the error probability depends exponentially on SNR, meaning that even small improvements in SNR lead to valuable improvements in the fidelity.

Sometimes it useful to plot similar data as a two-dimensional histogram in the $(\VI,\VQ)$ plane (Fig.~\ref{fig:noise}(h)) so that the two device states appear as two spots.
This makes it clear if the phase as well as the amplitude is changing.
In this figure, only the amplitude is varied, so the two spots lie on the same bearing from the origin.

\subsubsection{Sources of noise in realistic circuits}
\begin{figure}
	\centering
    \includegraphics{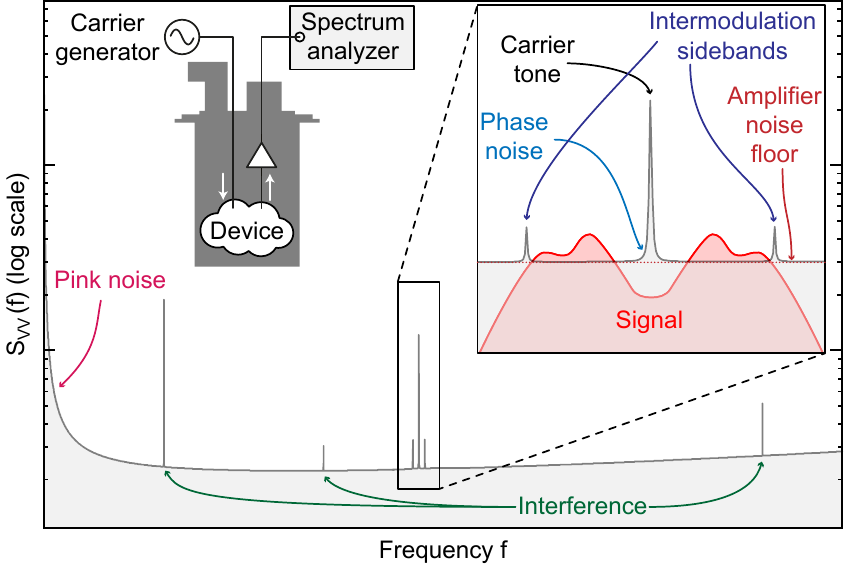}
	\caption{Cartoon showing different contributions to the spectral density in a reflectometry experiment. Left inset: experimental schematic. The device is illuminated by a carrier tone, and the emitted spectrum is measured. Right inset: Zoom-in near the carrier frequency. Also shown is a possible spectrum for the signal being measured. The smaller the overlap of this signal with the noise spectrum, the easier it will be to identify.
	\label{fig:realisticnoise}
	}
\end{figure}
Here are some types of noise encountered in an RF measurement, how you identify them, and what to do about them~\cite{Horowitz2015}. Figure~\ref{fig:realisticnoise} illustrates how some of them appear in the spectral density.
In general the system noise is a combination of contributions from the device, the tank circuit, and the amplifier chain~\cite{Muller2013}.
\begin{enumerate}
	\item \emph{Interference} from electronic instruments, power supplies, and radio transmitters appears as sharp peaks in the spectral density. It can be minimised by avoiding ground loops, by electromagnetic shielding, and by measuring at a frequency away from interference peaks. Often the most insidious interference comes from low-frequency signals, such as vibrations and power-line pickup, that create intermodulation sidebands near the carrier frequency.
	
	\item \emph{Pink noise} is a generic term for noise that is most intense at low frequency. Phenomenologically it is often found that $\SVVN(f) \propto 1/f$. A common cause is charge switchers in the device being measured. The cure for pink noise is to shift your signal away from zero frequency by using a carrier frequency above the relevant frequency band, scanning quickly, and/or making a lock-in measurement.
	
	Similar effects can also create a pair of spectral wings near the carrier frequency. These are often called phase noise\cite{Navid2004}. Again, the cure is to make your signal vary in a way that puts its frequency components outside the noisy range.
	\item \emph{Thermal noise} is the black-body radiation emitted by any dissipative circuit element. In cryogenic experiments, electromagnetic thermal noise coming away from the device is rarely a problem; thermal noise going towards the device must be suppressed with attenuators, filters, and circulators (Fig.~\ref{fig:amplifierchain}).
	The spectral density of thermal noise into a matched load is
	\begin{equation}
		\SVVN(f) = \frac{hfZ_0}{e^{hf/\kB T}-1}.
		\label{eq:thermalnoise}
	\end{equation}
	If $hf \ll \kB T$, which is often the case, then
	\begin{equation}
		\SVVN(f) \approx \kB T Z_0,
		\label{eq:RayleighJeans}
	\end{equation}
	which is where Eq.~\eqref{eq:noisetemperature} comes from (but see Footnote~\cite{foot:noisetemperature}).

	\item \emph{Shot noise} is broadband noise caused by a current flowing through a tunnel barrier.
	The spectral density of this current noise is given by Eq.~\eqref{equ:SII}.
	This current noise transforms to voltage noise at the amplifier input.
	Being fundamental, shot noise is generally unavoidable, but it is also usually small.
	\item \emph{Quantum noise} is the result of quantum fluctuations. 
	Under most conditions, an amplifier's noise temperature must satisfy the \emph{standard quantum limit} (SQL) for continuous measurements, which means~\cite{Clerk2010}
	\begin{equation}
		\TN\geq \frac{hf}{2\kB}.
		\label{eq:quantumNoiseTemperature}
	\end{equation}
	In electronic experiments it is very hard to reach this limit, let alone surpass it.
	It is discussed further in Section~\ref{sec:SQL}.
	\item \emph{Amplifier noise} is the noise added by the amplifiers. It includes the effects listed above, but also contributions from other physical processes inside the amplifiers, which are generically called technical noise. The noise from a commercial RF amplifier usually varies smoothly with frequency (Fig.~\ref{fig:AmpComparison}), leading to a nearly uniform spectral background which is hard to evade. In optimised experiments, technical amplifier noise usually dominates other sources. It can often be mitigated by buying a high-end amplifier and cooling it down.
\end{enumerate}

\subsection{Suppressing noise using cryogenic amplifiers}

To suppress noise, often the greatest single improvement is to cool down the primary amplifier.
Low temperature suppresses thermal noise and switching noise in semiconducting components.
It also makes it possible to use superconductors.
Since quantum electronic experiments are usually carried out in a dilution refrigerator, the required cold space is readily available.
Virtually all advanced high-frequency measurements in this field use cryogenic semiconductor amplifiers, and many now use superconducting amplifiers as well.

\subsubsection{Amplifier chains}
\begin{figure}
    \centering
    \includegraphics{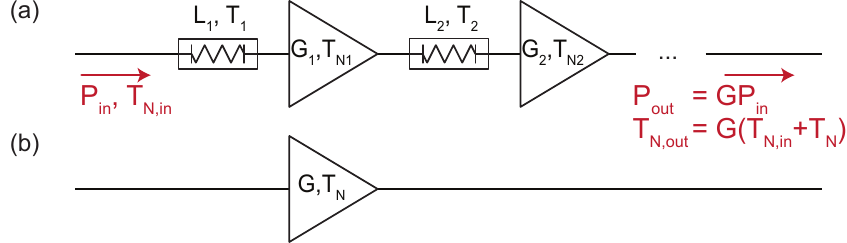}
    \caption{(a) An amplifier chain, consisting of a series of amplifiers, each with gain $G_i$ and noise temperature $T_{\mathrm{N}i}$, connected by transmission lines each with loss $L_i$ and at temperature $T_i$.
    (b) The corresponding equivalent amplifier, with gain and noise given by Eqs.~\eqref{eq:FriisGain} - \eqref{eq:FriisNoise}.
    If the chain receives a signal power $\Pin$ superimposed on thermal noise at temperature ${\TN}_\mathrm{,in}$, it will output signal power $\Pout$ superimposed on thermal noise at temperature ${\TN}_\mathrm{,out}$.
    \label{fig:friis}
    }
\end{figure}

To appreciate the benefit of a cryogenic primary amplifier, we need to know how the noise of a measurement changes when a series of amplifiers is cascaded as in Fig.~\ref{fig:friis}.
Each amplifier has a power gain ratio $G_i$ and a noise temperature ${\TN}_i$.
Furthermore we should take account of losses in the transmission lines leading to the amplifier inputs, each of which transmits a fraction $1/L_i$ of the power and is at physical temperature $T_i$.
Assuming there are no impedance mismatches, the entire chain behaves as a single amplifier~\cite{Pozar2012} with gain
\begin{equation}
	    G = \frac{G_1 G_2 \cdots}{L_1 L_2 \cdots}  \label{eq:FriisGain}
\end{equation}
and noise temperature
\begin{multline}
	    \TN = \left[(L_1-1)T_1 + L_1{\TN}_1\right] + \frac{L_1}{G_1} \left[(L_2-1)T_2 +L_2{\TN}_2\right] \\
	    +\frac{L_1L_2}{G_1G_2} \left[(L_3-1)T_3 +L_3{\TN}_3\right] \cdots 
\label{eq:FriisNoise}.
\end{multline}

The first term in each square bracket can usually be neglected, giving a noise temperature:
\begin{equation}
	\TN = L_1 {\TN}_1 + \frac{L_1 L_2}{G_1}{\TN}_2 + \frac{L_1 L_2 L_3}{G_1 G_2}{\TN}_3 + \cdots
	\label{eq:FriisNoise2}.
\end{equation}


Since all  of the gains appearing in Eq.~\eqref{eq:FriisNoise} are usually much greater than unity, it is clear that the overall noise is dominated by the first amplifier in the chain.
This is why a low-noise primary amplifier is so important.
Later amplifiers still contribute noise, but to a lesser extent.
Equation~\eqref{eq:FriisNoise2} also tells us that transmission loss before the amplifier should be minimised.
This points to another advantage of cryogenic amplifiers; they can be connected to the device by a short length of superconducting cable, which has extremely low loss.

\begin{figure}
    \centering
    \includegraphics{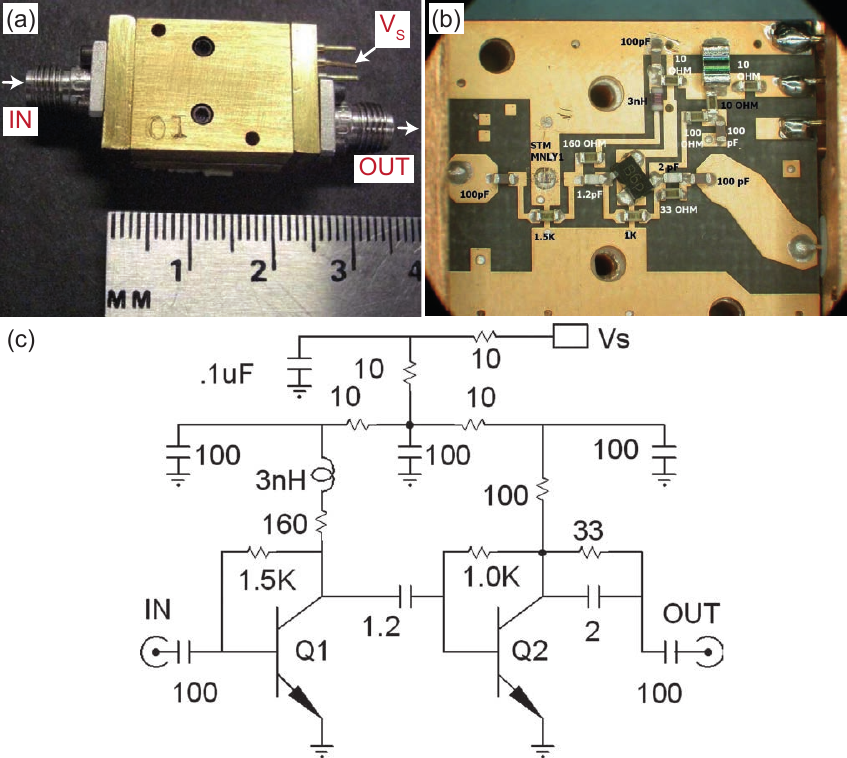}
    \caption{(a) A BJT cryogenic semiconductor amplifier in its package, with connectors marked for the RF input and output and for the dc power supply voltage $V_\mathrm{S}$.
    (b) Photograph of amplifier circuit board.
    (c) Circuit diagram. Adapted from Ref.~\onlinecite{Weinreb2009}.
    \label{fig:semiconAmp}
    }
\end{figure}

\subsubsection{Semiconductor amplifiers}
\label{sec:semiconAmp}

Packaged cryogenic semiconductor amplifiers (Fig.~\ref{fig:semiconAmp}) are commercially available and easy to use.
The active elements are usually SiGe bipolar junction transistors (BJTs)~\cite{Weinreb2009} or InGaAs/InAlAs/InP high-electron-mobility transistors (HEMTs)~\cite{Schleeh2012}.
Optimum amplifier design is a trade-off between noise, impedance, and stability.
For example, decreasing transistor size improves the high-frequency response by increasing the bandwidth, but also increases switching noise.
At present it appears that at microwave frequencies (above about 4~GHz), HEMTs generally work better.
At lower frequencies, BJTs are often preferred despite a sub-optimal noise temperature because of their good wideband input impedance matching, which prevents unwanted standing waves or, even worse, self-oscillations.
In both cases, the amplifier should be mounted at the 4~K stage of the refrigerator.

\subsubsection{Superconductor amplifiers}
\begin{figure}
    \centering
    \includegraphics{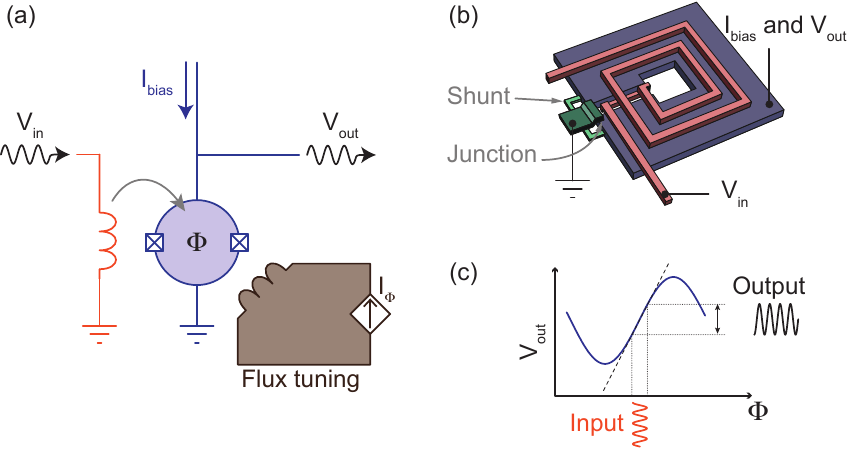}
    \caption{The SQUID microstrip amplifier~\cite{Muck2010} (a) Working principle.
    The active element is a SQUID (center) biased by a dc current $\Ibias$ greater than the critical current, which leads to voltage $\Vout$ at the output port.
    When a voltage $\Vin$ is applied at the amplifier input, it excites a current in the microstrip resonator.
    This modulates the flux $\Phi$ through the SQUID washer, which in turn modulates the critical current and therefore $\Vout$.
    A flux tuning loop adjusts the dc flux to the point of maximum response.
    (b) Geometry of washer (blue and green) and microstrip coil (pink), which is separated from the washer by an insulating layer (not shown).
    A pair of shunt resistors, not drawn in the circuit diagram, suppresses hysteresis.
    (c) Solid curve: $\Vout(\Phi)$ characteristic of an ideal SQUID.
    The modulation of the flux by the input signal and its effect on the output are sketched. 
    \label{fig:squidAmp}
    }
\end{figure}

The very quietest RF amplifiers are based on superconductors.
The active elements are Josephson junctions.
The simplest superconducting amplifier is the superconducting quantum interference device (SQUID) amplifier (Fig.~\ref{fig:squidAmp}).
This exploits the fact that the critical current of a dc SQUID\footnote{The name `dc SQUID' means that the SQUID has more than one junction and therefore shows interference when current-biased at zero frequency. Even when the device is operated at radio frequency, do not confuse it with an `rf SQUID', which is a loop with only one junction~\cite{Clarke2004}.} depends on the magnetic flux $\Phi$ enclosed between its two junctions~\cite{Muck2010}.
When the SQUID is biased above its critical current, changes in critical current lead to changes in the voltage across the terminals.
A small flux generated by the input signal therefore leads to a comparatively large output voltage.
To maximise the oscillating flux, the input coil is usually engineered as a resonator, for example the microstrip resonator shown in Fig.~\ref{fig:squidAmp}(b).

\begin{figure}
    \centering
    \includegraphics{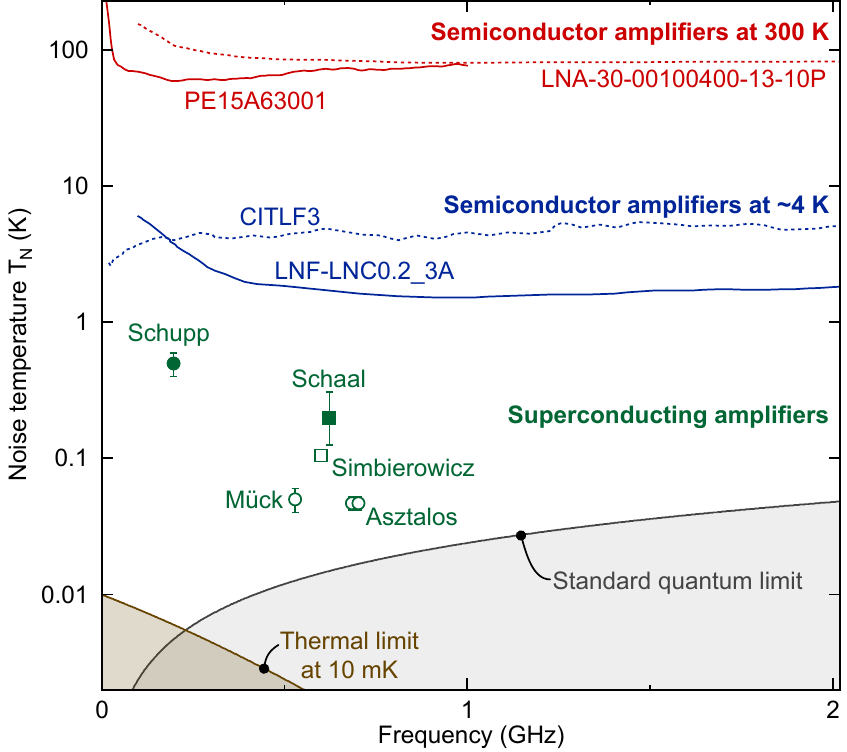}
    \caption{Noise temperature as a function of frequency for selected state-of-the-art amplifiers.
    Lines are examples of low-noise commercial amplifiers operating near room temperature and near 4~K.
    Symbols are superconducting amplifiers operating in dilution refrigerators.
    Amplifiers used in a reflectometry configuration are Schupp~\etal ~\cite{Schupp2020} using a SQUID, and Schaal~\etal~\cite{Schaal2020}, using a JPA.
    Lower-noise amplifiers not yet used for reflectometry include SQUID amplifiers (M\"{u}ck~\etal\cite{Muck2001}, Asztalos~\etal\cite{Asztalos2010}) and JPAs (Simbierowicz~\etal\cite{Simbierowicz2018}).
    The shaded regions lie beyond the standard quantum limit (Eq.~\eqref{eq:quantumNoiseTemperature}) and the thermal limit at 10~mK (Eq.~\eqref{eq:TNphysical}).
	Footnote~\onlinecite{foot:noisetemperature} explains why the thermal limit is not equal to the physical temperature.
    \label{fig:AmpComparison}
   }
\end{figure}

SQUID amplifiers achieve better sensitivity than semiconductor amplifiers, but are more difficult to operate.
The working bandwidth is small and not easily tunable, because it is set by the properties of the resonant coil.
The power handling is poor because of the SQUID's non-linearity, although this can be mitigated by injecting a cancellation tone to null out the carrier tone~\cite{Laird2020_CNT}.
SQUIDs must also be well-shielded from magnetic fields, even nominally constant ones such as from superconducting magnets in the same room.
Nevertheless, SQUID amplifiers hold the record for voltage sensitivity at low RF frequency (Fig.~\ref{fig:AmpComparison}) and have successfully been used for reflectometry~\cite{Court_2005,Schupp2020}.

\begin{figure}
    \centering
    \includegraphics{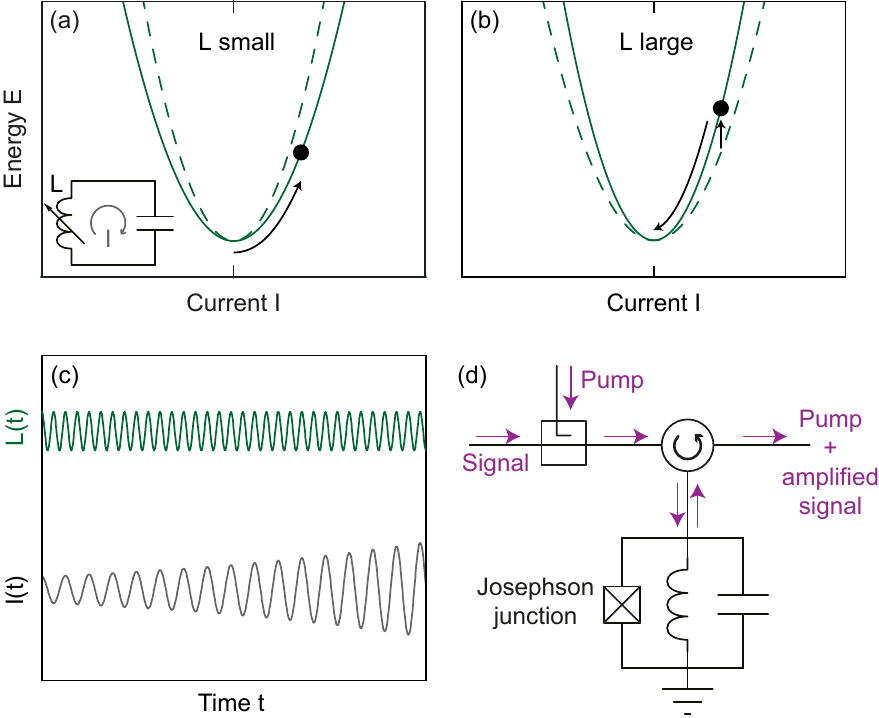}
    \caption{Parametric amplification. Panels (a) and (b) show energy as a function of current in an $LC$ resonator, whose circuit is shown in the inset. The state of the resonator is indicated by a dot.
    Twice per cycle, the resonator is switched between its high-$L$ and low-$L$ condition, so that the current increases under the low-$L$ condition (a) and decreases under the high-$L$ condition (b).
    In this way, the energy in the resonator increases in each cycle.
    (c) Sketch showing how $L$ is modulated at twice the resonance frequency causing the amplitude of $I$ to increase.
    (d) Simple Josephson parametric amplifier.
    The Josephson junction embedded in the $LC$ resonator is a non-linear element whose effective inductance is modulated by a pump tone. 
    To operate the amplifier, a circulator feeds the pump and the signal to the resonator, and routes the reflection containing the amplified signal towards a semiconductor postamplifier.~\cite{Aumentado2020}
    \label{fig:parAmp}
    }
\end{figure}

Another type of superconducting amplifier is the Josephson parametric amplifier (JPA)~\cite{Castellanos-Beltran2007,Yamamoto2008,Aumentado2020}
To understand the principle of parametric amplification, consider the $LC$ resonator shown in the inset of Fig.~\ref{fig:parAmp}(a).
The energy stored in the inductor depends quadratically on the instantaneous current $I(t)$ (Fig.~\ref{fig:parAmp}(a-b)).
Now suppose the inductance is changed twice per oscillation cycle, being increased when $I(t)$ is maximal and decreased when $I(t)$ is zero.
The effect is to increase the stored energy in each repetition, thus amplifying the current (Fig.~\ref{fig:parAmp}(c)).
In practice the inductance does not need to jump abruptly, but is modulated sinusoidally at twice the resonator frequency as shown
\footnote{This type of operation, in which the pump is at twice the signal frequency, is called degenerate parametric amplification. This is not the only way to operate a parametric amplifier, but it is the most intuitive.}

The simplest implementation of a JPA is shown in Fig.~\ref{fig:parAmp}(d).
The variable inductance is provided by a Josephson junction, whose inductance depends on the current according to~\cite{Devoret2004, Aumentado2020}
\begin{equation}
	\LJ(I)=\frac{\hbar}{2eI_0}\frac{1}{\sqrt{1-I^2/I_0^2}}
		\label{eq:LJdefinition}
\end{equation}
where $I_0$ is the critical current\footnote{This equation follows from the Josephson equations and the definition $V=L\frac{dI}{dt}$. Some authors~\cite{Feldman1975, Castellanos-Beltran2007} prefer to define the inductance $L'$ by $V=\frac{d}{dt}(L'I)$, which leads to $\LJ'=\frac{\hbar}{2eI_0}\frac{\sin^{-1}(I/I_0)}{I/I_0}$. This is equivalent to Eq.~\eqref{eq:LJdefinition} when $I \ll I_0$.}.
To modulate the inductance, $I(t)$ should be driven by an intense pump tone.
Since $\LJ(I)$ is an even function, pumping at the resonator frequency $f_{LC}$ generates the modulation at $2f_{LC}$ that Fig.~\ref{fig:parAmp}(c) requires.
More complex implementations of the JPA principle distribute the amplifier's non-linearity over a series of junctions.
Advanced JPAs, typically working at around 7~GHz, can reduce all other noise sources to the extent that intrinsic quantum noise given by Eq.~\eqref{eq:quantumNoiseTemperature} is the dominant remaining contribution~\cite{Castellanos-Beltran2008}.
In a reflectometry experiment~\cite{Schaal2020}, a JPA has attained a noise temperature of $\sim 200$~mK at 622~MHz, an order of magnitude better than a semiconductor amplifier.

Among the most advanced JPAs are travelling-wave parametric amplifiers (TWPAs), which replace the single resonator of Fig.~\ref{fig:parAmp}(d) by an array of cells through which the signal passes once~\cite{Macklin2015}.
As well as the convenience of operating in transmission instead of reflection, TWPAs allow good bandwidth and power handling compared with reflective JPAs, although fabrication is more difficult and the sensitivity is so far not quite as good.
To our knowledge no TWPA has yet been operated below about 4~GHz.
The advantages of different kinds of parametric amplifiers were recently reviewed by Aumentado~\cite{Aumentado2020}.

Although JPAs of various kinds now have excellent performance at microwave frequency and many experiments have operated close to the bounds set by quantum mechanics, radio-frequency JPAs are less well-developed.
This is illustrated by Fig.~\ref{fig:AmpComparison}, which compares the noise performance of different radio-frequency amplifiers.
The quietest amplifiers in this frequency range are SQUIDs, although both SQUIDs and JPAs are still some way from the standard quantum limit.

\begin{figure*}
    \centering
    \includegraphics{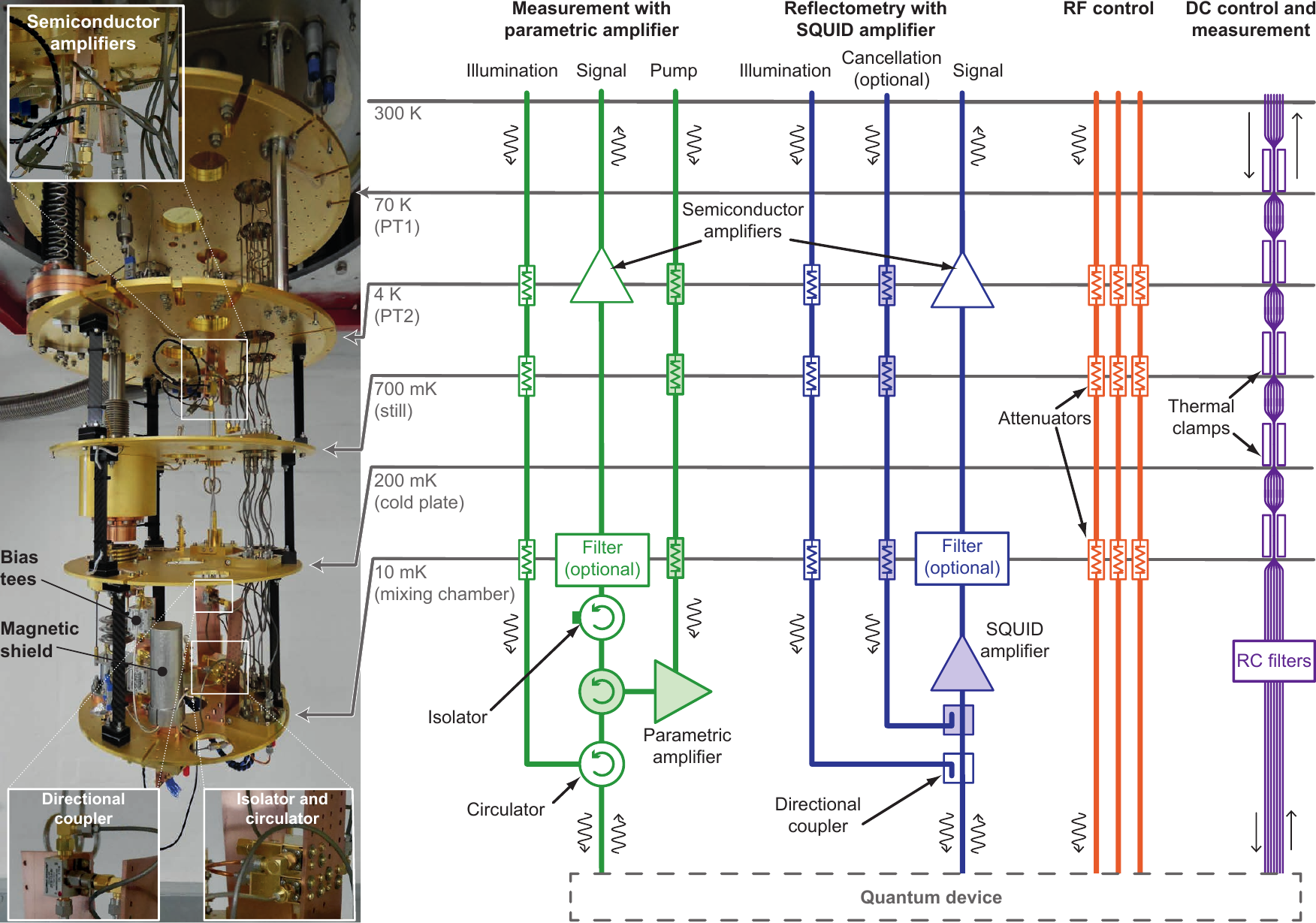}
    \caption{Right: Typical refrigerator wiring for high-frequency measurements. This setup includes (left to right):
    Measurement chain using reflection-mode parametric amplifier (based on Ref.~\onlinecite{Schaal2020});
    measurement chain using SQUID amplifier (based on Ref.~\onlinecite{Schupp2020});
    high-frequency control lines using coaxial cable;
    quasi-dc control and measurement lines using cryogenic loom.
    Such a setup can also be operated without superconducting amplifiers by omitting the shaded components.
    To protect the quantum device from thermal radiation, all lines are attenuated and/or filtered at various stages inside the refrigerator, and thermally clamped to minimise the heat load on the mixing chamber.
    Components drawn below the 10~mK line are in thermal contact with the mixing chamber plate, but not necessarily below it.
    Left: A Triton 200 refrigerator wired in a similar arrangement.
    This refrigerator is equipped with a flux-pumped parametric amplifier~\cite{Simbierowicz2018}, and therefore contains two bias tees (not drawn in the circuit diagram) through which the amplifier is biased.
    Selected components are labelled.
    \label{fig:amplifierchain}
    }
\end{figure*}

All cryogenic amplifier chains require careful engineering to operate with the best performance.
Figure~\ref{fig:amplifierchain} shows a typical wiring scheme.

\subsection{Opportunities and challenges}
\subsubsection{The standard quantum limit}
\label{sec:SQL}
How quiet can an amplifier be?
Quantum uncertainty limits the sensitivity of any continuous measurement, because the back-action induced at one time disturbs the observable's state a short time later.
To be precise, for an electromagnetic mode associated with a voltage
\begin{equation}
V(t)=\VI \cos( 2\pi f t) + \VQ \sin( 2\pi f t),
\label{eq:Vquadratures}
\end{equation}
a measurement of $\VI$ perturbs $\VQ$ and vice versa.
If an amplifier has large gain and is phase-preserving, meaning that it is equally sensitive to $\VI$ and $\VQ$ (which is the usual situation) this imposes a minimum noise given by Eq.~\eqref{eq:quantumNoiseTemperature}.
This is the standard quantum limit (SQL).

Most electronic amplifiers work far from this limit.
However, if an experiment is so sensitive that the SQL becomes a problem, then there is a way to evade it by combining two tricks.
The first trick is to make the observable of interest appear in only one quadrature of Eq.~\eqref{eq:Vquadratures}.
For example, to measure the reflected amplitude as in Fig.~\ref{fig:noise}, the phase can be defined so that the signal is entirely in the $\VI$ quadrature.
The second trick is that the parametric scheme shown in Fig.~\ref{fig:parAmp} only amplifies a signal with the correct phase relative to the pump; the complementary phase is attenuated.
By pumping in a way that amplifies only $\VI$, the observable can therefore be measured with arbitrary precision.
In the $(\VI,\VQ)$ plane (Fig.~\ref{fig:noise}(h)), the noise spots are squeezed along one axis at the price of spreading out along the other.
Squeezing measurements have been applied for precise measurements of microwave electromagnetic fields~\cite{Castellanos-Beltran2008} and thereby to electron spins~\cite{Bienfait2017} and superconducting qubits~\cite{Eddins2018}; the same strategy should work for spin qubits.

In a reflectometry experiment, this is possible because the axis of squeezing can be controlled by the phase between the carrier tone and the amplifier pump.
Remarkably, squeezing sometimes also helps measure incoherent emission, which has no defined phase.
In a measurement without squeezing, the sensitivity to such a signal is maximal at a cavity resonance but declines for frequencies on either side.
By injecting a squeezed signal into the cavity, the optimum frequency range can be extended while the optimal sensitivity stays the same~\cite{Malnou2019}.
This strategy is therefore useful when the possible frequency range of the target signal is greater than the linewidth of the cavity.
It was invented to search for dark matter, for which the frequency scan range must be very large~\cite{Backes2021}.
The subject of quantum limits on continuous measurements is an intricate one, reviewed in detail by Clerk \emph{et al.}~\cite{Clerk2010}.
There is probably room for future circumventions of the SQL, both by applying known schemes in new experiments and by devising even more ingenious ones.

\subsubsection{New types of quantum amplifier}
The first radio-frequency SET used a cryogenic HEMT amplifier with a noise temperature of $\TN=10$~K at 1.7~GHz, which at the time was the state of the art~\cite{Schoelkopf1998}.
As Fig.~\ref{fig:AmpComparison} shows, amplifiers have improved greatly, but there is still room to do better.
In the next few years, we hope that rf superconductor amplifiers become as widely available and user-friendly as semiconductor amplifiers are today.
As well as having low noise, they will also need to operate across a wide frequency range and handle comparatively large signals without saturating.
This will be particularly important when measuring many devices using frequency multiplexing, since the total input power scales with the number of devices.

As seen from Fig.~\ref{fig:AmpComparison}, there is still a need to extend the technology of quantum-limited microwave amplifiers down to rf frequencies.
However, at the lowest frequencies the SQL becomes less important than thermal noise.
In a 10~mK dilution refrigerator, thermal noise overtakes the SQL below 229~MHz.
While there is scope for much quieter amplifiers than exist today, there will be no particular benefit from reaching the SQL at this frequency unless there are equal advances in ultra-low-temperature electronics~\cite{Jones2020}.

Another need is for quantum-limited amplifiers that can operate in a magnetic field.
In spin quantum computing and for magnetic resonance, the device being measured necessarily operates in a field between a few tens of mT and a few~T.
Any amplifier based on alumina Josephson junctions must therefore be placed some distance away, which costs space and sacrifices part of the signal to transmission losses.
Interesting recent approaches that may one day overcome this problem include TWPAs based on kinetic inductance instead of Josephson inductance~\cite{Malnou2021}, and new Josephson junctions based on nanowires~\cite{Luthi2018} and graphene~\cite{Kroll2018} which can tolerate in-plane magnetic fields up to 1~T.
Alternatively, a parametric amplifier could be based on a different degree of freedom, such as mechanical motion~\cite{Ockeloen-Korppi2017}
or quantum capacitance~\cite{Cochrane2021}.
Eventually a single resonator, with many devices embedded within it, might serve as a readout cavity and a parametric amplifier cavity simultaneously~\cite{Eddins2019}.
Such a device would be the ultimate combination of sensitivity and density in future large-scale quantum circuits.

\section{Reading out multiple channels: The challenge of scaling up}

\label{sec:Multiplexing}
Many experiments at the frontier of nanoscale electronics require fast concurrent impedance measurements, for instance in quantum computers where the execution of error correcting codes potentially amounts to the correlated readout of a large number of qubits. However, addressing this challenge via brute force duplication of a measurement setup quickly becomes unwieldy, in terms of the physical footprint of the duplicate sub-systems, of their power dissipation, and of unwanted interaction between them \cite{Reilly_IEDM}. Duplicating all the necessary readout hardware for every parallel measurement is hardly a scalable approach. Multiplexing readout signals can dramatically improve the efficiency of these sub-systems by making use of total available bandwidth or duty cycle in the time domain. For measurements that must be performed simultaneously, frequency multiplexing is possible but requires a means of generating, amplifying, separating, and measuring signals across multiple frequencies. Conversely, time-domain multiplexing can be used for parallel-to-serial translation of measurement data. Both techniques can be combined \cite{ruffino2021integrated} to enable hardware-efficient readout of multiple devices at high frequencies.

\subsection{Frequency multiplexing}

\begin{figure}
     \includegraphics[width=\columnwidth]{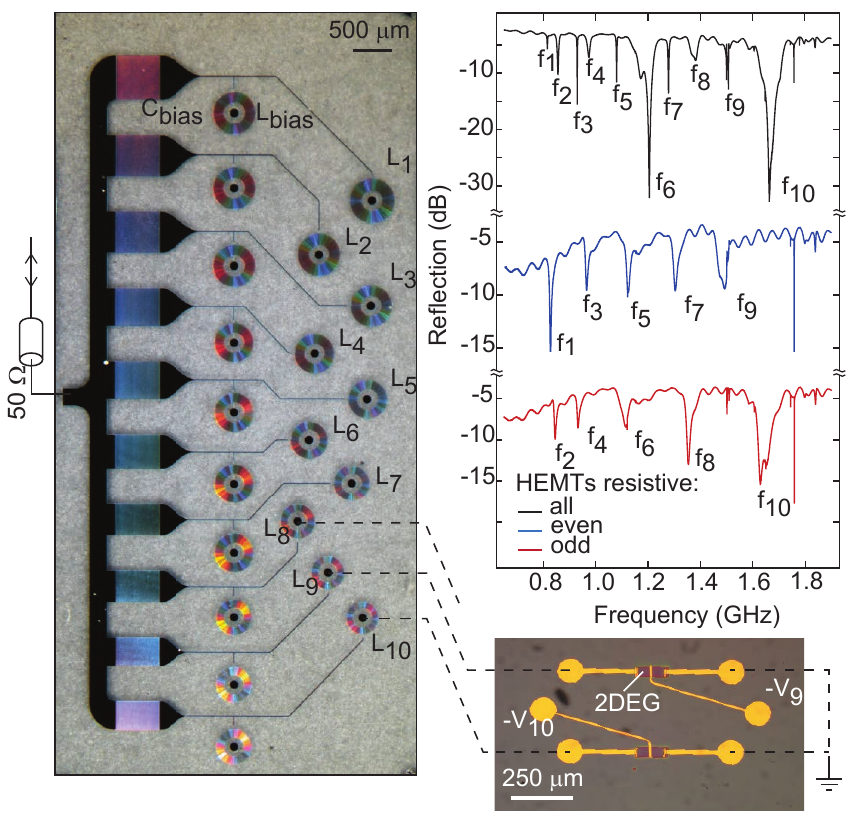}
    \caption{Frequency division multiplexing makes use of a bank of lithographically defined resonators and bias tees. Here, the resonators are lumped element circuits fabricated using niobium superconductor on a sapphire substrate. The frequency domain response of the resonators is shown on the right, using HEMT devices (shown lower right) as resistors to modulate the $Q$-factor of each resonator. Adapted from Ref.~\onlinecite{Reilly2014}.
    \label{fig:multiplexing}
    }
\end{figure}

The rf reflectometry technique is immediately amenable to the parallel readout of multiple devices or sensors by encoding each device with a unique frequency or channel. Such an approach, which is usually termed \textit{frequency division multiplexing (FDM)} at rf frequencies or equivalently, \textit{wavelength division multiplexing (WDM)} in the optics and photonics communities, is the mainstay of modern communication systems. Owing to the orthogonality of signals at different frequencies, the FDM technique enables the transmission of multiple frequency channels using a single transmission-line and amplification chain. Here we review the sub-components that make up a multi-channel system and describe how this approach enables efficient simultaneous readout of a large number of devices, for instance in the operation of a scaled-up qubit array.

\subsubsection{Multiplexed resonators}

Frequency multiplexing brings several new challenges in the design of the physical resonator structures used in an rf reflectometry setup.  Most importantly, the footprint of the $LC$ network can become critical since the signal feedline must branch into parallel lines that couple to each resonator simultaneously. At the frequency of one resonator the splitting of the feedline creates a `stub' in which a parallel length of line is terminated with an open circuit, i.e., it is terminated with an $LC$ network that is off-resonance. The stub allows interference of the standing wave reflected from the off-resonant open end of the parallel line with the signal feeding the resonator\cite{Pozar2012}. The phase accumulated by the parallel path, which depends on the length of the line, modifies the effective impedance of the network since it alters the ratio of current to voltage. This complication can be accounted for (or even exploited) in a few-channel system \cite{Buehler2004} but becomes increasingly challenging to address as the number of resonators and number of stubs is scaled up. 

A potential solution is to minaturize the entire network so that the total path length $l$ of any section is far smaller than the wavelength $\lambda$ of the signal\cite{Reilly2014}. A rule of thumb is to set $l < \lambda/20$, ensuring that the entire network is in the `near-field' regime where the inductive and capacitive contributions can be considered as lumped elements rather than a distributed circuit.

Defining the entire network lithographically enables a large number of resonators and feedlines to be integrated on a chip far smaller than the signal wavelength (recall, 1~GHz $\sim$ 25~cm). 
However, this requires superconducting materials, since miniaturised planar inductors made from normal metals such as copper have appreciable resistance.
An implementation of the on-chip superconducting approach, including both resonators and bias tees, is shown in Fig.~\ref{fig:multiplexing}. It is worth noting that the use of superconducting materials makes operation in large magnetic fields challenging. The need for magnetic field-compatible resonators has motivated recent approaches to mitigate adverse effects such as the penetration of flux into the superconductor~\cite{kroll2019}.

A further consideration with frequency multiplexing is that each resonator must operate at a unique frequency, lifting the freedom to choose the frequency where sensitivity is maximized. Rather than a narrowband system where each component has been selected to operate at a sweet spot, a multiplexed setup requires wideband sensitivity for the hardware components. In some instances the underlying device physics limits the possible operating frequencies. Examples include limits on tunnel rates, or energy scales at which high frequencies lead to back-action. Ultimately this limits the number of available channels owing to frequency crowding\cite{Reilly2014}.

Finally, we draw attention to the additional challenges caused by inductive or capacitive crosstalk between resonators.
One challenge is that nearby resonators can shift each other's frequencies, necessitating careful design of the entire network.
A second challenge is that an excitation applied to one resonator can leak to another resonator at a nearby frequency.
One mitigation is to design nearby resonators to have well-separated frequencies (i.e. allowing a guard band between their resonances).
Another is to include on-chip ground planes and grounding rings.
Fortunately such approaches are already widely used in the rf integrated circuit community. 

\subsubsection{Heterodyne techniques for frequency multiplexing}
\textit{Heterodyne} detection is yet another mitigation strategy, where the up- and down-conversion process has a character inherently amenable to multiplexing. The process can proceed as follows, utilizing the notation of Section \ref{Sec:Basics}, where we have already discussed the principle of heterodyne detection, where the signal is demodulated using $\fLO \neq \fin$, the input signal frequency. This results in two signals at frequencies $\fout- \fLO$ and $\fout + \fLO$, where the second term is usually filtered out.

\begin{figure}
    \includegraphics{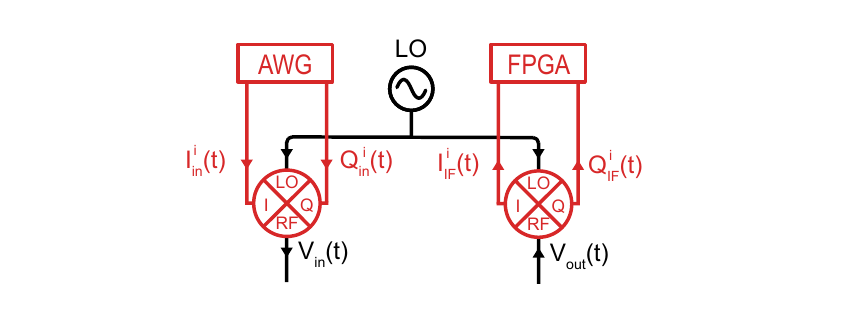}
   \caption{Up and down conversion using $IQ$ mixers~\cite{Sank2014}}
    \label{fig:Heterodyne}
\end{figure}

In the case of modulation, or up-conversion, we mix a local oscillator signal $\cos(\omega_\text{LO}  t$) and its quarter-phase shifted copy $-\sin(\omega_\text{LO}  t$) with a the modulating signals $I_\text{in}^i(t)=I_\text{in}^i\cos(\omega_i t)$ and $Q_\text{in}^i(t)=Q_\text{in}^i\sin(\omega_i t)$ respectively (for our purposes, we consider a signal from an arbitrary waveform generator (AWG)) and add them to form the input signal~\cite{Sank2014}
\begin{equation}
    \Vin(t)=\cos((\omega_\text{LO}+\omega_i)t+\phi). 
\end{equation}
at the up-converted angular frequency $\omega_\text{LO}+\omega_i$ (Fig.~\ref{fig:Heterodyne}). We ignore here the amplitude of the wave to focus on the frequency conversion.
This can be trivially extended to multiple frequencies, by asking the AWG to output modulation signals of the form $\omega_1$, $\omega_2$ and so on, such that we arrive at the desired number of upconverted frequencies, for example for $N$ qubits.

In the reverse process, during readout at multiple frequencies, the reflected signal
\begin{equation}
    \Vout(t)=\cos((\omega_\text{LO}+\omega_i)t+\phi)
\end{equation}
is mixed with the LO signal of frequency $\omega_\text{LO}$ and phase $\delta$ for demodulation or down-conversion.
This gives two outputs $I_\text{IF}^i(t)$ and $Q_\text{IF}^i(t)$ for the two quadratures~\cite{Sank2014}. We can represent these two signal as the real and imaginary parts of the complex signal  
\begin{align}
    &\VIF^i(t)=e^{i(\omega_i t+\phi-\delta)} \\
    &I_\text{IF}^i(t) =\frac{1}{2}\mathrm{Re}(\VIF^i(t)) \quad Q_\text{IF}^i(t) =\frac{1}{2}\mathrm{Im}(\VIF^i(t)).
\end{align}
This process therefore now results in the original $\omega_i$ term, which now carries the amplitude and phase information stemming from the physical phenomena we are measuring, for that particular subsystem, excited at that $\omega_i$, by the AWG. Both the up- and down-conversion process can be repeated for an arbitrary number of $\omega_i$, mixed into the same signal.

Readout of such a mixed signal, which resides at $\omega_i$ and not at DC, is typically accomplished via numerical post-processing, on-board on a field-programmable gate array (FPGA)~\cite{Sank2014}. The reflected and downconverted $I^i(t)$ and $Q^i(t)$, which contain $\omega_i$ components, are numerically mixed with the relevant $\omega_i$, which results in the $I^i$ and $Q^i$ information of each $\omega_i$ signal. For a particular $\omega_i$, this is done by multiplying the complex signal by $e^{-i\omega t}$ 
\begin{multline}
    \VIF^i(t)e^{-i\omega t}= I_\text{IF}^i(t)\cos(\omega_it)+Q_\text{IF}^i(t)\sin(\omega_it)\\
    +i(-I_\text{IF}^i(t)\sin(\omega_it)+Q_\text{IF}^i(t)\cos(\omega_it))
\end{multline}
and integrating, as follows:
\begin{align}
I^i=\sum_n I_\text{IF}^i(t)\cos(\omega_i t)+\sum_n Q_\text{IF}^i(t)\sin(\omega_i t)\\
Q^i=\sum_n Q_\text{IF}^i(t)\cos(\omega_i t)-\sum_n I_\text{IF}^i(t)\sin(\omega_i t)
\end{align}
where consecutive samples $n$ are digitally summed by the FPGA to remove the $2\omega_i$ components.
The above equation means that four integrals have to be performed numerically to find the result in the IQ plane. In practice, the same lookup table can be used to generate only two signals, a sine and a cosine, by offsetting the lookup by a quarter cycle in the table. These can then quickly be multiplied and summed with the signal to give the result, for each of our $\omega_i$.

\subsubsection{Constraints on amplifiers and related components}
Most cryogenic amplifiers used in rf reflectometry are designed to minimize the noise, maximise the gain, and achieve reasonable impedance matching. Optimizing all three parameters is difficult across a wide bandwidth. Semiconductor amplifiers that leverage feedback, for instance those based on SiGe transistors (see Section \ref{Sec:amplifiers}), achieve wideband operation at the price of increased noise from the feedback resistor. In comparison, amplifiers based on high electron mobility transistors (HEMTs) are typically configured to be open-loop and `noise matched', i.e., the $LC$ networks on the input and output of the transistor present an impedance that achieves the lowest noise and reasonable match. This is usually only possible across a narrow band. Frequency multiplexing is thus challenging for measurement setups that also require the lowest noise since encoding multiple parallel readout channels as a `comb' of frequencies is inherently wideband. Potentially, this limitation may be overcome by  making use of wideband superconducting amplifiers such as the traveling wave parametric amplifiers discussed in Section \ref{Sec:amplifiers}.

Beyond the bandwidth requirements, FDM brings two additional challenges for the amplification chain. Firstly, the total  power of signals at all frequency tones must be considered. If a system is to support 10 frequency channels, for instance, then the amplifier compression power must support an input power that is 10 times higher than for a single channel. Secondly, non-linearities in the transfer characteristics of the readout chain can lead to intermodular distortion in which signals at different frequencies are mixed (multiplied) to produce new frequency components, often overlapping other channels.

It is also worth mentioning the challenges associated with broadband transmission. Although cryogenic measurement setups are usually configured with substantial microwave filtering and attenuation to block radiation, FDM requires wideband transmission in order to accommodate all channels. Thus, experiments requiring the lowest electron temperature can be particularly difficult to combine with wideband frequency-multiplexing readout. 

\subsubsection{Digital approaches to signal generation and acquisition}

\begin{figure*}
    \includegraphics{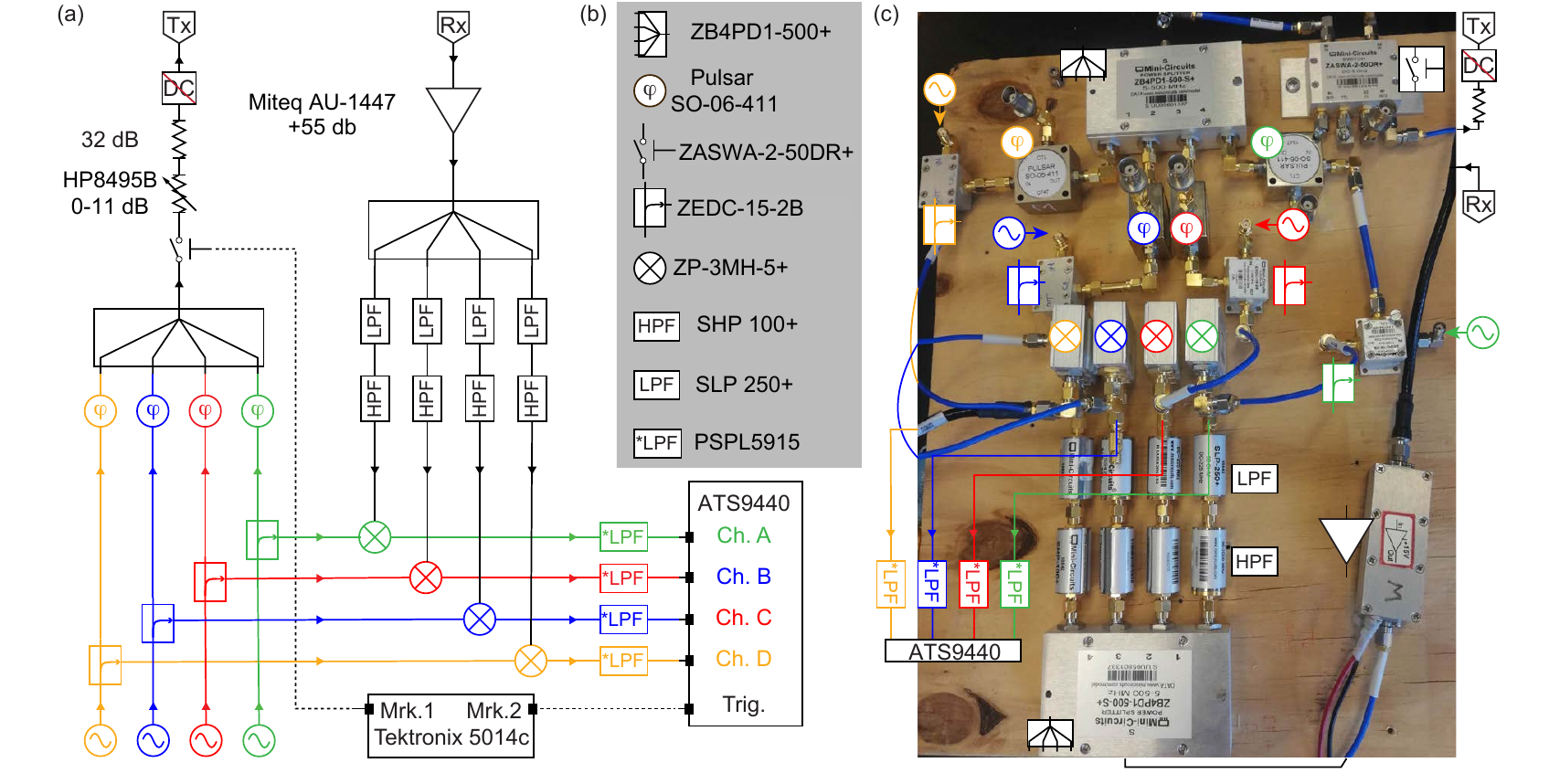}
   \caption{A typical modulation / demodulation setup built from analog components to enable frequency multiplexed readout, here for the simultaneous readout of four spin qubits.
   (a) Schematic of the circuit layout. 
   (b) Legend of specific components.
   (c) Photograph of the circuit layout.
   SMA cable "Rx" provides the undemodulated RF signals from the cryostat to an amplifier, before the signal is divided into four paths, each with its own filtering (LPF, HPF) and mixing with a local tone. Local tones (provided via directional couplers) carry the same frequency as the carriers (homodyne detection), thereby resulting in four dc signals that are detected by four independent channels of the digitizer (ATS9440).
   Figure adapted from Ref.~\onlinecite{Fedele2021}.
   }
    \label{fig:MultiDemod}
\end{figure*}

A key motivation for frequency division multiplexing is its potential to alleviate the burden posed by brute-force duplication of readout hardware. Although a single amplification chain can handle multiple frequency channels, demodulation hardware is still needed to create the baseband signals from which the device states are inferred. Conventional demodulation requires a separate frequency generator for each channel, as well as mixers, directional couplers, splitters, attenuators, and filters (see Fig.~\ref{fig:MultiDemod}). Using analog hardware for this purpose is cumbersome when the setup requires even a handful of frequency tones and quickly becomes unworkable for the large channel counts needed for scalable quantum computing. (Figure~\ref{fig:MultiDemod} shows the analog hardware required for demodulating four frequency channels.)

\begin{figure}
     \includegraphics[width=\columnwidth]{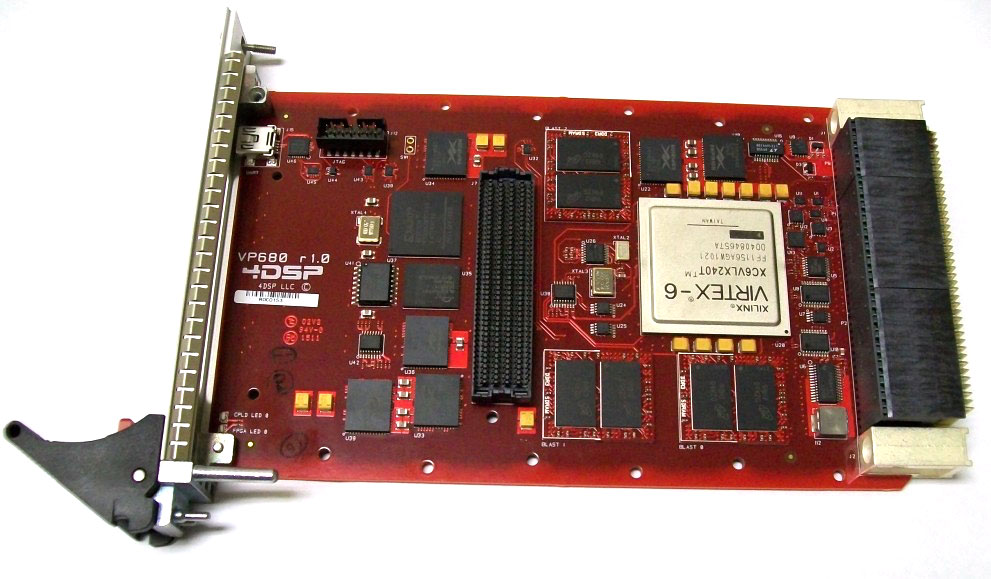}
    \caption{A fully integrated digital platform for generating and sampling rf signals combining FPGA for  digital signal processing with a ADC and DAC. Available from 4DSP Company.}
    \label{fig:DSP}
\end{figure}

Modern high-speed data converters and digital signal processing (DSP) can dramatically improve hardware efficiency when generating and detecting large numbers of frequency tones. Common approaches, such as the platform shown in Fig.~\ref{fig:DSP}, make use of a digital-to-analog converter (DAC) and an analog-to-digital converter (ADC), integrated with an FPGA that is accessible via a high-speed bus (such as the widely used PCIe platform).

This digital architecture first synthesizes a comb of frequencies in the digital domain, encoding frequency, phase, and amplitude of each tone. A wideband DAC then takes this stream of bits as input and generates the analog tones for transmission in a technique termed direct digital synthesis (DDS). On the receiver side, the collection of tones are amplified and then sampled at giga-sample per second clock rates using a wideband ADC. The digital output bit stream from the ADC generally feeds a bank of digital filters implemented in the FPGA, essentially performing a discrete Fourier transform. The platform writes to memory changes in amplitude and phase of carrier tones, referenced to the transmit signals generated via DDS. Adding or configuring new frequency tones is straightforward using a digital architecture for demodulation, in so far as the FPGA contains sufficient logic gates (and clock rate). 

Finally, we note that hybrid digital and analog architectures are now in widespread use. High-speed digitizers (ADCs) paired with analog mixers or frequency sources are particularly common.

\subsection{Time-division multiplexing}

 For many applications, the need for truly simultaneous measurements can be relaxed so that readout hardware can be used efficiently, switching between multiple devices sequentially or in an interleaved manner \cite{Reilly2015_Control,Schaalconditional2018,Gonzalez2019CMOS,jarratt2017,pauka2020}. The potential to share readout resources in this way is generally referred to as time-division multiplexing (TDM). Such schemes can be configured so that a subset of devices (or qubits) are being measured while others are being manipulated or prepared. In general however, measurement is usually the slowest task.

 Switches for implementing TDM are also not easy to come by. They need to operate at deep cryogenic temperatures, usually at the same temperature as the quantum devices, ($\lesssim 100$~mK), dissipating microwatts of power or less. For readout applications, such switches must also have extremely low insertion loss, since attenuation before the first stage amplifier degrades the SNR. A further requirement is a large on-off ratio (or isolation), which is important to minimize crosstalk. Finally, the impedance of the switch is critically important. 
 
Wide-band switches can be inserted in the readout chain in two places, depending on their attributes. For impedance-matched switches that have low insertion loss in the on state, it is possible to build switching networks that select distinct $LC$ resonators for the readout of a targeted device. 

An alternative and more scalable approach is to `recycle' the resonator by using the switch to connect it to each measured device in turn~\cite{Gonzalez2019CMOS}. This approach can dramatically reduce the footprint, since a single resonator structure reads out many devices. For such a configuration to be useful however, the switch should add minimal capacitance so as not to load the resonator.

\subsection{A look ahead: Limits to multiplexing approaches}
Multiplexing techniques provide a means to efficiently use all the available bandwidth or available time window to carry readout signals from multiple quantum devices and are key to the scale-up of quantum computing. However, it remains an open question how far these techniques can be extended and what new developments will be needed to enable parallel readout of millions of qubits. Below we discuss some of the likely constraints to scale-up and identify areas where new work is needed.

Above we discussed the requirements for low-noise amplifiers. Here we extend our discussion to include the entire readout chain. A scaled-up readout system must have ultra-wide bandwidth while preserving the noise, linearity, and power-handling capabilities of the state-of-the-art single-channel systems. To estimate some rough bandwidth requirements we note that applications of fast reflectometry typically require single-channel bandwidths of order a few MHz. Considering resonators constructed from lumped elements, a reasonable estimate is that 100 channels might occupy a total system bandwidth of 2 GHz, including frequency guard bands to suppress crosstalk.

In addition to the cryo-amplifiers, this estimate suggests that  non-reciprocal elements such as circulators or isolators must also exhibit wideband performance. 
Traditionally, non-reciprocal elements are implemented using interference of microwave signals confined to bulky ferrite resonators, a mechanism that is inherently narrowband. Alternative means of realizing non-reciprocity \cite{mahoney2017PRX,mahoney2017NatComm} will likely be required to enable scale-up of frequency multiplexing.

With the need for each quantum device to be paired with a resonator operating at a particular frequency, the physical dimensions of the resonators also pose a challenge to scale-up. As is well known from the development of monolithic microwave ICs, creating large inductors on a chip is difficult due to the significant loop areas required. For quantum applications, however, the use of cryogenic temperatures opens the prospect of leveraging the kinetic inductance associated with superconductors to create small-footprint inductors. Indeed, this is a well established technique in the astronomy community~\cite{Annunziata2010}.

Finally, we draw attention to requirements of the digital demodulation platform in a scaled-up system. Already, implementing the realtime digital synthesis and filtering sub-blocks of a handful of carriers requires some of the largest FPGAs available commercially. Likely, both the required algorithms and hardware can be optimized (effectively implementing the demodulation of highly multiplexed signals using optimized ASICs). Improvements in the performance of DACs and ADCs are also vital to enable multiplexed readout at scale. Again, the noise, linearity, and power-handling capability are key parameters that determine the suitability of data converters for readout applications. Recently, there have been several demonstrations of integrated circuits that provide a compact alternative to distributed readout chains~\cite{Prabowo2021, Ruffinoieee2021}.

\newcommand{\com}[1]{{\color{blue} #1}}
\newcommand{\comFK}[1]{{\color{magenta} #1}}

\section{Spin qubits}
\label{sec:Singlet-triplet qubits}

A leading application of radio-frequency reflectometry for quantum information processing is readout of spin qubits in QDs.
Semiconducting spin qubits comprise different qubit encodings (most commonly single-spin single-dot~\cite{nowack2007}, singlet-triplet double-dot~\cite{Petta2005}, and exchange-only triple-dot encodings~\cite{Gossard2010_Coherent}) and implementations in various semiconducting materials (most prominently GaAs, Si and Ge structures). 
A recent review of spin qubits is given by Ref.~\onlinecite{Chatterjee2021}, whereas details of GaAs and silicon spin qubits were previously reviewed in Ref.~\onlinecite{Hanson2007} and Ref.~\onlinecite{Zwanenburg2013, SCHAAL2021265, gonzalezzalba2020scaling}, respectively.  
In the following, we explain the main rf techniques (Section \ref{sec:rf readout of spin qubit}) to detect spin in QDs, how to perform and interpret single-shot readout (Section \ref{Single-shot}) and we highlight the state-of-the-art experiments involving high-frequency singlet-triplet measurements (Section \ref{state-of-the art experiments}).

\subsection{rf readout of spin qubit}
\label{sec:rf readout of spin qubit}

The first step to reading out a spin qubit is to create an electrical signal that depends on the qubit state. 
The most common way to do it relies on spin-dependent tunneling mechanisms known as spin-to-charge conversion~\cite{Elzerman2004,Ono2002,Petta2005}. The qubit state can then be deduced either from the measurement of a charge sensor or by dispersive readout.

\subsubsection{Spin readout using a charge sensor}
\begin{figure}
	\includegraphics{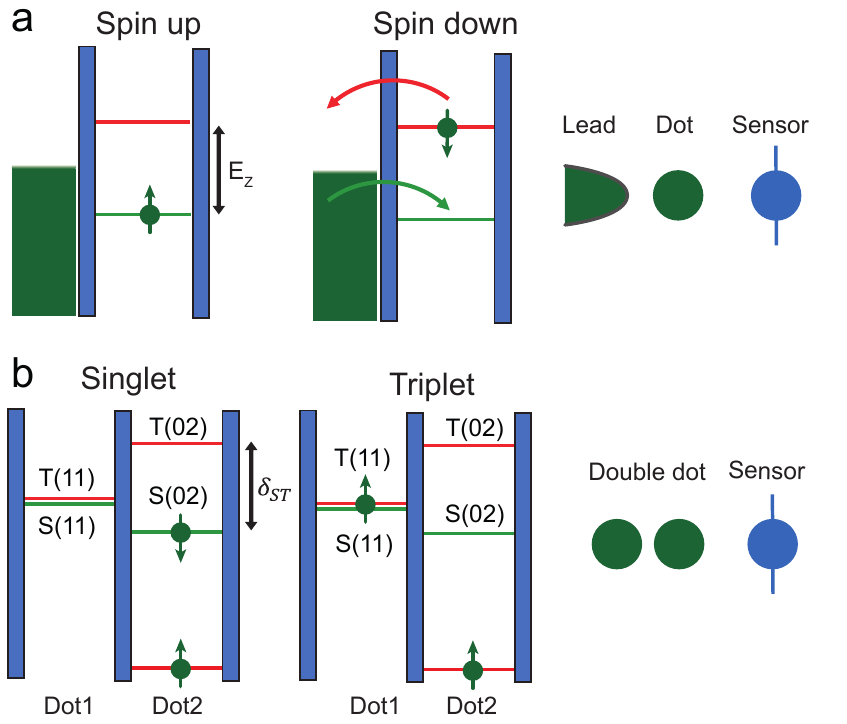}   	
	\caption{
		Schemes for spin-to-charge conversion.
		(a) Single-spin readout using a charge sensor.
		In a magnetic field, the spin levels in a QD are separated by Zeeman energy~$E_\mathrm{Z}$. (This figure is drawn assuming a negative $g$-factor, as in GaAs.)
		If these two levels straddle the Fermi level in a nearby lead, the higher-energy state can decay by electron tunneling.
		This gives rise to a transient change in the electric field seen by the sensor, and therefore in its resistance.
		(b) Singlet-triplet readout using a charge sensor.
		For two electrons in the same dot, there is a splitting~$\delta_\text{ST}$ between the singlet and triplet levels.
		In a DQD with the level alignment shown, a singlet spin state therefore favors a (02) charge occupation while a triplet state favors (11).These two configurations are distinguished by the sensor.
	}
	\label{fig:SpinToCharge}
\end{figure}

Fig.~\ref{fig:SpinToCharge} shows two main mechanisms for accomplishing spin-to-charge conversion, which rely respectively on energy-selective and on spin-selective tunneling.
In both cases the information about the spin is correlated with a specific charge-tunneling event or a static dot charge occupation that can be detected using a nearby charge sensor. 

To perform energy selective spin readout, or Zeeman readout~\cite{Elzerman2004}, the $\{ \uparrow,\downarrow \}$-spin states of a charge confined in a QD are separated in energy using a large magnetic field, and the QD potential is tuned such that only spin-$\downarrow$ electrons are allowed to tunnel off the QD, whereas spin-$\uparrow$  electrons will remain confined within the QD potential (Fig.~\ref{fig:SpinToCharge}(a)). Due the large energy separation between the two spin-states, when a tunneling event occurs the charge in the QD is quickly replaced by a charge with the opposite spin. 
A fast charge sensor can therefore detect the interval between these two tunnelling events during which the QD is empty, and thereby identify the initial spin state.

Although conceptually simple, this readout method presents some challenges. First, it requires the energy splitting of the spin states to be larger than the electron thermal energy, which demands low temperatures and large magnetic fields. Second, the precise tuning of the QD energy levels can be very sensitive to charge noise and fluctuations of the dot-electrostatic potential. 
However, as demonstrated by Ref.~\onlinecite{Vandersypen2005}, if the tunneling rates of the two spin states with the electron reservoir are very different, both these conditions can be relaxed resulting in a more robust readout-mechanism. 

Another popular method for spin-to-charge conversion, typically used in DQD systems, uses current rectification due to Pauli spin blockade (Fig.~\ref{fig:SpinToCharge}(b)).
Consider two electrons confined in a DQD. The combination of two-particle charge and spin degrees of freedom can be classified respectively as separated and joint singlets, S(11) and S(02), and separated and joint triplets, T(11) and T(02). The latter two each has a degeneracy of three which is broken by a magnetic field. 
Pauli selection rules forbid the existence of two fermions with the same quantum numbers, forcing the second electron to a higher orbital state in the T(02) configuration which is separated from S(02) by an energy $\delta_\text{ST}$~\cite{Tarucha2001}. On the other hand, S(11) and T(11) are quasi-degenerate since the spatial separation of the participant spins results in a vanishing small $\delta_\text{ST}$.

Because Pauli exclusion raises the energy of the T(02) state compared to the T(11) and S(11) states, spin conservation requires the T(11 state to remain blocked while the singlet S(11) is allowed to tunnel to the state S(02).
A charge sensor can then detect the difference between these two static charge configurations, either T(11) or S(02). Note how the spin state is now correlated to the charge configuration.

\subsubsection{Dispersive spin readout}
\begin{figure}
	\includegraphics{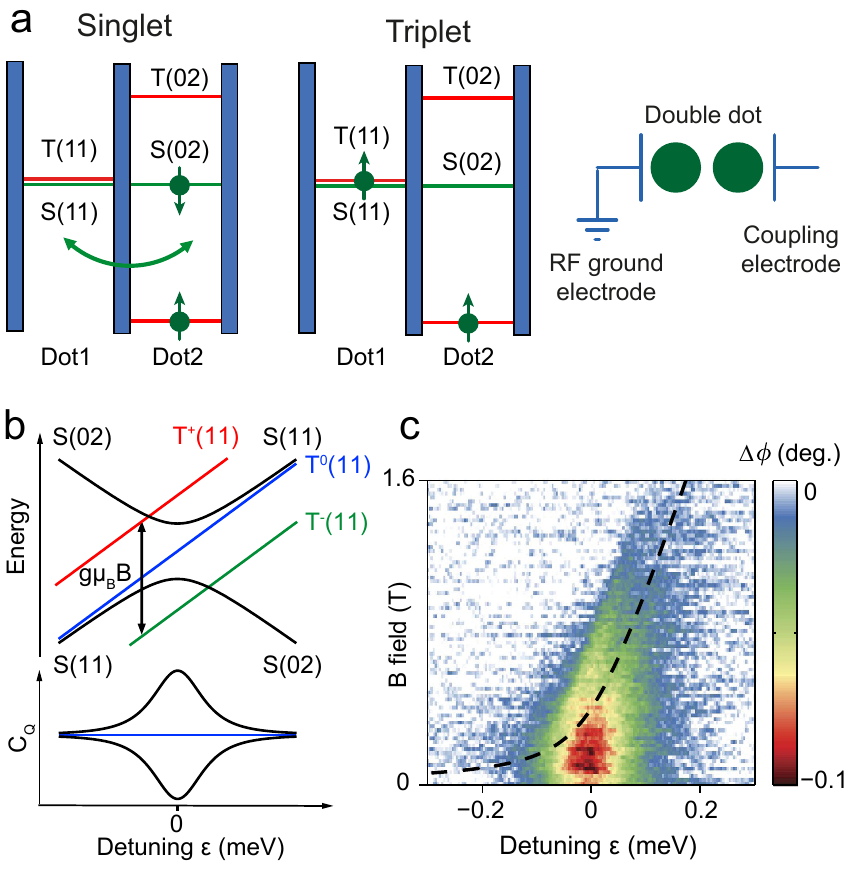}
	\caption{Dispersive singlet-triplet readout.
	(a) Energy diagram of the four charge-spin configurations of two electrons in a DQD at $B=0$. The brackets (left, right) give the charge occupation of each QD for each state. The green arrows symbolise the tunneling between the singlet states. The polarizability of a DQD depends on whether a charge can tunnel in response to a small electric field.
	With the level alignment shown, this only happens for the singlet state.
	This gives rise to a spin-dependent admittance across the DQD, which can be measured using an rf resonator attached to a coupling electrode (sketched on the right).
    (b) Top: Energy levels of the two-electron configurations as a function of detuning $\varepsilon$ (top). The magnetic field separates the three triplet states T$^-(11)$, T$^0(11)$ and T$^+(11)$ (We ignore the higher-energy T(02) states for simplicity.). Bottom: Corresponding quantum capacitance $\CQ$. The quantum capacitances of T$^-(11)$, T$^0(11)$ and T$^+(11)$ overlap.
    (c) Dispersive measurement of a double quantum dot in Pauli spin blockade as a function of detuning $\varepsilon$ and magnetic field B. The dashed line indicates the degeneracy of the lowest energy singlet and triplet states. Panel (c) adapted from Ref.~\onlinecite{Simmons2015}.}
	\label{fig:Dispersive:SpinReadout}
\end{figure}

Spin readout via Pauli spin blockade can also be measured dispersively without a charge sensor.
In this case, the DQD is configured so that the S(11) and S(02) configurations are degenerate ~(Fig.~\ref{fig:Dispersive:SpinReadout}(a)), with the weighting of these two configurations depending on the electric field.
As long as the singlet (triplet) coupling $\Delta_\text{S(T)} < \delta_\text{ST}$, the system is free to tunnel between the S(11) and S(02) charge states, whereas a system in the T(11) cannot tunnel to the T(02) state unless extra energy is provided.

As we saw in Section~\ref{sec:dispersiveQD}, a double quantum dot presents a quantum capacitance 
\begin{equation}\label{spinQ}
	\CQ=(e\alpha')^2\frac{dP_2}{d\varepsilon},
\end{equation}
which depends on how the charge distribution among QDs reacts to a change in detuning induced by the rf signal.
Hence, tunneling between singlets manifests itself as a quantum capacitance, allowing these the singlet and triplet spin configurations to be distinguished. In Fig.~\ref{fig:Dispersive:SpinReadout}(b), we plot the two-electron spectrum as a function of detuning. The plot includes the singlet eigenenergies (Eq.~\ref{eigenenergies}) and the uncoupled triplet energies $E_{\text{T}0}=\varepsilon/2$, $E_{\text{T}\pm}=\varepsilon/2 \pm g\mu_\text{B}B$ where $\mu_\text{B}$ is the Bohr magneton, $g$ the electron g-factor and $B$ the external magnetic field. In the low temperature limit, Eq.~\eqref{spinQ} can be conveniently generalized t~\cite{Gonzalez2020Quintet},
\begin{equation}
	\CQ=-(e\alpha')^2\sum_\text{i}\frac{\partial^2E_i}{\partial\varepsilon^2}P_i
\end{equation}
where $E_i$ and $P_i$ are the eigenenergies and their occupation probabilities, respectively. As we see, at zero detuning, the singlet ground (S-) and excited states (S+) present a quantum capacitance
\begin{equation}
	C_{\text{Q,S}\pm}=\pm\frac{(e\alpha')^2}{2\Delta_\text{S}},
\end{equation}
whereas the triplets have zero quantum capacitance. The difference in quantum capacitance can be determined using reflectometry when the system is biased at zero detuning. At $B=0$~T, the overall ground state is a singlet ground state and electrons are free to tunnel between the S(11) and S(02) states, resulting in a net phase shift of the resonator, see Fig~\ref{fig:Dispersive:SpinReadout}(c). For magnetic fields $g\mu_\text{B}B>\Delta_\text{S}/2$, T$^-(11)$ becomes the ground state and the phase shift tends to zero. The signal vanishes asymmetrically from the (11) side tracking the position in $\varepsilon-B$ space of the singlet-triplet crossing.

\textit{In-situ} dispersive spin readout has been achieved in double quantum dots in InAs~\cite{Schroer2012}, GaAs~\cite{Petersson2010, Reilly2013} and Si~\cite{Gonzalez-Zalba2015_Dispersive, urdampilleta2015, Heijden2018, DeFranceschi2019}. Furthermore, dispersive Pauli spin blockade has been used for single-shot spin readout~\cite{Simmons2018, West2019, Zheng2019}.

\subsection{Single-shot readout}
\label{Single-shot}

Fault-tolerant quantum computing requires the state of individual qubits to be read out in single-shot mode,
meaning that the state of a single qubit, i.e. 0 or 1, must be determined from one iteration of the measurement.
For error correction to be scalable, the fidelity of this process, i.e. the probability to correctly identify the qubit state, must be well above a threshold determined by the error-correction protocol~\cite{Fowler2012}.
The acceptable error rate for measurements (so-called `class-1 errors'~\cite{Fowler2012}) depends on the fidelity of other gate operations but is likely to be around 0.1\%, meaning that fast readout needs to attain a fidelity of 99.9\% or better.
Furthermore, this process should happen within a single repetition time of the error-correction cycle, which means within the qubit coherence time.
This is one of the most demanding and important applications of fast readout, and requires sufficient sensitivity to detect a small signal and sufficient bandwidth to respond within the qubit coherence time.

Unfortunately, the short relaxation lifetime $T_1$ of the state being measured often makes single-shot measurements challenging, and if the signal-to-noise ratio is too small, the state cannot accurately be determined within this time.
Electron spin lifetimes can be greater than 1~s in gate-defined quantum dots~\cite{Amasha2008, CirianoTejel2021} or 30~s in donor-based devices~\cite{Watson2017}, but are typically of the order of 1~ms or less in qubit devices~\cite{Pakkiam2018, West2019, Urdampilleta2019, Zheng2019}.

\begin{figure}
    \includegraphics[width=\columnwidth]{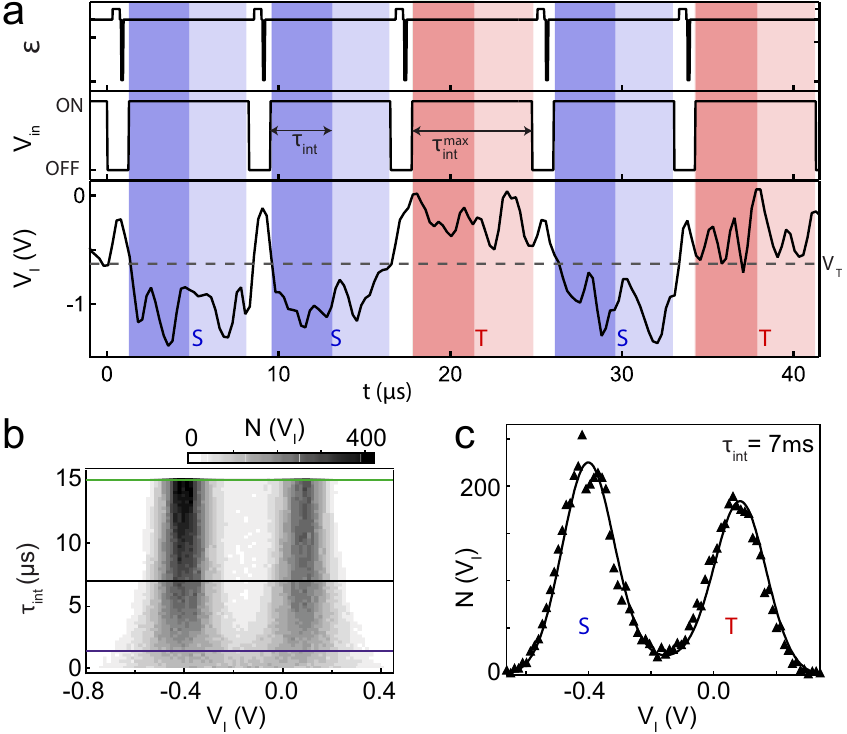}	
    \caption{
    (a) Series of single-shot measurements of a spin qubit~\cite{}. The unshaded portion of each cycle marks the interval during which a spin superposition is generated and projectively converted to a charge state; the shaded portion marks when the qubit is measured.
    Top: Detuning $\varepsilon$ as a function of time. The cycle shown generates an approximately equal mixture of the two states.
    Middle: Level of the illumination signal.    
    Bottom: Demodulated reflected signal $\VI$.
    Each iteration is identified as singlet $S$ or triplet $T$ depending whether the average level is below or above the threshold $V_\mathrm{T}$.
The integration time $\tauM$ is adjusted up to a maximum value $\tau_\mathrm{M}^\mathrm{max}$.
(b) Histogram of average readout signal for different choices of $\tauM$. Here $N(\VI)$ is the number of counts in each bin of the histogram.
The purple line at $\tauM=1.5~\mu$s marks a choice for which the two states are insufficiently distinct; the green line at $\tauM=15~\mu$s marks a choice for which decay from $T$ to $S$ has significantly degraded the fidelity.
(c) Histogram at $\tauM=7~\mu$s (black line in (b)), for which the fidelity is maximal.
Figures adapted from Ref.~\onlinecite{Gossard2009}.
}

    \label{fig:SingleShot}    
\end{figure}

An example of a single-shot spin qubit measurement is shown in Fig.~\ref{fig:SingleShot}~\cite{Gossard2009}.
The qubit in this case is a singlet-triplet qubit~\cite{Petta2005} measured using an rf-QPC charge sensor in the scheme of Fig.~\ref{fig:SpinToCharge}(b).
The qubit is controlled by rapidly adjusting the detuning $\varepsilon(t)$ in a cycle that generates an approximately equal mixture of the two states (Fig.~\ref{fig:SingleShot}(a)).
To read out the state at the end of each cycle, $\varepsilon$ is held constant and the illumination tone $\Vin$ is turned on.
This leads to a demodulated signal $\VI(t)$ whose average value during the readout step is low or high depending which state was generated.

The optimum integration time $\tauM$ is long enough to minimise electrical noise but short enough that the qubit usually has not decayed during the measurement.
Figure~\ref{fig:SingleShot}(b) illustrates this trade-off by plotting a histogram of averaged $\VI$ values for different choices of $\tauM$.
Similar to Fig.~\ref{fig:noise}(e-f), the distribution shows two peaks at $\VIF^\mathrm{T}$ and $\VIF^\mathrm{S}$ corresponding to the two qubit states, with each becoming narrower as the integration time increases.
However, unlike in Fig.~\ref{fig:noise}, the two peaks do not have equal weighting; the right-hand peak becomes weaker as T states are given more time to decay, leading them to be misidentified as S.

The optimum value of $\tauM$ is chosen by maximising the fidelity (see Eq.~\eqref{eq:footFidelity})
\begin{equation}
\mathcal{F}\equiv	 \frac{\FS}{2} +\frac{\FT}{2}
\end{equation}
where $\FS$ and $\FT$ are the fidelity associated with identifying with success the S or T states, respectively.
It leads to the histogram in Fig.~\ref{fig:SingleShot}(c).
Typically, a threshold voltage $V_T$ is chosen between the two peaks, with outcomes below threshold interpreted as S and outomes above threshold as T. $\FS$ and $\FT$  are 
\begin{equation}
    \FS=1-\int_{V_T}^\infty n_\mathrm{S}(V)dV, \quad  \FT=1-\int_{-\infty}^{V_T} n_\mathrm{T}(V)dV
\end{equation}
where $n_\mathrm{S}$ and $n_\mathrm{T}$ are respectively the Singlet and Triplet probability density.
Here $n_\mathrm{S}$ can modeled as a noise-broadened Gaussian~\cite{Gossard2009} with standard deviation $\sigma$ and centered on $\VIF^\mathrm{S}$:
\begin{equation}
    \nS(\VIF)=(1-\left \langle \PT \right \rangle) e^{-\frac{(\VIF-\VIF^\mathrm{S})^2}{2\sigma^2}}\frac{1}{\sqrt{2\pi}\sigma}
\end{equation}
where $\EE{\PT}$ is the Triplet probability over all the experiment. The Triplet outcomes $\nT$, need to take into account relaxation during $\taum$:
\begin{equation}
    \begin{split}
    &\nT(\VIF)= e^{-\frac{\taum}{T_1}}\left \langle \PT \right \rangle e^{-\frac{(\VIF-\VIF^\mathrm{T})^2}{2\sigma^2}}\frac{1}{\sqrt{2\pi}\sigma} \\
    & +\int_{\VIF^\mathrm{S}}^{\VIF^\mathrm{T}}\frac{\taum}{T_1}\frac{\left \langle \PT \right \rangle}{\Delta \VIF} e^{-\frac{(V-\VIF^\mathrm{S})}{ \Delta \VIF}\frac{\taum}{T_1}}e^{-\frac{(\VIF-V)^2}{2\sigma^2}}\frac{dV}{\sqrt{2\pi}\sigma},
    \end{split}
\end{equation}
where $\Delta \VIF=\VIF^\mathrm{T}-\VIF^\mathrm{S}$.

In this experiment with a qubit relaxation time $T_1 = 34~\mu$s, the maximum fidelity is $\mathcal{F}\approx95\%$ for $\tauM=7~\mu$s.
Experiments since then have reached higher values (see Supplementary Table~SI).
The optimum strategy for identifying the qubit state from the voltage record, which is more sophisticated than the simple average used in Fig.~\ref{fig:SingleShot}, is discussed in Supplementary Section~S3 B 4.
Currently the record fidelity for reading out a singlet-triplet qubit is $99.86\%$~\cite{Harvey-Collard2018}, or $99.5~\%$ in a short array~\cite{Nakajima2017}.

Single-shot measurements of a single spin using energy-selective readout, require a different fidelity analysis. For energy-selective readout, charge sensors are necessary. The experiment needs to detect the reflected voltage signal (or current if dc charge sensors are used) occurring between the two charge-tunneling events in Fig.~\ref{fig:SpinToCharge}(a). The bandwidth needs to be sufficiently large to resolve the transient during which the electron resides outside the QD. The important parameters are therefore the tunneling in and out times, the integration time per point, the relaxation time and the voltage threshold to define whether a measurement outcome is called a spin up or down. The optimization of these parameters and the evaluation of the corresponding readout fidelity are now commonly performed using Monte Carlo simulations~\cite{Simmons2019Benchmarking}. Currently the highest fidelity reported for single-spin qubits using energy-selective readout is $99.8~\%$ in 65~ms for a p-donor in silicon~\cite{Watson2017}, and $97~\%$ in $1.5~\mu$s using an rf-SET~\cite{Keith2019}.

\subsection{Examples of state-of-the art experiments}
\label{state-of-the art experiments}

\subsubsection{Readout of four qubits with charge sensors}

\begin{figure}
    \includegraphics{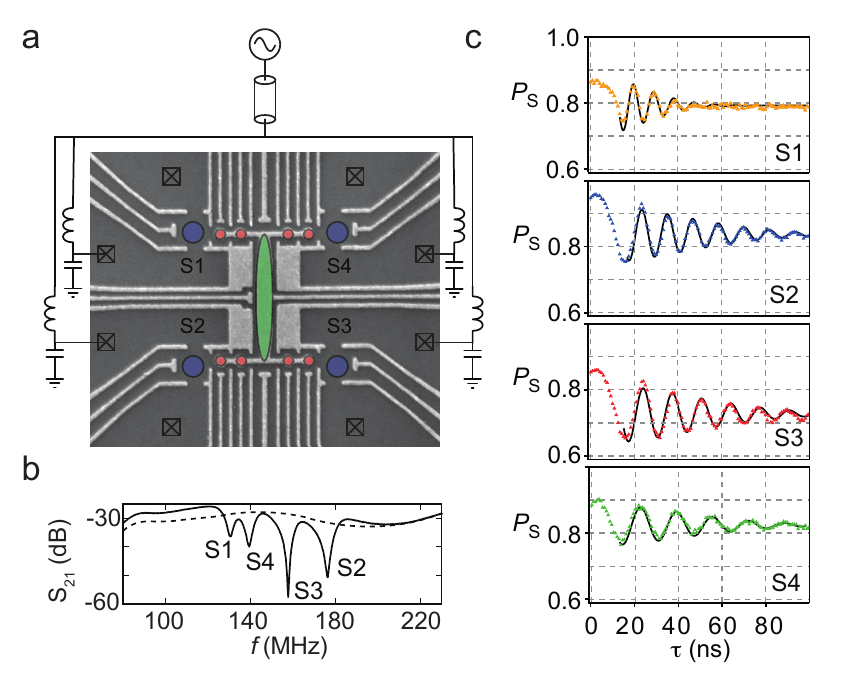}
    \caption{(a) Four singlet-triplet qubits (red double dots) with proximal sensor dots (blue dots) implemented in a GaAs heterostructure, allowing simultaneous four-qubit operation and four-qubit single-shot readout via frequency multiplexing using a commercial PCB sample holder system.
    (b) Each charge sensor is wirebonded to a tank circuit with unique resonance frequency, allowing simultaneous readout of all four sensors (S1-4) via frequency multiplexing. 
    (c) Simultaneous exchange rotations of all four qubits induced by suitable detuning pulses (not shown). Each data point represents the average of many single-shot outcomes obtained for each qubit. 
    Figure adapted from Ref.~\onlinecite{Fedele2021}.
    \label{fig:spinqubitreadoutFederico}
    }
\end{figure}

While in state-of-the-art silicon devices most qubits are operated one at a time, GaAs devices have recently allowed the simultaneous operation (and readout) of up to four singlet-triplet qubits~\cite{Fedele2021}. The device shown in Fig.~\ref{fig:spinqubitreadoutFederico}(a) employs a multi-electron coupler (elongated QD highlighted in green) to space four DQDs sufficiently far apart to allow individual qubit manipulation with minimal cross-talk between electrodes. Each DQD implements one singlet-triplet qubit, with proximal charge-sensing QDs that are read out simultaneously by frequency-multiplexed reflectometry. 
(The high-frequency PCB sample holder for this experiment is commercially available from QDevil\footnote{High-frequency PCB sample holders are available from QDevil (www.qdevil.com) \label{footnoteQDevil}}.) 
In this experiment, one contact lead of each sensor is wirebonded to an SMD resonator (with a unique inductance) on the PCB sample holder, and all four resonators are capacitively coupled to one reflectometry channel of the cryostat. The measured reflectometry signal (Fig.~\ref{fig:spinqubitreadoutFederico}(b)) shows four dips sufficiently spaced in frequency to allow qubit-resolved single-shot readout using separate carrier signals injected via the same line. This work demonstrates not only that all four qubits can be rotated simultaneously with similar speed (Fig.~\ref{fig:spinqubitreadoutFederico}(c) shows a $\pi$ rotation within 15~ns), but also that all four qubits can be read out in single-shot mode simultaneously (in Fig.~\ref{fig:spinqubitreadoutFederico}(e), each data point is an average of 512 single-shot outcomes). 

The ability to combine time-domain and frequency-domain multiplexing means that reflectometry will likely continue to play an important readout tool as qubit devices are scaled to 100-qubit processors or even beyond 1000 qubits. 

\subsubsection{Spin readout with superconducting on-chip microwave resonators}

\begin{figure}
    \includegraphics{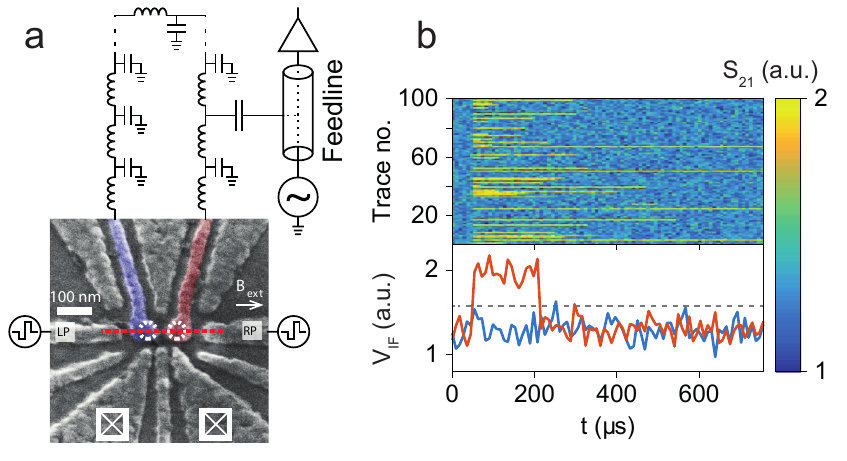}
    \caption{
    (a) A silicon double quantum dot (dashed white circles) is capacitively coupled to an on-chip superconducting resonator (shown here as a series of LC elements), which is monitored via a transmission line. Detuning pulses applied to LP and RP configure the double dot in its readout position where the charge associated with spin-singlet states can tunnel while spin-triplet states are Pauli blocked. 
    (b) The tunnel and dispersive capacitance associated with the singlet state yield an enhancement of the transmission amplitude $|S_{21}|$, corresponding to a single-shot readout fidelity >98\% in 6~$\mu$s. Figures adapted from Ref.~\onlinecite{Vandersypen2019}.
    \label{fig:spinqubitreadoutVandersypen}
    }
\end{figure}

In Fig.~\ref{fig:spinqubitreadoutVandersypen}(a), a silicon DQD is capacitively coupled to a 5.7-GHz on-chip superconducting resonator that is capacitively coupled to transmission line~\cite{Vandersypen2019}. Transmission measurements $S_{21}$ can distinguish singlet and triplet states with high sensitivity and temporal response, as exemplified by several single-shot readout traces in Fig.~\ref{fig:spinqubitreadoutVandersypen}(b). In this experiment, a random spin configuration is repeatedly initialized and measured after 50 $\mu$s. Singlet and triplet states are then distinguished by a different response in the resonator transmission. In the top panel, blue pixels are associated with triplet statesand  yellow pixels are associated with singlet states. Two individual linecuts in the bottom panel illustrate the difference in the resonator response. In this example, the resonator quality factor (2600) yields a maximum bandwith of 2~MHz. In conjunction with an estimated spin relaxation time of 0.16~ms, this allowed single-shot read-out of the two-electron spin state with an average fidelity of >98\% with an integration time of $6~\mu$s. 

Because the carrier frequency of 5.7 GHz exceeds the interdot tunnel coupling (2 GHz), the system is not in the adiabatic limit during readout, i.e. in addition to the quantum capacitance associated with the curvature of the dispersion relation, there are significant contributions from the tunnelling capacitance. See Refs.~\cite{Petta2017,Vandersypen2018Coupling} to understand the coupling between the spin of the electrons in the DQDs and the photons in the resonator.

\newcommand{\com}[1]{{\color{blue} #1}}
\newcommand{\comFK}[1]{{\color{magenta} #1}}
\section{Rapid detection of important quantum phenomena}
\label{sec:Sample}

So far, our examples have focused on the technical aspects of high-frequency reflectometry, and were chosen to illustrate important variations and optimizations rather then the breadth of physical insights that can be gained with this technique. In this section, we describe different condensed-matter experiments that have already benefited from this powerful measurement tool. Our selection is by no means exhaustive, and intends to inspire new applications in diverse subject areas. 

\subsection{Noise-protected superconducting qubits}
\label{Noise-protected superconducting qubits}

\begin{figure*}
    \includegraphics[width=\textwidth]{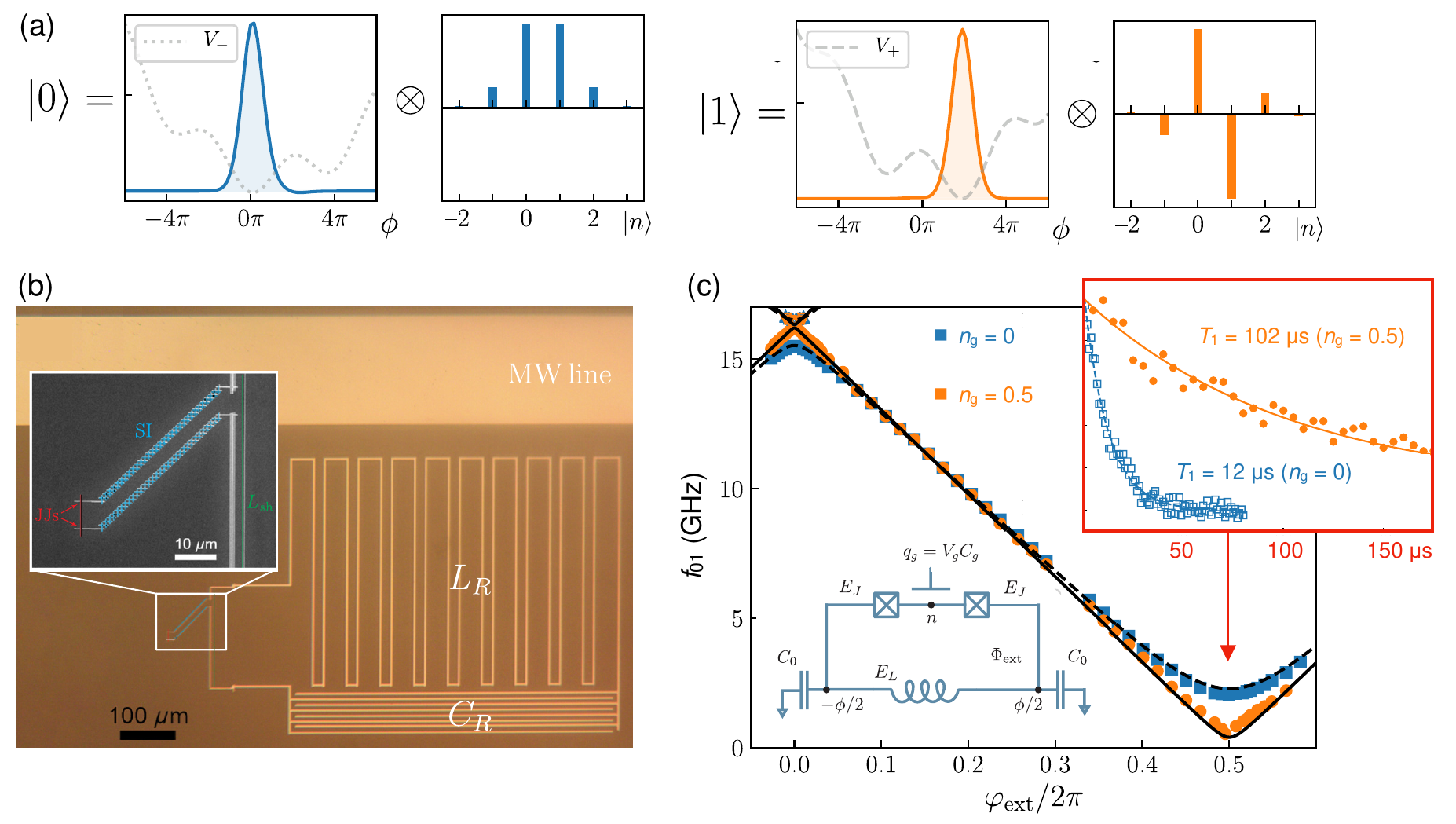}
    \caption{A protected superconducting qubit, here implemented as a bifluxon qubit. 
    (a) For a suitable symmetric operating point, the two qubit states $|0\rangle$ and $|1\rangle$ become nearly degenerate in energy. Visualization of their flux-like and and charge-like wavefunctions shows the origin of their protection: the flux-like parts (in the basis of the phase difference $\phi$ across the superinductor, see inset of (c)) are localized in different minima of the fluxonium potential $V$, and the charge-like parts are symmetric ($|0\rangle$) and antisymmetric ($|1\rangle$) (in the basis of number of Cooper pairs on the Cooper-pair box, $n$). 
    (b) Experimental implementation of the bifluxon qubit (inset), readout resonator ($L_R$, $C_R$), and microstrip transmission line (MW), all fabricated in a single multi-angle Al-evaporation process. The Cooper-pair box (red in inset) is defined by two small Josephson junctions, whereas the superinductor (SI) is implemented as an array of 122 larger Josephson junctions (blue in inset). During one cooldown of such a chip multiple qubits can be read out via the same microwave line, by using different resonance frequencies for each readout resonator. 
    (c) Experimental signatures of the protection include a decrease of the qubit splitting, $f_\mathrm{01}$, as the symmetry point is approached ($\varphi_\mathrm{ext}=\pi, n_g=0.5$), and an increase of the qubit relaxation time $T_1$ (red inset). To operate this qubit and take measurements like this, the gate voltage is pulsed away from $n_g=0.5$ to temporarily break the protection of the qubit and make it interact with the control and readout signals (not shown).
    Figures adapted from Ref.~\onlinecite{Kalashnikov2020}.}
    \label{fig:ExampleProtectedQubit}
\end{figure*}

The coupling between superconducting qubits and superconducting microwave resonators plays a key role in current quantum processors for enabling coherent two-qubit gates and efficient qubit readout. A key challenge is maintaining high control and readout fidelities while increasing the number of qubits. As an alternative to conventional superconducting qubits (such as Xmons, transmons, etc.\cite{KjaergaardReview2020}), new superconducting circuits are being studied that combine inductors, capacitors, and Josephson junctions in such a way that the resulting two-level system is more robust to environmental noise. The reduced error rates associated with such qubits may then possibly simplify the scaling towards larger qubit arrays. The key idea behind such noise-protected superconducting qubits is to simultaneously suppress qubit relaxation and qubit dephasing by creating special symmetries in the effective circuit ("error correction by hardware engineering")~\cite{Bell2014,Kalashnikov2020,Larsen2020,Gyenis2021,Gyenis2021b}.
For qubits protected by the topologies of the underlying system Hamiltonian, see the following section on Majorana qubits.
The hope of error correction on the hardware level is to eventually engineer circuits comprising frustrated chains of Josephson junctions~\cite{Ioffe2002,Doucot2002} such that not only quantum memory is protected, but also gate operations~\cite{Kitaev2006, Brooks2013}.

To achieve a good quantum memory, both bit-flip errors (qubit relaxation) and phase errors (qubit dephasing) need to be suppressed. While energy relaxation can be suppressed by decreasing the wavefunction overlap between the two qubit states (for example by \emph{localizing} qubit states in distinct minima of the qubit potential, as in the "heavy fluxonium" qubit~\cite{Lin2018}), and dephasing can be exponentially suppressed by \emph{delocalizing} the qubit wavefunction (for example in charge space, as in the "transmon" qubit~\cite{Koch2007}), the simultaneous suppression of both errors requires more complicated "few-body" systems such as the $0$-$\pi$ qubit~\cite{Brooks2013}. 
In such qubits, multiple Josephson junctions are connected by superconducting loops in such a way that two nearly degenerate ground states emerge that are localized in distinct minima of a superconducting phase difference, namely at zero phase and at $\pi$. The exponentially small qubit splitting combined with a robustness to weak local perturbations should make $0$-$\pi$ qubits highly resistant to decoherence arising from local noise. Theoretically, such qubits also offer routes towards topologically protected gate operations, although this has not yet been demonstrated experimentally. 

Experimentally, the characterization of protected qubits is complicated by its very protection: near its (protective) symmetry point, not only does the qubit splitting become impractically small (preventing the typical microwave techniques such as two-tone spectroscopy), but also its coupling to the control pulses (thereby preventing straightforward Hahn echo experiments to study dephasing characteristics, for example). To maintain the ability to control and read out such qubits, they have been intentionally mistuned from the protected symmetry such that noise protection can be quantified~\cite{Larsen2020,Gyenis2021}. (In the pioneering devices in Ref.~\onlinecite{Bell2014}, the asymmetry naturally arose from imperfections in the Josephson rhombus arrays.)

An interesting alternative to $0$-$\pi$ qubits protected by the parity of Cooper pairs (as in Refs.~\onlinecite{Larsen2020,Gyenis2021}), are $0$-$\pi$ qubits protected by the parity of flux quanta, as in the bifluxon qubit~\cite{Kalashnikov2020}. Here, a superconducting loop comprises a Cooper-pair box and a superinductor~\cite{Bell2012}, suppressing tunneling of flux quanta between the outside and inside of the loop (and errors that would be generated by such tunneling) for a particular charge tuning of the Cooper-pair box. 
In Figure~\ref{fig:ExampleProtectedQubit}, the gate-induced charge of the Cooper-pair box, $n_\mathrm{g}$, is controlled by a gate voltage, and the applied flux through the loop ($\Phi_\mathrm{ext}$, controlled by an external magnetic field) induces a phase winding $\varphi_\mathrm{ext}$ along the loop. At $\varphi_\mathrm{ext}=\pi$, the qubit splitting, $f_\mathrm{01}$, is observed to drop dramatically at the symmetric operating point $n_\mathrm{g}=0.5$ (Fig.~\ref{fig:ExampleProtectedQubit}(c)), while the qubit relaxation time, $T_1$, increases significantly (red inset). The underlying protection at that symmetry point can be understood by visualizing the wavefunctions of the qubit states as tensor products of a flux-like and charge-like part (Fig.~\ref{fig:ExampleProtectedQubit}(a)), and noting that the flux-like part of $|0\rangle$ is localized in a different minimum than $|1\rangle$ while the charge-like parts are symmetric ($|0\rangle$) and antisymmetric ($|1\rangle$) (when expressed in the basis of number of Cooper pairs $|n\rangle$ on the Cooper-pair box). 

The readout of the bifluxon qubit shown in Figure~\ref{fig:ExampleProtectedQubit} is performed by inductively coupling a readout resonator (consisting of $L_R$ and $C_R$ elements as shown on the micrograph) to a microwave transmission line (MW line). The properties of the readout resonator change if the state of the qubit changes. The Cooper-pair box (red in the inset) is connected via two Josephson junctions to a superinductor (blue), which is implemented as an array of 122 larger Josephson junctions. 
The bifluxon qubit, readout resonator, and microstrip transmission lines are fabricated in a single multiangle electron-beam deposition of aluminum through a liftoff mask.
In the transmission measurements, the microwave signals travel along the microstrip line and couple to the readout resonators of up to five different bifluxon qubits located on the same chip. By using different resonant frequencies of the readout resonators, the qubits can be individually addressed and characterized in the same cooldown. In this case, frequency-multiplexing is not essential to the operation of the device, but is simply an experimental trick to increase the chances of finding one device with suitable device characteristics (symmetry of the two small Josephson junctions, in this case).

\subsection{Topological superconductivity and Majorana devices}
\label{subsec:Majorana devices}

\begin{figure*}
    \includegraphics[width=\textwidth]{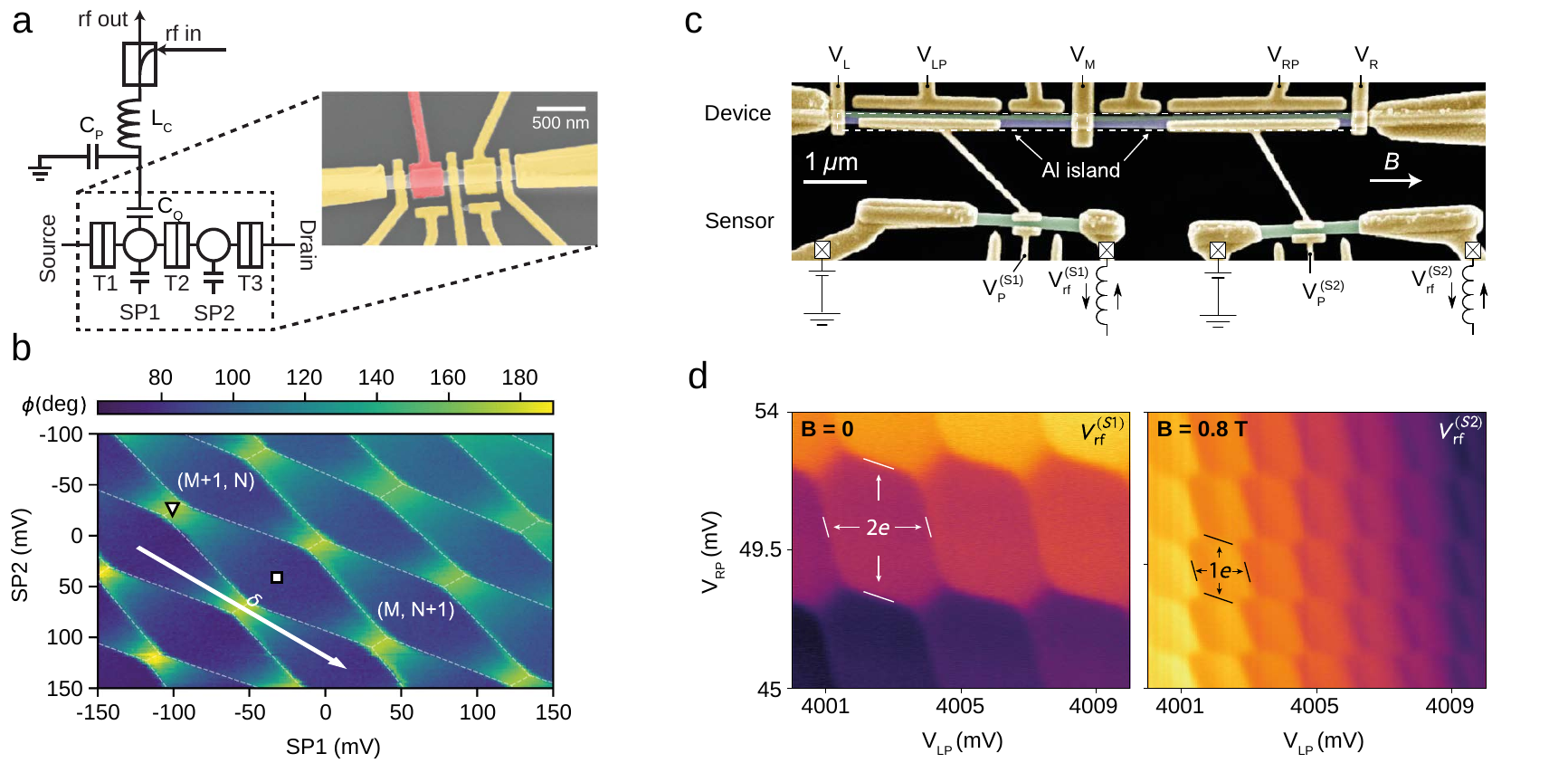}
\caption{Nanowire devices with high-frequency reflectometry readout to investigate Majorana modes. 
(a) SEM of an InAs nanowire double dot to demonstrate rapid detection of interdot tunneling (here without superconductivity and without Majorana modes). One of the quantum dots is capacitively coupled (via the red gate electrode) to a 0.4-GHz superconducting resonator ($\Lc,\Cp$) that is probed by reflectometry. Interdot tunneling (modeled by $\CQ$) results in a substantial  shift of the resonance frequency, and therefore of reflected carrier phase $\phi$. 
(b) Charge stability diagram measured by sweeping the two plunger gates, SP1 vs SP2. Tunneling at charge degeneracy ($\triangledown$) can be distinguished from Coulomb blockade ($\square$) by sampling the reflected carrier phase $\phi$ for a short time (in this work, achieving a SNR of 2 within 1 $\mu$s for a carrier power of -109 dBm and 5 GHz interdot tunneling). 
(c) Superconducting double dot fabricated from a hybrid InAs/Al nanowire (Device) suitable for investigating Majorana modes. The charge occupation is measured by two InAs nanowire quantum dots (Sensor) that are monitored using different reflectometry frequencies (40 and 60 MHz in this work), each yielding a reflectometry signal (namely $V_\mathrm{rf}^{(S1)}$ and $V_\mathrm{rf}^{(S2)}$ in panel d). 
(d) Charge stability diagram measured at $B=0$ and $B=0.8$~T via the rf response of the right and left sensor respectively, for weak interdot tunneling. By studying the transition from 2$e$-periodic Coulomb valleys at $B=0$ to 1$e$-periodic Coulomb valleys at finite field, one can in principle measure the interaction energy within Majorana pairs~\cite{Albrecht2016}. 
Panels (a,b) adapted from Ref.~\onlinecite{Watson2019}. Panels (c,d) adapted from~\onlinecite{Marcus2019}.
}
    \label{fig:ExampleMajorana}
\end{figure*}

One keenly studied sub-field of condensed-matter physics is that of topological materials~\cite{Ando2015, Armitage2018}, whose coherence and time-dependent properties are largely unexplored despite their potential applications in inherently fault-tolerant quantum computation. To date most experiments on topological systems, such as those seeking non-abelian Majorana bound states in nanowires or 2DEGs, focus on transport signatures, even though topologically-protected quantum computing~\cite{Nayak2008} will require time-domain control such as braiding and single-shot parity measurements. 
Several considerations drive the development of fast parity-to-charge detectors: first, readout times can be as low as microseconds, thereby potentially mitigating quasiparticle poisoning of Majorana modes that occurs on longer time scales. Second, quantum non-demolition measurements become possible with high SNR, which is crucial for measurement-based quantum computation based on topological superconductivity. 
Third, no matter how long the coherence times of protected qubits ultimately may be, in order to operate many qubit cycles within reasonable timescales, fast qubit readout of charge or current will be beneficial.

If a topological superconductor hosting two physically separated Majorana zero modes is sufficiently small, then the Coulomb charging energy associated with such a ``Majorana island" can be utilized to create two degenerate ground states that differ in their total occupation number by one electron. (This is in contrast to trivial superconducting islands, where the addition or removal of one electron would require an energy related to the superconducting gap.) Liang Fu realized that the injection of an electron into one Majorana mode, and simultaneous extraction of an electron from the other mode, constitutes a non-local phase-coherent electron transfer~\cite{Fu2010}. This prediction of Majorana-assisted electron teleportation inspired numerous other proposals that suggest \emph{conductance measurements} and \emph{charge sensing} of Majorana islands as a tool to study non-local and non-abelian properties of Majorana modes. For example, in the proposal by David Aasen et al., superconducting double-dot devices in which the various tunnel couplings (i.e. Josephson couplings and Majorana couplings) can be controlled by gate voltages play a central role, allowing parity-to-charge conversion for charge sensing experiments that are targeted towards the detection of Majorana fusion rules (which are unique to non-Abelian anyons) and towards the coherent operation of a prototype topological qubit~\cite{Aasen2016}. Not surprisingly, the ability of reflectometry to reveal conductance changes or charge changes of quantum devices is therefore relevant for studying topological superconductivity.  

For example, Majorana readout based on conductance measurements has been proposed for so-called Majorana box qubits. The simplest box qubit consists of an island of topological superconductor hosting four Majorana bound states at four different locations~\cite{Plugge2017}. Two of them are coupled via controllable tunnel barriers to a semiconducting region, such that they can participate in conductance measurements. 
Because the combination of two Majorana operators ($\gamma_{i}$) constitutes one fermionic operator, the even parity state of the four-Majorana box is two-fold degenerate, thereby encoding one qubit. Importantly, it can be arranged such that the transmission phase of the two Majorana states that participate in transport depends on the state of the qubit. Interferometric measurement of the device conductance would therefore allow \emph{readout} of the state of the box qubit. 
The functionalities can be extended to qubit \emph{control} by hosting six Majorana bound states on the topological island, and by connecting three of them to gate-controlled semiconducting quantum dots. Conductance measurements that involve pairs of these three Majorana states would then implement readout of different Pauli operators ($\hat{x}=i\gamma_1\gamma_2$, $\hat{y}=i\gamma_3\gamma_1$, $\hat{z}=i\gamma_2\gamma_3$). Although never realized in practice, such gate-controlled measurements in the time domain would resemble, from an operational viewpoint, certain gate-controlled experiments in the field of spin qubits, with rf reflectometry providing useful tools for accurately and quickly detecting conductance changes.

Motivated by the intriguing roles of Josephson couplings and Majorana couplings between topological superconductors, readout schemes that potentially detect the associated dispersive shifts of rf resonators or superconducting microwave cavities have been proposed. In the specific case of reflectometry-based readout, Ref.\cite{Maman2020a} considers a popular scheme for Majorana readout, where zero modes ($\gamma_{1}$ and $\gamma_{2}$) are coupled to an auxiliary quantum dot. This setup can be used to read out the joint parity of the two Majorana zero modes by tunnel coupling to an auxiliary quantum dot. Calculations of the parity-dependent capacitances of the coupled system (which depend on the on-site energy of the readout dot and the complex-valued but tunable tunnel coupling between the dot and $\gamma_{i}$) and of the ensuing reflectances are given in Ref.\cite{Maman2020a}, along with the estimated readout fidelity.  

Experimentally, rf readout has been applied to Majorana-type devices. Examples in this direction include measurements of the tunneling rates in InAs nanowire devices, which are an early test bed to search for Majorana bound states. Another example is the development of superconducting resonators that can withstand magnetic fields in the range of 1-2~T where Majorana bound states occur. 

Two studies carried out in InAs nanowires are shown in Fig.~\ref{fig:ExampleMajorana}. The first study~\cite{Watson2019} (Fig.~\ref{fig:ExampleMajorana} (a-b)) utilises dispersive charge sensing, relying on a quantum dot charge sensor controlled with a top gate connected to a standard off-chip lumped-element resonant circuit. The resulting dispersive shift of the resonance frequency was significant ($\approx$1~MHz, of the order of the resonator linewidth), corresponding to a  detected phase shift of the reflectometry signal of nearly 180 degrees. The experimental demonstration to resolve the tunneling-dependent quantum capacitance $\CQ$ paves the way to detecting coherent Majorana couplings between topological islands. 

A second study~\cite{Marcus2019} (Fig.~\ref{fig:ExampleMajorana} (c-d)) used a proximal nanowire charge sensor instead of gate-based reflectometry, with high readout SNR reported up to 1~T. In this study, the evolution of Coulomb blockade regions could be studied without a current flow through the device.  There was a transition from 2$e$ periodicity at zero magnetic field to 1$e$ periodicity at finite axial magnetic fields, as experimentally observed in Fig.~\ref{fig:ExampleMajorana}(d). When accompanied by the hard superconducting gap remaining, this change in periodicity has been theorized as an indicator for transitioning between trivial and topological superconductivity. 

Lastly, depending on the bound state occupation, the fermion parity of a nanowire Josephson junction can be even or odd. Dynamic polarization of this even/odd parity and its single-shot detection has been very recently demonstrated, but for Andreev bound states, with up to 94\% fidelity, with measurements performed via a superconducting LC resonator\cite{Wesdorp2021}.

While these experiments show that fast and high-SNR measurements based on reflectometry potentially allow the identification of topological properties, they have not yet been applied to complex experiments such as braiding or the demonstration of fractional statistics.

\subsection{Noise experiments}
\label{sec:noiseMeasrement}

\begin{figure}
    \includegraphics{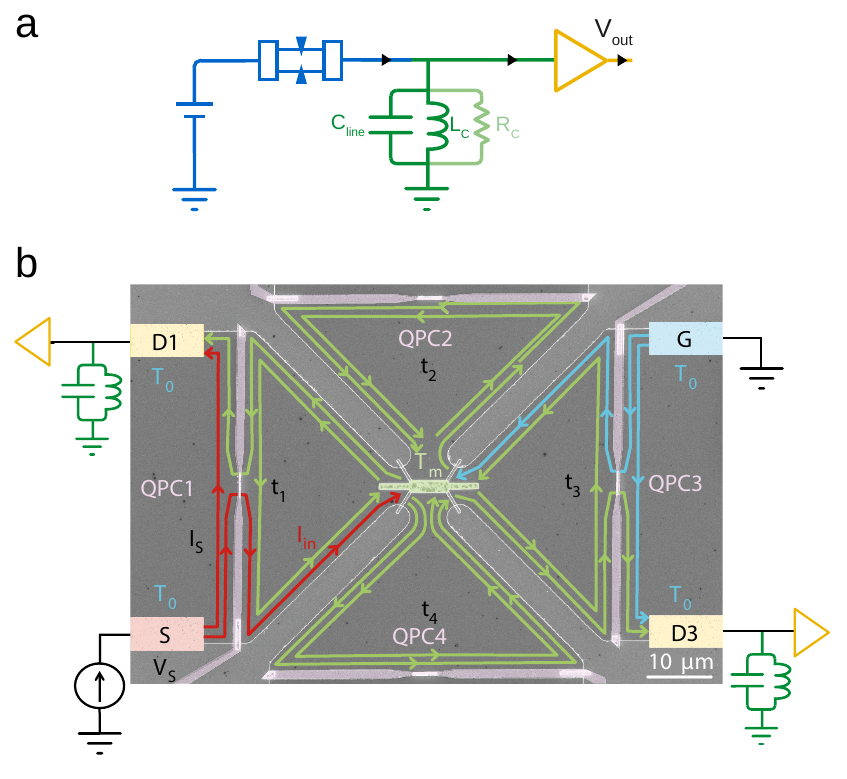}
    \caption{
    (a) Schematic of a shot noise measurement circuit~\cite{Mahalu1998}. 
    The tank circuit is formed from the parasitic transmission line capacitance $\Cline$, the resistance $Rc$ of the device in parallel with the real part of the amplifier input impedance, and an added inductor $\Lc$.
    In this circuit, the amplifier is sensitive to device noise within a bandwidth of order $1/\Rc \Cline$ centered on the frequency $\fr = 1/2\pi\Lc\Cline$.
    By measuring the voltage noise integrated across this bandwidth, the spectral density $S_{II}(\fr)$ can be inferred.
    (b) An SEM micrograph of a four-arm device fabricated in GaAs, with a small floating contact lead at its center (green; the depleting gates underneath are not visible), a quantum point contact (QPC) in each arm (an air-bridge shorts the two sides of the split-gate) and source (S1), drains (D1, D2) and ground (G) contacts. 
    The device is placed in the quantum Hall regime with filling factor $\nu = 2$ by setting the required magnetic field. QPC2 and QPC4 are fully pinched off while QPC1 and QPC3 transmit only the outmost ballistic chiral quantum Hall edge mode  (t$_i$ is the transmission coefficient of QPC$_i$).
    The Source current ($I_\mathrm{S}$, red) impinges on QPC1, which transmits $I_\mathrm{in}$ that is absorbed in the floating contact. 
    Edge modes (green) at temperature $T_\mathrm{m}$, leave the floating contact into the four arms (in arms 2 and 4 they are fully reflected). Cold edge modes, at temperature $T_\mathrm{0}$ (blue) arrive from the grounded contacts. $LC$ circuits at each drain transmit the signal at $\fr=740$~kHz with a bandwidth $\Delta f= 10-30$~kHz. Panel (b) adapted from Ref.~\onlinecite{Banerjee2017}.
    }
    \label{fig:ShotNoise}
\end{figure}

Electrical noise itself is a valuable source of information associated with different physical phenomena~\cite{landauer1998noise,Buttikker2000}. 
Shot noise, which results from the discreteness of charge carriers, leads to current fluctuations with spectral density $\SII=F2qI$ (see Eq.~\eqref{equ:SII}), where $I$ is the average current and $q$ is the charge $q$ of the particles (or quasiparticles) carrying that current. 
By measuring $\SII$ in a situation in which the Fano factor $F$ is known, the charge $q$  can be deduced.
Usually the current is carried by electrons with charge $q=e$, but in some correlated states the excitations are quasiparticles with a fractional charge.
A particularly clear example is the $\nu=1/3$ fractional quantum Hall state, for which the carrier charge is~\cite{Mahalu1998,Mahalu1998Nature} $q=e/3$

The experiment that confirmed this fact measured the shot noise generated by a tunnel barrier between two regions of two-dimensional electron gas in the fractional quantum Hall state.
The tunnel barrier acts as a broadband noise source.
It is measured using a cryogenic voltage amplifier as in Fig.~\ref{fig:ShotNoise}(a).
The current noise is transduced to a voltage noise by the real part of the amplifier input impedance in parallel with the sample impedance, together represented by the resistor $R_\mathrm{C}$.
As with the rf-SET, the bandwidth of the measurement is limited by the capacitance $\Cline$ of the transmission line, meaning that if the amplifier is directly connected to the noise source, it can only detect current noise up to a frequency $\sim 1/R_\mathrm{C} \Cline$, which would be  $\sim 30$~kHz in this experiment.
In this frequency range, pink noise due to background charges usually overwhelms the shot noise of interest, making precise measurements impossible.
This problem is circumvented by inserting an inductor in parallel with the line capacitance, thus forming a resonant tank circuit and shifting the range of frequencies over which the experiment is sensitive up to the tank circuit's resonant frequency, which in this case is $\sim 4$~MHz.
The sensitive bandwidth is unchanged.
Comparing the shot noise measured in this way with Eq.~\eqref{equ:SII} implies a quasiparticle charge $q=e/3$~\cite{Mahalu1998}.

While measuring shot noise gives information about the charge carriers in a device, measuring thermal noise gives information about their temperature.
This means that noise measurements can be used to study how heat flows in quantum devices.
One example is the measurement of quantised heat transport by anyonic carriers~\cite{Banerjee2017}.
As shown in Fig.~\ref{fig:ShotNoise}(b), resonant circuits at the drain electrodes of a 2DEG gated by two QPCs, QPC1 and QPC2, are used to filter the chiral ballistic 1D edge modes that transmit the carriers under high magnetic field. These edge modes are at an equilibrium temperature $T_\mathrm{N}$, where the power dissipated in the central floating contact is equal to the power carried by phonons and the chiral edge modes. Since the phononic heat contribution is negligible compared to the strongly-interacting electronic contribution for $T<35$~mK, the temperature $T_\mathrm{N}$ can be determined from thermal noise measurements in one of the arms. These thermal noise measurements, carried out as above, use cold amplifiers to measure thermal voltage fluctuations.

\subsection{Micro- and nanomechanical resonators}
\label{sec:NanomechanicalResonator}

\begin{figure*}
    \centering
    \includegraphics{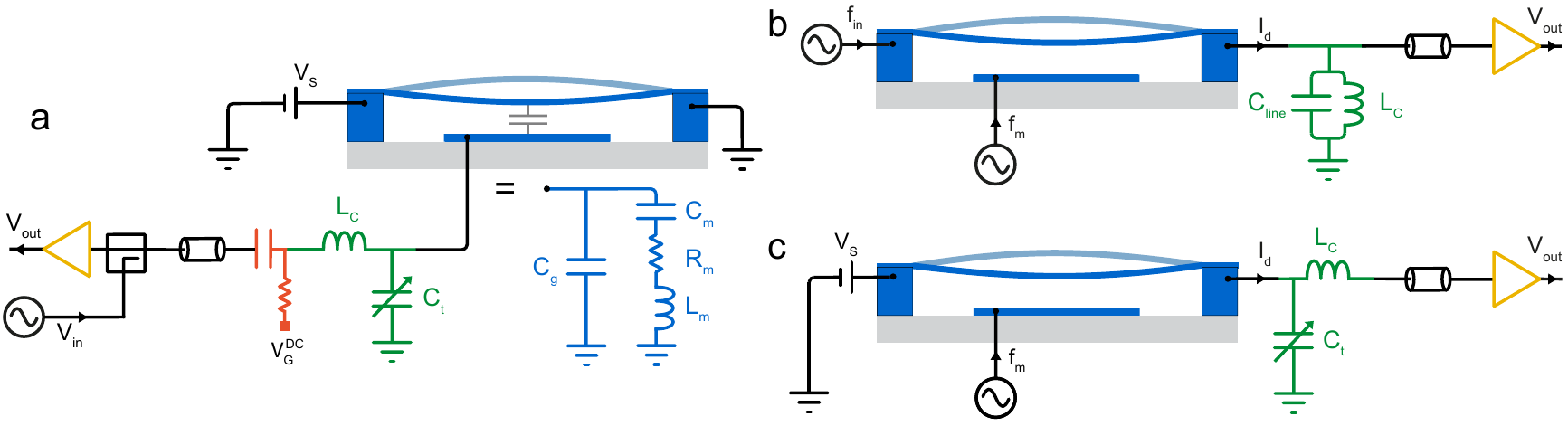}
    \caption{
    (a)	Schematic of the measurement of the motion of a carbon nanotube mechanical resonator using gate reflectometry~\cite{ Schwab2007,Laird2016_CNT}. An $LC$ resonator made of an inductance $\Lc$ and a tunable capacitance $\Ct$ is attached to the gate of the device. The nanomechanical resonator is represented using the van Dyke–Butterworth model~\cite{vanDyke1928} by a static capacitor $\Cg$ in parallel with an RLC circuit with equivalent elements  $\Cm$, $\Rm$ and $\Lm$.
    (b) Schematic of the measurement of the motion of a carbon nanotube mechanical resonator with the  two source method. In this example~\cite{Bachtold2018}  the $LC$ resonator comprises an inductor $\Lc = 66~\micro$H in parallel with the capacitance of the line $\Cline=242$~pF to transmit the current noise $I_d$ generated motion of the mechanical resonator.
    (c) Schematic of the measurement of the motion of a carbon nanotube mechanical resonator~\cite{Laird2018} using a tuneable capacitance $\Ct=3.82$~pF and an inductor $\Lc = 223$~nH to bring the $LC$ resonator in resonance with the mechanical motion to capture the current noise $I_d$ generated directly at the mechanical resonance frequency.
    }
    \label{fig:MEMS}
\end{figure*}

Due to their small size and mass, nanomechanical resonators have high mechanical resonance frequencies $\fm$ of the order of hundreds of MHz. It is therefore natural to turn to radio-frequency measurement techniques to measure their motion. Moreover, radio-frequency measurements can provide additional information; an example is the demonstration of coherent mechanical oscillations in carbon nanotubes using IQ demodulation\cite{Laird2020_CNT,Bachtold2019}.
Nanomechanical resonators made of carbon nanotubes, graphene or aluminium sheets find exciting applications in sensing \cite{Bachtold2012,Bachtold2013} and qubit readout \cite{Schwab2003,LaHaye2009}. The measurement of the mechanical vibrations of carbon nanotubes and nanowires have been used to reveal fundamental properties that can be difficult to probe with transport\cite{Hone2016,Ilani2019,Suh2019}. Current challenges for exploring the foundations of quantum mechanics~\cite{Marquardt2014Review} include measuring mechanical resonators in their quantum ground states of motion~\cite{OConnell2010Apr} or generating a macroscopic quantum superposition of states~\cite{Laird2018Displacemon,Steele2020flux}. Some recent proposals also suggest using QDs coupled to a mechanical oscillator as mechanical qubits~\cite{Pistolesi2021}. 

Gate reflectometry can be used to monitor the motion of nanomechanical resonators by sensing the change of capacitance between the mechanical resonator and a metallic gate electrode connected to the rf cavity.  The sample is illuminated by an input signal at frequency $\fin$ via the gate electrode. The reflected signal contains sidebands at frequencies
\begin{equation}
 	\fout=\fin \pm \fm.
\end{equation}
that transduce the displacement of the sample.
This techniques works well with large devices, such as metallised SiN membranes~\cite{Ares2020Membrane}, for which the capacitance varies considerably with the motion.

It is more challenging to measure smaller devices (such as carbon nanotubes or aluminium nanosheets) where the variation of capacitance with the displacement is small compared to the static capacitance. One solution is to operate on resonance with the mechanical resonator $\fin \approx \fm$ while applying a static voltage between the mechanical resonator and a gate electrode~\cite{Schwab2007}. In this case, the mechanical resonator impedance can be expressed using the van~Dyke–Butterworth~\cite{vanDyke1928} model as a static capacitance $\Cg$ in parallel with an LCR circuit~\cite{Schwab2007} ($\Lm$, $\Cm$ and $\Rm$)  (Fig. \ref{fig:MEMS}(a)) with equivalent impedance:
\begin{equation}
    \frac{1}{\Zm}=j\omega \Cg +\frac{1}{j\omega \Lm + \frac{1}{j\omega \Cm}+\Rm}. 
\end{equation}
When $\fin$ is out of resonance with the mechanical resonance frequency, $\Zm$ is large and dominates by $1/j\omega \Cg$. On resonance, the impedance drops to $\Zm^{-1} \approx \Rm$ allowing detection of the motion. This technique has been employed to detect the motion of an aluminium drum~\cite{Schwab2007} and of carbon nanotubes~\cite{Laird2016_CNT}, although it requires tuning the frequency of the rf resonator to $\fr\approx \fm$ which can be achieved with \textit{in-situ} tuneable cicuits (see Section~\ref{sec:insitu}).

In gated semiconducting mechanical resonators, the motion can be transduced into a current modulation $I_d$ emitted by the device. This current is measured with a cryogenic amplifier, similar to the shot noise measurements described in Section~\ref{sec:noiseMeasrement}, using LCR circuits with resonance frequency $\fr$. In the \textit{two source method}~\cite{Bachtold2018,Ilani2019,Bachtold2019}, the mechanical resonator is biased with an ac voltage at frequency  $\fin=\fr \pm \fm$ while the mechanical motion is excited by a second source at frequency $\fm$ (Fig.~\ref{fig:MEMS}(b)). These two frequencies mix such that the current noise spectrum $I_d$ has a sideband at $\fr$ that is transmitted by the LCR resonator. Alternatively, the LCR circuit can be tuned into resonance with the mechanical resonance frequency $\fr \approx \fm$ using \textit{in-situ} tunable elements~\cite{Laird2018,Laird2020_CNT} (Fig.~\ref{fig:MEMS}(c)).

\subsection{Fast thermometry}
\label{fast thermometry}

\begin{figure*}
    \includegraphics[width=0.9\textwidth]{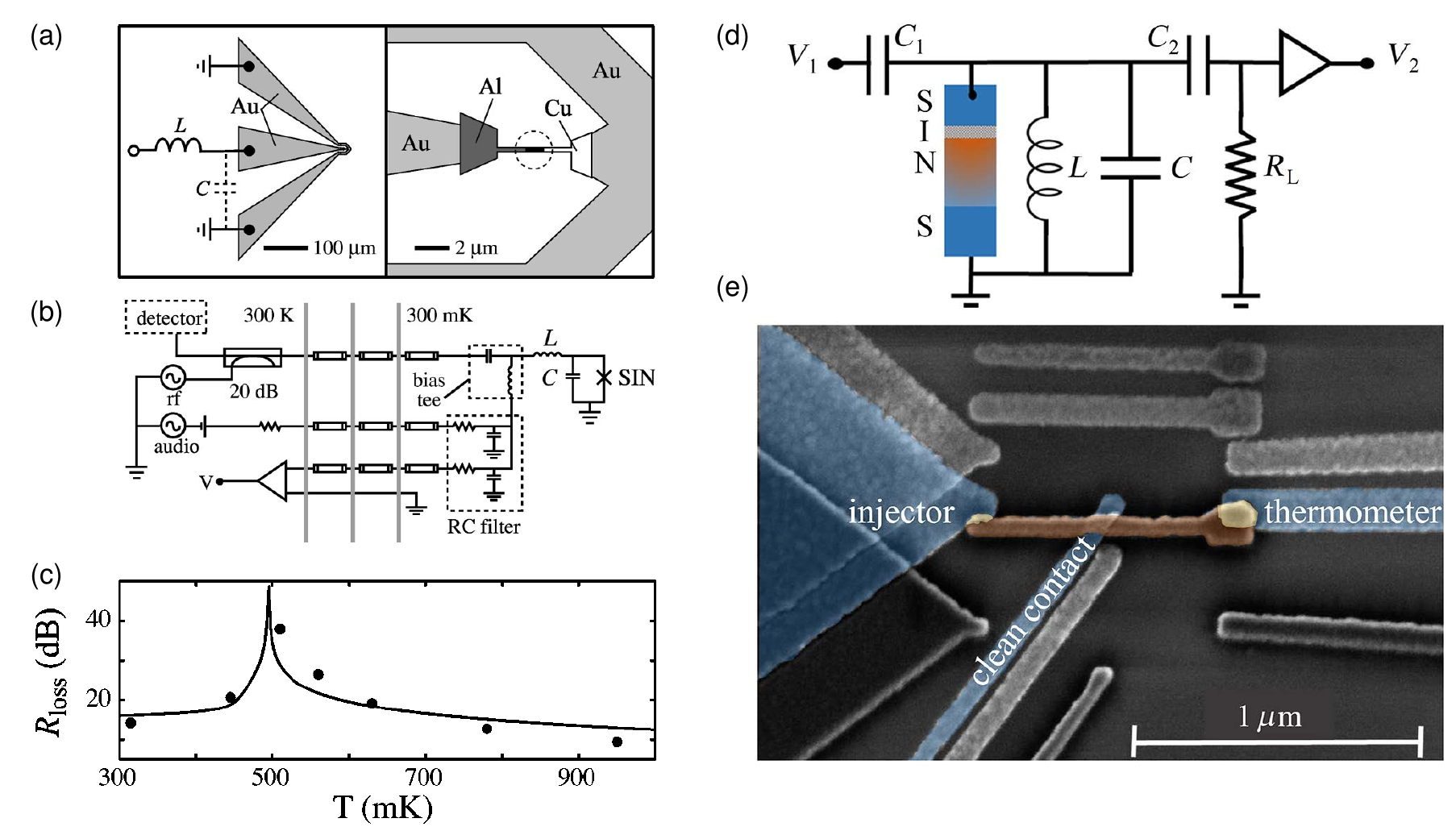}
    \caption{Fast and nanoscale thermometry and calorimetry. 
    (a) To demonstrate nanoscale radio-frequency thermometry, an LC resonator is loaded by a temperature-sensitive Al-oxide-Cu SIN junction (dashed circle). 
    (b) Bias tees allow the application of ac bias voltages, to keep the junction at a temperature-dependent operating point on its IV characteristics, while the remaining reflectometry setup in this work was kept at room temperature. 
    (c) The temperature dependence of the junction's operating point yields the matching condition for the LC tank circuit, evident here by a peak in the reflectometry return loss at 510 mK. Near matching, the demodulated reflectometry signal is sensitive to temperature changes.
(d) SINS junction, in which the proximitized normal metal electrode gives rise to a zero-bias conductance anomaly that makes this thermometer useful for ultrasensitive calorimetry~\cite{Pekola2018}. Importantly, this junction can be read out with reflectometry without the need to apply large (and invasive) bias voltages. 
(e) Implementation of the proximitized normal electrode by establishing a clean contact between the copper electrode (orange) and an aluminum electrode (blue). The overlapping region (yellow) with another aluminum electrode (thermometer) constitutes the SIN junction. The tunnel junction on the left (injector) was used to inject heat on the copper island, thereby manipulating its temperature in a well controlled manner. Panels (a),(b),(c) adapted from Ref.~\onlinecite{Cleland2003}. Panels (d),(e) adapted from Ref.~\onlinecite{Pekola2018}.} \label{fig:PekolaPlusCleandThermometry}
\end{figure*}

Already in the early days of dilution refrigerator experiments, it was noticed that high-frequency measurements provide practical solutions for the challenging task of implementing reliable subkelvin thermometry. For example, Johnson noise thermometry utilizes the fluctuation-dissipation theorem, on the principle that the voltage fluctuations of a resistor (i.e. the mean square noise voltage within some suitable bandwidth) are proportional to the resistor's temperature. At low temperatures, the minuscule voltage fluctuations of a resistor can be elegantly converted to frequency fluctuations using the ac Josephson effect (the factor $2e/h$ corresponds to an attractive conversion factor of $\approx$ 484~MHz/$\mu$V), yielding reliable thermometry in the 10~mK range with a measurement noise temperature (response time) of 0.05~mK (50~ms) as early as 1973~\cite{Webb1973}. While this technique is not based on reflectometry, it loosely fits into the category of high-frequency emission measurements (in this case, via a 19~MHz tank circuit connected to a SQUID amplifier). Another early  high-frequency technique (important for microkelvin applications) involves the detection of $^{125}$Pt nuclear spin susceptibilities (which follow a Curie law) using pulsed nuclear magnetic resonance techniques, allowing not only determination of nuclar spin temperatures, but also (via the Korringa law) the temperature of the electrons~\cite{Aalto1973}.

More recently, efforts have been made to study the effective temperature and time scales associated with different degrees of freedom in small systems (phonon temperature in mechanical resonators, electron temperature in isolated quantum devices, photon temperatures in microwave cavities, etc.). Quantum devices that involve superconductor–insulator–normal metal (SIN) junctions are of particular interest, as they allow fundamental insights (interplay of high thermal conductance associated with normal metals with low thermal conductance associated with superconductors) as well as technological applications (such as SINIS on-chip coolers~\cite{Nguyen2016}). 

\begin{figure*}
    \includegraphics[width=\textwidth]{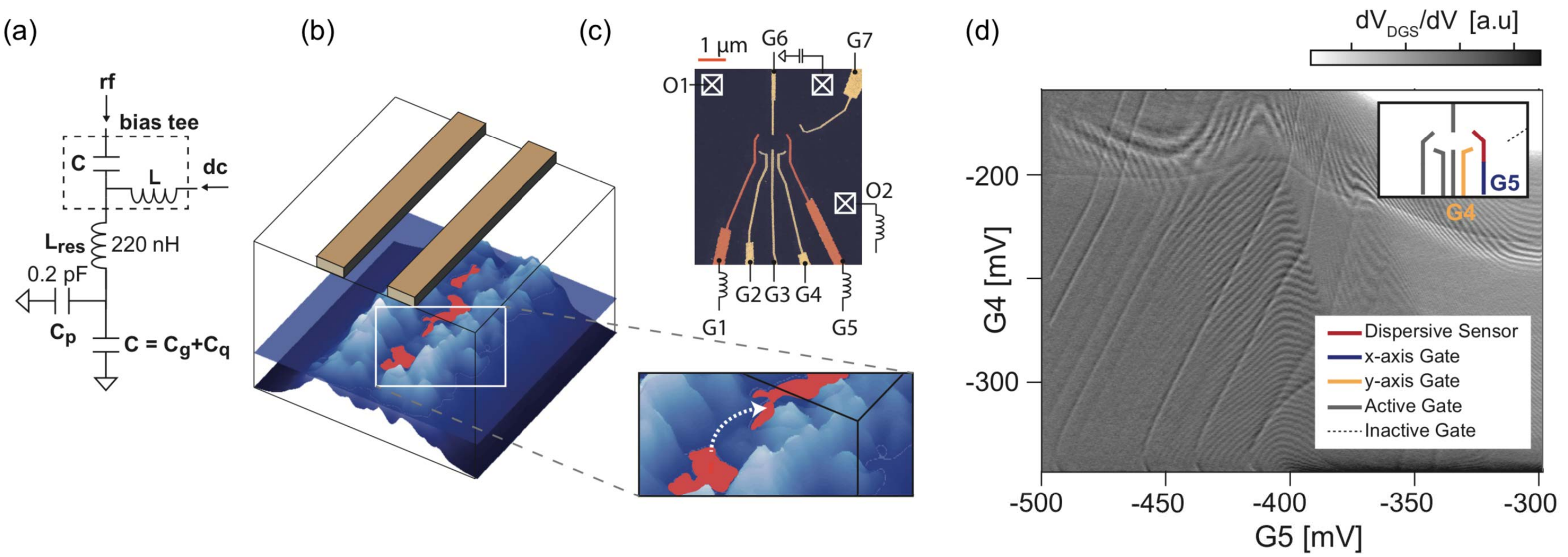}
\caption{Gate-sensing of charge pockets in the semiconductor environment of GaAs QPC devices.
(a) Schematic of the reflectometry circuit, indicating the the rf response is sensitive to geometrical capacitances ($C_\mathrm{g}$, set by the geometry of gate electrodes) and quantum capacitances ($C_\mathrm{q}$, sensitive to the density of states in the semiconductor). 
(b) Visualization of electrostatic disorder potentials (blue) affecting the localization of the 2DEG in charge pockets (red areas). 
(c) Charge-pocket phenomena occuring near different gate electrodes (and not necessarily contributing to traditional conductance measurements of the different QPCs of this device) can be detected by performing frequency-multiplexed reflectometry of different gate electrodes, for example G1 and G5. 
(d) Example of the G5 reflectometry signal as a function of gate voltages applied to G5 and G4. To increase the visibility of quasi-periodic Coulomb oscillations with varying slopes and periods, associated with different charge puddles, the derivative of the demodulated voltage (wrt gate-voltage G4) is plotted. The precise pattern of Coulomb-blockade oscillations was sensitive to the exact voltages applied to neighboring gate electrodes (not shown). Figures adapted from Ref.~\onlinecite{Reilly2019Pocket}.
    }
    \label{fig:Croot_charge_puddle_physics}
\end{figure*}

Figure~\ref{fig:PekolaPlusCleandThermometry}(a) shows a nanoscale SIN junction (implemented by the aluminum - aluminum oxide - copper junction in the dashed circle, which was created by a multi-angle shadow evaporation technique) that is wirebonded to an inductor $\Lc=390$~nH~\cite{Cleland2003}. In conjunction with the stray capacitance $\Cp=0.6$~pF of the bond wire, it forms a 338~MHz resonator that goes through matching as the SIN junction is cooled and temporarily reaches the matching resistance of 20~k$\Omega$. As a consequence, the return loss (measured by reflectometry, and plotted in Fig.~\ref{fig:PekolaPlusCleandThermometry}(c) shows a pronounced peak near 500~mK. The effective temperature (bandwidth) of this high-bandwidth "reflectometry thermometer" was 0.3~mK/Hz$^{1/2}$ (10~MHz), but adjustments to the choice of circuit parameters and improvements to device and reflectometry setup allow application to other temperature ranges and to thermodynamic and calorimetric studies of mesoscopic nanostructures and far-infrared detectors. 
To put these numbers into the context of calorimetry, Ref.~\onlinecite{Pekola2018} notes that a temperature noise of 10~$\mu$K/Hz$^{1/2}$ would be desired to detect the heat quantum associated with, say, a 1-K microwave photon. 
An improved modification is shown in Fig.~\ref{fig:PekolaPlusCleandThermometry}(d), where the normal metal contact has been replaced by a proximitized normal metal~\cite{Pekola2018}. In contrast to the SIN junction, this results in a zero-bias conductance feature in the IV characteristics of the junction, thereby obliterating the need for large biasing voltages (which constitute a source of heating). Accordingly, this thermometer was demonstrated to function at temperatures as low as 25~mK, with a sensitivity and noise performance almost sufficient to detect heat quanta relevant for superconducting qubit circuits (1-K photon conversions). 

For semiconducting quantum devices, primary thermometry (measured via reflectometry) is possible by employing that the cyclic electron tunneling between a discrete state in a QD and an electron reservoir depends on the thermal distribution function of the reservoir. By embedding the plunger gate electrode of a quantum dot in an rf resonator, the reflectometry carrier induces cyclic tunneling and dispersively senses the tunnelling response~\cite{Gonzalez-Zalba2018_thermo}. Interestingly, in certain regimes, the width of the tunneling capacitance along the detuning of the dot (with respect to the Fermi level of the reservoir) depends only on temperature, thereby making this a primary thermometer for the electron reservoir (if the detuning lever arm is known). 
This lever arm is usually measured using a source-drain bias across the thermometer device, but it can also be calibrated by measuring the width of the tunneling peak at known temperature.
This allows the temperature to be measured, even when the circuit is galvanically isolated~\cite{Chawner2021}.
Alternatively, the  thermal distribution of the reservoir can be deduced by conductance measurement of the quantum dot from which the temperature is deduced. This process can be considerably accelerated using rf-readout of the QD resistance~\cite{Blanchet2021}.

\subsection{Sensing the semiconductor environment}
\label{sensing Environement}

In semiconducting devices with reduced dimensionality for the effective carriers, physical intuition from bulk systems has been shown again and again to break down with the emergence of quantum phenomena. For the kinetic energies, a drastic modification of the density of states by spatial quantization appears already in the simplest, non-interacting treatment of single-particle states (such as van Hove singularities in quasi 1D systems). For the electrostatic potentials, distinctly different length scales associated with electrostatic screening appear. For example, in 1D systems, the familiar exponential decay of potentials in 2D interfaces, or depletion lengths, are replaced by very long-range (logarithmic) tails in the charge distribution that affect the physics and engineering of p-n junctions, n-i junctions, and metal-semiconductor heterojunctions (Schottky barriers)~\cite{Tersoff1999, Odintsov2000}. Moreover, the categorical classification of any 3D fundamental particle as either a boson or a fermion (originating from the strictly integer and half-integer eigenvalues associated with rotational symmetry in 3-dimensional space) no longer holds for emergent quasiparticles in 2D and 1D systems. Here, the reduced spatial symmetries allow other exchange statistics, including abelian and non-abelian exchange statistics associated with \emph{anyons} that are neither bosons nor fermions. 
For instance, in 2D systems, capacitance measurements play an important role for establishing localization of normally or fractionally charged quasiparticles in GaAs quantum Hall systems~\cite{Ilani2004, Martin2004, Venkatachalam2011}, and electron-hole puddles~\cite{Martin2008} and correlated instulators~\cite{cao2018correlated} in graphene structures.  

In 1D systems, electron interaction gives rise to such exotic phenomena as spin–charge separation and the emergence of correlated-electron insulators and Wigner crystals (facilitated by the ineffective screening of the long range Coulomb interaction in 1D), which have been traditionally been measured by transport~\cite{Bockrath1999, Auslaender2005, Deshpande2010, Pecker2013} but are also conducive to charge sensing~\cite{Ilani2019Wigner}. 

Not surprisingly, reflectometry of semiconductors using dispersive gate sensing reveals information complementary to traditional transport measurements. For example, conductance measurements of quantum point contacts~\cite{Kouwenhoven1988, Wharam1988} often show a mysterious anomaly at $0.7 \times 2e^2/h$ that is now thought to comprise interaction and scattering effects in the presence of a smeared van Hove singularity in the local density of states at the bottom of the lowest one-dimensional subband~\cite{Bauer2013} or perhaps Kondo correlations~\cite{Iqbal2013}. 

When applying dispersive gate sensing to similar quantum point devices, a surprising richness of features appeared in gate voltage space that persists even below the threshold for non-zero conductance~\cite{Reilly2019Pocket}. In Fig.~\ref{fig:Croot_charge_puddle_physics}(c), we show a clever use of frequency multiplexing that allows reflectometry of various gate electrodes (and contact leads) of a top-gated GaAs heterostructure. This way, the reflectometry features of various gate electrodes could be compared for the same chip (an example for the reflectometry signal from gate G5 is shown in panel (c). The appearance of many (quasiperiodic) oscillations with varying slope in gate-voltage space were attributed not to physics associated with the region of the quantum point contacts (which dominates the signal in transport measurements), but to regions near the extended gate electrodes that, in the presence of spatial disorder in the potential landscape, form charge puddles. Follow-up work suggests that reflectometry features in the pinched-off regime (i.e. zero QPC conductance) may also have contributions from asymmetric capacitive couplings between the reflectometry gate and the source and drain reservoirs, whereas the  non-zero conductance staircase associated with QPC behaviour can in fact show up clearly in reflectometry~\cite{Reilly2020QPC}. 

\subsection{SQUID magnetometer}
\label{Section9SQUID}

Superconducting quantum interference devices (SQUIDs) are used as extremely sensitive magnetometers, among other applications, in quantum sensing and quantum technologies. For example, SQUIDs are employed to measure cosmic radiation~\cite{Ullom2017} or as particle detectors~\cite{Enss2017}.

SQUIDs falls into two categories~\cite{Clarke2004}: the dc SQUID and the rf SQUID. dc SQUIDs are made of two junctions in parallel in a superconducting loop. When the dc SQUID pick-up a small magnetic flux it generates a screening current along the loop that maintains the total flux to a multiple of the flux quantum $\Phi_0=h/2e$. As we have already seen, dc SQUIDs can be employed as ultra-low noise amplifiers. 

\begin{figure}
    \includegraphics[width=\columnwidth]{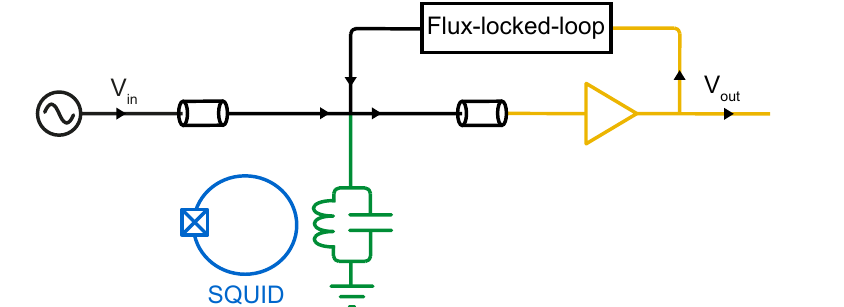}
    \caption{rf SQUID setup inspired by Ref.~\onlinecite{Clarke2004}}
    \label{fig:SQUID}
\end{figure}

We are interested here in the second type: the radio-frequency SQUID (rf SQUID). These are made of only one Josephson junction in a superconducting loop that is inductively coupled to a $LC$ resonator formed by an inductance and a capacitance in parallel (Fig.~\ref{fig:SQUID}). Because they are made of only one Josephson junction, rf SQUIDs are easier to fabricate. But their sensitivity is limited by the readout setup which is a motivation for further optimisation~\cite{Clarke2004}. 

The readout of an rf SQUID is a transmission-type measurement with the $LC$ resonator inductively coupled to the SQUID (Fig. \ref{fig:SQUID}). A radio-frequency signal $\Vin$ is sent to the $LC$ resonator and the SQUID while the transmitted signal is amplified to become the output signal $\Vout$. The magnetic flux picked up by the SQUID changes the phase across the Josephson junction, which then modifies the impedance of the SQUID and the resonator. This translates into the transmitted signal phase and amplitude. 
The rf SQUID measurement setup generally integrates a flux-locked-loop feedback system to maintain the SQUID at its sweet spot where its sensitivity is best. This is especially important when the bandwidth of the LC resonator is low. 

We distinguish two types of rf SQUID, based on device parameters which influence the readout method, depending on $\beta_\text{rf}=2\pi L I_0/\Phi_0$ \cite{Clarke2004} where $L$ is the loop inductance and $I_0$ is the critical current of the Josephson junction. If $\beta_\text{rf}>1$ the SQUID is hysteretic and the transmission measurement is dissipative. The radio-frequency signal causes the Josephson junctions to switch periodically between two quantum states causing dissipation of the signal. This dissipation reduces the quality factor of the $LC$ resonator and affects the amplitude of the transmitted signal, similar to a resistance measurement. This switch causes intrinsic noise that limits the sensitivity of a hysteretic rf SQUID.

If $\beta_\text{rf}<1$, the SQUID is non-hysteretic and the measurement is dispersive. The inductor of the tank circuit is parametrically coupled to the SQUID. The magnetic flux in the SQUID loop modifies the total inductance of the circuit, which changes its resonance frequency. This regime gives less intrinsic noise than the hysteretic SQUID.

\section{Conclusion and Outlook}

After two decades of developments, the high-speed electrical readout of quantum devices is allowing us to advance quantum computing and several other fields of research. In this Review, we focused on reflectometry circuits to perform high-speed sensitive measurements of quantum devices. In the development of radio-frequency readout techniques, circuit quantum electrodynamics and its application to the readout of superconducting qubits has been a source of inspiration \cite{William2019SuperQubits}. 

One of the main driving forces for the advancement of radio-frequency technologies has been the rise of charge and spin qubits. The need for fast, sensitive and scalable readout of charge and spin states has promoted the development of single-shot readout techniques, the integration of superconducting components and the search for circuit multiplexing approaches described in this review. Careful engineering of the rf circuits made the difference. The optimisation of matching circuits, amplifier chains and PCB designs and materials are a few of the strategies discussed. 

Fast measurements are not only directed at the readout of charge and spin states, but also to the tuning and characterisation of quantum devices \cite{Petta2015,Schupp2020,Pfaf2021}. 
Video-mode measurements demonstrate the potential of rf circuits for tuning quantum dot devices. The limiting factor is no longer the speed of the measurements but the ability of humans to analyse and interpret the data. The integration of machine learning techniques is allowing us to tackle such limitations~\cite{Ares2020Expert,Ares2020fine-tuning,Taylor2020}.

Rf readout has also allowed for sensitive measurements of temperature and motion at the nanoscale, with applications such as the thermalisation of quantum circuits and other aspects of non-equilibrium and quantum thermodynamics. It is also used to sense the semiconductor environment of quantum devices, investigate Majoranas and dark matter, and probe other phenomena in the solid state.
We expect rf-based techniques to enable yet new types of experiments. The use of pulsed magnetic fields~\cite{Chuck2016,Nicolin2017} is an example of a technique that asks for the fast readout capabilities that rf reflectometry can offer. 
Rf readout could also be a key tool for the exploration of different mechanisms of electron transfer. Kondo physics, commonly probed by transport measurements, was found to be ‘transparent’ in a cavity quantum electrodynamic architecture~\cite{Desjardins2017}. In the same way, rf techniques can be used to explore many-body correlations. 

We hope this review is a guide for students and researchers to explore the full potential of rf readout. Our optimisation guidelines, focused not only on rf components but also on the specifics of high-frequency lines, noise floors and amplifier chains, provide a starting point to advance rf reflectometry and to use it to further a broad variety of research fields.

\begin{acknowledgments}

DJR acknowledges Microsoft Corporation and the ARC Centre of Excellence for Engineered Quantum Systems.
FK acknowledges the Danish National Research Foundation, Independent Research Fund Denmark, Novo Nordisk Foundation grant No. NNF20OC0060019, and the European Commission (grant agreements 951852 and 856526). 
MFGZ acknowledges support from Quantum Motion Technologies, the European Commission (grant agreement 951852), the Innovate UK Industry Strategy Challenge Fund, and the UKRI via a Future Leaders Fellowship (grant number MR/V023284/1).
EAL acknowledges the ERC (grant agreement 818751), the EU H2020 European Microkelvin Platform (grant agreement 824109), the Joy Welch Educational Charitable Trust. We thank the UK Science and Technology Facilities Council (STFC) for funding this work through support for the Quantum Sensors for the Hidden Sector (QSHS) collaboration under grants ST/T006102/1, ST/T006242/1, ST/T006145/1, ST/T006277/1, ST/T006625/1, ST/T006811/1, ST/T006102/1 and ST/T006099/1.
NA acknowledges the Royal Society (URF R1 191150), an EPSRC Platform Grant (EP/R029229/1), grant number FQXi-IAF19-01 from the Foundational Questions Institute Fund, a donor advised fund of Silicon Valley Community Foundation, and the ERC (grant agreement 948932).

\end{acknowledgments}

\subsection*{Competing Interests} The authors have no conflict of interest to declare.

\subsection*{Correspondence}
Correspondence and requests for materials should be addressed to Edward A. Laird at e.a.laird@lancaster.ac.uk, Fernando Gonzalez-Zalba at fernando@quantummotion.tech, and/or Natalia Ares at natalia.ares@eng.ox.ac.uk.


\end{document}


\parbox[c]{\textwidth}{\protect \centering \Large \MakeUppercase{Supplementary Material}}
\rule{\textwidth}{1pt}
\vspace{0.4cm}

\maketitle

\tableofcontents

\vspace{0.3cm}
\hrule

\section{Scattering parameters for a transmission measurement}
\label{sec:TransmissionScattering}

\begin{figure}
	\centering
	\includegraphics{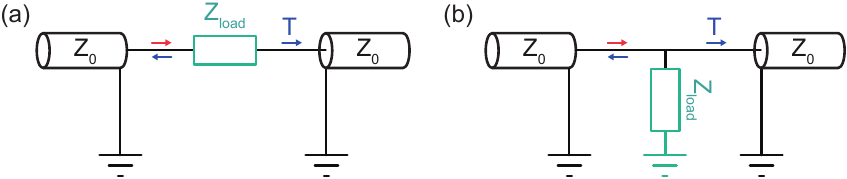}
	\caption{The simplest geometries for a transmission measurement. (a) Transmission through an impedance. (b) Transmission past an impedance.
		\label{fig:transmission}
	}
\end{figure}

Just as the load impedance determines the reflection amplitude in a reflectometry experiment (Eq.~(10)), it also determines the scattering amplitudes in a transmission experiment.
The two simplest transmission circuits are shown in Fig.~\ref{fig:transmission}.
The amplitude for transmission through an impedance $\Zload$ (Fig.~\ref{fig:transmission}(a)) is
\begin{equation}
T = \frac{2\Zo}{2\Zo+\Zload}.
\end{equation}
The amplitude for transmission past an impedance $\Zload$ (Fig.~\ref{fig:transmission}(b)) is
\begin{equation}
T = \frac{2\Zload}{2\Zload+\Zo}.
\end{equation}

Examples of rf-SETs measured in a transmission configuration are Refs.~\onlinecite{Hirayama2000,fujisawa2000_transmission}.

\section{The series equivalent of a reflectrometry resonator; derivation of Equation (42) of the main text}

Here, we demonstrate the approximate equivalence between the reflectometry circuit of Fig.~7(a), described by Eq.~(40), and its series model in Fig.~7(b), described by Eq.~(42). We do this by showing that they have the same impedance near resonance.

First, we write explicitly the real and imaginary parts of Eq.~(40):
\begin{equation}
    \Zload =j\omega \Lc + \RL+\frac{\Req}{1+\omega^2\Req^2 \Cc^2}-j\omega \frac{\Req^2\Cc}{1+\omega^2\Req^2\Cc^2}
\end{equation}
In the limit $\omega \Req \Cc \gg 1$, which is true for most applications, we obtain:
\begin{equation}
    \Zload =
    \RL+\frac{1}{\omega^2 \Req \Cc^2} + j\omega \Lc +\frac{1}{j\omega\Cc}.
    \label{eq:LoadReIm}
\end{equation}
The resonant angular frequency $\omegar=1/\sqrt{\Lc\Cc}$ can be found by setting the imaginary part of Eq.~\eqref{eq:LoadReIm} equal to zero. 

Finally, to see the equivalence between the reflectometry circuit on resonance and a standard RLC circuit, substitute the resonant frequency into Eq.~\eqref{eq:LoadReIm}, to find the effective resistance: 
\begin{equation}
    \Reff=\frac{\Lc}{\Cc\Req} +\RL,
\end{equation}
which implies that near the resonance frequency, the reflectometry circuit behaves like an series RLC circuit with impedance 
\begin{equation}
    \Zload = \Reff + j\omega \Lc+\frac{1}{j\omega \Cc}.
\end{equation}

\section{Using spectral densities}

In this section, we summarise how to calculate and use a spectral density, with a focus on quantitative experimental analysis.
Two excellent explanations of how to understand and use spectral densities are the review article by Clerk \etal~\cite{Clerk2010}, written from a theoretical physics perspective, and the textbook by Press \etal~\cite{Press2007}, written from a computer science perspective.
Unfortunately nomenclature differs in many ways between these two fields, and both differ from the conventions of electronic engineering, represented e.g.\ by the textbook of Horowitz and Hill~\cite{Horowitz2015}.
Infuriating scaling factors proliferate, and some of them are infinite. Here we present a self-consistent pedagogical treatment, written from an experimentalist's perspective and including brief derivations and examples, of how to calculate a spectral density and use it to estimate uncertainty in a measurement.

The spectral density $\SVV[f]$ represents the intensity of a signal $V(t)$ near frequency $f$.
This representation involves some choices.
We make the following choices in order to make our spectral densities consistent with what appears on the screen of your spectrum analyser:
\begin{enumerate}
\item
The signal $V(t)$ is assumed to be real and classical.
\item
The spectral density of a voltage signal is defined by Eq.~(134), giving units $\mathrm{V}^2\mathrm{s}$.
Some authors~\cite{Press2007} call this ``power spectral density per unit time.''
\item
The spectral density is one-sided, which means that it is defined for both positive and negative $f$ but is normalised so that $\int_0^\infty \SVV[f]\,df = \langle V^2(t) \rangle$.
\end{enumerate}

With these conventions, we will show how to calculate a spectral density in different situations, and how to use it for its most valuable purpose, which is to derive uncertainties in measured quantities.

\floatstyle{boxed}
\newfloat{Textbox}{b!}{lop}[section]

\begin{Textbox*}
\centering
\noindent\textsf{\textbf{Recognizing and converting between definitions of the spectral density}}\\
\vspace{0.2cm}
\begin{minipage}[t]{0.48\textwidth}
\justifying
Here's our cheat sheet for converting between conventions for the classical spectral density. It covers most of the definitions we have encountered.

\begin{itemize}
	\item[(a)]
	This Review follows the \textbf{one-sided convention} common among experimentalists, in which the factors in Eqs.~(\ref{eq:SVVfourier}-\ref{eq:windowedFT}) are chosen so that
	\begin{equation}
		\EE{V^2(t)} = \int_0^\infty \SVV[f] \, df.
	\end{equation}
	In this convention the Wiener-Khinchin theorem, i.e.\ the inverse of Eq.~\eqref{eq:WKtheorem}, is
	\begin{equation}
		\SVV[f] = 2\intinfty \EE{V(0)V(\tau)} e^{-2\pi i f \tau} \, d \tau.
		\label{eq:WKbox}
	\end{equation}
	
	The noise density $e_\mathrm{n}$ used by electrical engineers~\cite{Horowitz2015} is
	\begin{equation}
		e_\mathrm{n}^2[f] = \SVVN[f].
	\end{equation}
	\item[(b)]
	In the \textbf{one-sided convention using angular frequency}, the spectral density $\SVV'$ satisfies
	\begin{align}
		\EE{V^2(t)} 	&= \int_0^\infty S'_{VV}[\omega] \, \frac{d\omega}{2\pi}	\\
		S'_{VV}[\omega] 	&= 2\intinfty \EE{V(0)V(\tau)} e^{- i \omega \tau} \, d \tau.
	\end{align}
	To convert from our convention, use
	\begin{equation}
		S'_{VV}[\omega] = \SVV\left[\frac{\omega}{2\pi}\right].
	\end{equation}
	\item[(c)]
	In the \textbf{two-sided convention using frequency},
	\begin{align}
		\EE{V^2(t)}	&= \intinfty S'_{VV}[f] \, df,	\\
		S'_{VV}[f]		&= \intinfty \EE{V(0)V(\tau)} e^{-2\pi i f \tau} \, d \tau.
	\end{align}
	and the conversion is
	\begin{equation}
		S'_{VV}[f] = \frac{1}{2}\SVV[f].
	\end{equation}
\end{itemize}
\end{minipage}
\hspace{0.5cm}
\begin{minipage}[t]{0.48\textwidth}
\justifying
\begin{itemize}
	\item[(d)]
	In the \textbf{two-sided convention using angular fre-}
	\textbf{quency}, which is common among theorists~\cite{Clerk2010},
	\begin{align}
		\EE{V^2(t)}	&= \intinfty S'_{VV}[\omega] \, \frac{d\omega}{2\pi},	\\
		S'_{VV}[\omega]		&= \intinfty \EE{V(0)V(\tau)} e^{- i \omega \tau} \, d \tau.
	\end{align}
	with
	\begin{equation}
		S'_{VV}[\omega] = \frac{1}{2}\SVV\left[\frac{\omega}{2\pi}\right].
	\end{equation}
	
	\item[(e)]
	In the \textbf{two-sided convention using angular frequency and normalised over $\omega$},
	\begin{align}
		\EE{V^2(t)}	&= \intinfty S'_{VV}[\omega] \, d\omega,	\\
		S'_{VV}[\omega]		&= \frac{1}{2\pi}\intinfty \EE{V(0)V(\tau)} e^{-i \omega \tau} \, d \tau.
	\end{align}
	with
	\begin{equation}
		S'_{VV}[\omega] = \frac{1}{4\pi}\,\SVV\left[\frac{\omega}{2\pi}\right].
	\end{equation}
	\item[(f)]
	In the \textbf{one-sided computer science convention~\cite{Press2007}}, the ``power spectral density'' $P_V[f]$ is defined such that
	\begin{equation}
	    \SVV[f] = \lim_{T \rightarrow \infty} \Braket{\frac{1}{T} \, P_V[f]} 
	\end{equation}
	where $T$ is the measurement duration,
	meaning that
	\begin{equation}
	    \lceil V^2(t)\rceil = \frac{1}{T} \int_0^\infty P_V[f] \, df. 
	\end{equation}
	where $\lceil \cdot \rceil$ denotes a time average.
	Confusingly, $P_V[f]$ has units $\mathrm{V}^2 \, \mathrm{s}^2$, which means it's neither a power nor a density per unit frequency.
\end{itemize}
A final freedom is the sign of the exponent in Eq.~\eqref{eq:WKbox}. Fortunately, if $V(t)$ is real, both choices give the same~$S_{VV}[f]$.

Most papers containing spectral densities either state their convention as one-sided or two-sided, or else define $\SVV$ by an equation similar to Eq.~\eqref{eq:WKbox} by which their convention is implied.
However, some contain more subtle clues, or even no clues at all.
If anything here was of service to you, we implore you to play your part in ending this misery: Whenever you use a spectral density, \emph{say clearly how it is defined.}

\end{minipage}
\end{Textbox*}

\subsection{How to calculate a spectral density}

Suppose our experiment is generating a voltage~$V(t)$.
How do we calculate its spectral density $S_{VV}[f]$?
We will answer this question by presenting the definition of $S_{VV}[f]$ in terms of a Fourier integral.
Under nearly all practical conditions, this definition implies Eq.~(134) of the main text.
We will prove this statement and discuss when it holds.
We  then explain how to estimate the Fourier integral in different situations.

\subsubsection{Definition of $\SVV[f]$ in terms of a Fourier integral}

The one-sided spectral density is defined as 
\begin{equation}
    \SVV(f)~\text{or}~S_{VV}[f]   \equiv 2 \lim_{T\rightarrow \infty} \langle |\VT[f]|^2 \rangle
    \label{eq:SVVfourier}
\end{equation}
where 
\begin{equation}
    \VT[f]    \equiv \frac{1}{\sqrt{T}}\int_{-T/2}^{T/2} V(t) e^{-2\pi i f t} dt.
    \label{eq:windowedFT}
\end{equation}
is the \emph{windowed Fourier transform}\footnote{The sign of the exponent in our Fourier transforms is chosen so that voltage $V$ and current $I$ are related by $V[f]=Z[f]I[f]$ with the conventional definition~\cite{Horowitz2015} of impedance $Z[f]$.
For clarity, this Supplementary uses square brackets for quantities in frequency space, for example $\SVV[f]$.}.
Since $V(t)$ is real, we have $S_{VV}[f]=S_{VV}[-f]$. 

As noted, there is more than one way to define the spectral density.
The most common conventions are summarised in the box overleaf.

\subsubsection{When the two expressions for $\SVV[f]$ are equivalent}
\label{sec:SVVequivalenceproof}

We take Eq.~\eqref{eq:SVVfourier} to define the spectral density, but Eq.~(134) is more intuitive.
Here we explain when the first expression implies the second.

Suppose $V(t)$ is stationary, which means that its statistical properties are independent of time.
(We return shortly to the question of when this is true.)
Then its spectral density, defined by Eq.~\eqref{eq:SVVfourier}, obeys the \emph{Wiener-Khinchin theorem}\footnote{For a proof of Eq.~\eqref{eq:WKtheorem}, see Ref.~\onlinecite{Clerk2010}.}, which states that $\SVV[f]$ is related to the autocorrelation function through a Fourier transform:
\begin{equation}
	\EE{V(t)V(t')} = \frac{1}{2} \int_{-\infty}^\infty df \, e^{2\pi i f (t-t')} \SVV[f].
	\label{eq:WKtheorem}
\end{equation}

Now apply this to the filtered voltage $\VF(t)$ of Eq.~(134), from which all spectral components of $V(t)$ have been removed except those within a small bandwidth $\Bf$ of $f$.
The Wiener-Khinchin theorem now gives
\begin{equation}
	\EE{\VF(t)\VF(t')} = \int_{f-\Bf/2}^{f+\Bf/2} df' \, e^{2\pi i f' (t-t')} \SVV[f'],
\end{equation}
where we have also used that $\SVV[f]=\SVV[-f]$. 
Setting $t'=t$ and dividing both sides by $\Bf$ leads to
\begin{align}
	\frac{\EE{\mathbb{V}^2(t)}}{\Bf}	&= \frac{1}{\Bf}\int_{f-\Bf/2}^{f+\Bf/2} df' \, \SVV[f']
\end{align}
In the limit $\Bf \rightarrow 0$ this becomes\footnote{We need to assume here that $\SVV[f]$ is well-approximated by its average over a small range. This is obviously true if $\SVV[f]$ is continuous, and in fact Eq.~\eqref{eq:SVVdefnSupp2} is also true if $\SVV[f]$ is a delta function.}
\begin{align}
\SVV[f] &= \lim_{\Bf \rightarrow 0} \frac{\EE{\VF^2(t)}}{\Bf}
	\label{eq:SVVdefnSupp1}\\
		&=\lim_{\Bf \rightarrow 0} \frac{\EEE{\VF^2(t)}}{\Bf}.
	\label{eq:SVVdefnSupp2}
\end{align}
where the second equality follows because $V(t)$ is stationary and therefore the time average $\lceil \cdot \rceil$ does not change the right hand side.
This is identical to Eq.~(134) in the main text.

What about a non-stationary $V(t)$?
For example, $V(t) = A \cos (2\pi f_0 t)$ is clearly non-stationary because its variance depends on time as $\EE{V^2(t)} \propto \cos^2 (2 \pi f_0 t)$.
Does Eq.~\eqref{eq:SVVdefnSupp2} hold for such an observable?
Although we cannot use our argument based on Eq.~\eqref{eq:WKtheorem}, we show below Eq.~\eqref{eq:SofFourierSeries2} that any signal that can be represented as a Fourier series nevertheless obeys Eq.~\eqref{eq:SVVdefnSupp2}.
Thus we have proved Eq.~(134) in the main text, provided that $V(t)$ is either stationary or a Fourier series.

These two cases cover many observables that are encountered experimentally
\footnote{
	An example of a voltage that is not stationary and cannot be represented by a Fourier series is $V(t) = At$.
	If you have this in your experiment and you cannot correct for it, then you have a problem.
}.
The reason that most observables, especially noise, are stationary is time translation invariance; once an experiment has been running for a long time, its behavior should not depend on when it was turned on.
As we shall see in Section~\ref{sec:measurementuncertainty}, this is an extremely useful property when estimating measurement uncertainty.
Unfortunately it is not always true, even for noise: an obvious counterexample is a constant drift in experimental parameters.
Such non-stationary noise is not accurately described by a spectral density, and indeed the right-hand side of Eq.~\eqref{eq:SVVfourier} may not be mathematically defined.

\subsubsection{Evaluating the Fourier integral}
Equation~\eqref{eq:SVVfourier} defines the spectral density $\SVV[f]$, but is not directly useful for calculating it in a real experiment, where we cannot wait for infinite $T$ and we may not have access to multiple iterations.
In that case we should use the following approximation to Eq.~\eqref{eq:SVVfourier}:
\begin{equation}
    S_{VV}[f]   \approx 2 |\VT[f]|^2.
    \label{eq:SVVfourierApprox}
\end{equation}
with $\VT[f]$ given by Eq.~\eqref{eq:windowedFT}.

Often Eq.~\eqref{eq:SVVfourierApprox} is still insufficient because we do not have a continuous record $V(t)$, but instead a series of samples $V(t_k)$ taken at regular instants $t_k$ separated by a sampling interval $\Delta$.
Now we must be careful, because frequency components separated by the Nyquist frequency $1/2\Delta$ are indistinguishable in the sampled record.
A high-frequency component of $V(t)$ may therefore appear spuriously at a lower frequency in the calculated spectrum, an effect known as aliasing.
For this reason, before digitising any signal, it should be filtered using a low-pass filter with a cutoff below the Nyquist frequency.
If this has been done, the spectral density is~\cite{Press2007}:
\begin{equation}
	S_{VV}[f] \approx 2 \frac{\Delta^2}{T} \langle|\VD[f]|^2\rangle
	\label{eq:SVVDiscreteApprox}
\end{equation}
where the \emph{discrete Fourier transform} of $V(t_k)$ is:
\begin{equation}
	\VD[f] \equiv \sum_{k=0}^{N-1} V(t_k) e^{-2\pi i k f \Delta}.
\end{equation}
If only one iteration of the measurement is available, we must omit the expectation value in Eq.~\eqref{eq:SVVDiscreteApprox}.

Lastly, we may need to calculate the spectral density of a mathematical function $V(t)$ that is known for all values of $t$.
If $V(t)$ is stationary, then Eq.~\eqref{eq:WKtheorem} leads to:
\begin{equation}
	\SVV[f] = 2\int_{\infty}^\infty \EE{V[f]V[f']}\,df'
\end{equation}
where $V[f]\equiv \int_{-\infty}^\infty V(t) e^{-2\pi i f t} dt$ is the conventional Fourier transform.

If not, then Eq.~\eqref{eq:SVVfourier} needs to be evaluated directly.
A useful case is the Fourier series
\begin{equation}
V(t) = \sum_n A_n \cos (2\pi f_n t) + B_n \sin (2\pi f_n t)
\label{eq:VofFourierSeries}
\end{equation}
with $A_n$ and $B_n$ real.
The corresponding spectral density is
\begin{equation}
	\SVV[f] = \sum_n (\EE{A_n^2} + \EE{B_n^2})\, \frac{\delta[f+f_n] + \delta[f-f_n])}{2}.
	\label{eq:SofFourierSeries}
\end{equation}
If the voltage in Eq.~\eqref{eq:VofFourierSeries} is filtered around a single frequency $f$, then clearly 
\begin{equation}
	\EEE{\VF^2}=\frac{1}{2}(\EE{A_n^2}+\EE{B_n^2})(\delta[f+f_n] + \delta[f-f_n]),
	\label{eq:SofFourierSeries2}
\end{equation}
in agreement with Eq.~\eqref{eq:SofFourierSeries} and Eq.~\eqref{eq:SVVdefnSupp2}.

\subsection{How to derive a measurement uncertainty from the spectral density}
\label{sec:measurementuncertainty}

\begin{figure}
   \centering
   \includegraphics[width=86mm]{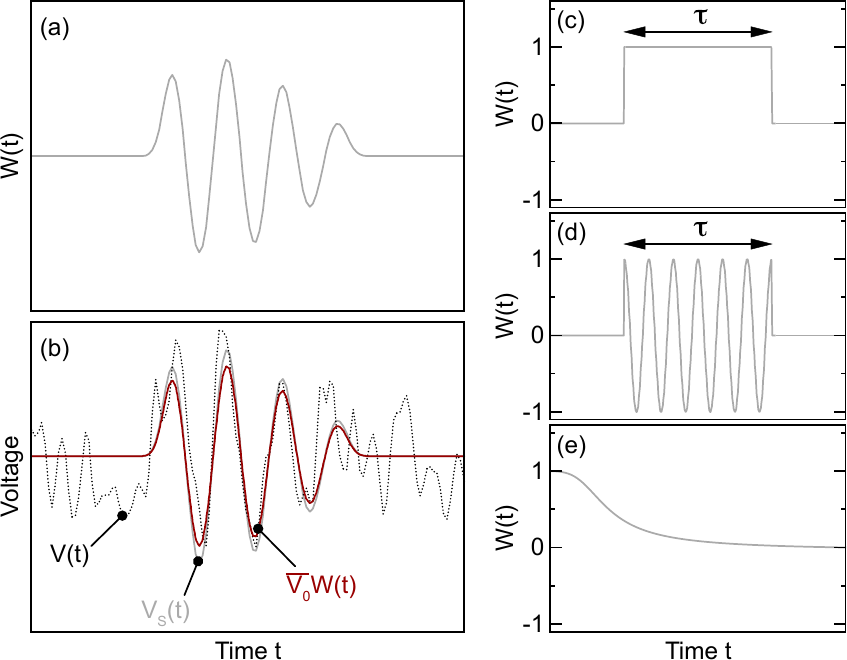} 
   \caption{Using a window function to estimate an observable.
   (a) Example of a window function,  proportional to the noise-free signal.
   (b) Typical measured signal $V(t)$, including noise, arising from an underlying signal $\VS(t)$.
   Applying Eq.~\eqref{eq:Vbarestimate} leads to an estimate of the signal amplitude $\Vbar$ and a reconstructed signal $\Vbar W(t)$.
   (c-e) Examples of weighting functions for (c) averaging a dc voltage (Eq.~\eqref{eq:Wtophat}); (d) estimating the amplitude of an oscillating voltage (Eq.~\eqref{eq:Woscillating}); (e) high-fidelity qubit readout~\cite{Gambetta2007}.
   \label{fig:SuppNoiseModel}}
\end{figure}

\subsubsection{Uncertainty in measuring a voltage}

As stated in Section~VI~A~3, a valuable property of the spectral density is that it determines the uncertainty of a measurement in the presence of noise.
Let us explain how this is done.
In general, electrical measurements transduce the observable of interest (for example qubit state, displacement, temperature, or impedance) into a voltage $\VS(t)$ contaminated by noise $\VN(t)$.
From a record of $V(t) = \VS(t)+\VN(t)$, acquired over a duration $\tau$, it is our task to extract the observable with an associated uncertainty or error bar.

A general model of this process is shown in Fig.~\ref{fig:SuppNoiseModel}.
We expect the signal to be
\begin{equation}
\VS(t) = V_0 \, W(t),
\label{eq:V0W}
\end{equation}
where $V_0$ is proportional to the observable and $W(t)$ is a weighting function.
Figure~\ref{fig:SuppNoiseModel}(a) shows an example of such a weighting function. For example, if we are measuring a constant voltage then
\begin{equation}
\label{eq:Wtophat}
W(t)=
\begin{cases}
	1 & \text{if}\  0<t<\tau	\\
	0 & \text{otherwise}
\end{cases}
\end{equation}
as in Fig.~\ref{fig:SuppNoiseModel}(c).

The optimal estimate $\Vbar$ can be derived using a least-squares fit~\cite{Press2007}.
In other words, we choose $\overline{V_0}$ to minimise the integrated squared difference between the model and the data.
This implies that
\begin{equation}
	\frac{\partial}{\partial \Vbar} \intinfty (V(t)-\Vbar\,W(t))^2\,dt = 0
	\label{eq:Vbarestimate}
\end{equation}
where
\begin{equation}
	V(t) = V_0 W(t) + \VN(t)
	\label{eq:VtSuppDefn}
\end{equation}
is the measured voltage trace including noise.
Solving Eq.~\eqref{eq:Vbarestimate} gives
\begin{equation}
	\Vbar=\frac{1}{\tauW}\intinfty V(t) \, W(t)\, dt,
	\label{eq:VbarSoln}
\end{equation}
where 
\begin{equation}
	\tauW \equiv \intinfty W^2(t)\, dt
\end{equation}
is a normalisation factor which can be thought of as the weighted duration of the measurement.
Equation.~\eqref{eq:Vbarestimate} provides an optimal estimate of $V_0$ in the sense that the expectation value of $\Vbar$ over many iterations is  the true value:
\begin{align}
	\langle \Vbar \rangle &= \frac{1}{\tauW}\intinfty \langle V_0 W(t) + \VN(t)\rangle \, W(t)\, dt \\
		&= V_0,
	\label{eq:VbarOptimality}
\end{align}
since $\langle \VN(t) \rangle=0$.
(If not, $\VN(t)$ is a correctable offset rather than noise).

Figure~\ref{fig:SuppNoiseModel}(b) shows an example of a ``true'' signal $\VS(t)$ associated with the weighting function in Fig.~\ref{fig:SuppNoiseModel}(a), and one realisation of a measured signal $V(t)$.
Applying Eq.~\eqref{eq:VbarSoln} to generate an estimate $\Vbar$ leads to a reconstructed signal $\Vbar W(t)$ which fairly accurately matches the ``true'' signal.

As this figure suggests and Eq.~\eqref{eq:VbarOptimality} confirms, the procedure estimates the correct $\Vbar$ on average.
However, the value derived from any individual voltage trace has an uncertainty.
This uncertainty is determined by the variance over a large number of estimates, each incorporating a different realisation of the random noise.
To calculate this, we evaluate
\begin{widetext}
\begin{align}
\langle \Vbar^2 \rangle 
&= \frac{1}{\tauW^2} \left\langle \intinfty  V(t) W(t) \, dt \intinfty  V(t') W(t') \, dt' \right\rangle \\
&= \frac{1}{\tauW^2} \iintinfty dt \, dt' \, \langle V(t)V(t')\rangle \, W(t)W(t') \\
&= \frac{1}{\tauW^2} \iintinfty dt \, dt' \, \langle (V_0 W(t) + \VN(t)) (V_0 W(t') + \VN(t'))\rangle \, W(t)W(t') \\
&= \frac{1}{\tauW^2} \iintinfty dt \, dt' \, \langle V_0^2 W(t)W(t') + V_0 \VN(t) W(t') + V_0 \VN(t') W(t)+ \VN(t)\VN(t')
\rangle \, W(t)W(t') \label{eq:VbarSquared4} \\
&= \frac{1}{\tauW^2} \iintinfty dt \, dt' \, ( V_0^2 W(t)W(t') + V_0 \langle\VN(t)\rangle W(t') + V_0 \langle\VN(t')\rangle W(t)+ \langle\VN(t)\VN(t') \rangle \, W(t)W(t')). \label{eq:VbarSquared5}
\end{align}
\end{widetext}
Here the first line is a substitution from Eq.~\eqref{eq:VbarSoln}, the second line follows by rearrangement, the third line by substituting from Eq.~\eqref{eq:VtSuppDefn}, the fourth line by expanding the brackets, and the fifth line follows because the expectation values need to be taken only over combinations of $\VN(t)$, which are the only stochastic terms.
Since the expectation value of $\VN(t)$ is zero, Eq.~\eqref{eq:VbarSquared5} simplifies to:
\begin{align}
\begin{split}
\langle \Vbar^2 \rangle
&= \frac{1}{\tauW^2} \iintinfty dt \, dt' \, V_0^2 W^2(t)W^2(t')\\
 &\qquad\qquad + \langle\VN(t)\VN(t') \rangle \, W(t)W(t') \\
\end{split}
\\[2ex]
&= V_0^2 + \frac{1}{\tauW^2} \iintinfty dt \, dt' \, \langle\VN(t)\VN(t') \rangle \, W(t)W(t').
\label{eq:VbarSquared7}
\end{align}

To proceed further, we need to assume that $\VN(t)$ is stationary.
Our justification is discussed at the end of Section~\ref{sec:SVVequivalenceproof}. 
If we do this, we can evaluate Eq.~\eqref{eq:VbarSquared7} using the Wiener-Khinchin theorem (Eq.~\eqref{eq:WKtheorem}). 
The double integral becomes
\begin{align}
&\iintinfty dt \, dt' \int_{-\infty}^\infty df\, e^{2\pi i f (t-t')} \SVVN[f] W(t)W(t') \nonumber\\
&=\int_{-\infty}^\infty df \, \SVVN[f] \int_{-\infty}^\infty dt\,e^{2\pi i f t} W(t) \int_{-\infty}^\infty dt\,e^{-2\pi i f t'} W(t')
\\[2ex]
&=\int_{-\infty}^\infty df \, \SVVN[f] \, W[-f]W[f] \\[2ex]
&=\int_{-\infty}^\infty df \, \SVVN[f] \, |W[f]|^2
\label{eq:SecondTerm3}
\end{align}
where $W[f]$ is the Fourier transform of $W(t)$.
The first equation follows by rearrangement, the second equation follows from the definition of the Fourier transform, and the third equation follows because $W(t)$ is real and therefore $W[-f]=(W[f])^*$.

Finally, Eqs.~\eqref{eq:VbarOptimality}, \eqref{eq:VbarSquared7}, and~\eqref{eq:SecondTerm3} can be combined to give a compact expression for the variance of the estimate $\Vbar$:
\begin{align}
\var(\Vbar) 	&\equiv \langle \Vbar^2 \rangle - \langle \Vbar \rangle ^2\\
				&=\frac{1}{\tauW^2}\int_{0}^\infty df \, \SVVN[f] \, |W[f]|^2 .	\label{eq:variance}
\end{align}
The uncertainty in the measured parameter $V_0$ is 
\begin{equation}
\sigma(\Vbar) = \sqrt{\var(\Vbar)}.
\label{eq:uncertainty}
\end{equation}

Equation~\eqref{eq:variance} is intuitive because the uncertainty is determined by the overlap between the noise spectral density $\SVVN[f]$ and the spectral weighting of the expected signal $|W[f]|^2$.
This is the fundamental relationship between the spectral density of stationary noise and the corresponding measurement uncertainty.

\subsubsubsection{Example 1: Uncertainty from a measurement with fixed duration}

Calculating the uncertainty is now a matter of choosing the appropriate weighting function $W[f]$ in Eq.~\eqref{eq:variance}.
For example, consider the measurement described by Eqs.~(135a) and~(135b) in the main text, in which~$\Vbar$ must be estimated from a measurement of fixed duration~$\tau$.
If we are measuring a constant voltage, i.e.\ using~$W(t)$ given by Eq.~\eqref{eq:Wtophat}, then we find:
\begin{align}
\tauW 		&= \tau \\
|W[f]|^2		&= \tau^2 \left( \frac{\sin (\pi \tau f)}{\pi \tau f}\right)^2
\end{align}
and therefore
\begin{align}
\var(\Vbar) 	&= \int_{0}^{\infty} df\, \left( \frac{\sin (\pi \tau f)}{\pi \tau f}\right)^2 \SVVN[f] \\
				&\approx \frac{1}{2\tau} \SVVN[0],	\label{eq:sigmaUncertainty1}
\end{align}
where the approximation holds provided that $\SVVN[f]$ is smooth near the origin where $|W[f]|^2$ is large.
This is Eq.~(124a) in the main text.

If we are measuring an oscillating voltage such as Eq.~(139) in the main text, then the appropriate window function is
\begin{equation}
\label{eq:Woscillating}
W(t)=
\begin{cases}
	\cos (2\pi f_0 t) & \text{if}\  0<t<\tau	\\
	0 & \text{otherwise},
\end{cases}
\end{equation}
as in Fig.~\ref{fig:SuppNoiseModel}(b).
If we can average over many cycles of the oscillation, i.e. $f_0 \tau \gg 1$, then
\begin{align}
\tauW 		&\approx \frac{\tau}{2} \\
|W[f]|^2	&\approx \frac{\tau^2}{4} \left(\delta[f-f_0] + \delta[f+f_0]\right)
\end{align}
and therefore
\begin{equation}
\var(\Vbar) \approx \frac{1}{\tau}\SVVN[f_0]
\label{eq:sigmaUncertainty}
\end{equation}
This leads to Eq.~(140b).

\subsubsubsection{Example 2: Uncertainty from a measurement using a frequency filter}

Another common situation is that we have filtered the voltage record using a filter with amplitude transmission~$F[f]$.
The filtered record can be regarded as a measurement of the underlying signal $\VS(t)$.
What is the uncertainty of this measurement?

If the Fourier transform of the original voltage is~$V[f]$, the Fourier transform of the filtered signal is
\begin{equation}
	\VF[f]=F[f] V[f],
\end{equation}
or equivalently
\begin{equation}
	\VF(t) = \intinfty V(u) F(t-u) \, du
	\label{eq:VfilterConvolution}
\end{equation}
where $u$ is a time interval and $F(u)$ is the inverse Fourier transform of $F[f]$.
(Obviously a causal filter has $F(t-u)=0$ for $t < u$.)
This is the process that generates the low-pass filtered traces in Fig.~24(d).

Equation~\eqref{eq:VfilterConvolution} has the same form as Eq.~\eqref{eq:VbarSoln}, except that $W(u)$ has been replaced by a new weighting function $\tauW F(t-u)$.
The filtered voltage $\VF(t)$ is thus an estimate of $\VS(t)$.
Although the estimate may not be optimal in the sense of Eq.~\eqref{eq:VbarOptimality}, a sensibly chosen filter often gets pretty close, meaning that the error is dominated by fluctuations due to $\VN(t)$ rather than by distortion of $\VS(t)$ due to the filter.

Provided this is true, then the measurement uncertainty can be calculated by the same procedure as led to Eq.~\eqref{eq:variance}, giving
\begin{align}
	\var(\VF(t)-\VF_\mathrm{S}(t))	&= \int_0^\infty df\, \SVVN[f] \, |F[f]|^2		\\
									& \approx \Bf\, \SVVN[f_0]	\label{eq:BfSVVN}
\end{align}
where the approximation holds provided the noise spectrum is smooth across the filter passband.
Here $\VF_\mathrm{S}(t) \approx \VS(t)$ is the filtered signal voltage, $f_0$ is the center frequency of the filter, and
\begin{equation}
	\Bf \equiv \int_{0}^{\infty} |F[f]|^2 \, df
	\label{eq:varianceFilter}
\end{equation}
is its \emph{equivalent noise bandwidth}.
As above, the uncertainty is the square root of Eq.~\eqref{eq:varianceFilter}.

In terms of the windowing function in the time domain associated with a filter in the frequency domain, the equivalent noise bandwidth can be written
\footnote{It is tempting to associate $\tauW$ with the ``time constant'' of the filter.
	The temptation should be resisted, because this name is usually reserved for the $RC$ time constant of a particular filter implementation.
	Reference~\onlinecite{ZurichPrinciplesOfLockindetection2016} tabulates the equivalent noise bandwidth  in terms of the $RC$ time constant for filters of different order.
	This bandwidth can be converted to $\tauW$ using~Eq.~\eqref{eq:BfTauW}.
	For example, a first-order low-pass $RC$ filter has equivalent noise bandwidth $\Bf= 1/4RC$ and therefore $\tauW=2RC$.}
\begin{equation}
	\Bf = \frac{1}{2\tauW}.
	\label{eq:BfTauW}
\end{equation}
In other words, a top-hat window of duration $\tau$ admits the same amount of white noise as a brick-wall filter of bandwidth $1/2\tau$.

\subsubsubsection{Example 3: Single-shot readout}
\label{sec:suppSingleShot}

Suppose we are trying to determine the state of a qubit.
Unlike the situation in Fig.~24, we do not simply need to distinguish two levels of the readout signal, because the qubit can decay during the measurement.
The best way to determine the state in this situation is explained in Ref.~\onlinecite{Gambetta2007}.

At first sight, we might choose to apply Eq.~\eqref{eq:VbarSoln} with an exponentially decaying weighting function $W(t)$, to match the expected decay profile of the qubit.
This is indeed the optimal way to determine the average qubit state, but this is not the same as optimising single-shot fidelity; to achieve high fidelity it is necessary (among other things) to identify the small number of experimental runs in which the qubit decays rapidly from its excited state.
The optimum $W(t)$ must be determined numerically using the known signal-to-noise ratio and qubit relaxation time~\cite{Gambetta2007}; an example is shown in Fig.~\ref{fig:SuppNoiseModel}(e).
In fact, it is possible to do even better than this by  applying a non-linear filter\cite{Gambetta2007} not described by Eq.~\eqref{eq:VbarSoln}.

\subsubsubsection{Example 4: Uncertainty in a combined measurement of more than one observable}

Suppose that we are trying to extract more than one observable from a signal.
For example, if
\begin{equation}
\VS(t) = V_\mathrm{R} \cos (2\pi \fc t + \varphi)
\label{eq:estimateAmplitudePhase}
\end{equation}
we may want to estimate both the amplitude $V_\mathrm{R}$ and the phase $\varphi$.

We approach this problem by explaining how to do a linear fit and calculate its uncertainty.
Suppose we generalise Eq.~\eqref{eq:V0W} by writing
\begin{equation}
\VS(t) = \sum_k V_k W_k(t)
\label{eq:VSlinearfit}
\end{equation}
where $V_k$ are the observables we want to estimate and $W_k(t)$ are their corresponding weightings.
Then the same process that led to Eq.~\eqref{eq:VbarSoln} leads to the matrix equation
\begin{equation}
\sum_j \alpha_{kj} \overline{V}_j = \beta_k
\end{equation}
where
\begin{align}
\alpha_{kj} 	&\equiv 	\intinfty W_k(t) W_j(t) \, dt	\label{eq:defineAlpha}\\
\beta_k		&\equiv	\intinfty W_k(t) V(t) \, dt.
\end{align}
Thus the optimal estimate is
\begin{equation}
\overline{V}_j = \sum_k C_{jk} \beta_k
\end{equation}
where $\mathbf{C}$ is the covariance matrix, defined as the inverse of Eq~\eqref{eq:defineAlpha}:
\begin{equation}
\mathbf{C} = \boldsymbol{\alpha}^{-1}.
\label{eq:defineC}
\end{equation}
By a similar process that led to Eq.~(\ref{eq:variance}), the variance of the estimate, which by Eq.~\eqref{eq:uncertainty} determines the uncertainty in $\overline{V}_j$, is
\begin{equation}
\var(\overline{V}_j) = \sum_{k,l} C_{jk} C_{jl} \int_0^\infty \SVVN[f] \, W_k^*[f] \, W_l[f]\, df.
\end{equation}
If the noise spectral density is white over the frequency range of the signal, then this simplifies to
\begin{equation}
\var(\overline{V}_j) = \frac{1}{2}\SVVN C_{jj}.
\label{eq:sigmaMultiDimensional}
\end{equation}

Let us apply Eq.~\eqref{eq:sigmaMultiDimensional} to the observables in~Eq.~\eqref{eq:estimateAmplitudePhase}.
Equation~\eqref{eq:estimateAmplitudePhase} is not of the form of~Eq.~\eqref{eq:VSlinearfit} because it is not linear in the observable~$\varphi$.
However, we will assume the common situation in which the fit function varies linearly with changes in the fit parameters over the range of uncertainty.
For example, if we were trying to measure the amplitude and phase of a segment of signal from Fig.~24(b), the corresponding location in $(\VI, \VQ)$ space lies near the spots in Fig.~24(h), and the relative uncertainty, given by the separation of the spots, is small.
We therefore convert the problem to a linear fit by writing
\begin{align}
V_\text{R} &= V_{\text{R}0} + \delta V_\text{R}\\
\varphi &= \varphi_0 + \delta \varphi
\end{align}
where $V_{\text{R}0}$ and $\varphi_0$ are known approximate values, and $\delta R$ and $\delta \varphi$ are the unknown deviations.
Expanding in $\delta R$ and $\delta \varphi$ leads to
\begin{equation}
\begin{split}
\VS(t)-V_{\mathrm{R}0} \cos (2\pi\fc t + \varphi_0)  &\approx\\
&\hspace{-2.7cm} \delta V_\mathrm{R}\cos (2\pi\fc t + \varphi_0) - \delta \varphi  V_{\mathrm{R}0} \sin (2\pi\fc t + \varphi_0). \\
\end{split}
\label{eq:estimateAmplitudePhaseLinearised}
\end{equation}

Clearly, fitting the left-hand side is equivalent to fitting $\VS(t)$, and estimating $\delta V_\mathrm{R}$ and $\delta \varphi$ is equivalent to estimating $V_\mathrm{R}$ and $\varphi$.
The right-hand side of Eq.~\eqref{eq:estimateAmplitudePhaseLinearised} is of the form of Eq.~\eqref{eq:VSlinearfit}, with
\begin{align}
V_1	 	&=	\delta V_\mathrm{R}	\\
V_2		&=	\delta \varphi \\
W_1(t)	&=	\cos (2\pi\fc t + \varphi_0)		\label{eq:W1}\\
W_2(t)	&=	-V_{\mathrm{R}0} \sin (2\pi\fc t + \varphi_0).	\label{eq:W2}
\end{align}
If we measure this signal for a time $\tau$ extending over many cycles, Eq.~\eqref{eq:defineAlpha} leads to 
\begin{equation}
\boldsymbol{\alpha} =   
	\begin{pmatrix}
    		\tau/2 & 0 \\
      		0 &V_{\mathrm{R}0}^2 \tau/2 \\
  	\end{pmatrix}.
\end{equation}
The resulting covariance matrix (Eq.~\eqref{eq:defineC}) is
\begin{equation}
\mathbf{C} =   
\begin{pmatrix}
      2/\tau & 0 \\
      0 & 2/\tau V_{\mathrm{R}0}^2 \\
  \end{pmatrix}.
\end{equation}
Substituting into Eq.~\eqref{eq:sigmaMultiDimensional} finally gives the uncertainties in the observables $V_\mathrm{R}$ and $\varphi$:
\begin{align}
	\sigma(\overline{V_\mathrm{R}})		&= \sqrt{\frac{\SVVN[\fc]}{\tau}}	\\
	\sigma(\overline{\varphi})	&= \sqrt{\frac{\SVVN[\fc]}{V_{\mathrm{R}0}^2\tau}	}
\end{align}
where the noise spectral density is evaluated at $\fc$ because that is the noise frequency which overlaps with the weighting functions (Eqs.~(\ref{eq:W1}-\ref{eq:W2})).

As noted, this procedure requires the uncertainty in the fit parameters to be small enough for the fit function to be linearised.
If this is not true, the uncertainty must be determined in some other way and does not in general have a simple relation to the noise spectral density.

\subsubsection{Uncertainty in measuring power}

Equation~\eqref{eq:variance} can be applied to a measurement of voltage and, via Eq.~(141), to any observable on which the voltage depends linearly.
However, a common situation in which the model of Fig.~\ref{fig:SuppNoiseModel} no longer holds is when the observable is proportional to the signal power, for example when measuring thermal noise.
We can still estimate the uncertainty using Eqs.~\eqref{eq:VbarSoln} and \eqref{eq:sigmaUncertainty}, but we need to use the spectral density of the power instead of the voltage
\footnote{Unfortunately it is wrong to use Eq.~(141) with $X$ being the power. The reason is that $\partial P/\partial V = 2 V/Z_0$ is not constant over the range of the noise.
}.

\begin{figure}
	\centering
	\includegraphics[width=86mm]{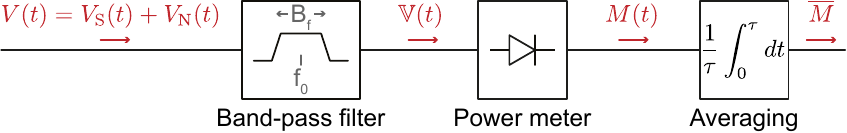} 
	\caption{Model of the process for estimating power in a signal $V(t)$.
	A real power meter, whose response depends on the input frequency, is modelled as the combination of a band-pass-filter whose output is the filtered voltage $\mathbb{V}(t)$ followed by an ideal power meter whose output is $M(t) \equiv \mathbb{V}^2(t)$.
	The average of the meter output gives $\overline{M}$, which is the optimal estimate of $\EE{ \mathbb{V}^2 }$.
	\label{fig:Radiometer}}
\end{figure}

To do this, assume that the signal $V(t)$ whose power content we are estimating is stationary.
We model the estimation process by assuming that we have a power meter whose output~$M(t)$ is equal to the square of the incident voltage within its detection bandwidth:
\begin{equation}
	M(t) \equiv \mathbb{V}^2(t).
\end{equation}
This may represent a real power meter, or $M(t)$ may be calculated from the digitised $V(t)$.
As in Section~\ref{sec:measurementuncertainty}, we must estimate the power from a record of $M(t)$ acquired over a time $\tau$.
A model of this process
\footnote{You may ask what happens if you don't filter the voltage before the power meter. The answer is that you cannot make that choice. Any power meter, including one realised in software, must have a limited bandwidth; otherwise, it would need to respond instantaneously to any input.}
is shown in Fig.~\ref{fig:Radiometer}.

We need the spectral density $\SMM[f]$.
To calculate it, we first evaluate the autocorrelation function of $M(t)$.
This is done with the help of Isserlis' theorem
\footnote{
Isserlis' theorem (also known as the Wick probability theorem) is proved in several places online, and for the valiant in Ref.~\onlinecite{Janson1997}.
Here's a proof of the special case Eq.~\eqref{eq:Isserlis}, pitched at the level of this Review.

Define $X=\VF(t_1)$ and $Y=\VF(t_2)$.
Each is due to the combination of many independent noise sources, so obeys a Gaussian distribution, as does the linear combination $aX + bY$ for any values of $a$ and $b$.
(In statistical terminology, $X$ and $Y$ follow a multivariate normal distribution.)
The variance of the combination is
\begin{align}
	\sigma^2	&\equiv \EE{(aX + bY)^2} \\ 
				&= a^2 \EE{X^2} + 2ab \EE{XY} + b^2 \EE{Y^2}. \label{eq:IsserlisProof1}
\end{align}

Now consider
\begin{align}
	\EE{(aX + bY)^4} 	&= 3\sigma^4	\\
						&= a^4 \EE{X^4} + 6 a^2 b^2 \EE{X^2 Y^2} + b^4 \EE{Y^4}
\end{align}
where the first line follows from the properties of the univariate Gaussian and the second line follows by expanding the bracket and using that any expectation value containing an odd number of terms vanishes.
Substituting $\sigma^2$ from Eq.~\eqref{eq:IsserlisProof1} and balancing the $a^2 b^2$ terms on each side gives:
\begin{equation}
	3 (2 \EE{X^2} \EE{Y^2} + 4 \EE{XY}^2 ) = 6 \EE{X^2 Y^2},
\end{equation}
from which (since the statistical properties of $X$ and $Y$ are identical):
\begin{equation}
	\EE{X^2 Y^2} = \EE{X^2}^2 + 2\EE{XY}^2
\end{equation}
as required.
}, which states that
\begin{align}
	\langle M(t) M(t') \rangle 	&\equiv \langle \VF^2(t) \VF^2(t') \rangle \\
								&= \langle \VF^2(t)\rangle \langle \VF^2(t') \rangle
	+ 2\, \langle \VF(t)\VF(t')\rangle^2 \\
								&= \langle \VF^2(0)\rangle^2
								+ 2 \, \langle \VF(0)\VF(t-t')\rangle^2.
	\label{eq:Isserlis}
\end{align}
The theorem holds provided that $\VF(t)$ obeys a multivariate normal distribution, which it should do because it is a sum of many independent contributions to the noise.
Using Eq.~\eqref{eq:Isserlis} in combination wih the Wiener-Khinchin theorem (Eq.~\eqref{eq:WKbox}) gives
\begin{equation}
		\SMM[f] = 2 \left\{\langle \VF^2(0)\rangle^2 \delta[f] + 2 \int_{-\infty}^{\infty} dt\,
			e^{-2\pi i f t}
			\langle \VF(0)\VF(t)\rangle^2 \right\}.
			\label{eq:SMMcalculated}
\end{equation}
To evaluate the second term we again use the Wiener-Khinchin theorem, this time for the correlator~$\langle \VF(0)\VF(t)\rangle$:
\begin{align}
	&\int_{-\infty}^{\infty} dt\,	e^{-2\pi i f t}	\langle \VF(0)\VF(t)\rangle^2 \nonumber\\
	&=
	\frac{1}{4} \iiint_{-\infty}^\infty dt\ df_1\ df_2\  \SVFVF[f_1] \SVFVF[f_2] e^{2\pi i (f_1+f_2-f)t} \\[1ex]
	&=
	\frac{1}{4} \iint_{-\infty}^\infty df_1\ df_2\  \SVFVF[f_1] \SVFVF[f_2] \delta[f_1+f_2-f] \\[1ex]
	&=
	\frac{1}{4} \int_{-\infty}^\infty df_1\  \SVFVF[f_1] \SVFVF[f_1-f].
	\label{eq:Mcorrelator3}
\end{align}
We now make the approximation that $f$ is small enough that $\SVFVF[f_1-f] \approx \SVFVF[f_1]$.
This is valid because in the final evaluation of the uncertainty, which comes from an equation analogous to Eq.~\eqref{eq:SecondTerm3}, the noise spectral density is multiplied by the Fourier transform of the weighting function corresponding to the final averaging step in Fig.~\ref{fig:Radiometer}.
By choosing a weighting function that varies slowly (e.g. by averaging over a long time~$\tau$), we suppress high-frequency components
\footnote{To be precise, we need $\Bf \tau \gg 1$, where $\Bf$ is the bandwidth of the sharpest feature in $\SVFVF$. Often this is the bandwidth of the power detector.}
of $W[f]$.
Applying this approximation to Eq.~\eqref{eq:Mcorrelator3} and substituting into Eq.~\eqref{eq:SMMcalculated} gives
\begin{equation}
			\SMM[f] \approx 2 \langle \VF^2(0)\rangle^2 \delta[f] + \int_{-\infty}^{\infty} df_1\, \SVFVF^2[f_1].
\end{equation}
The first term, which is proportional to the average power, contains the signal; the second term is the noise~$\SMM^\mathrm{N}[f]$.

We now use analogs of Eqs.~\eqref{eq:VbarSoln} and~\eqref{eq:variance} to calculate the expectation value and variance of $\overline{M}$.
For simplicity, assume that the expected power is independent of time so that the appropriate weighting function is Eq.~\eqref{eq:Wtophat}.
This leads (via Eq.~\eqref{eq:VbarSoln}) to:
\begin{align}
	\langle \overline M \rangle &= \langle \VF^2(0) \rangle \\
								&= \frac{1}{2} \int_{-\infty}^{\infty} \SVFVF[f]\,df \\
								&= \int_0^\infty |F[f]|^2 \SVV[f] \, df  \label{eq:MbarSoln3}
\end{align}
and (via Eq.~\eqref{eq:variance}) to:
\begin{align}
	\mathrm{var}(\overline{M})	&= \frac{1}{2\tau} \, \lim_{f\rightarrow 0}\SMM^\mathrm{N}[f] \\
								&= \frac{1}{\tau} \int_0^\infty |F[f]|^4 \SVV^2[f] \, df, \label{eq:MvarSoln2}
\end{align}
where $F[f]$ is the amplitude transmission of the filter before the power meter.
Obviously the power estimate is related to $\overline M$ by
\begin{equation}
	\overline P = \frac{\overline M}{\Zo}.
\end{equation}

Let us approximate that $\SVV[f]$ is white, i.e. independent of frequency within the detection bandwidth, and that the filter transmits either all the signal or none of it.
In that case Eqs.~\eqref{eq:MbarSoln3} and~\eqref{eq:MvarSoln2} combine into a single expression for the signal-to-noise ratio:
\begin{equation}
	\frac{\sqrt{\mathrm{var}(\overline{M})}}{\overline{M}} \approx \frac{1}{\sqrt{\tau \Bf}}.
	\label{eq:Dicke1}
\end{equation}
where $\Bf$ is the detection bandwidth.
Equation~\eqref{eq:Dicke1} holds for any observable proportional to the power.
Another way to express Eq.~\eqref{eq:Dicke1} is as an uncertainty in estimating the spectral density, once the noise is fully characterised:
\begin{equation}
	\sqrt{\mathrm{var}(\SVVS[f_0])} = \frac{\SVVS[f_0] + \SVVN[f_0]}{\sqrt{\tau \Bf}},
\end{equation}
where $f_0$ is the center of the power meter's detection bandwidth.
This is the famous \emph{radiometer equation}, derived by Dicke~\cite{Dicke1946} for microwave thermometers.

Another form of the radiometer equation, useful for dark-matter searches~\cite{Asztalos2010}, is as the amplitude signal-to-noise ratio in the power meter's output when it is fed a weak narrowband signal, for which $\SVVS \ll \SVVN$.
Then
\begin{align}
	\frac{\mathrm{Signal}}{\mathrm{Noise}} 
	&\equiv \frac{\overline{M}-\langle \VF_\mathrm{N}^2\rangle}{\sqrt{\mathrm{var}(\overline{M})}} \\
	&=\frac{\VS^2}{\SVVN \sqrt{\Bf/\tau}} \\
	&=\frac{P_\mathrm{S}}{\kB \TN} \sqrt{\frac{\tau}{\Bf}},
\end{align}
where $P_\mathrm{S}\equiv\VS^2/\Zo$ is the signal power.
This is the signal-to-noise ratio with which a signal power $P_\mathrm{S}$ can be measured within an acquisition time $\tau$.

\begin{figure*}[t]
	\centering
	\includegraphics{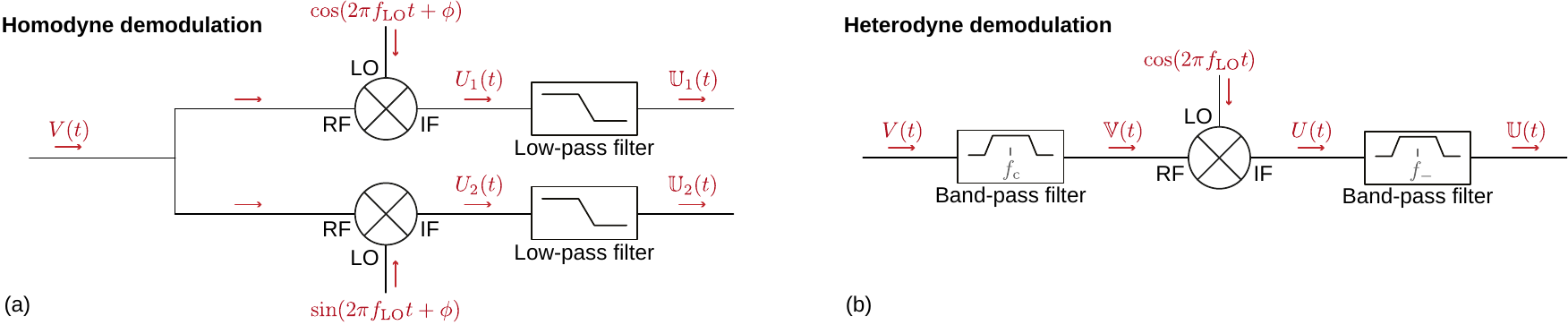} 
	\caption{Circuits for generating demodulated filtered voltages.
	(a) Homodyne circuit, to generate voltages as in Eqs.~(\ref{eq:Ufiltered1}-\ref{eq:Ufiltered2}).
	(b) Heterodyne circuit, to generate a voltage as in Eq.~\eqref{eq:Uheterodyne}.
	\label{fig:Demodulation}}
\end{figure*}

\subsection{Effect of demodulation on the spectral density}

As shown in Fig.~2, a high-frequency measurement nearly always involves demodulation of the signal by mixing it with a local oscillator.
As one would expect, when done properly this does not affect the accuracy of any measurement based on this signal.
We will now justify this statement by calculating the signal and noise spectral density after demodulation.

Suppose we have a voltage of the form
\begin{equation}
V(t) = A(t) \cos (2\pi \fc t) + B(t) \sin (2\pi \fc t) + \VN(t),
\label{eq:Vcarrier}
\end{equation}
where $\fc$ is the carrier frequency and $\VN(t)$ is stationary noise.
We want to estimate the two slowly varying
\footnote{If $A(t)$ and $B(t)$ do not vary slowly compared to $\fc$, then the partition of $V(t)$ into two quadratures need not be unique.
For example $V(t)= \sin (2\pi\fc t) \cos (2\pi \fc t)$ cannot be partitioned in this way.}
 quadratures $A(t)$ and $B(t)$, assumed for simplicity to be uncorrelated.
 An example of a voltage described by~Eq.~\eqref{eq:Vcarrier} is the reflected signal from a coherently illuminated circuit when both the real and imaginary parts of the reflection coefficient are changing.
 
In principle we can estimate $A(t)$ and $B(t)$ directly from $V(t)$. If our measurement duration $\tau$ is longer than $1/\fc$ but shorter than the timescale over which $A$ and $B$ vary, then by Eq.~\eqref{eq:sigmaUncertainty} the uncertainties are
\begin{align}
\sigma(A)		= \sigma(B)
			&= \sqrt{\var(\overline{A})}	 \\
			&= \sqrt{\frac{\SVVN[\fc]}{\tau}}	\label{eq:sigmaModulated}
\end{align}
provided that $\SVVN[f]$ varies smoothly near $\fc$.

If our measurement includes a demodulation step, then we must estimate $A(t)$ and $B(t)$ from the demodulated voltage.
Whether the demodulation is homodyne (with $\fLO=\fc$) or heterodyne (with $\fLO \neq \fc$), the estimates should have the same uncertainty as Eq.~\eqref{eq:sigmaModulated}.

\subsubsection{Homodyne demodulation}

\begin{figure}
	\centering
	\includegraphics[width=86mm]{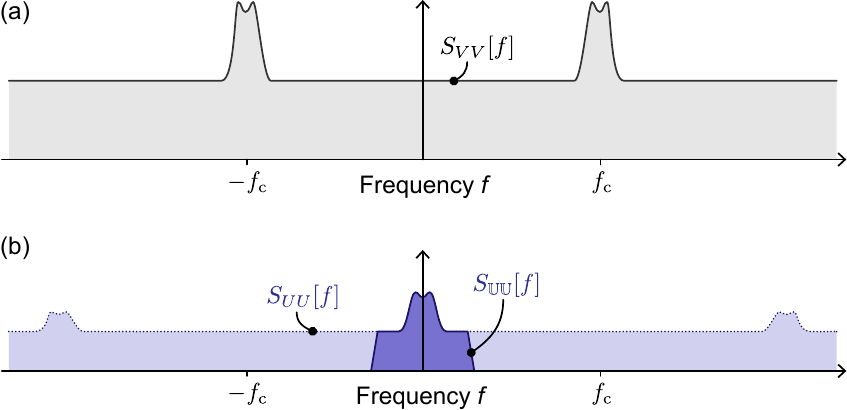} 
	\caption{The effect of homodyne demodulation on the spectral density.
	(a) Cartoon of the spectral density at the mixer rf input.
	The signal is concentrated near frequency $\fc$; the noise is white.
	(b) Cartoon of the spectral density at one of the outputs, before and after low-pass filtering.
	\label{fig:DemodulationSpectra}}
\end{figure}

In a homodyne setup (Fig.~\ref{fig:Demodulation}(a)), we need to demodulate with two quadratures in order to extract both $A(t)$ and $B(t)$.
This generates the two output voltages
\begin{align}
	U_1(t)	&\equiv V(t) \cos (2\pi \fLO t)	    \label{eq:U1}\\
	U_2(t)	&\equiv V(t) \sin (2\pi \fLO t)		\label{eq:U2}
\end{align}
where $\fLO$ is the local oscillator frequency.
(For simplicity we have omitted a prefactor $\sqrt{2/\LC}$, where $\LC$ is the mixer conversion loss
\footnote{We follow here the definition of Ref.~\onlinecite{Pozar2012}, according to which the mixer conversion loss $\LC$ (when expressed in linear units instead of in dB) is the ratio of rf input power to IF output power.
Conversion loss is sometimes defined~\cite{MiniCircuitsMixerDefinitions} as the ratio of rf input power to power in one IF sideband; by this definition the conversion loss is $\LC'=2\LC$.}.)
Application of Eq.~\eqref{eq:SVVfourier} shows that the noise spectral density in both mixer outputs is related to the noise spectral density in $V(t)$ by
\begin{equation}
	\SUU^\mathrm{N}[f] = \frac{1}{4} (\SVVN[f-\fLO] + \SVVN[f+\fLO]).
	\label{eq:SUU}
\end{equation}
This is illustrated in Fig.~\ref{fig:DemodulationSpectra}.

To extract $A(t)$ and $B(t)$, the demodulated voltages $U_1(t)$ and $U_2(t)$ are low-pass filtered to generate  voltages $\UF_1(t)$ and $\UF_2(t)$.
The filter cut-off should be chosen to pass all components of $A(t)$ and $B(t)$ but reject components near $2\fc$.
It then follows from Eqs.~\eqref{eq:Vcarrier} and (\ref{eq:U1}-\ref{eq:U2}) that the filtered demodulated voltages are
\begin{align}
\UF_1(t)	&= \frac{A(t)}{2} + \UF_1^\mathrm{N}(t)		\label{eq:Ufiltered1}\\
\UF_2(t)	&= \frac{B(t)}{2} + \UF_2^\mathrm{N}(t),		\label{eq:Ufiltered2}
\end{align}
showing as expected that the outputs of the homodyne circuit contain the two signal quadratures of $V(t)$, plus noise.

The spectral density of both noise components $\UF_1^\mathrm{N}(t)$ and $\UF_2^\mathrm{N}(t)$ is
\begin{equation}
	\SUFUF^\mathrm{N}[f] = \frac{1}{2} \SVVN[f+\fc].
	\label{eq:SUFUF}
\end{equation}
Since $\UF_{1,2}^\mathrm{N}(t)$ is stationary
\footnote{
This isn't obvious, because $U_1(t)$ and $U_2(t)$ are clearly non-stationary.
To apply Eq.~\eqref{eq:WKtheorem} and therefore Eq.~\eqref{eq:sigmaUncertainty1}, we need to show that
$\EE{\UFN_1(t)\UFN_1(t')}$
is invariant under a common translation of $t$ and $t'$.
To do this, write
\begin{equation}
	\qquad \EE{\UFN_1(t)\UFN_1(t')} = \iint_0^{\Bf} df \, df' e^{2 \pi i (ft+f't')} \EE{U_1^\mathrm{N}(f) U_2^\mathrm{N}(f')}
\end{equation}
and use that 
\begin{align}
	U_1^\mathrm{N}[f] &= \frac{\VN[f-\fLO] + \VN[f+\fLO]}{2}	\\
	U_2^\mathrm{N}[f] &= \frac{\VN[f-\fLO] - \VN[f+\fLO]}{2i}	\\
	\EE{\VN(f_1)\VN(f_2)} &= \frac{1}{2}\SVVN[f_1]\,\delta[f_1+f_2].
\end{align}
This eventually leads to 
\begin{equation}
	\begin{split}
		\qquad\EE{\UFN_1(t)\UFN_1(t')} = &\frac{1}{8}\int_0^{\Bf} df \,e^{2 \pi i f(t-t')}\\
		&\qquad\times \left(\SVVN[f+\fLO] + \SVVN[f-\fLO]\right),\\
	\end{split}
\end{equation}
where $\Bf$ is the cutoff of the low-pass filter.
This expression depends only on $t-t'$ as required.
},
we can apply Eq.~\eqref{eq:sigmaUncertainty1}, obtaining
\begin{align}
\sigma(A)	= \sigma(B) 	&= \sqrt{\var(\overline{A})}				\\
					&= \sqrt{\frac{2\SUFUF^\mathrm{N}[0]}{\tau}}	\\
					&= \sqrt{\frac{\SVVN[\fc]}{\tau}},
\end{align}
in agreement with Eq.~\eqref{eq:sigmaModulated}.
This confirms that the same information is present in the homodyne outputs as was contained in the input.

Another way to express this result is to say that the sensitivity when measuring the demodulated filtered noise voltage (defined above Eq.~(140)) is related to the sensitivity when measuring the voltage at the mixer input by
\begin{equation}
\sqrt{\SUFUF^\mathrm{N}[0]} = \sqrt{\frac{\SVVN\fc]}{2}}
\end{equation}
but that this does not degrade the accuracy of the measurement because the signal power in each quadrature is decreased by a factor 2.

\subsubsection{Heterodyne demodulation}
In a heterodyne setup (Fig.~\ref{fig:Demodulation}(b)) the entire signal information is contained in the output of a single mixer.
We find
\begin{equation}
	\UF(t) = \frac{A(t)}{2} \cos (2\pi f_-t) + \frac{B(t)}{2} \sin (2\pi f_-t) + \UF_\mathrm{N}(t)	
	\label{eq:Uheterodyne}
\end{equation}
with $\UF_\mathrm{N}(t)$ described by the spectral density
\begin{equation}
	\SUFUF^\mathrm{N}[f_-] = \frac{1}{4}\SVVN[\fc]	
	\label{eq:SUUheterodyne}
\end{equation}
where $f_- \equiv \fc-\fLO$.
Once again, this leads to the measurement uncertainties given by Eq.~\eqref{eq:sigmaModulated}, thus confirming that heterodyne demodulation, like homodyne demodulation, preserves the information in the original signal.

\subsection{The sideband method of determining measurement sensitivity; derivation of Equation~(124)}
The sensitivity of a reflectometry measurement can in principle be determined from Eq.~(136).
However, this requires knowledge of the proportionality constant $\left|\frac{\partial V}{\partial X}\right|$, which depends on many details of the circuit.
It is usually better to use Eq.~(124), which we will now derive.

Suppose we want to find the sensitivity to charge $Q$ on an SET.
We modulate this charge in a known way, so that
\begin{equation}
    Q(t) = \sqrt{2} \, \Qrms \cos (2\pi \fM t),
    \label{eq:Qmod}
\end{equation}
where $\fM \ll \fc$ is the modulation frequency and $\Qrms$ is the rms modulation amplitude.
Since the reflected signal is
\begin{equation}
    \VS(t)=\Vc \, \mathrm{Re} \left(e^{i(2\pi \fc t + \varphic)} \Gamma (t)\right),
\end{equation}
where $\Vc$ and $\varphic$ are the amplitude and phase of the carrier, and since for weak modulation we have
\begin{equation}
    \Gamma(t) = \Gamma_0 + \frac{\partial \Gamma}{\partial Q} Q(t),
\end{equation}
this leads to
\begin{equation}
\begin{split}
    \VS(t)    &=\Vc\,\mathrm{Re} \left(\Gamma_0 \, e^{i(2\pi \fc t + \varphic)}\right)     \\
            &\qquad + \Vc Q(t) \, \mathrm{Re}\left( \frac{\partial \Gamma}{\partial Q} \, e^{i(2\pi \fc t + \varphic)} \right).
\end{split}
\end{equation}
Without loss of generality we assume $\Gamma_0=0$ and choose $\varphic=-\mathrm{arg}\left(\frac{\partial \Gamma}{\partial Q}\right)$.
The signal voltage is then
\begin{equation}
    \VS(t) = \Vc \left| \frac{\partial \Gamma}{\partial Q} \right| Q(t) \cos(2\pi \fc t).
    \label{eq:VtforSidebands}
\end{equation}

We cannot yet use Eq.~(136) because the $\cos(2\pi \fc t)$ term makes $\partial \VS / \partial Q$ non-constant.
However, we can define
\begin{align}
    \mathbb{U}(t)   &\equiv \mathsf{LPF}\left\{V(t) \cos (2\pi \fc t)\right\}  \\
                    &= \frac{\Vc}{2} \left| \frac{\partial \Gamma}{\partial Q} \right| Q(t)
\end{align}
where $\mathsf{LPF}\{\cdot\}$ denotes a low-pass filter.
This gives us the proportionality we need to use Eq.~(136), which leads to:
\begin{align}
    \SQQN[f]    &= \frac{4}{\Vc^2 |\partial \Gamma /\partial Q|^2} {\SUFUFN}[f]    \\
                &= \frac{2}{\Vc^2 |\partial \Gamma /\partial Q|^2} \SVVN[f+\fc]
                \label{eq:SQQN}
\end{align}
where the second line follows from Eq.~\eqref{eq:SUFUF}.

We now substitute Eq.~\eqref{eq:Qmod} into Eq.~\eqref{eq:VtforSidebands}, leading to
\begin{equation}
    \VS(t) = \frac{\Vc \Qrms}{\sqrt{2}} \left|\frac{\partial \Gamma}{\partial Q}\right| (\cos (2 \pi f_+ t) + \cos (2 \pi f_- t)),
\end{equation}
where now $f_\pm \equiv \fc \pm \fM$.
The corresponding spectral density is
\begin{equation}
    \SVVS[f] = \frac{\Vc^2 \Qrms^2}{4} \left| \frac{\partial \Gamma}{\partial Q} \right|^2 (\delta[f-f_+] + \delta [f-f_-]).
\end{equation}
This describes the sidebands that appear in a spectrum such as Fig.~22(c).

Now we are ready to derive Eq.~(124).
First, we note that since a spectral analyser obviously cannot resolve a delta function but instead measures the average of $\SVV[f]$ over the resolution bandwidth $\Rf$, the apparent spectral density of the signal at the peak of a sideband is
\begin{align}
    s_{VV}^\mathrm{S}[f_\pm]  = \frac{\Vc^2 \Qrms^2}{4 \Rf} \left| \frac{\partial \Gamma}{\partial Q} \right|^2.
\end{align}
We then use this apparent spectral density to calculate the power SNR, with the noise spectral density taken from Eq.~\eqref{eq:SQQN}:
\begin{align}
    \mathrm{SNR}    &\equiv \frac{s_{VV}^\mathrm{S}[f_\pm]}{\SVVN[f_\pm]}   \\
                    &= \frac{\Qrms^2}{2\Rf \, \SQQN[\fm]},
\end{align}
which by further rearrangement gives the charge sensitivity
\begin{equation}
    \sqrt{\SQQN[\fm]} = \frac{\Qrms}{\sqrt{2 \Rf \cdot \mathrm{SNR}}}.
    \label{eq:SQQNlinear}
\end{equation}
All scaling factors, such as $|\partial \Gamma /\partial Q|$, have dropped out of this equation, meaning that it can be used without knowing details of the reflectometry chain.

The final step is to re-express SNR in dB, after which Eq.~\eqref{eq:SQQNlinear} becomes
\begin{equation}
    \sqrt{\SQQN[\fm]} = \frac{\Qrms}{\sqrt{2 \Rf} \, 10^{\mathrm{SNR_{dB}}/20}}.
    \label{eq:SQQNlog}
\end{equation}
This is Eq.~(124).
Clearly it can be applied to any other measured quantity instead of charge $Q$.

Conveniently, $\mathrm{SNR_{dB}}$ can be read off as a peak height as in Fig.~22(c), provided that it is large enough; if not, then the peak height overestimates $\mathrm{SNR_{dB}}$ because it fails to account for the contribution of the noise to the total sideband power.

Finally we comment on the relationship between sensitivity measured in the frequency domain (Eq.~(124)) and in the time domain (Eq.~(126)).
If all relevant noise sources are white and no additional noise is introduced by demodulating or digitising the signal, then these two equations will give the same result.
In practice, the frequency-domain method often gives a slightly better apparent sensitivity because $\fm$ can be chosen away from noise spurs.
When the target signal is nearly monochromatic, the frequency-domain method is often appropriate; for broadband signals such as for qubit readout, the frequency-domain result can be used as a lower bound but the time-domain result is usually more representative.
Ultimately it is Eq.~\eqref{eq:variance} that determines which components of the noise corrupt a measurement.

  \clearpage
  \onecolumngrid
  
\section{Charge detection table}

 \rowcolors{2}{gray!25}{white}

  \begin{table*}[h]
  \begin{tabular}{ccccccccc}
   \hline\hline
        Paper                               & Technique & System    & $\fr$ & $\Qr$ & $\Sqq$ & $\taumin$  &  $\mathcal{F}$ ($\tauM$) & Special features \\
    \hline\hline
    Schoelkopf 1998 \cite{Schoelkopf1998}   &  Resistive    & SET Al/AlOx   &  1700 &   6   &  12   &       &   &       \\
    Fujusawa 2000 \cite{fujisawa2000_transmission} & Resistive & SET GaAs   &  680  &   10  &  500  &       &   &  No cryo amp.  \\
    Fujusawa 2000 \cite{Hirayama2000}       &   Resistive   & SET GaAs      &  700  &   4   &  36   &       &   &       \\
    Aassime 2001 \cite{Delsing2001}         &   Resistive   & SET Al/AlOx   &  331  &   18  &  6.3  &       &   &       \\
    Aassime 2001 \cite{Schoelkopf2001}      &   Resistive   & SET Al/AlOx   &  332  &   24  &  3.2  &       &   &       \\
    Lehnert 2003 \cite{Schoelkopf2003}      &   Resistive   & SET Al/AlOx   &  500  &   10  &  40   &       &   &       \\
    Lu 2003 \cite{Rimberg2003}              &   Resistive   & SET Al/AlOx   &  1091 &       &  24   &       &   &       \\
    Roschier 2004 \cite{Schoelkopf2004}     &   Resistive   & SET Al/AlOx   &  471.2&       &  38   &       &   &       \\
    Brenning 2006 \cite{Delsing2006}        &   Resistive   & SET Al/AlOx   &  345  &   11  &  0.9  &       &   &       \\
    Angus 2007 \cite{Clark2007}	            &   Resistive   & SET Si        &  340  &   20  &  7.2  &       &   &       \\
    Ares 2016\cite{Laird2016matching}       &   Resistive   & SET GaAs      &  211  &       &  1650 &       &   &       \\
    Schupp 2020\cite{Schupp2020}        &   Resistive   & SET GaAs      &  200  &   15  &  60   &       &   & SQUID amp. \\
    Qin 2006 \cite{Williams2006}            &  QPC CS       & DQD GaAs      & 810   &  10   &  2000 &       &   &       \\
    Cassidy 2007 \cite{Smith2007}           &  QPC CS       & DQD GaAs      & 332   &  8    &  200  &       &   &       \\
    Reilly 2007 \cite{Gossard2007}          &  QPC CS       & DQD GaAs      & 220   & 15    &  1000 &       &   &       \\
    Barthel 2009 \cite{Gossard2009}	        &  QPC CS       & DQD GaAs      &       &       &  600  &       &   90 (6~$\micro$s) &    \\
    Mason 2010 \cite{Kycia2010}             &  QPC CS       & DQD GaAs      &   763 &       &  146  &       &   &   Superconducting $\Lc$ \\
    House 2016 \cite{Simmons2016}           &  SET CS       & DQD Si:P      & 283.6 &   45  &       & 55~ns &   &    \\
    Volk 2019 \cite{Kuemmeth2019}           &  SET CS       & DQD Si/SiGe   &  136  &       &  1500 & 2.1 $\micro s$& &    \\
    Keith 2019 \cite{Simmons2019SingleShot} &  SET CS       & DQD Si:P      &  223  &   40  &    50 &       & 97 (1.5 $\micro$s)    &          \\ 
    Noiri 2020 \cite{Tarucha2020}           &  SET CS       & DQD Si        & 206.7 &       &   & 22~ns*   &  99.99 (1.8 $\micro$s)&    \\
    Connors 2020 \cite{Nichol2020}          &  SET CS       & DQD SiGe      &       &       &       &       & 99.9 (1 $\micro $s )  &     \\
    Petersson 2010 \cite{Petersson2010}     &  Disp.        & DQD GaAs      &  385  &   8   & 200   &       &   &      \\     
    Stehlik 2015 \cite{Petta2015}           &  Disp.        & DQD InAs NW   &  7881 & 3000  &       &  7~ns &   &   QED cavit, JPA \\
    Colless 2013 \cite{Reilly2013}          & Disp.         & DQD  GaAs     & 704   & 70    & 6300  & 5~$\micro$s   &                   &      \\
    Gonzalez 2015  \cite{Gonzalez-Zalba2015_limits} & Disp. & DQD Si     & 335   & 42    &	37  &       &                   &  	    \\
    Pakkiam 2018 \cite{Simmons2018}         & Disp.         & DQD Si        & 339.6 &  266  &       &       & 82.3 (300~$\micro$s) & Superconducting $\Lc$  \\
    Ahmed 2018 \cite{Gonzalez-Zalba2018_rfgate}& Disp.      & DQD Si        & 616   & 790   &	1.3 &       &   & Superconducting $\Lc$  \\
    West 2019 \cite{Dzurak2019}             & Disp.         & DQD Si        & 266.9 &   38  &       & 2.6~ms& 73 (2 ms)&      \\
    Schaal 2019 \cite{Morton2020_JPA}       & Disp.         & DQD Si        & 621.9 & 966   &       & 80~ns &   & Superconducting $\Lc$, JPA \\
    Zheng 2019 \cite{Vandersypen2019}       & Disp.         & DQD Si        & 5711.6& 2600  &  400  & 170~ns& 98 (6 $\micro$s)& QED cavity \\
    Ibberson 2021 \cite{Gonzalez-Zalba2021Interaction}& Disp. & DQD Si  & 1880  &       & 100   & 10~ns  &    & Superconducting $\Lc$, waveguide \\
    House 2016 \cite{Simmons2016}           & Disp. CS      & DQD Si:P      & 244.8 & 100   &       &550~ns &                   & \\
    Urdampilleta 2019 \cite{Urdampilleta2019}& Disp. CS     & DQD Si        &       &  234  &  58   &       & 99 (1 ms) &      \\
    Schaal 2019 \cite{Morton2020_JPA}       & Disp. CS      & DQD Si        & 621.9 & 966   & 0.25  &       &     & Superconducting $\Lc$, JPA \\
    Bohuslavsky 2020 \cite{Kuemmeth2020quadruple}& Disp. CS & DQD Si   & 191   &       & & 17 $\micro s$* &                   & \\
    Chanrion 2020 \cite{Urdampilleta2020}   & Disp. CS      & DQD Si        & 286   &   70  &  2100 &       &                   & \\
  
 \end{tabular}

 \caption{Table referencing the sensitivity $\Sqq$ (in $\micro\eHz$), from available sources, over reading the charge occupation of single electron transistors (SETs) or double quantum dots (DQDs) of various sorts.
 In the SET experiments, $\Sqq$  refers to measuring the charge occupation of the SET itself which are obtain by detecting its variation of resistance with the rf setup (Resistive). 
 In the DQD experiments, $\Sqq$ refer to measuring the charge occupation of the two quantum dots using either a radio-frequency charge sensor a quantum point contact charge sensor (QPC CS), a SET charge sensor (SET CS), in-situ dispersive readout (Disp.) or dispersive charge sensing (Disp. CS).
 Next to the charge sensitivity $\Se$ and the minimum integration time $\taumin$ to reach $\text{SNR} = 1$, we display the resonance frequency $\fr$ (in MHz), the total quality factor $\Qr$ and in the case of single shot readout of spin qubit: the fidelity (in \%) $\mathcal{F}$ (and corresponding integration time $\taum$). 
The last column contain special features of the resonators or the setups. \\
* These papers do not report $\taumin$ directly; instead they report the SNR at another value of $\taum$, and we assume $\taumin= \taum^2/\text{SNR}_x^2$.}
\label{table:ChargeSensitivity}
 \end{table*}
 
 \newpage
 
 \twocolumngrid
 
 \section{Component table}
 \vspace{-0.7cm}
\rowcolors{2}{gray!25}{white}
  \begin{table}[H]
  \begin{tabular}{p{14mm} p{49mm} p{20mm}}
  \hline \hline
    	& Name                              &    Reference  \\
    	\hline \hline
   	\multicolumn{3}{l}{Resistor}\\    
    1~k$\Ohm$       &   TE RP73D1J1K0BTDG  & \onlinecite{Gonzalez-Zalba2019tunable} \\
    10~k$\Ohm$      &   TE RP73D1J10KBTDG  & \onlinecite{Gonzalez-Zalba2019tunable}\\
    10~k$\Ohm$      &   ERA3APB103V                     &	\\
    100~k$\Ohm$     &   TE RP73D1J100KBTDG &
    \onlinecite{Gonzalez-Zalba2019tunable}
       	 \global\rownum=1\relax \\
       	\hline
   	\multicolumn{3}{l}{Capacitor} \global\rownum=2\relax \\
    1~pF            &   KEMET BR06C109BAGAC             & \onlinecite{Gonzalez-Zalba2019tunable}\\
    100~pF          &   Murata GRM1885C1H101JA01        & \onlinecite{Gonzalez-Zalba2019tunable}\\
    100~pF          &   CC0603JRNPO9BN101               &   \\
    1~nF            &   Murata GRM1885C1H102JA01        & \onlinecite{Gonzalez-Zalba2019tunable}\\
    10~nF           &   KEMET C0603C103J3GACTU          & \onlinecite{Gonzalez-Zalba2019tunable}\\
    10~nF           &   TDK  CGA3E2C0G1H103J080AA       &  \global\rownum=1\relax \\
       	\hline
   	\multicolumn{3}{l}{Inductor} \global\rownum=2\relax \\
    270~nH          &   TDK B82498F3271J001             & \onlinecite{Gonzalez-Zalba2019tunable}\\
    390~nH          &   EPCOS  B82498B3391J             &   \\
    470~nH          &   B82498B3471J                    &    \\
    560~nH          &   TDK B82498F3561J001             & \onlinecite{Gonzalez-Zalba2019tunable}\\
    820~nH          &   Coilcraft 1206CS-821XJL         & \onlinecite{Kuemmeth2020quadruple} \\
    820~nH          &   Coilcraft 1206CS-821XJE         & \onlinecite{Simmons2018} \\
    1200 nH         &   Coilcraft 1206CS122XJEB         &  \onlinecite{Kuemmeth2019} \\ 
    \hline
	\multicolumn{3}{l}{Varicap diode}\\
    0.7~pF          &   MA46H200                        & \onlinecite{Gonzalez-Zalba2019tunable}\\
    11~pF           &   MACOM MA46H204-1056             & \onlinecite{Laird2016matching}, \onlinecite{Simmons2016},
    \onlinecite{Gonzalez-Zalba2019tunable}\\
     \end{tabular}
     \caption{List of components for PCB board resonator used in various experiments. PCB sample holders stuffed with SMD inductors for reflectometry are also commercially available from QDevil (www.qdevil.com).}
    \label{table:Components}

 \end{table}

  \twocolumngrid
  
\section*{Bibliography}
\vspace{-0.6cm}